\documentclass[12pt]{article}
\usepackage{amsmath,amssymb}
\usepackage[english]{babel}
\usepackage[cp866]{inputenc}
\usepackage{hyperref}

\numberwithin{equation}{section}

\newcommand{\bq}{\begin{eqnarray}}
\newcommand{\eq}{\end{eqnarray}}

\newcommand{\bbq}{\begin{equation*}}
\newcommand{\eeq}{\end{equation*}}

\newcommand{\ra}{\rightarrow}

\newcommand{\lm}{\Lambda_2}
\newcommand{\la}{\Lambda_Q}

\newcommand{\ov}{\overline}

\newcommand{\lt}{\tilde\Lambda}
\newcommand{\wh}{\widehat}
\newcommand{\wt}{\widetilde}
\newcommand{\wmu}{\widetilde{\mu}_{\rm x}}

\newcommand{\qq}{{\ov Q}Q}
\newcommand{\Qo}{({\ov Q}Q)_1}
\newcommand{\Qt}{({\ov Q}Q)_2}

\newcommand{\nd}{{\ov N}_c}

\newcommand{\mx}{\mu_{\rm x}}

\newcommand{\bb}{2N_c-N_F}
\newcommand{\bt}{{\rm b}_2}
\newcommand{\w}{{\cal W}}
\newcommand{\no}{{\rm n}_1}
\newcommand{\nt}{{\rm n}_2}
\newcommand{\tm}{\widetilde m}

\begin{document}

\begin{center}{\bf \large Softly broken $\mathbf{{\cal N}=2}$ SQCD: mass spectra in vacua with unbroken \\ $\mathbf{Z_{\bb}}$ symmetry}\end{center}
\vspace{1cm}
\begin{center}\bf {Victor L. Chernyak $^{a,\, b}$ } \end{center}
\begin{center}(e-mail: v.l.chernyak@inp.nsk.su) \end{center}
\begin{center} a)\,\, Novosibirsk State University, 630090 Novosibirsk, Russia \end{center}
\begin{center} b)\,\, Budker Institute of Nuclear Physics SB RAS, 630090 Novosibirsk, Russia
\end{center}
\vspace{1cm}
\begin{center}{\bf Abstract} \end{center}
\vspace{1cm}

Considered are ${\cal N}=2\,\, SU(N_c)$ or $U(N_c)$ SQCD with $N_F<2N_c-1$ equal mass quark flavors.  ${\cal N}=2$ supersymmetry is softly broken down to ${\cal N}=1$ by the mass term $\mu_{\rm x}{\rm Tr}\,(X^2)$ of colored adjoint scalar partners of gluons, $\mu_{\rm x}\ll\Lambda_2$ (\,$\Lambda_2$ is the scale factor of the $SU(N_c)$ gauge coupling).

There is a large number of different types of vacua in these theories with both unbroken and spontaneously broken global flavor symmetry, $U(N_F)\rightarrow U({\rm n}_1)\times U({\rm n}_2)$. We consider in this paper the large subset of these vacua with the unbroken non-trivial $Z_{2N_c-N_F\geq 2}$ discrete symmetry, at different hierarchies between the Lagrangian parameters $m\gtrless\Lambda_2,\,\, \mu_{\rm x}\gtrless m$. The forms of low energy Lagrangians, quantum numbers of light particles and mass spectra are described in the main text for all these vacua.

The calculations of power corrections to the leading terms of the low energy quark and dyon condensates are presented in two important Appendices. The results agree with also presented in these Appendices {\it independent} calculations of these condensates using roots of the Seiberg-Witten spectral curve. This agreement confirms in a non-trivial way a self-consistency of the whole approach.

Our results differ essentially from corresponding results in e.g. recent related papers arXiv:1304.0822, arXiv:1403.6086 and arXiv:1704.06201 of M.Shifman and A.Yung (and in a number of their previous numerous papers on this subject), and we explain in the text the reasons for these differences. (See also the extended critique of a number of results of these authors in section 8 of arXiv:1308.5863).
\vspace*{1cm}

\begin{center} To Arkady Vainshtein on his 75-th Anniversary \end{center}

\newpage

\tableofcontents
\numberwithin{equation}{section}

\section{Introduction}

\hspace*{4mm} The masses and quantum numbers of BPS particles in ${\cal N}=2\,\, SU(2)$ SYM and $SU(2)$ SQCD with $N_F=1...4$ quark flavors at small $\mx\neq 0$ were found in seminal papers of N.Seiberg and E.Witten \cite{SW1,SW2}. They presented in particular the corresponding spectral curves from which the masses and quantum numbers of these BPS particles can be calculated at $\mx\ra 0$. The forms of these curves have been generalized then, in particular, to ${\cal N}=2\,\, SU(N_c)$ SYM \cite{DS} and ${\cal N}=2\,\, SU(N_c)$ SQCD with $1\leq N_F\leq 2N_c$ quark flavors \cite{KL,AF,HO,Shap,APS}.

In what follows we deal mainly with ${\cal N}=2\,\, SU(N_c)$ SQCD with $N_F<2N_c-1$ flavors of equal mass quarks $Q^i_a,\,{\ov Q}_i^{\,a},\,i=1...N_F,\,\,a=1...N_c$. The Lagrangian of this UV free theory, broken down to ${\cal N}=1$ by the mass term $\mx{\rm Tr}\,(X^{\rm adj}_{SU(N_c)})^2$ of adjoint scalars, can be written at the sufficiently high scale $\mu\gg\max\{\lm, m\}$ as (the gluon exponents in Kahler terms are implied in \eqref{(1.1)} and everywhere below)
\bbq
K=\frac{2}{g^2(\mu)}{\rm Tr}\,\Bigl [(X^{\rm adj}_{SU(N_c)})^\dagger X^{\rm adj}_{SU(N_c)}\Bigr ]_{N_c}+{\rm Tr}\,(Q^\dagger Q+{\ov Q}^{\,\dagger} {\ov Q})_{N_c}\,, \quad X^{\rm adj}_{SU(N_c)}=T^A X^A\,,\quad A=1,\,...,N_c^2-1\,,
\eeq
\bq
{\cal W}_{\rm matter}=\mx{\rm Tr}\,(X^{\rm adj}_{SU(N_c)})_{N_c}^2 +{\rm Tr}\,\Bigl (m\,{\ov Q} Q-{\ov Q}\sqrt{2} X^{\rm adj}_{SU(N_c)} Q \Bigr )_{N_c}\,,\quad {\rm Tr}\, (T^{A_1} T^{A_2})=\frac{1}{2}\,\delta^{A_1 A_2}\,,\label{(1.1)}
\eq
with the scale factor $\lm$ of the gauge coupling $g(\mu)$, and ${\rm Tr}$ in \eqref{(1.1)} is over all colors and flavors. The "softly broken"\, ${\cal N}=2$ theory means $0<\mx\ll\lm$. Considering $\mx$ as a soft background field, there is the unbroken $U(1)$ R-symmetry with the charges of fields and parameters:
$r_{\lambda}=r_{\theta}=r_{Q}=r_{\ov Q}=1,\, r_X=r_m=r_{\lm}=0,\, r_{\mx}=2$.

The spectral curve corresponding to \eqref{(1.1)} can be written at $N_F < 2N_c-1$ e.g. in the form \cite{APS}
\bq
y^2=\prod_{i=1}^{N_c}(z+\phi_i)^2-4\lm^{\bb}(z+m)^{N_F}\,,\quad \sum_{i=1}^{N_c}\phi_i=0{\rm\,\,\, in\,\,\, SU(N_c)}\,, \label{(1.2)}
\eq
where $\{\phi_i\}$ is a set of gauge invariant co-ordinates on the moduli space. Note that, as it is, this curve can be used for the calculation of BPS particle masses at $\mx\ra 0$ only.

In principle, it is possible to find  at $\mx\ra 0$ from the curve \eqref{(1.2)} the quantum numbers and masses of all massive and massless BPS particles in different multiple vacua of this theory. But in practice it is very difficult to do this for general values of $N_c$ and $N_F$, especially as for quantum numbers . The important property of the $SU(N_c)$ curve \eqref{(1.2)} which will be used in the text below is that there are maximum $N_c-1$ double roots. And the vacua we will deal with below at $\mx\neq 0$ are just these vacua. In other words, {\it in all $SU(N_c)$ vacua we will deal with, the curve} \eqref{(1.2)} {\it has $N_c-1$ double roots and two single roots}.

The first detailed attempt has been made in \cite{APS} (only the case $m=0$ was considered) to classify the vacua of \eqref{(1.1)} for general values of  $N_c$ and $N_F$, to find quantum numbers of light BPS particles (massless at $\mx\ra 0$) in these vacua  and forms of low energy Lagrangians. It was found in particular that in all vacua the $SU(N_c)$ gauge symmetry is broken spontaneously at the scale $\mu\sim\lm$ by higgsed adjoint scalars, $\langle X^{\rm adj}_{SU(N_c)}\rangle\sim\lm$.

That this should happen can be understood qualitatively as follows \cite{ch6}. The perturbative NSVZ $\beta$-function of the (effectively) massless unbroken ${\cal N}=2$ SQCD is exactly one loop \cite{NSVZ1,NSVZ2}. The theory \eqref{(1.1)} is UV free at $(2N_c-N_F)>0$. At small $\mx$ and $m\ll\lm$, and if the whole matrix $\langle X^{adj}_{SU(N_c)}\rangle\ll\lm$ (i.e. in the effectively massless at the scale $\mu\sim\lm$ unbroken $\,{\cal N}=2$ SQCD), the coupling $g^2(\mu)$ is well defined at $\mu\gg\lm$, has a pole at $\mu=\lm$ and becomes negative at $\mu<\lm$. To avoid this unphysical behavior, {\it the field $X$ is necessarily higgsed breaking the $SU(N_c)$ group, with (at least some) components $\langle X^A\rangle\sim\lm$}. (This becomes especially clear in the limit of the unbroken ${\cal N}=2$ theory at $\mx\ra 0$ and small $m\ll\lm$, when all chiral quark condensates and the gluino condensate, i.e. their corresponding mean values $\langle...\rangle$, approach zero, so that all quarks are not higgsed in any case, and all particles are very light at $\langle X^{\rm adj}_{SU(N_c)}\rangle\ll\lm$).

Remind that exact values of $\langle X^A\rangle$ account for possible nonperturbative instanton contributions. And it is worth also to note that, in the limit of unbroken ${\cal N}=2$ SQCD with small $m$ and $\mx\ra 0$, if there were no some components $\langle X^A\rangle\sim\lm$ (i.e. all $\langle X^A\rangle\ll\lm$), {\it the nonperturbative instanton contributions also would be non operative at the scale $\lm$ without both the corresponding fermion masses $\sim\lm$ and the infrared cut off $\,\,\rho\lesssim\, 1/\lm$ supplied by some $\langle X^A\rangle\sim\lm$. Therefore, the problem with $g^2(\mu<\lm)<0$ will survive in this case. Moreover, for the same reasons, if there remains the non-Abelian subgroup unbroken at the scale $\lm$, it has to be IR free (or at least conformal)}.

It was found in \cite{APS} that (at $m=0$) and $\mx\ra 0$ there are two qualitatively different branches of vacua. On the baryonic  branch the lower energy gauge group at $\mu<\lm$ is $SU(N_F-N_c)\times U^{\bb}(1)$, while on non-baryonic branches these are $SU(\no)\times U^{N_c-\no}(1),\,\, \no\leq N_F/2$.

Besides, at $m\neq 0$ and small $\mx\neq 0$, the global flavor symmetry $U(N_F)$ of \eqref{(1.1)} is unbroken or broken spontaneously as $U(N_F)\ra U(\no)\times U(\nt),\, 1\leq\no\leq [N_f/2]$ in various vacua \cite{CKM}.

As was also first pointed out in \cite{APS}, there is the residual $Z_{2N_c-N_F}$ discrete R-symmetry in theory \eqref{(1.1)}. {\it This symmetry will play a crucial role in what follows, as well as a knowledge of multiplicities of all different types of vacua}, see e.g. section 3 in \cite{ch4} and/or section 4 in \cite{ch5}. The charges of fields and parameters in the superpotential of \eqref{(1.1)} under $Z_{\bb}=\exp\{i\pi/(\bb)\}$ transformation are: $q_{\lambda}= q_{\rm\theta}=1,\,\, q_{X}=q_{\rm m}=2,\,\, q_Q=q_{\,\ov Q}=q_{\lm}=0,\,\, q_{\mx}=-2$. If this non-trivial at $\bb\geq 2$ discrete $Z_{\bb}$ symmetry is broken spontaneously in some vacua, there will appear then the factor $\bb$ in their multiplicity. Therefore, if multiplicities of different types of vacua are known, we can see explicitly whether this $Z_{\bb}$ symmetry is broken spontaneously or not in these vacua. As a result, it is seen the following at $0<N_F<2N_c-1,\, m\ll\lm$ (in the language of \cite{ch6,ch4}).

I) In $SU(N_c)$ theories at $m\ll\lm$ ($\,\langle ({\ov Q} Q)_{1,2}\rangle_{N_c}$ and $\langle S\rangle_{N_c}=\langle \lambda\lambda/32\pi^2\rangle$ are the quark and gluino condensates summed over all their colors\,). The discrete $Z_{\bb\geq 2}$ symmetry is broken spontaneously.-

1) In L (large) vacua with the unbroken global flavor symmetry $U(N_F)$, the multiplicity $2N_c-N_F$ and with $\langle\qq\rangle_{N_c}\sim\mx\lm,\,\,\langle S\rangle_{N_c}\sim\mx\lm^2,\,\,\langle{\rm Tr} (X^{adj}_{SU(N_c)})^2\rangle\sim\lm^2$.\\
2) In Lt (\,L-type\,) vacua with the spontaneously broken flavor symmetry, $U(N_F)\ra U(\no)\times U(\nt),\,\,1\leq\no\leq N_F/2$, the multiplicity $(2N_c-N_F)C^{\,\no}_{N_F},\,C^{\,\no}_{N_F}=(N_F!/\no!\,\nt!)\,,$ and with $\langle\Qo\rangle_{N_c}\sim\langle\Qt\rangle_{N_c}\sim\mx\lm,\,\,\langle S\rangle_{N_c}\sim\mx\lm^2,\,\, \langle{\rm Tr} (X^{adj}_{SU(N_c)})^2\rangle\sim\lm^2$.\\
3) In special vacua with $U(N_F)\ra U(\no=N_F-N_c)\times U(\nt=N_c)$, the multiplicity $(2N_c-N_F)C^
{\,N_F-N_c}_{N_F}$, and with $\langle\Qo\rangle_{N_c}\sim\mx m,\,\, \langle\Qt\rangle_{N_c}=\mx\lm,\,\, \langle S\rangle_{N_c}\sim\mx m\lm,\,\, \langle{\rm Tr} (X^{adj}_{SU(N_c)})^2\rangle\sim m\lm$.

While $Z_{\bb\geq 2}$ is unbroken. - \\
4) In br2 (br=breaking) vacua with $U(N_F)\ra U(\no)\times U(\nt),\,\, \nt>N_c$, the multiplicity $(N_F-N_c-\no)C^{\,\no}_{N_F}$, and with the dominant condensate $\langle\Qt\rangle_{N_c}\sim\mx m\gg \langle\Qo\rangle_{N_c}\sim\mx m (m/\lm)^{\frac{2N_c-N_F}{\nt-N_c}}$, $\langle S\rangle_{N_c}\sim\mx m^2 (m/\lm)^{(2N_c-N_F)/(\nt-N_c)},\,\, \langle{\rm Tr} (X^{adj}_{SU(N_c)})^2\rangle\sim m^2$.\\
5) In S (small) vacua with unbroken $U(N_F)$, the multiplicity $N_F-N_c$ and with $\langle\qq\rangle_{N_c}
\sim\mx m,\,\, \langle S\rangle_{N_c}\sim\mx m^2 (m/\lm)^{(2N_c-N_F)/(N_F-N_c)},\,\, \langle{\rm Tr} (X^{adj}_{SU(N_c)})^2\rangle\sim m^2$ (see section 3 in \cite{ch4} or section 4 in \cite{ch5} for all details).

II). $Z_{\bb\geq 2}$ is unbroken in all multiple vacua of $SU(N_c)$ theories at $m\gg\lm$ (see Part II in the table of contents).

III) There are additional vs (very special) vacua with e.g. $N_c\leq N_F\leq 2N_c-1$, with the spontaneously broken global flavor symmetry $U(N_F)\ra U(\no=N_F-N_c)\times U(\nt=N_c),\,\, \langle\Qt\rangle_{N_c}=\mx m,\,\, \langle\Qo\rangle_{N_c}=0,\,\,\,\langle S\rangle_{N_c}=0,\,\, \langle{\rm Tr} (X^{adj}_{SU(N_c)})^2\rangle=0$, and the multiplicity $C^{N_F-N_c}_{N_F}$ in $U(N_c)$ theories at all $m\gtrless\lm$ (such vacua are absent in $SU(N_c)$ theories, see Appendix in \cite{ch4}).  $Z_{\bb\geq 2}$ is unbroken in these vs-vacua at all $m\gtrless\lm$.\\

It was stated in \cite{APS} that {\it in all vacua} of $SU(N_c)$ \eqref{(1.1)}, baryonic and non-baryonic, all light particles in low energy Lagrangians with $m=0$ and small $\mx$, i.e. quarks, gluons and scalars in non-Abelian sectors and corresponding particles in Abelian ones, are pure magnetic.

Really, the final purpose of \cite{APS} was an attempt to derive the Seiberg duality for ${\cal N}=1$ SQCD \cite{S2,IS}. The idea was as follows. On the one hand, the electric ${\cal N}=2$ SQCD in \eqref{(1.1)} with small both $\mx$ and $m$ flows clearly to the standard ${\cal N}=1$ electric SQCD with $SU(N_c)$ gauge group, $N_F$ flavors of light quarks and without adjoint scalars $X^{\rm adj}$ at $m={\rm const},\,\,\mx\ra \infty$ and fixed $\la^{3N_c-N_F}=\lm^{2N_c-N_F}\mx^{N_c}$. On the other hand, the authors of \cite{APS} expected that, starting in \eqref{(1.1)} at $\mx\ll\lm$ from vacua of the baryonic branch with the $SU(\nd=N_F-N)$ low energy gauge group {\it by itself} and increasing then $\mx\gg\lm$, they will obtain Seiberg's dual ${\cal N}=1$ magnetic SQCD with $SU(\nd=N_F-N_c)$ gauge group and $N_F$ flavors of massless dual magnetic quarks. (But, in any case, $N^2_F$ light Seiberg's mesons $M^i_j\ra ({\ov Q}_j Q^i)$ of dual ${\cal N}=1\,\, SU(\nd)$ SQCD \cite{S2} were missing in this approach, see section 8 in \cite{ch6} for a more detailed critique of \cite{APS}\,).

The results from \cite{APS} were generalized then in \cite{CKM} to nonzero mass quarks, with emphasis on properties of flavor symmetry breaking in different vacua. Besides, multiplicities of various vacua were discussed in this paper in some details. (A complete detailed description of different vacua multiplicities and values of quark and gluino condensates, $\langle{\ov Q}_j Q^i\rangle_{N_c}$ and $\langle S\rangle_{N_c}$, in various vacua and at different hierarchies between $\mx,\,m$ and $\lm$ (or $\la^{3N_c-N_F}=\lm^{2N_c-N_F}
\mx^{N_c}$) see in \cite{ch4,ch6,ch5}\,).

But further detailed investigations of light particles quantum numbers in theory \eqref{(1.1)}, in some simplest examples with small values of $N_c$ and $N_F$, did not confirm the statement \cite{APS} that all these particles, and in particular light quarks, are pure magnetic.

Specifically, the $SU(N_c=3)$ vacua of \eqref{(1.1)}, with $N_F=4,\,\, \no=2,\, \nt=N_F-\no=2$ (in the language of \cite{ch4}, these are Lt vacua with the gluino condensate $\langle S\rangle_{N_c}\sim\mx\lm^2$ and spontaneously broken non-trivial $Z_{\bb=2}$ symmetry at $m\ll\lm$), and with $N_F=5,\,\, \no=2,\, \nt=3$ (these are special vacua with $\langle S\rangle_{N_c}\sim\mx m\lm$ in the language of \cite{ch4}, $Z_{\bb=1}$ symmetry acts trivially in this case and gives no restrictions on the form of $\langle X^{\rm adj}_{SU(3)}\rangle$), were considered in details at $\mx\ra 0$ and $m\lessgtr\lm$ in \cite{MY}. Besides, the vs (very special) $U(N_c=3)=SU(N_c=3)\times U^{(0)}(1)$ vacua with $N_F=5,\,\, \no=2,\, \nt=3,\, \langle S\rangle_{N_c}=0$ were considered earlier in \cite{SY3} (these are $D^{\,i}_a,\,{\ov D}^{\,a}_i,\,\, a=1,\,2,\,\,i=1...4 (5)$ (massless at $\mx\ra 0$) form $SU(2)$ doublet of dyons with the $SU(3)$ fundamental electric charges and with the hybrid magnetic charges which are fundamental with respect to $SU(2)$ but adjoint with respect to $SU(3)$.

We can propose e.g. the following picture for these UV free $SU(N_c=3)$ theories with $N_F=4$ or $N_F=5$ at $m\ll\lm,\,\mx\ra 0$. To avoid $g^2(\mu<\lm)<0$ and without restrictions from the non-trivial unbroken $Z_{2N_c-N_F}$ symmetry, the electric gauge group $SU(3)$ is higgsed in these vacua by $\langle X^{\rm adj}_{SU(3)}\rangle\sim diag (\lm, \lm, -2\lm)$, i.e. $SU(3)\ra SU(2)\times U(1)$, so that all original pure electrically charged quarks acquire masses $\sim\lm$. But, due to strong coupling $a(\mu\sim\lm)=N_c g^2(\mu\sim\lm)/8\pi^2\sim 1$ and very specific properties of the enhanced ${\cal N}=2$ SUSY, the heavy flavored electric quarks $Q^{\,i}_a,\,{\ov Q}^{\,a}_i,\,\, a=1,\,2,\,\,i=1...4 (5)$ (i.e. $SU(2)$ color doublet) with equal masses $\sim\lm$ combined with two heavy unflavored pure magnetic monopoles with the same equal masses $\sim\lm$ and with above described hybrid magnetic charges, and formed the $SU(2)$ doublet of light flavored BPS dyons $D^{\,i}_a,\,{\ov D}^{\,a}_i$, massless at $\mx\ra 0$.

These massless at $\mx\ra 0$ quark-like dyons $D^{\,i}_a,\,{\ov D}^{\,a}_i$ coupled then with massless $SU(2)$ adjoint gluons and scalars (which transformed from pure electric to dyonic ones, and this is a most surprising phenomenon) and formed the dyonic ${\cal N}=2\, SU(2)$ non-Abelian group (conformal at $N_F=4$ and IR free at $N_F=5$).

As follows from the curve \eqref{(1.2)} and the formulae like \eqref{(2.2.10)} from \cite{SY5}, the $SU(2)$ dyons are higgsed simultaneously at small $\mx\neq 0$ e.g. as $\langle{\ov D}^{\,a}_i\rangle=\langle D_{\,a}^i\rangle\sim \delta^i_a (\mx\lm)^{1/2},\, a,i=1,2$ (and for this to be possible in the ground state they have to be massless at $\mx=0$ and mutually local).  As a result, the whole color $SU(2)$ group is broken and $(\no=2)^2=4$ long ${\cal N}=2$  multiplets of massive gluons with masses $\sim g_D (\mx\lm)^{1/2}$ are formed. Besides, the flavor $U(N_F)$ symmetry is broken spontaneously as $U(N_F)\ra U(\no=2)\times U(N_F-2)$ and $2(N_F-2)$ massless Nambu-Goldstone multiplets appear (in essence, these are non-higgsed dyons).

These higgsed $SU(2)$ dyons $D^{\,i}_a, {\ov D}^{\,a}_i,\, i,a=1,2$ are mutually non-local with all heavy original quarks with $SU(3)$ colors and with the $SU(2)$ doublet of pure magnetic monopoles, all with masses $\sim\lm$. Therefore, when these dyons $D^{\,i}_a, {\ov D}^a_i$ are higgsed at small $\mx\neq 0$ as $\langle D^{\,i}_a\rangle=\langle{\ov D}_i^{\,a}\rangle\sim\delta^i_a\,(\mx\lm)^{1/2}$, this results in a weak confinement of these heavy charged particles, the string tension is $\sigma^{1/2}\sim (\mx\lm)^{1/2}\ll\lm$. (The confinement is weak in a sense that the string tension is much smaller than particle masses). These heavy particles form hadrons with masses $\sim\lm$. In the case of $U(N_c=3)$ vs-vacua with $N_F=5$ and $\langle S\rangle_{N_c}=0$ from \cite{SY3}, there are in addition the massless at $\mx\ra 0$ {\it flavorless} $SU(2)$ singlet BPS dyons $D_3, {\ov D}_{3}$ with $\langle{\ov D}_{3}\rangle=\langle D_3\rangle\sim (\mx\lm)^{1/2}$ at $\mx\neq 0$. Now the additional long ${\cal N}=2\,\,  U(1)$ multiplet of massive gluon with the mass $\sim g_D (\mx\lm)^{1/2}$ is formed.\\

Besides, it was shown explicitly in \cite{SY3,MY} at $\mx\ra 0$ on examples considered, how vacua with some massless at $m\gg\lm$ and $\mx\ra 0$ pure electric original quarks $Q^i$ can evolve at $m\ll\lm$ into vacua with only massless dyons $D^i$. E.g., in the vicinity of the Argyres-Duglas point $m=m_0\sim\lm$, there are vacua in which {\it additional} corresponding particles become light. At $m=m_0$ the vacua collide, all these particles become massless, and the lower energy regime at $\mu<\lm$ is conformal. In particular, both some flavored pure electric quarks $Q^i$ and flavored dyonic quarks $D^i$ are massless at this point (as well as some pure magnetic monopoles). The masses of these quarks with {\it fixed quantum numbers} behave continuously but non-analytically in $m/\lm$ at this special point $m= m_0$. E.g., in selected at $m\lessgtr m_0$ vacua colliding at $m=m_0$ with other vacua, the mass of some light pure electric quarks $Q^i$ stays at zero at $m\geq m_0$, then begins to grow at $m<m_0$ and becomes $\sim\lm$ at $m\ll\lm$, while the mass of light composite dyonic quarks $D^i$ is zero in these vacua at $m\leq m_0$, then begins to grow at $m>m_0$ and becomes large at $m\gg m_0$. {\it Equivalently}, tracing e.g. the evolution with $m/\lm$ in the small vicinity of $m=m_0$ of {\it quantum numbers} of, separately, massless or nearly massless flavored quark-like particles at $m\lessgtr m_0$, one can say that their quantum numbers jump at the collision point $m=m_0$. I.e.: a) the massless particles at $m > m_0$ are pure electric quarks $Q^i$, while they are dyonic quarks $D^i$ at $m < m_0$; b) the very light (but not massless) flavored quark-like particles are the dyonic quarks $D^i$ at $m > m_0$, while they are  pure electric quarks $Q^i$ at $m < m_0$. These two views of the non-analytic behavior at the collision point $m= m_0$ when traversing the small vicinity of the collision point, i.e. either looking at the non-analytic but continuous evolution of small masses of particles with fixed quantum numbers, or looking on the jumps of quantum numbers of, separately, massless particles and of nearly massless particles, are clearly {\it equivalent and indistinguishable} because all these particles simultaneously become massless at the collision point $m=m_0$. These are two different projections (or two sides) of the one phenomenon.

The examples of these conformal regimes for the ${\cal N}=2\,\,SU(2)$ theory were described earlier in \cite{APSW,BF,GVY} (but it is clear beforehand that to avoid unphysical $g^2(\mu<\lm)<0$ in UV free ${\cal N}=2\,\,SU(2)$ with $N_F=1,2,3$ and $m\ll\lm$, the whole gauge group is higgsed necessarily at the scale $\mu\sim\lm$ by $\langle X^{\rm adj}_{SU(2)}\rangle\sim\lm,\, SU(2)\ra U(1)$ \cite{SW1,SW2}, so that all original pure electrically charged particles are indeed heavy, with masses~ $\sim\lm$).\\

The properties of light particles with masses $\ll\lm$ in theory \eqref{(1.1)} (really, in the $U(N_c)$ theory with especially added $U^{(0)}(1)$ multiplet) for general values of $N_c,\, N_F$ and possible forms of low energy Lagrangians at scales $\mu<\lm$ in different vacua were considered later by M.Shifman and A.Yung in a series of papers, see e.g. the recent papers \cite{SY1,SY2,SY6} and references therein. At $m\ll\lm$, for vacua of baryonic branch with the lower energy ${\cal N}=2\,\,\, SU(\nd=N_F-N_c)\times U^{2N_c-N_F+1}(1)$ gauge group at $\mu<\lm$ \cite{APS}, it was first proposed naturally by these authors in \cite{SY1} that all light charged particles of ${\cal N}=2\,\, SU(\nd)$ are the original pure electric particles, as they have not received masses $\sim\lm$ directly from higgsed $\langle X^{\rm adj}_{SU(N_c)}\rangle$ (this is forbidden in these vacua by the unbroken non-trivial $Z_{\bb\geq 2}$ symmetry \cite{APS}). But they changed later their mind to the opposite
\footnote{\,
Because not $SU(N_c)$ but $U(N_c)$ group was considered in \cite{SY1,SY2}, this their statement concerns really br2 and very special vacua of $U(N_c)$, see table of contents, sections 2.2 and 6 (these are respectively zero vacua and $r=N_c$ vacua in the language of \cite{SY1,SY2}).
}
and, extrapolating freely by analogy their previous results for $r=N_c=3,\, N_F=5,\, \no=2$ very special vacua in \cite{SY3} (see also \cite{SY4}) to all br2 and very special vacua, stated in \cite{SY2} that this $SU(\nd)$ group is not pure electric but dyonic. This means that at $m\ll\lm$ all original pure electrically charged particles of $SU(\nd)$ have received large masses $\sim\lm$, but now not directly from $\langle X^{\rm adj}_{SU(N_c)}\rangle\sim\lm$ (this is forbidden by the non-trivial unbroken $Z_{\bb\geq 2}$ symmetry), but from some mysterious "outside" sources and decoupled at $\mu<\lm$, while the same gauge group $SU(\nd)$ of light composite dyonic solitons was formed. Unfortunately, it was overlooked in \cite{SY2} that, in distinction with their example with $U(N_c=3),\,\,N_F=5,\,\,\no=2$ in \cite{SY3} with the trivial $Z_{2N_c-N_F=1}$ symmetry giving no restrictions on the form of $\langle X^{adj}_{SU(N_c)}\rangle$, the appearance e.g. in the superpotential at $\mu\sim\lm$ of mass terms of original $SU(\nd)$ quarks like $\sim\lm{\rm Tr}\,({\ov Q}Q)_{\nd}$ or terms like $\sim\lm{\rm Tr\,}(X^2_{SU(\nd)})$ for scalars is forbidden in vacua with the non-trivial unbroken $Z_{2N_c-N_F\geq 2}$ symmetry, independently of whether these terms originated directly from $\langle X^{adj}_{SU(N_c)}\rangle$ or from unrecognized "outside".

The authors of \cite{APS} understood that, at $m\ll\lm$ and small $\mx$, the quarks $Q^i$ from $SU(\nd)$ cannot acquire heavy masses $\sim\lm$ in vacua with unbroken non-trivial $Z_{2N_c-N_F\geq 2}$ symmetry and remain light (i.e. with masses $\ll\lm$). But they argued then that a number of vacua with light (massless at $\mx\ra 0$) pure electric quarks at $m\gg\lm$, when evolved to $m\ll\lm$, have only light quarks with nonzero magnetic charges. And they considered this as a literal evidence that phases with light pure electrically and pure magnetically charged quarks are not different and can transform freely into each other by some unidentified mechanism. And they supposed that in vacua of the baryonic branch {\it just this mysterious mechanism transformed at $m\ll\lm$ all light pure electric particles from $SU(\nd)$ into light pure magnetic ones}.

Remind that, as described above on the examples of vacua with spontaneously broken or trivial $Z_{2N_c-N_F}$ symmetry, even in such vacua at $m\ll\lm$ the  electric quarks do not literally transform by themselves into light dyonic ones. The mechanism is different, see above. Therefore, we will consider in what follows that such mysterious mechanism which transforms literally at $m\ll\lm$ light electric quarks {\it in vacua with unbroken non-trivial $Z_{2N_c-N_F\geq 2}$ symmetry} into light magnetic (or dyonic) ones does not exist really.~
\footnote{\,
It is clear from \cite{SY1,SY2} that these authors also disagree with pure magnetic $SU(\nd)$ quarks from \cite{APS}.
}
\vspace*{5mm}

Our purpose in this paper is similar to \cite{SY1,SY2}, but we limit ourselves at $m\ll\lm$ only to vacua of baryonic branch with the non-trivial unbroken discrete $Z_{\bb\geq 2}$ symmetry. As will be seen from the text below, this $Z_{\bb}$ symmetry is strong enough and helps greatly.

We introduce also the two following assumptions of general character.\,-

{\bf Assumption A}: {\it in unbroken ${\cal N}=2$ (i.e. at $\mx\ra 0$) and at least in considered vacua with the non-trivial unbroken $Z_{\bb\geq 2}$ symmetry, the original pure electrically charged particles receive additional contributions to their masses only from higgsed} $\langle X\rangle\neq 0$ (remind that $\langle X\rangle$ includes in general all nonperturbative instanton contributions). In other words, at $\mx\ra 0$ {\it they are BPS particles} in vacua considered. For instance, if light quarks $Q^i$ with $m\ll\lm$ do not receive contributions to their masses from some $\langle X^{A}\rangle\neq 0$, they remain as they were, i.e. the light electric quarks with masses $m$. The opposite is not true in this very special ${\cal N}=2$ theory in the strong coupling (and nonperturbative) regime $a(\mu\sim\lm)=N_c g^2(\mu\sim\lm)/8\pi^2=O(1)$. The original pure electric quarks which have received masses $\sim\lm$ from some $\langle X^{A}\rangle\sim\lm$, can combine e.g. with some heavy composite pure magnetic monopoles, also with masses $\sim\lm$, and form massless at $\mx\ra 0$ composite BPS dyons. And all that.

{\bf Assumption B}: {\it there are no {\it extra} massless particles at $\mx\neq 0,\, m\neq 0$ in considered vacua of the $SU(N_c)$ theory} \eqref{(1.1)} (i.e. {\it in addition} to Nambu-Goldstone multiplets originating from spontaneously broken global flavor symmetry $U(N_F)\ra U(\no)\times U(\nt)$\,). This more particular assumption will help us to clarify the quantum numbers of massless at $\mx\ra 0$ charged BPS dyons $D_j$, see section 2.1\,.

Besides, to calculate mass spectra in some ${\cal N}=1$ SQCD theories obtained from ${\cal N}=2$ after increasing masses $\mx$ of adjoint scalar multiplets and decoupling them as heavy ones at scales $\mu<\mx^{\rm pole}$, we use the dynamical scenario introduced in \cite{ch3}. {\it But it is really needed in this paper only for those few cases with $3N_c/2<N_F<2N_c-1$ when so obtained ${\cal N}=1$ SQCD enters at lower energies the strong coupling conformal regime with} $a_*=O(1)$, see section 2.4 below. Remind that this scenario \cite{ch3} assumes for these cases that at those even lower scales where this conformal regime is broken when quarks are either higgsed or decouple as heavy, no {\it additional} parametrically light composite solitons are formed at these scales in this ${\cal N}=1$ SQCD without colored scalar fields $X^{\rm adj}$.
\footnote{\,
It is worth noting that the appearance of additional light solitons will influence the 't Hooft triangles of this ${\cal N}=1$ theory. \label{(f3)}
}

Let us emphasize that this dynamical scenario satisfies all those checks of Seiberg's duality hypothesis
for ${\cal N}=1$ SQCD which were used in \cite{S2}. In other cases, if this ${\cal N}=1$ SQCD stays in the IR free weak coupling logarithmic regime, its dynamics is simple and clear and there is no need for any additional dynamical assumptions.\\

Below in sections 2-8 are presented our results (see the table of contents) for quantum numbers of light particles, the forms of low energy Lagrangians and mass spectra in various vacua of \eqref{(1.1)} with the unbroken discrete $Z_{\bb\geq 2}$ symmetry, at different hierarchies between lagrangian parameters $m,\,\mx$ and $\lm$.

For further convenience, we present here also the equations for the quark condensates (with account of Konishi anomalies \cite{Konishi}) in the $SU(N_c)$ theory \eqref{(1.1)} in vacua with the spontaneously broken global flavor symmetry $U(N_F)\ra U(\no)\times U(\nt)$, following from the well known  effective superpotential $\w^{\,\rm eff}_{\rm tot}$ of ${\cal N}=1$ SQCD which contains {\it only} quark bilinears $\Pi^i_j=({\ov Q}_j Q^i)_{N_c}$. Because, at all $\mx\lessgtr\lm$, the quark and gluino condensates, $\langle (\qq)_{1,2}\rangle_{N_c}$ and $\langle S\rangle_{N_c}$, considered as functions of $\lm,\, m$ and $\mx$, depend trivially on $\mx$, i.e. $\sim\mx$, we can take first sufficiently large $\mx\gg\lm$ and integrate out all heavy scalars in \eqref{(1.1)}, obtaining the superpotential of (modified only by addition of the 4-quark term) ${\cal N}=1$ $\,SU(N_c)$ SQCD. Proceeding then as in section 3 of \cite{ch4} (see also section 4 in \cite{ch5} or section 8 of \cite{ch6} for the corresponding effective superpotential in ${\cal N}=1$ $\,SU(N_c)$ SQCD), $\w^{\,\rm eff}_{\rm tot}$  looks as ($N_F=\no+\nt,\,\,1\leq\no\leq [N_F/2],\,\, \nt\geq\no$)
\bq
\w^{\,\rm eff}_{\rm tot}(\Pi)=m\,{\rm Tr}\,({\ov Q} Q)_{N_c}-\frac{1}{2\mx}\Biggl [ \,\sum_{i,j=1}^{N_F} ({\ov Q}_j Q^i)_{N_c}({\ov Q}_{\,i} Q^j)_{N_c}-\frac{1}{N_c}\Bigl ({\rm Tr}\,({\ov Q} Q)_{N_c}\Bigr )^2\Biggr ]-\nd S_{N_c} \,, \label{(1.3)}
\eq
\bbq
S_{N_c}=\Bigl(\frac{\det (\qq)_{N_c}}{\la^{3N_c-N_F}=\lm^{2N_c-N_F}\mx^{N_c}}\Bigr )^{\frac{1}{N_F-N_c}}\,,
\quad \langle S\rangle_{N_c} =\Bigl(\frac{\langle\det\qq\rangle=\langle\Qo\rangle^{\no}\langle\Qt
\rangle^{\nt}}{\lm^{2N_c-N_F}\mx^{N_c}}\Bigr )^{\frac{1}{N_F-N_c}}_{N_c}\,.
\eeq
In particular, in those vacua where the four-quark terms in \eqref{(1.3)} are small and can be neglected, this effective superpotential reproduces the standard effective superpotential of the ordinary ${\cal N}=1$ SQCD. But, as it is, it is valid also both at small $\mx$ and in all others vacua of ${\cal N}=2$ SQCD. It is worth only to recall that \eqref{(1.3)} is {\it not} a genuine low energy superpotential, it can be used {\it only} for finding the values of mean vacuum values $\langle{\ov Q}_j Q^i\rangle_{N_c}$ and $\langle S\rangle_{N_c}$. The genuine low energy superpotentials in each vacuum are given below in the text.

From  \eqref{(1.3)} (and also directly from \eqref{(1.1)})
\bq
\langle \Qo+\Qt-\frac{1}{N_c}{\rm Tr}\,(\qq)\rangle_{N_c}=\mx m\,,\quad \langle S \rangle_{N_c}=\frac{\langle \Qo\rangle_{N_c}\langle\Qt\rangle_{N_c}}{\mx}\,,\label{(1.4)}
\eq
\bbq
\langle\,\sum_{a=1}^{N_c}{\ov Q}^{\,a}_j Q^i_a\,\rangle=\delta^i_j\,\langle\Qo\rangle_{N_c}\,,\,\,i,j=
1\,...\,\no,\,\, \langle\,\sum_{a=1}^{N_c}{\ov Q}^{\,a}_j Q^i_a\,\rangle=\delta^i_j\,
\langle\Qt\rangle_{N_c}\,,\,\,i,j=\no+1\,...\,N_F\,,
\eeq
\bbq
\langle {\rm Tr}\,({\ov Q} Q \rangle_{N_c}=\no\langle\Qo\rangle_{N_c}+\nt\langle\Qt\rangle_{N_c}\,,
\eeq
\bbq
m_Q\langle\sum_{a=1}^{N_c}{\ov Q}^a_i Q^i_a\rangle-\sum_{A=1}^{N_c^2-1}\sum_{a,b=1}^{N_c}\langle{\ov Q}^b_i \sqrt{2} X^A (T^A)^a_b Q^i_a\rangle=\langle S\rangle_{N_c}\,,\quad \rm{no\,\, summation\,\, over\,\, flavor\,\,  here}
\eeq
\bbq
\mx\langle {\rm Tr}\, (\sqrt{2}\,X^{\rm adj}_{SU(N_c)})^2\rangle=(2N_c-N_F)\langle S \rangle_{N_c}+m\,\langle {\rm Tr}\,({\ov Q} Q) \rangle_{N_c}=2\langle\w^{\,\rm eff}_{\rm tot}\rangle\,,
\eeq
where $\langle S\rangle_{N_c}=\langle\lambda\lambda/32\pi^2\rangle_{N_c}$ is the gluino condensate summed over all $N_c^2-1$ colors.\\

The organization of this paper is as follows (see also the table of contents). In $SU(N_c)$ gauge theories with $m\ll\lm$, the br2 vacua with the broken global flavor symmetry, $U(N_F)\ra U(\no)\times U(\nt),\, \nt>N_c$, and the unbroken non-trivial discrete symmetry $Z_{\bb\geq 2}$ are considered in section 2.1\,. We discuss in detail in this section 2.1 the quantum numbers of massless at $\mx\ra 0$ BPS particles in these vacua, the low energy Lagrangians and mass spectra at smallest $0<\mx\ll\Bigl (\Lambda^{SU(\nd-{\rm n}_1)}_{{\cal N}=2\,\, SYM}\Bigr )^2/\lm$. The mass spectra in these br2 vacua at larger values of $\mx$ are described in sections 2.3 and 2.4. These results serve then as a basis for a description of similar regimes in sections 6-8. The br2 vacua of the $U(N_c)$ theory at smallest $0<\mx\ll\Bigl (\Lambda^{SU(\nd-\no)}_
{{\cal N}=2\,\, SYM}\Bigr )^2/\lm$ are considered in section 2.2. Besides, the S vacua with the unbroken flavor symmetry $U(N_F)$ (i.e. $\no=0$) in the $SU(N_c)$ theory at $m\ll\lm$ are considered in section 3.

The light particle quantum numbers, the low energy Lagrangians and mass spectra in a large number of vacua of $SU(N_c)$ or $U(N_c)$ theories with $m\gg\lm$ are considered in sections 4 and 5.

Mass spectra in specific additional vs (very special) vacua with $\langle S\rangle_{N_c}=0,\, N_c\leq N_F< 2N_c-1$ present in $U(N_c)$ theory (but absent in $SU(N_c)$, see Appendix in \cite{ch4}\,), are considered in section 6 at $m\lessgtr\lm$ and various values of $0<\mx\ll\lm$.

The mass spectra are calculated in section 7 in br2 vacua of $SU(N_c)$ theory with $U(N_F)\ra U(\no)\times U(\nt),\, \nt<N_c,\,\,m\gg\lm$ at various values of $\mx$.

And finally, the mass spectra are described in section 8 for vacua with $0<N_F< N_c-1$ and $m\gg\lm\,$.\\

Calculations of power corrections to the leading terms of the low energy quark and dyon condensates are presented in two important Appendices. The results agree with also presented in these Appendices {\it independent} calculations of these condensates using roots of the Seiberg-Witten spectral curve. This agreement confirms in a non-trivial way a self-consistency of the whole approach.\\

This article supersedes arXiv:1603.04255.

\addcontentsline{toc}{section}
{\hspace*{4cm}\bf Part I.\,\, Small quark masses, $\Large\mathbf{m\ll\lm}$ }

\vspace*{3mm}

\begin{center}{\hspace*{1cm}\bf\Large Part I.\,\, Small quark masses, $\Large\mathbf{m\ll\lm}$ } \end{center}

\section{ Broken flavor symmetry,\,\, br2 vacua}
\numberwithin{equation}{section}

\hspace{4mm} From \eqref{(1.3)},\eqref{(1.4)} for the $SU(N_c)$ theory, the quark condensates (summed over all $N_c$ colors) at $N_c+1<N_F<2N_c-1$ in the considered br2 vacua (i.e. breaking 2, with the dominant condensate $\langle({\ov Q}Q)_2\rangle_{N_c}$) with $\nt>N_c\,,\,1\leq\no<\nd=(N_F-N_c)$ in theory \eqref{(1.1)} (these are vacua of the baryonic branch in the language \cite{APS} or zero vacua in \cite{SY1,SY2}\,) look as, see section 3 in \cite{ch4} or section 4 in \cite{ch5}, (the leading terms only)
\footnote{\,
Here and everywhere below\,: $A\approx B$ has to be understood as an equality neglecting smaller power corrections, and $A\ll B$ has to be understood as $|A|\ll |B|$. \label{(f4)}
}
\bq
\langle({\ov Q}Q)_2\rangle_{N_c}=\mx m_1+\frac{N_c-\no}{\nt-N_c}\langle({\ov Q}Q)_1\rangle_{N_c}\approx \mx m_1\,,\quad \langle({\ov Q}Q)_1\rangle_{N_c}\approx\mx m_1\Bigl (\frac{m_1}{\lm}
\Bigr)^{\frac{\bb}{{\rm n}_2-N_c}},\label{(2.1)}
\eq
\bbq
\frac{\langle ({\ov Q}Q)_1\rangle_{N_c}}{\langle({\ov Q}Q)_2\rangle_{N_c}}\approx\Bigl (\frac{m_1}{\lm}
\Bigr)^{\frac{2N_c-N_F}{{\rm n}_2-N_c}}\ll 1\,,\,\,  N_c+1< N_F < 2N_c-1\,,\,\,  m_1=\frac{N_c}{N_c-n_2}\,m\,,\,\, \nd\equiv N_F-N_c,
\eeq
\bbq
\langle S\rangle_{N_c}=\frac{\langle\Qo\rangle_{N_c}\langle\Qt\rangle_{N_c}}{\mx}\approx\mx m_1^2\Bigl (\frac{m_1}{\lm}\Bigr)^{\frac{\bb}{{\rm n}_2-N_c}}\ll\mx m^2,\quad \nt > N_c\,,
\eeq
and this shows that {\it the multiplicity of these vacua} is $N_{\rm br2}=(\nt-N_c)\, C^{\,\nt}_{N_F}=(\nd-n_1)\,C^{\,\no}_{N_F}$, the factor $C^{\,\no}_{N_F}=N_F!/(\no!\,\nt!)$ originates from the spontaneous breaking $U(N_F)\ra U(\no)\times U(\nt)$ of the global flavor symmetry of \eqref{(1.1)}.

Because the multiplicity $N_{\rm br2}$ of these br2 vacua in \eqref{(2.1)} does not contain the factor $\bb$, this shows that the non-trivial at $\bb\geq 2$ discrete symmetry $Z_{\bb}$ {\it is not broken spontaneously in these vacua}.

\numberwithin{equation}{subsection}
\subsection{$SU(N_c)$, smallest $\mx$}

\hspace{4mm} We describe first the overall qualitative picture of various stages (in order of decreasing energy scale) of the gauge and flavor symmetry breaking in these br2 vacua. As will be shown below, it leads in a practically unique way to a right multiplicity of these vacua with unbroken discrete $Z_{\bb\geq 2}$ symmetry, and it is non-trivial to achieve this. The value of $\mx$ is taken in this section to be very small, $0<\mx<\langle\Lambda^{SU(\nd-\no)}_{{\cal N}=2\,\, SYM}\rangle^2/\lm$, see \eqref{(2.1.4)} below.\\

{\bf 1)}\, As was explained in Introduction, at small $m$ and $\mx$, the adjoint field $X^{\rm adj}_
{SU(N_c)}$ higgses necessarily the UV free $SU(N_c)$ group at the largest scale $\mu\sim\lm$ to avoid $g^2(\mu<\lm)<0$. In the case considered this first stage looks as: $\,SU(N_c)\ra SU(N_F-N_c)\times U^{(1)}(1)\times U(1)^{\bb-1}$ \cite{APS}, while all other components of $\langle X^A \rangle\lesssim m$ are much smaller, see below \eqref{(2.1.3)},\eqref{(2.1.6)},
\bq
\langle\sqrt{2}\, X^{\rm adj}_{SU(\bb)}\,\rangle =C_{\bb}\lm\,{\rm diag}(\,\underbrace{\,0}_{\nd}\,; \underbrace{\,\omega^0,\,\,\omega^1,\,...\,,\,\omega^{\bb-1}}_{\bb}\,)\,,\quad \omega=\exp\{\frac{2\pi i}{\bb}\},\, \,\,\,  \label{(2.1.1)}
\eq
\bbq
\sqrt{2}\, X^{(1)}_{U(1)}=a_{1}\,{\rm diag}(\,\underbrace{\,1}_{\nd}\,;\, \underbrace{\,c_1}_{\bb}\,),\quad c_1=-\,\frac{\nd}{\bb}\,,\quad \langle a_{1}\rangle=\frac{1}{c_1}\, m\,,\quad C_{\bb}=O(1)\,.
\eeq
This pattern of symmetry breaking is required by the unbroken $Z_{\bb\geq 2}$ discrete symmetry. Remind that  charges under $Z_{\bb}=\exp\{\pi i/(\bb)\}$ transformation are: $q_{\lambda}=q_{\theta}=1,\,\,q_{X}=q_{\rm m}=2,\,\,q_Q=q_{\ov Q}=q_{\lm}=0,\,\,\mx=-2$, so that $\langle a_{1}\rangle$ respects $Z_{\bb}$ symmetry, while the form of $\langle\sqrt{2}\, X^{\rm adj}_{SU(\bb)}\,\rangle$ ensures the right behavior under $Z_{\bb }$ transformation up to interchanging terms in \eqref{(2.1.1)} (the Weyl symmetry).

As a result, all original $SU(2N_c-N_F)$ charged electric particles acquire large masses $\sim\lm$ and decouple at lower energies $\mu<\lm/(\rm several)$. But due to strong coupling, $a(\mu\sim\lm)=N_c g^2(\mu\sim\lm)/2\pi\sim 1$ and the enhanced ${\cal N}=2$ SUSY, the light composite BPS dyons $D_j, {\ov D}_j,\,\, j=1...2N_c-N_F$ (massless at $\mx\ra 0$) are formed in this sector at the scale $\mu\sim\lm$, their number is required by the unbroken $Z_{\bb\geq 2}$ symmetry which operates interchanging them among one another. This is seen also from the spectral curve \eqref{(1.2)}: it has $\bb$ unequal double roots $e_j\approx \omega^{\,j-1}\lm,\,\, j=1...\bb,$ corresponding to these $\bb$ massless BPS solitons \cite{APS,CKM}. The charges of these dyons $D_j$ are discussed below in detail. They are all flavor singlets, their $U^{(1)}(1)$ and $SU(N_c)$ baryon charges which are $SU(\bb)$ singlets are pure electric, while they are coupled with the whole set of $U^{\bb-1}(1)$ independent light Abelian (massless at $\mx\ra 0$) multiplets in \eqref{(2.1.2)}.

{\it All original electric particles with $SU(\nd)$ colors remain light}, i.e. with masses $\ll\lm$. This is a consequence of the non-trivial unbroken $Z_{\bb\geq 2}$ symmetry and their BPS properties. As for quarks with $SU(\nd)$ colors, the appearance in the superpotential of the mass terms like $\sim\lm {\rm Tr\,}({\ov Q} C_Q Q)_{\nd}$ with the constants $C^i_Q$ such that this term will be  $Z_{\bb\geq 2}$ invariant is impossible, independently of the source of such terms. Therefore, the unbroken $Z_{\bb\geq 2}$ is sufficient by itself to forbid $SU(\nd)$ quark masses $\sim\lm$. Similarly, it also forbids (as well as the unbroken $U(1)$ R-symmetry) the  "outside" contributions in the superpotential of mass terms of adjoint scalars like $\lm (X^{A}_{SU(N_c)})^2$ or $m (X^{A}_{SU(N_c)})^2$.

But we will need below somewhat stronger assumption ${\bf "A"}$ formulated in Introduction that all $SU(N_c)$ quarks are BPS particles, i.e. they receive additional contributions to their masses {\it only directly} from the couplings with $\langle X^{(adj)}_{SU(N_c)}\rangle$, see below. This forbids also the appearance of {\it any "outside"} (i.e. originating not from the direct couplings with $\langle X^{(adj)}_{SU(N_c)}\rangle$ contributions to their masses, e.g. the additional contributions in the superpotential $c_m m {\rm Tr\,}({\ov Q} Q)_{\nd},\, c_m=O(1)$, although these are not forbidden by the unbroken $Z_{\bb\geq 2}$. Besides, this BPS assumption ${\bf "A"}$ concerns all original electric particles of $SU(N_c)$. On the other hand, it is natural to expect that all original particles of the $SU(N_c)$ theory have the same properties, in the sense that in considered isolated vacua at $\mx\ra 0$ either they all are BPS or they all are not BPS. And because the "outside"\, contributions $\sim\lm$ to masses in the superpotential are forbiden for $SU(\nd)$ quarks and scalars, and $\sim m$ are forbidden for scalars, then all "outside"\, contributions to masses are forbidden for all original electric particles, i.e. they all are indeed the BPS particles. (The absence of such additional contributions is confirmed by {\it two independent} calculations of $SU(\nd)$ quark condensates, see \eqref{(2.1.14)} and the text just below it).\\

The $SU(\nd)$ gauge symmetry can still be considered as unbroken at scales $m\ll\mu<\lm$ (it is broken only at the scale $\sim m$ due to higgsed $X^{\rm adj}_{SU(\nd)}$, see below). Therefore, after integrating out all heavy particles with masses $\sim\lm$, the general form of the superpotential at the scale $\mu_{\rm cut}=\lm/{(\rm several)}$ can be written as (the dots in \eqref{(2.1.2)} denote smaller power corrections)
\bq
\wh{\w}_{\rm matter}=\w_{SU(\nd)}+\w_D+\w_{a_1}+\dots\,,\label{(2.1.2)}
\eq
\bbq
\w_{SU(\nd)}={\rm Tr}\,\Bigl [{\ov Q}\Bigl (m-a_1-\sqrt{2}X^{\rm adj}_{SU(\nd)}\Bigr ) Q\Bigr ]_{\nd}+\wmu {\rm Tr}\,(X^{\rm adj}_{SU(\nd)})^2\,,\quad \wmu=\mx(1+\delta_2)\,,
\eeq
\bbq
\w_{D}=(m-c_1 a_1)\sum_{j=1}^{\bb}{\ov D}_j D_j-\sum_{j=1}^{\bb} a_{D,j}{\ov D}_j D_j-\mx\lm\sum_{j=1}^{\bb}\omega^{j-1}a_{D,j}+
\eeq
\bbq
+\mx L \sum_{j=1}^{\bb} a_{D,j}+O\Bigl (\mx a_D^2 \Bigr )\,,
\eeq
\bbq
\w_{a_1}=\frac{\mx}{2}(1+\delta_1)\frac{\nd N_c}{\bb}\,a^2_1+\mx N_c \delta_3\,a_1(m-c_1 a_1)+\mx N_c\delta_4 (m-c_1 a_1)^2\,,
\eeq
where $a_{D,j}$ are the light neutral scalars, $L$ is the Lagrange multiplier, $\langle L\rangle=O(m)$, while all $\delta_i=O(1)$.

These terms with $\delta_i$  in the superpotential \eqref{(2.1.2)} originate finally from {\it the combined effect} of\,: a) the quantum loops with heavy particles with masses $\sim\lm$, integrated over the energy range $[\mu_{\rm cut}=\lm/(\rm several)]<\mu<(\rm several)\lm$, in the strong coupling regime $a(\mu\sim\lm)
\sim 1$ and in the background of lighter fields;\, b) the unsuppressed non-perturbative instanton contributions from the broken color subgroup $SU(\bb)\ra U^{\bb-1}(1)$ which are operating at the scale $\mu\sim\lm$. Without instantons, different pure perturbative loop contributions to the superpotential cancel in the sum, but the additional instanton contributions spoil this cancelation. See also the footnote \ref{(f6)}. Compare with the absence of such terms in sections 4.1,\,4.2, see \eqref{(4.1.4)},\eqref{(4.2.3)}, when $\langle X^{adj}_{SU(N_c)}\rangle\sim m\gg\lm$ is higgsed in the weak coupling regime, resulting in $SU(N_c)\ra SU(\no)\times U(1)\times SU(N_c-\no)$ with unbroken at this scale the non-Abelian  ${\cal N}=2\,\, SU(N_c-\no)$ SYM part. The instanton contributions are then power suppressed. They originate only from this ${\cal N}=2$ SYM part and operate in the lower energy strong coupling region at the scale $\mu\sim\langle\Lambda^{SU(N_{c}-{\rm n}_1)}_{{\cal N}=2\,\,SYM}\rangle\ll m$, after the color breaking $SU(N_c-\no)\ra U^{N_c-\no-1}(1)$ by higgsed $\langle X^{adj}_{SU(N_c-\no)}\rangle\sim\langle\Lambda^{SU(N_{c}-{\rm n}_1)}_{{\cal N}=2\,\,SYM}\rangle$, see \eqref{(4.1.2)}).

The way the parameter "m" enters \eqref{(2.1.2)} can be understood as follows. This parameter "m" in the original superpotential $\w_{\rm matter}$ in \eqref{(1.1)} can be considered as a soft scalar $U^{(0)}(1)$ background field coupled universally to the $SU(N_c)$ singlet scalar electric baryon current, $\delta\w_B=m J_B$, and this current looks at the scale $\mu\gg\lm$ as $J_B={\rm Tr}\,({\ov Q} Q)_{N_c}$. But at the lower energy scale $\mu\ll\lm$ only lighter particles with masses $\ll\lm$ and with a nonzero electric baryon charge contribute to $J_B$, these are ${\rm Tr}\,({\ov Q} Q)_{\nd}$ and $\sum_{j=1}^{\bb}{\ov D}_j D_j$ (see bellow). Moreover, the field $a_{1}$  plays a similar role, but its couplings are different for two different color sectors of $SU(N_c)$, see \eqref{(2.1.1)},\eqref{(2.1.2)}.

The terms with $\delta_i,\,\, i=1\,...\,4\,,\,\,\delta_i=O(1)$, arose in \eqref{(2.1.2)} from integrating out heaviest original fields with masses $\sim \lm$ in the soft background of lighter $X^{\rm adj}_{SU(\nd)},\,\, a_1$ and $(m-c_1 a_1)$ fields. These heavy fields are\,: charged $SU(\bb)$ adjoints with masses
$ (\Lambda_j-\Lambda_i),\,\, \Lambda_j\sim\omega^{\,j-1}\lm$, the heavy quarks with $SU(\bb)$ colors and masses $\Lambda_j-(m-c_1 a_1)$, and heavy adjoint hybrids $SU(N_c)/[SU(\nd)\times SU(\bb)\times U^{(1)}(1)]$ with masses $\Lambda_j-[\sqrt{2}X^{\rm adj}_{SU(\nd)}+(1-c_1)a_1]$. Note that terms like $\mx\lm^2,\,\,\mx a_1\lm$ and $\mx (m-c_1 a_1)\lm$ cannot appear in \eqref{(2.1.2)}, they are forbidden by the unbroken $Z_{\bb\geq 2}$ symmetry. \\

The above described picture is more or less typical, the only really non-trivial point concerns the quantum numbers of $2N_c-N_F$ dyons $D_j$. Therefore, we describe now some necessary requirements to them.

a) First, the $SU(\nd)$ part of \eqref{(2.1.2)}, {\it due to its own internal dynamics}, ensures finally the right pattern of the spontaneous flavor symmetry breaking $U(N_F)\ra U(\no)\times U(\nt)$ and the right multiplicity of vacua, $N_{\rm br2}=(\nd-\no) C^{\,\no}_{N_F}$, see below. Therefore, the quantum numbers of these $\bb$ dyons $D_j$ should not spoil the right dynamics in the $SU(\nd)$ sector.

b) These quantum numbers should be such that the lower energy Lagrangian respects the discrete $Z_{\bb\geq 2}$ symmetry. This fixes, in particular, the number $\bb$ of these dyons because $Z_{\bb\geq 2}$ transformations interchange them among one another. Notice also that the quantum numbers of these dyons $D_i$ formed at the scale $\mu\sim\lm$ cannot know about properties of further color or flavor breaking in the $SU(\nd)$ color part at much lower scales $\mu\sim m\ll\lm$ or $\mu\sim (\mx m)^{1/2}\ll m$, i.e. they do not know the number $\no$.

c) According to M.Shifman and A.Yung, see  e.g. \cite{SY6}, the massive diagonal quarks ${\ov Q}^i_i,\, Q^i_i,\, i=1...2N_c-N_F$ with $SU(2N_c-N_F)$ colors and masses $\sim\lm$ combine  at the scale $\mu\sim\lm$  with $SU(2N_c-N_F)$ appropriate massive flavorless magnetic (anti)monopoles and form $2N_c-N_F$ massless dyons ${\ov D}_i, D_i$ in \eqref{(2.1.2)}. But for quarks with equal own masses $"m"$ the non-Abelian flavor symmetry $SU(N_F)$ is {\it not broken} by higgsed $\langle X_{SU(\bb)}\rangle\sim\lm$ at the scale $\mu\sim\lm$, and {\it flavorless monopoles cannot distinguish between quarks with different $SU(N_F)$ flavors. Therefore, such dyons will really have the same flavor as quarks, i.e. they will form the (anti)fundamental representation of global $SU(N_F)$}.

But these $\bb$ dyons  ${\ov D}_i, D_i$ in \eqref{(2.1.2)} have to be $SU(N_F)$ flavor singlets. Otherwise, because they all are higgsed at the scale  $\mu\sim (\mx\lm)^{1/2}\neq 0$, this will result in a wrong pattern of the spontaneous non-Abelian flavor symmetry breaking $SU(N_F)\ra SU(2N_c-N_F)\times SU(2N_F-2N_c)$ instead of the right one $SU(N_F)\ra SU(\no)\times SU(\nt)$. Besides, there will be $N_F (\bb)$ these dyons $D$, this is clearly too much.

d) Their magnetic quantum numbers should be such that they are mutually local with respect to all original pure electric particles in the $SU(\nd)$ sector. Otherwise, in particular, because all these dyons and all original pure electric quarks from $SU(\nd)$ have nonzero $U^{(1)}(1)$ charges, and because all dyons are higgsed, $\langle{\ov D}_j\rangle=\langle D_j\rangle\sim (\mx\lm)^{1/2}$, if the  magnetic $U^{(1)}(1)$ charge of these dyons were nonzero, this would lead to confinement of all quarks from the $SU(\nd)$ sector with the string tension $\sigma^{1/2}\sim (\mx\lm)^{1/2}$. E.g., see below, all $N_F$ flavors of confined light electric quarks with $SU(\no)$ colors would decouple then in any case at the scale $\mu<(\mx\lm)^{1/2}$, the flavor symmetry $U(N_F)$ would remain unbroken in the whole $SU(\nd)$ sector, while the remained ${\cal N}=2 \,\,SU(\no)$ SYM would give the additional wrong factor $\no$ in the multiplicity. All this is clearly unacceptable. This excludes the variant with the nonzero $U^{(1)}(1)$ magnetic charge (and also with non-zero $SU(\nd)$ magnetic charges) of these dyons. Similarly, when considering the $U(N_c)=SU(N_c)\times U^{(0)}(1)$ theory in section 2.2 below, this also requires the $SU(N_c)$ baryon charge of these dyons to be pure electric. For the same reasons, the magnetic charges of these dyons cannot be $SU(N_c)/[SU(\nd)\times SU(\bb)$ hybrids. On the whole, to be mutually local with the whole $SU(\nd)$ part, the magnetic parts of all charges of these dyons have to be $SU(\bb)$ root-like (i.e. $SU(\bb)$ adjoints most naturally).

e) If these $\bb$ dyons $D_j$ had zero $U^{(1)}(1)$ charge this would correspond to the first term of $\w_D$ in \eqref{(2.1.2)},\eqref{(2.1.5)} of the form $m\sum_{j}^{\bb}{\ov D}_j D_j$. They would not be coupled then with the $U^{(1)}(1)$ multiplet \eqref{(2.1.1)} and this would lead to extra massless particles at $\mx\neq 0$, and even to the internal inconsistency, see below.

The case with zero $SU(N_c)$ baryon charge of dyons, i.e. with the first term of $\w_{D}$ of the form $[\,-c_1 a_1\sum_{j=1}^{\bb}{\ov D}_j D_j\,]$ in \eqref{(2.1.2)}, \eqref{(2.1.5)}, will lead to wrong values of quark condensates, see below.

Finally, about the case when this first term of $\w_{D}$ in \eqref{(2.1.2)} equals zero. The whole dyon sector will be then completely decoupled from the $SU(\nd)\times U^{(1)}(1)$ sector, so that $\langle{\ov D}_j D_j\rangle$ will not include corrections containing $\langle\Lambda^{SU(\nd-\no)}_{{\cal N}=2\,\, SYM}\rangle$. And this will be wrong, because the values of  $\langle{\ov D}_j D_j\rangle$ obtained {\it independently} from the roots of the curve \eqref{(1.2)} contain such corrections, see \eqref{(B.15)}. Besides, all $\bt=\bb$ dyons ${\ov D}_j, D_j$ are higgsed at $\mx\neq 0$ (this is a consequence of the unbroken $Z_{\bb\geq 2}$ symmetry which operates interchanging them among one another). But they will be coupled then only maximum with $\bb-1$ independent multiplets $U^{\bb-1}(1)$.  Then {\it additional} massless Nambu-Goldstone bosons will remain at $\mx\neq 0$ resulting in the additional (infinite) factor in the multiplicity of these vacua. Moreover, counting degrees of freedom in the $SU(\nd)\times U^{(1)}(1)$ sector (see below), it is seen that one ${\cal N}=1$ photon multiplet will remain exactly massless (while its ${\cal N}=2$ scalar partner has very small mass $\sim\mx$ only due to breaking ${\cal N}=2\ra {\cal N}=1$ at small $\mx\neq 0$).

Clearly, these considerations with extra massless particles concern as well the cases when even one out of $U^{2N_c-N_F-1}(1)$ Abelian photon multiplets is not coupled with the whole set of $2N_c-N_F$ dyons. Due to the unbroken $Z_{2N_c-N_F}$ discrete symmetry, all $2N_c-N_F-1$ Abelian multiplets will not be
coupled then. In any case (not even speaking about other problems), at least all $U^{2N_c-N_F-1}(1)$ photons will remain massless at $\mx\neq 0$. But since all dyons are higgsed (this follows from the curve \eqref{(1.2)}, see \eqref{(B.15)}),  there will be a large number of extra Numbu-Goldstone particles. According to our assumption "B" (see Introduction), all variants with extra massless particles at $\mx\neq 0,\, m\neq 0$  in $SU(N_c)$ theories (in addition to normal Nambu-Goldstone particles originating from the spontaneous breaking of the global flavor symmetry) are excluded.

f) The regime in the $SU(\bb)$ color sector at low energies is IR free. As follows from the curve \eqref{(1.2)}, see \eqref{(B.15)}), all $\bb$ these BPS dyons are higgsed simultaneously in the ground state at small $\mx\neq 0$. And for this to be possible, {\it they all have to be massless at $\mx\ra 0$ and to be  mutually local} (so that $\langle a_{D,j}\rangle=0$, see \eqref{(2.1.5)},\eqref{(2.1.6)} below).

g) No one of $U^{\bb-1}(1)$ charges of these light BPS dyons $D_j$ can be {\it pure electric}. Indeed, since all $SU(\bb)$ adjoint $\langle a_j\rangle\sim\lm$ in \eqref{(2.1.1)}, all BPS solitons with some pure electric $U^{\bb-1}(1)$ charges will have large masses $\sim\lm$ in this case.

Therefore, as a result, these mutually local $\bb$ BPS dyons $D_j, {\ov D}_j$, massless at $\mx\ra 0$, should have nonzero: i) {\it pure electric} both $SU(N_c)$ baryon and $U^{(1)}(1)$ charges; ii) on the whole, all $\bb-1$ nonzero independent Abelian charges of the broken $SU(\bb)$ subgroup, with their $SU(\bb)$ adjoint magnetic parts. Then they will be mutually local with the whole $SU(\nd)$ sector, and coupled with $"m"$ and with the electric $U^{(1)}(1)$ multiplet, and with all $U^{\bb-1}(1)$ Abelian multiplets, see\eqref{(2.1.1)},\eqref{(2.1.2)}. There will be no massless particles at $\mx\neq 0,\, m\neq 0$ (except for $2\nt\no$ Nambu-Goldstone multiplets), and the multiplicity and pattern of flavor symmetry breaking will be right, see below.\\

{\bf 2)}\, The second stage is the breaking $SU(\nd)\ra SU(n_1)\times  U^{(2)}(1)\times SU(\nd-n_1)$ at the lower scale $\mu\sim m\ll\lm$ {\it in the weak coupling regime}, see \eqref{(2.1.6)} below, $1\leq \no<\nd=N_F-N_c$, (this stage is qualitatively similar to those in section 4.1)
\bq
\sqrt{2}\, X^{(2)}_{U(1)}=a_{2}\,{\rm diag}(\,\underbrace{\,1}_{\no}\,;\, \underbrace{\,c_2}_{\nd-\no}\,;\,\underbrace{0}_{\bb}\,),\quad c_2=-\,\frac{\no}{\nd-\no}\,,\quad \langle a_{2}\rangle=\frac{N_c}{\nd}\, m\,.\label{(2.1.3)}
\eq

As a result, all original electric quarks with $SU(\nd-\no)$ colors and $N_F$ flavors have masses $m_2=\langle m- a_1-c_2 a_2\rangle=m\,N_c/(\nd-\no)$, and all original electrically charged $SU(\nd)$ adjoint hybrids $SU(\nd)/[SU(\no)\times SU(\nd-\no)\times U(1)]$ also have masses $(1-c_2)\langle a_2\rangle= m_2$, and so all these particles decouple as heavy at scales $\mu\lesssim m$. But the original electric quarks with $SU(\no)$ colors and $N_F$ flavors have masses $\langle m- a_1- a_2\rangle=0$, they remain massless at $\mx\ra 0$ and survive at $\mu\lesssim m$. Therefore, there remain two lighter non-Abelian sectors $SU(\no)\times SU(\nd-\no)$. As for the first one, it is the IR free ${\cal N}=2\,\, SU(\no)$ SQCD with $N_F$ flavors of original electric quarks massless at $\mx\ra 0\,,\,\, N_F>2\no,\,\, 1\leq\no<\nd\,$. As for the second one, it is the ${\cal N}=2\,\, SU(\nd-\no)$ SYM with the scale factor $\Lambda^{SU(\nd-{\rm n}_1)}_{{\cal N}=2\,\, SYM}$ of its gauge coupling, compare with $\langle S\rangle_{N_c}$ in \eqref{(2.1)},
\bq
\Bigl (\Lambda^{SU(\nd-\no)}_{{\cal N}=2\,\, SYM}\Bigr )^{2(\nd-\no)}=\frac{\Lambda_{SU(\nd)}^{2\nd-N_F}
(m-a_1-c_2 a_2)^{N_F}}{[\,(1-c_2)a_2\,]^{\,2\no}}\,,\label{(2.1.4)}
\eq
\bbq
\langle\Lambda^{SU(\nd-\no)}_{{\cal N}=2\,\, SYM}\rangle^2=m^2_2\Bigl (\frac{m_2}{\Lambda_{SU(\nd)}}\Bigr )^{\frac{2N_c-N_F}{\nd-\no}}\ll m^2\,,\quad\langle S\rangle_{\nd-\no}=\mx(1+\delta_2)\langle\Lambda^{SU(\nd-\no)}_{{\cal N}=2\, SYM}\rangle^2\,,
\eeq
\bbq
m_2=\langle m-a_1-c_2 a_2\rangle=(1-c_2)\langle a_2\rangle=\frac{N_c}{\nt-N_c}\,m=-m_1\,,
\eeq
where $\Lambda_{SU(\nd)}$ is the scale factor of the $SU(\nd)$ gauge coupling at the scale $\mu=\lm/(\rm several)$ in \eqref{(2.1.2)}, after integrating out heaviest particles with masses $\sim\lm$ (it will be determined below in \eqref{(2.1.15)}).\\

{\bf 3)}\, At the third stage, to avoid unphysical $g^2(\mu<\langle\Lambda^{SU(\nd-\no)}_{{\cal N}=2\,\, SYM}\rangle)<0$ of  UV free $SU(\nd-\no)\,\, {\cal N}=2$ SYM, the field $X^{adj}_{SU(\nd-\no)}$ is higgsed, $\langle X^{adj}_{SU(\nd-\no)}\rangle\sim \langle\Lambda^{SU(\nd-{\rm n}_1)}_{{\cal N}=2\,\, SYM}\rangle$, breaking $SU(\nd-\no)$ in a well known way \cite{DS}, $SU(\nd-\no)\ra U^{\nd-\no-1}(1)$. All original pure electrically charged adjoint gluons and scalars of ${\cal N}=2\,\,SU(\nd-\no)$ SYM acquire masses $\sim\langle\Lambda^{SU(\nd-{\rm n}_1)}_{{\cal N}=2\,\, SYM}\rangle$ and decouple at lower energies $\mu<\langle\Lambda^{SU(\nd-{\rm n}_1)}_{{\cal N}=2\,\, SYM}\rangle$. Instead, $\nd-\no-1$ light composite pure magnetic monopoles (massless at $\mx\ra 0$), $M_i,\,{\ov M}_i,\,\, i=1,...,\nd-\no-1$, with their $SU(\nd-\no)$ adjoint magnetic charges are formed at this scale $\sim\langle\Lambda^{SU(\nd-{\rm n}_1)}_{{\cal N}=2\,\, SYM}\rangle$. The factor $\nd-\no$ in the overall multiplicity of considered br2 vacua originates from the multiplicity of vacua of this ${\cal N}=2\,\,SU(\nd-\no)$ SYM. Besides, the appearance of two single roots with $(e^{+}-e^{-})\sim \langle\Lambda^{SU(\nd-\no)}_{{\cal N}=2\,\, SYM}\rangle$, of the curve \eqref{(1.2)} in these br2 vacua is connected just with this ${\cal N}=2\,\,SU(\nd-\no)$ SYM \cite{DS}. Other $\nd-\no-1$ double roots originating from this ${\cal N}=2$ SYM sector are unequal double roots corresponding to $\nd-\no-1$ pure magnetic monopoles $M_{\rm n}$, massless at $\mx\ra 0$.\\

{\bf 4)}\,  All $\bb$ dyons ${\ov D}_j, D_j$ are higgsed at the scale $\sim (\mx\lm)^{1/2}$ in the weak coupling regime:  $\,\,\langle D_j\rangle=\langle {\ov D}_j\rangle\sim (\mx\lm)^{1/2}\gg (\mx m)^{1/2}$. As a result, $\bb$ long ${\cal N}=2\,\, U(1)$ multiplets of massive photons are formed (including  $U^{(1)}(1)$ with its scalar $a_1$), all with masses $[\,g_D(\mu\sim(\mx\lm)^{1/2})\ll 1\,](\mx\lm)^{1/2}$ (for simplicity, we ignore below logarithmically small factors $\sim g$ in similar cases, these are implied where needed). No massless particles remain in this sector at $\mx\neq 0$. And there remain no particles at all in this sector at lower energies.

The $(\bb)$-set of these higgsed dyons with the $SU(2N_c-N_F)$ adjoint magnetic charges is mutually non-local with all original $SU(2N_c-N_F)$ pure electrically charged particles with largest masses $\sim\lm$. Therefore, all these heaviest electric particles are weakly confined, the string tension is $\sigma^{1/2}_D\sim (\mx\lm)^{1/2}$. This confinement is weak in the sense that the tension of the confining string is much smaller than particle masses, $\sigma^{1/2}_{D}\sim (\mx\lm)^{1/2}\ll\lm\,$. All these heaviest confined particles form a large number of hadrons with masses $\sim\lm$.\\

{\bf 5)}\, $\no$ out of $N_F$ electric quarks of IR free ${\cal N}=2\,\, SU(\no)$ SQCD are higgsed at the scale $\mu\sim (\mx m)^{1/2}$ in the weak coupling region, with $\langle{\ov Q}_k^{\,a}
\rangle=\langle Q^k_a\rangle\sim \delta^k_a\,(\mx m)^{1/2},\,\, a=1,...,\no,\,\, k=1,...,N_F$. As a result, the whole electric group $SU(\no)$ is broken and $\no^2$ long ${\cal N}=2$ multiplets of massive electric gluons are formed (including  $U^{(2)}(1)$ with its scalar $a_2$), all with masses $\sim (\mx m)^{1/2}\ll (\mx\lm)^{1/2}$. The global flavor symmetry is broken spontaneously {\it only at this stage}, $U(N_F)\ra U(\no)\times U(\nt)$, and $2\no\nt$ (complex) massless Nambu-Goldstone multiplets remain in this sector at lower energies (in essence, these are quarks $Q^k_a,\, {\ov Q}^{\,a}_{k},\,a=1,...,\no,\,\, k=\no+1,...,N_F$ ). This is a reason for the origin of the factor $C^{\,\no}_{N_F}$ in multiplicity of these br2 vacua. No additional massless particles remain in this sector at $\mx\neq 0,\, m\neq 0$.\\

{\bf 6)}\,  $\nd-\no-1$ magnetic monopoles are higgsed at the lowest scale, $\langle{\ov M}_{\rm n}\rangle=\langle M_{\rm n}\rangle\sim (\mx\langle\Lambda^{SU(\nd-{\rm n}_1)}_{{\cal N}=2\,\, SYM}\rangle)_{,}^{1/2}$ and $\nd-{\rm n}_1-1$ long ${\cal N}=2$ multiplets of massive $U^{\nd-\no-1}(1)$ photons are formed, all with masses $\sim (\mx\langle\Lambda^{SU(\nd-{\rm n}_1)}_{{\cal N}=2\,\, SYM}\rangle)^{1/2}\ll (\mx m)^{1/2}$. All original pure electrically charged particles with non-singlet $SU(\nd-{\rm n}_1)$ charges and masses either $\sim m$ or $\sim\langle\Lambda^{SU(\nd-{\rm n}_1)}_{{\cal N}=2\,\, SYM}\rangle$ are weakly confined, the string tension is $\sigma^{1/2}_{\rm SYM}\sim (\mx\langle\Lambda^{SU(\nd-{\rm n}_1)}_{{\cal N}=2\,\, SYM}\rangle)^{1/2}$. No massless particles remain in this sector at $\mx\neq 0,\, m\neq 0$.\\

As a result of all described above, the lowest energy superpotential at the scale $\mu=\langle\Lambda^
{SU(\nd-\no)}_{{\cal N}=2\,\, SYM}\rangle/(\rm several)$, and e.g. at smallest $0<\mx\ll\langle\Lambda^
{SU(\nd-\no)}_{{\cal N}=2\,\, SYM}\rangle^2/\lm$, can be written in these br2 vacua as
\bq
\w^{\,\rm low}_{\rm tot}=\w^{\,(SYM)}_{SU(\nd-\no)}+\w^{\,\rm low}_{\rm matter}+\dots\,,\quad
\w^{\,\rm low}_{\rm matter}=\w_{SU(\no)}+\w_{D}+\w_{a_1,\,a_2}\,,\label{(2.1.5)}
\eq
\bbq
\w^{\,(SYM)}_{SU(\nd-\no)}=(\nd-\no)\wmu\Bigl (\Lambda^{SU(\nd-{\rm n}_1)}_{{\cal N}=2\,\, SYM}\Bigr )^2+\w^{\,(M)}_{SU(\nd-\no)}\,,\quad \wmu=\mx(1+\delta_2)\,,
\eeq
\bbq
\w^{\,(M)}_{SU(\nd-\no)}= - \sum_{n=1}^{\nd-\no-1}  a_{M, \rm n}\Biggl [\, {\ov M}_{\rm n} M_{\rm n}+\wmu\Lambda^{SU(\nd-\no)}_{{\cal N}=2\,\,SYM}\Biggl (1+O\Bigl (\frac{\langle\Lambda^{SU(\nd-\no)}_{{\cal N}=2\,\,SYM}\rangle}{m}\Bigr )\Biggr )\,d_{\rm n}\,\Biggr ]\,,
\eeq
\bbq
\w_{SU(\no)}=(m-a_1-a_2)\,{\rm Tr}\,({\ov Q} Q)_{\no}-\,{\rm Tr}\,({\ov Q}\sqrt{2} X^{\rm adj}_{SU(\no )} Q)_{\no}+\wmu{\rm Tr}\,(X^{\rm adj}_{SU(\no)})^2\,,
\eeq
\bbq
\w_{D}=(m-c_1 a_1)\sum_{j=1}^{\bb}{\ov D}_j D_j-\sum_{j=1}^{\bb} a_{D,j}\,{\ov D}_j D_j\,-
\,\mx\lm\sum_{j=1}^{\bb}\omega^{j-1}\,a_{D,j}+\mx L \sum_{j=1}^{\bb} a_{D,j}\,,
\eeq
\bbq
\w_{a_1,\,a_2}=\frac{\mx}{2}(1+\delta_1)\frac{\nd N_c}{\bb} a_1^2+\frac{\wmu}{2}\frac{\no \nd}{\nd-\no} a_2^2+\mx N_c \delta_3 a_1(m-c_1 a_1)+\mx N_c\delta_4 (m-c_1 a_1)^2+O\Bigl (\mx a_{D,j}^2 \Bigr ),
\eeq
where coefficients $d_i=O(1)$ in \eqref{(2.1.5)} are known from \cite{DS}), and dots in \eqref{(2.1.5)} denote smaller power suppressed corrections (these are always implied and dots are omitted below in the text), . It is useful to notice that the unbroken $Z_{2N_c-N_F\geq 2}$ symmetry restricts strongly their possible values. The matter is that e.g. power suppressed corrections in $\w_{a_1}$ like $\sim\mx a_1^2(a_1/\lm)^{N}$ originate from $SU(N_c)\ra SU(\nd)\times U^{(1)}(1)\times SU(2N_c-N_F)$ at the scale $\sim\lm$ and they do not know the number $\no$ originating only at much lower scales $\mu\sim m\ll\lm$. Then the unbroken $Z_{2N_c-N_F\geq 2}$ symmetry requires that $N=(2N_c-N_F) k,\, k=1,2,3...$.

Remind that charges of fields and parameters entering \eqref{(2.1.5)} under $Z_{\bb}=\exp\{i\pi/(2N_c-N_F)\}$ transformation are\,: $q_{\lambda}=q_{\theta}=1,\,\, q_{X_{SU(\no)}^{\rm adj}}=q_{a_1}=q_{a_2}=q_{a_{D,j}}=q_{a_{M,\rm n}}=q_{\rm m}=q_{L}=2,\,\,q_{Q}=q_{\ov Q}=q_{D_j}=q_{{\ov D}_j}=q_{M_{\rm n}}=q_{{\ov M}_{\rm n}}=q_{\lm}=0,\,\, q_{\mx}=-2$. The non-trivial $Z_{\bb\geq 2}$ transformations change only numerations of dual scalars $a_{D,j}$ and dyons $D_j, {\ov D}_j$ in \eqref{(2.1.5)}, so that $\int d^2\theta\,\w_{\rm tot}^{\,\rm low}$ is $Z_{\bb}$-invariant.

All (massless at $\mx\ra 0$) $\bt=2N_c-N_F$ dyons $D_j$, $\no<\nd$ quarks $Q^k$ and $\nd-\no-1$ monopoles $M_{\rm n}$ in \eqref{(2.1.5)} are higgsed at $\mx\neq 0$. As a result, we obtain from \eqref{(2.1.5)} (neglecting power corrections)
\bq
\langle a_1\rangle=\frac{1}{c_1}\,m=-\frac{\bb}{\nd}\,m\,,\,\, \langle a_2\rangle=\langle m-a_1\rangle=
\frac{N_c}{\nd}\,m\,,\,\, \langle X^{\rm adj}_{SU({\rm n}_1)}\rangle=\langle a_{D,\,j}\rangle=\langle a_
{M,\rm n}\rangle=0\,,\label{(2.1.6)}
\eq
\bbq
\langle{\ov M}_{\rm n} M_{\rm n}\rangle=\langle{\ov M}_{\rm n}\rangle\langle M_{\rm n}\rangle\approx - \wmu\langle\Lambda^{SU(\nd-\no)}_{{\cal N}=2\,\, SYM}\rangle\,d_{\rm n}\,,\,\, d_{\rm n}=O(1),\quad \langle{\ov D}_j D_j\rangle=\langle{\ov D}_j\rangle\langle D_j\rangle\approx -\mx\lm\,\omega^{j-1}\,,
\eeq
\bbq
\langle\Qo\rangle_{\no}=\langle{\ov Q}^1_1\rangle\langle Q^1_1\rangle\approx\wmu\frac{\nd}{\nd-\no}\langle a_2\rangle\approx \wmu\frac{N_c}{\nd-\no}\,m,\,\, \langle\Qt\rangle_{\no}=\sum_{a=1}^{\no}\langle{\ov Q}^{\,a}_2\rangle\langle Q^2_a\rangle=0,\,\,\langle S\rangle_{\no}=0
\eeq
(the dyon condensates are dominated by terms $\sim\mx\lm$ plus smaller terms $\sim\mx m$, see Appendix B).\\

We have to explain at this point the meaning of notations used e.g. in \eqref{(2.1.5)},\eqref{(2.1.6)} for the superpotential and mean values written for the theory with the upper cutoff $(\mx m)^{1/2}\ll\mu_{uv}=\langle\Lambda^{SU(\nd-\no)}_{{\cal N}=2\,\, SYM}\rangle/(\rm several)\ll m$. By definition, for any operator $O,\,\,\langle O\rangle$ denotes its {\it total mean vacuum value} integrated from sufficiently high energies (formally, from the very high UV cutoff $M_{UV}$ for the elementary fields and from $\mu_{\rm sol}$ for composite solitons formed at the scale $\sim\mu_{\rm sol}$) down to $\mu^{\rm lowest}_{\rm cut}=0$, i.e. $\langle O\rangle\equiv\langle O\rangle^{M_{UV}\,{\rm or}\,\mu_{\rm sol}}_{\mu^{\rm lowest}_{\rm cut}=0}$. Therefore, we have, strictly speaking, to write e.g. for the lower energy superpotential in \eqref{(2.1.5)}: $\,a_{1,2}=\langle a_{1,2}\rangle^{M_{UV}}_{\mu^{\rm low}_{\rm cut}=\mu_{uv}}+(\hat a_{1,2})^{\mu_{uv}}$, with $\mu^{\rm low}_{\rm cut}=\mu_{uv}=\langle\Lambda^{SU(\nd-\no)}_{{\cal N}=2\,\, SYM}\rangle/(\rm several)$, and the remaining soft parts $(\hat a_{1,2})^{\mu_{uv}}$ of operators $a_{1,2}$ with the upper energy cutoff $\mu_{uv}\ll m$. But in the case considered, with higgsed $\langle a_{1,2}\rangle\sim m\ll\lm$, these their total mean values {\it originate and saturate} in the range of scales $m/(\rm several)<\mu<(\rm several) m$, because otherwise all $SU(\nd)$ quarks and all dyons would have masses $\sim m$. They all will decouple then as heavy at the scale $\mu<m/(\rm several)$ and this will be clearly wrong for these br2-vacua (the multiplicity will be wrong, the flavor symmetry will remain unbroken, the number of charged BPS particles massless at $\mx\ra 0$ will be wrong, etc.).  Therefore, $\langle a_{1,2}\rangle^{M_{UV}}_{\mu_{uv}}=\langle a_{1,2}\rangle^{M_{UV}}_{\mu^{\rm lowest}_{\rm cut}=0}\equiv\langle a_{1,2}\rangle=\langle a_{1,2}\rangle^{(\rm several) m}_{m/(\rm several)}$, and $\langle{\hat a}\rangle^{\mu_{uv}}_{\mu^{\rm lowest}_{\rm cut}=0}=0$. For this reason, only in order to simplify all notations (here and in most cases below in the text) all is written as in \eqref{(2.1.5)},\eqref{(2.1.6)}. We hope that the meaning of operators and mean values entering all formulas will be clear in each case from the accompanying text.\\

Now, in a few words, it is not difficult to check that the variant of \eqref{(2.1.5)} with the first term of $\w_D$ of the form $m\sum_{j=1}^{\bb}{\ov D}_j D_j$ (i.e. with zero $U^{(1)}(1)$ charge of dyons) results in $\langle\partial\w^{\,\rm low}_{\rm matter}/\partial L\rangle=\mx m (\bb)/2\neq 0$, and this is the internal inconsistency. Another way, with such a form of the first term of $\w_D$, one obtains from $\langle{\partial\w_D}/{\partial{\ov D}_j}\rangle=0$ : $\langle a_{D,j}\rangle=m,\,\, \sum_{j} \langle a_{D,j}\rangle=(\bb) m\neq 0$, and this is wrong.

The $\bb$ unequal double roots $e^{(D)}_j\approx\omega^{j-1}\lm$ of the curve \eqref{(1.2)} correspond to BPS dyons formed at the scale $\mu\sim\lm$. Together with $\no$ equal double roots $e^{(Q)}_k= - m$ of original pure electric quarks (higgsed at small $\mx\neq 0$) from ${\cal N}=2 \,\, SU(\no)$ SQCD, and $\nd-\no-1$ unequal double roots of $SU(\nd-\no)$ adjoint pure magnetic monopoles from ${\cal N}=2 \,\,  SU(\nd-\no)$ SYM (formed at the scale $\mu\sim\langle\Lambda^{SU(\nd-\no)}_{{\cal N}=2\,\, SYM}\rangle$), they constitute the total set of $N_c-1$ double roots of the curve \eqref{(1.2)}. Therefore, clearly, there will be no additional massless at $\mx\ra 0$ charged solitons because there will be then too many double roots of the curve \eqref{(1.2)}.

The short additional explanations will be useful at this point. As was pointed out in Introduction, the spectral curve \eqref{(1.2)} can be used, in general, to determine the masses of BPS charged particles (and the multiplicities and charges of massless charged BPS particles in particular) in the formal limit $\mx\ra 0$ only. But in cases with equal mass quarks considered in this paper, in br2 vacua considered in this section (and in many other vacua) some additional infrared regularization is needed to really have only $N_c-1$ exactly massless at $\mx\ra 0$ charged BPS particles and corresponding them $N_c-1$ well defined {\it unequal double roots} of the $SU(N_c)$ curve \eqref{(1.2)}. We have in mind that all charged particles of the ${\cal N}=2\,\, SU(\no)$ subgroup will be massless at $\mx\ra 0$ for equal mass quarks, this corresponds to $\no$ {\it equal double roots} $e^{(Q)}_k= - m,\,\, k=1...\no$ of the curve \eqref{(1.2)}. To have really only $\no$ exactly massless at $\mx\ra 0$ charged BPS particles in this $SU(\no)$ sector, we have to split slightly the quark masses. For instance, in br2 vacua of this section it is sufficient to split slightly the masses of quarks from $SU(\no)$ with the first $\no$ flavors. I.e.\,: $m_k= m+{\delta m}_k,\,\, k=1...N_F,\,\, {\delta m}_k\neq 0,\,\,  {\delta m}_k\sim {\delta m}_n, \,\, \sum_{k=1}^{\no}{\delta m}_k=0$, for $k, n=1...\no$\,; while ${\delta m}_k=0$ for $k=\no+1\, ...\, N_F$. The mass splittings ${\delta m}_k,\,\, k=1...\no$ are arbitrary small but fixed to be unequal and nonzero. In this case, $\langle X^{\rm adj}_{SU({\rm n}_1)}\rangle$ in \eqref{(2.1.6)} will be nonzero, $\langle{\sqrt 2} X^{\rm adj}_{SU({\rm n}_1)}\rangle={\rm diag}(\,{\delta m}_{1}\,...\,{\delta m}_{\no}\,)$, and $SU(\no)\ra U^{\no-1}(1)$. The massless charged particles at $\mx\ra 0$ will be only $\no$ quarks $Q^{i=a}_a,\, {\ov Q}^{\,a}_{j=a},\, a=1...\no$ (higgsed at $0<\mx\ll {\delta m}_k$), while all other charged particles of $SU(\no)$ (including all previously massless Nambu-Goldstone particles) will acquire nonzero masses $O({\delta m})$. And the $SU(N_c)$ curve \eqref{(1.2)} will have $\no$ slightly split quark double roots $e^{(Q)}_k= - m_k,\,\, k=1...\no$, and $N_c-1$ unequal double roots on the whole. And this shows now most clearly that {\it there are no other charged BPS solitons massless at $\mx\ra 0$}. This infrared regularization will be always implied in the text below when we will speak about only $N_c-1$ double roots of the $SU(N_c)$ curve \eqref{(1.2)} at $\mx\ra 0$ (or about $N_c$ double roots of the $U(N_c)$ curve \eqref{(1.2)} in vs vacua of section~ 6).\\

Our purpose now is to calculate $\delta_2$ in order to find the leading terms of quark condensates $\langle{\rm Tr}\,{\ov Q} Q\rangle_{\no}$ and the monopole condensates $\langle M_i\rangle$ in \eqref{(2.1.6)}, and to calculate $\Lambda_{SU(\nd)}$ in \eqref{(2.1.4)}. For this, on account of leading terms $\sim\mx m^2$ and the leading power correction $\sim\langle S\rangle_{SU(\nd-\no)}$, we write, see \eqref{(1.1)},\eqref{(2.1.5)},\eqref{(2.1.4)},
\bq
\langle\w^{\,\rm low}_{\rm tot}\rangle=\langle\w^{\,(SYM)}_{SU(\nd-\no)}\rangle+\langle\w^{\,\rm low}_{\rm matter}\rangle\,,\quad \wmu=\mx(1+\delta_2)\,,\label{(2.1.7)}
\eq
\bbq
\langle\w^{\,(SYM)}_{SU(\nd-\no)}\rangle=\wmu\langle{\rm Tr\,}(X^{adj}_{SU(\nd-\no)})^2\rangle=(\nd-\no)
\langle S\rangle_{SU(\nd-\no)}=(\nd-\no)\wmu\langle\Lambda^{SU(\nd-\no)}_{{\cal N}=2\,\, SYM}\rangle^2\,,
\eeq
where $\langle\w^{\,(SYM)}_{SU(\nd-\no)}\rangle$ is the contribution of the ${\cal N}=2\,\,SU(\nd-\no)$ SYM.

First, to determine $\delta_{1,2}=O(1)$ in \eqref{(2.1.5)}, it will be sufficient to keep only the leading term $\langle\w^{\,\rm low}_{\rm matter}\rangle\sim\mx m^2$ in \eqref{(2.1.7)},\eqref{(2.1.5)} and to neglect even the leading power correction $\langle\w^{\,(SYM)}_{SU(\nd-\no)}\rangle$,
\bq
\langle\w^{\,\rm low}_{\rm tot}\rangle\approx\langle\w^{\,\rm low}_{\rm matter}\rangle=\frac{\mx}{2}(1+\delta_1)\frac{\nd N_c}{\bt}\Bigl [ \langle a_1\rangle^2=\frac{\bt^2}{\nd^{\,2}}\,m^2\Bigr ]+\frac{\mx}{2}(1+\delta_2)\frac{{\rm n}_1\nd}{\nd-{\rm n}_1}\Bigl [\langle a_2\rangle^2=\frac{N_c^{\,2}}{\nd^{\,2}}\,m^2 \Bigr ].\quad\,\,\label{(2.1.8)}
\eq

On the other hand, the exact total effective superpotential $\w^{\,\rm eff}_{\rm tot}$ which accounts explicitly for all anomalies and contains {\it only} quark bilinears $\Pi^i_j=({\ov Q}_j Q^i)_{N_c}$ looks as, see \cite{ch4,ch6} and \eqref{(1.3)},\eqref{(1.4)},
\bq
\w^{\,\rm eff}_{\rm tot}(\Pi)=m\,{\rm Tr}\,({\ov Q} Q)_{N_c}-\frac{1}{2\mx}\Biggl [ \,\sum_{i,j=1}^{N_F} ({\ov Q}_j Q^i)_{N_c}({\ov Q}_{\,i} Q^j)_{N_c}-\frac{1}{N_c}\Bigl ({\rm Tr}\,({\ov Q} Q)_{N_c}\Bigr )^2\Biggr ]-\nd S_{N_c} \,, \label{(2.1.9)}
\eq
\bbq
\langle S\rangle_{N_c}=\Bigl(\frac{\langle\det\qq\rangle=\langle\Qo\rangle^{\no}\langle\Qt
\rangle^{\nt}}{\lm^{2N_c-N_F}\mx^{N_c}}\Bigr )^{\frac{1}{N_F-N_c}}_{N_c}\,.
\eeq

Kipping only the leading terms $\sim\mx m^2$ in \eqref{(2.1.9)} and using \eqref{(2.1)} obtained from \eqref{(2.1.9)},\eqref{(1.4)},
\bq
\langle\w^{\,\rm eff}_{\rm tot}\rangle\approx\frac{1}{2}\frac{\nt N_c}{N_c-\nt}\mx m^2\approx -\,\frac{1}{2}\frac{\nt N_c}{\nd-\no}\mx m^2\,.\label{(2.1.10)}
\eq

From $\langle\w^{\,\rm low}_{\rm tot}\rangle=\langle\w^{\,\rm eff}_{\rm tot}\rangle$ in \eqref{(2.1.8)},\eqref{(2.1.10)},
\bq
-2\nd N_c=\nd\bt\delta_1+\no(N_c\delta_2-\bt\delta_1)\,.\label{(2.1.11)}
\eq
But all coefficients $\delta_{i}=O(1),\,\, i=1...4,$ in \eqref{(2.1.2)} originate from the large scale $\sim\lm$, from integrating out the heaviest fields with masses $\sim\lm$, and they do not know the number $\no$ which originates only from the behavior of $X^{\rm adj}_{SU(\nd)}$ in \eqref{(2.1.2)} at the much lower scale $\sim\langle a_2\rangle\sim m\ll\lm$, so that
\bq
N_c\delta_2=\bt\delta_1\,,\quad \delta_1=-\,\frac{2N_c}{\bt}\,,\quad \delta_2= -2\,,\quad \bt=2N_c-N_F\,.\label{(2.1.12)}
\eq
(Besides, it is shown in Appendix B that $\delta_3=0$ in \eqref{(2.1.2)},\eqref{(2.1.5)}, see \eqref{(B.4)}).\\

Therefore, we obtain from $\langle\partial\w^{\,\rm low}_{\rm matter}/\partial a_2\rangle=0$, see \eqref{(2.1.5)},\eqref{(2.1.6)},\eqref{(2.1.12)}, for the leading term of $\langle\Qo\rangle_{\no}\rangle$ in br2 vacua, compare with \eqref{(2.1)},
\bq
\langle{\rm Tr}\,({\ov Q} Q)\rangle_{\no}=\no\langle\Qo\rangle_{\no}+\nt\langle\Qt\rangle_{\no}=
\no\langle\Qo\rangle_{\no}\approx\frac{{\rm n}_1 N_c}{N_c-\nt}\,\mx m\approx\no\mx m_1\,,\label{(2.1.13)}
\eq
\bbq
\langle\Qo\rangle^{SU(N_c)}_{\no}=\sum_{a=1}^{\no}\langle{\ov Q}_1^{\,a} Q^1_a\rangle=\langle{\ov Q}_1^{\,1} \rangle\langle Q^1_1\rangle\approx\mx m_1\approx\langle\Qt\rangle^{SU(N_c)}_{N_c},\,\, \langle\Qt\rangle_{\no}=\sum_{a=1}^{\no}\langle{\ov Q}_2^{\,a} \rangle\langle Q^2_a\rangle=0.
\eeq

The whole group $SU(\no)\times U^{(2)}(1)$ is higgsed by $\no$ electric quarks (the power correction from the SYM part in \eqref{(2.1.5)} is accounted for in \eqref{(2.1.14)}, see \eqref{(A.24)})
\bq
\frac{1}{\mx}\langle{\ov Q}_k^{\,a}\rangle\langle Q^k_a\rangle\approx\delta^k_a\, m_1 \Bigl [\, 1+\frac{\bb}{\nt-N_c}\Bigl (\frac{m_1}{\lm}\Bigr )^{\frac{2N_c-N_F}{\nt-N_c}}\,\Bigr ]\,,\quad \wmu\langle S\rangle_{\no}=\langle\Qo\rangle_{\no}\langle\Qt\rangle_{\no}=0,\quad \label{(2.1.14)}
\eq
\bbq
a=1...\no\,,\quad k=1...N_F\,,\quad m_1=\frac{N_c}{N_c-\nt}\, m\,,
\eeq
and $\no^2$ long ${\cal N}=2$ multiplets of massive gluons with masses $\sim (\mx m)^{1/2}$ are formed (including $U^{(2)}(1)$ with its scalar $a_2$). It is worth to remind that at sufficiently small  $\mx$ the ${\cal N}=2$ SUSY remains unbroken at the level of particle masses $O(\sqrt{\mx})$, and only small corrections $O(\mx)$ to masses of corresponding scalar superfields break ${\cal N}=2$ to ${\cal N}=1$, see e.g. \cite{VY2}.
\footnote{\,
In comparison with the standard ${\cal N}=1$ masses $\sim (\mx m_1)^{1/2}$ of $\no^2$ gluons and $\no^2$ of their ${\cal N}=1$ (real) scalar superpartners originating from D-terms, the same masses of additional $2\no^2$ (complex) scalar superpartners of long ${\cal N}=2$ gluon multiplets originate here from F-terms in \eqref{(2.1.5)}: $\sim\sum_{a,b=1}^{\no} X^a_b \Pi^b_a$, where $\no^2$ of $X^a_b,\, \langle X^a_b\rangle=0$, include $(\no^2-1)$ of $X^{adj}_{SU(\no)}$ and one ${\hat a}_2=a_2-\langle a_2\rangle$, and $\no^2$ "pions" are $\Pi^b_a=\sum_{i=1}^{\no}({\ov Q}^{\,b}_i Q_a^i),\, \langle\Pi^b_a\rangle\approx \delta^b_a\,\mx m_1$. The Kahler terms of all $\no^2$ pions $\Pi^b_a$ look as $K_{\Pi}=2\,{\rm Tr}\,\sqrt{\Pi^\dagger \Pi}\,$.\\
}

The value of the quark condensate in \eqref{(2.1.14)} is verified by two independent calculations using
values of double roots of the curve \eqref{(1.2)} and \eqref{(2.2.10)}, see \eqref{(A.31)},\eqref{(B.14)}, and {\it these calculations with roots of the curve \eqref{(1.2)} are valid only for charged ${\cal N}=2$ BPS particles massless at $\mx\ra 0$}. This right value in \eqref{(2.1.14)} is of real importance because it is a check of the main assumption "A" formulated in Introduction, i.e. the BPS properties of original quarks in \eqref{(1.1)} at $\mx\ra 0$. Indeed, if original quarks were not BPS particles, then their mass terms in \eqref{(2.1.2)},\eqref{(2.1.5)} would receive e.g. the {\it additional contributions} $c_{a_1} a_1 {\rm Tr}(\qq)_{\nd},\,\, c_{a_1}=O(1)$ and/or $c_m m{\rm Tr}(\qq)_{\no},\,\, c_m=O(1)$, from integrated out heavier particles (or from elsewhere). The value of $\langle a_2\rangle$ in \eqref{(2.1.6)} will be then changed, and this will spoil the correct value of the quark condensate in \eqref{(2.1.14)}. (And the same for the $U(N_c)$ theory in section 2.2 below, see \eqref{(2.2.5)},\eqref{(2.2.8)},\eqref{(A.27)},\eqref{(A.29)}). On the other hand, the additional contributions $\sim\mx\delta_i$ into masses of adjoint scalars in \eqref{(2.1.2)} do not contradict the ${\cal N}=2$ BPS properties of these scalars as these additional contributions to their masses are only due to the explicit breaking ${\cal N}=2\ra {\cal N}=1$ by $\mx\neq 0$.\\

As it should be, there remain $2\no\nt$ {\it exactly massless} Nambu-Goldstone multiplets $({\ov Q}_2 Q^1)_{\no}, ({\ov Q}_1 Q^2)_{\no}$ (these are in essence the quarks ${\ov Q}$ and  $Q$ with $\nt=N_F-\no$ flavors and $SU(\no)$ colors). The factor $\nd-\no$ in the multiplicity $N_{\rm br2}=(\nd-\no)C^{\,\no}_{N_F}$ of these vacua originates from $SU(\nd-\no)$ SYM, while $C^{\,\no}_{N_F}$ is due to the spontaneous $U(N_F)$ global flavor symmetry breaking in the $SU(\no)$ sector.\\

Clearly, higgsed flavorless scalars $\langle X^{adj}_{SU(\bb)}\rangle^{(\rm several)\lm}_{\lm/(\rm several)}\sim\lm$ in \eqref{(2.1.1)} {\it operating on the scale $\mu\sim\lm$ do not break by themselves the global flavor $SU(N_F)$ symmetry}. In particular, largest $|\rm masses|\sim\lm$ of all $N_F$ quarks with $SU(\bb)$ colors are the same. At the same time, e.g. the non-Abelian color $SU(\bb)$ is broken at this scale by this higgsed $X^{adj}_{SU(\bb)}$ down to Abelian one,  and masses $\sim\lm$ of heavy flavorless $SU(\bb)$ adjoint gluons (and scalars) differ greatly. {\it There is no color-flavor locking in this $SU(\bb)$ color sector}. We would like to emphasize (see below in this section) that the global $SU(N_F)$ is really broken spontaneously {\it only} at lower energies $\sim (\mx m)^{1/2}\ll\lm$ in the $SU(\no)$ color sector due to higgsed light quarks with $N_F$ flavors and $SU(\no)$ colors, see \eqref{(2.1.13)},\eqref{(2.1.14)}.\\

Now, in short about the variant of \eqref{(2.1.5)} with the first term of $\w_D$ of the form $[\,-c_1 a_1\sum_{j=1}^{\bt}{\ov D}_j D_j\,]$, i.e. with zero $SU(N_c)$ baryon charge of dyons $D_j$. With such a form of the first term of $\w_D$, one obtains from $\langle{\partial\w_D}/{\partial{\ov D}_j}\rangle=0$ : $\langle a_{D,j}\rangle= -c_1\langle a_1\rangle,\,\, \sum_{j} \langle a_{D,j}\rangle=0=(\bb)( -c_1\langle a_1\rangle)\ra \langle a_1\rangle=0,\, \langle a_2\rangle= m$. Then, instead of \eqref{(2.1.12)},\eqref{(2.1.13)} we will obtain
\bbq
(1+\delta_2)=-\frac{N^2_c}{\nd^{\,2}}\,\, \ra\,\,  \langle\Qo\rangle^{SU(N_c)}_{\no}\approx\mx(1+\delta_2)
\frac{\nd}{\nd-\no}\langle a_2\rangle\approx -\frac{N_c^2}{\nd(\nd-\no)}\mx m\approx\frac{N_c}{\nd}\mx m_1\,,
\eeq
and this disagrees with the {\it independent} calculation of $\langle\Qo\rangle^{SU(N_c)}_{\no}\approx \mx m_1$ in this $SU(N_c)$ theory using the roots of the curve \eqref{(1.2)}, see \eqref{(A.31)},\eqref{(B.14)}.

And finally, to determine $\Lambda_{SU(\nd)}$ in \eqref{(2.1.4)}, we account now for the leading power corrections $\langle\,\delta \w^{\,\rm eff}_{\rm tot}\rangle\sim\langle S\rangle_{N_c}\sim\langle\,\delta \w_{\rm tot}^{\,\rm low}\rangle\sim\langle S\rangle_{\nd-\no}$, see \eqref{(2.1)},\eqref{(2.1.4)},\eqref{(2.1.9)},\eqref{(2.1.12)},
\bbq
\langle\,\delta \w^{\,\rm eff}_{\rm tot}\rangle=(N_c-\nt)\langle S\rangle_{N_c}\approx (N_c-\nt)\mx\, m^2_1\Bigl (\frac{m_1}{\Lambda_2}\Bigr )^{\frac{2 N_c-N_F}{\nt-N_c}}\,,\quad m_1=\frac{N_c}{N_c-\nt}\,m\,,
\eeq
\bbq
\langle\,\delta \w^{\,\rm low}_{\rm tot}\rangle=\langle\w^{\,(SYM)}_{SU(\nd-\no)}\rangle=\wmu\langle\,{\rm Tr\,} ( X^{adj}_{SU(\nd-\no)} )^2\,\rangle=(\nd-\no)\langle S\rangle_{SU(\nd-\no)}= (\nd-\no)\wmu\langle\Lambda^{SU(\nd-\no)}_{{\cal N}=2\,\, SYM}\rangle^2
\eeq
\bbq
\approx(N_c-\nt)\mx m^{2}_2\Bigl (\frac{m_2}{\Lambda_{SU(\nd)}}\Bigr )^{\frac{2N_c-N_F}{\nt-N_c}},\quad \wmu=\mx(1+\delta_2)=-\mx\,,\quad m_2=\frac{N_c}
{\nd-\no}\,m = - m_1\,,
\eeq
\bq
\langle\,\delta \w^{\,\rm eff}_{\rm tot}\rangle=\langle\,\delta \w^{\,\rm low}_{\rm tot}\rangle\quad\ra\quad \Lambda_{SU(\nd)}=-\lm\,.\label{(2.1.15)}
\eq

The whole spectrum of masses in the br2 vacua considered and, in particular, the spectrum of nonzero masses arising from \eqref{(2.1.5)} at small $\mx\ll\langle\Lambda^{SU(\nd-\no)}_{{\cal N}=2\,\, SYM}\rangle$ was described above in this section. It is in accordance with $N_c-1$ double roots of the curve \eqref{(1.2)} in these br2-vacua at $\mx\ra 0$. Of them: $\,2N_c-N_F$ unequal roots corresponding to $2N_c-N_F$ dyons $D_j$, then $\no$ equal roots corresponding to $\no$ pure electric quarks $Q^k$ of $SU(\no)$ (higgsed at $\mx\neq 0$), and finally $\nd-\no-1$ unequal roots corresponding to $\nd-\no-1$ pure magnetic monopoles $M_i$ of $SU(\nd-\no)\,\, {\cal N}=2$ SYM. Remind that two single roots $(e^{+}-e^{-})\sim\langle\Lambda^{SU(\nd-
{\rm n}_1)}_{{\cal N}=2\,\, SYM}\rangle$ of the curve \eqref{(1.2)} originate in these br2 vacua from the ${\cal N}=2\,\, SU(\nd-\no)$ SYM sector.

As it is seen from the above, {\it the only exactly massless at $m\neq 0,\, \mx\neq 0$ particles} in the Lagrangian \eqref{(2.1.5)} are $2 \no\nt$ (complex) Nambu-Goldstone multiplets originating from the spontaneous breaking of the global flavor symmetry, $U(N_F)\ra U(\no)\times U(\nt)$ in the $SU(\no)$ color sector.

The calculation of the leading power correction $\sim\mx\langle\Lambda^{SU(\nd-{\rm n}_1)}_{{\cal N}=2\,\, SYM}\rangle^2/m$ to the value of $\langle\Qo\rangle^{SU(N_c)}_{\no}$ in \eqref{(2.1.14)} is presented in Appendices A and B, and power corrections to the dyon condensates in \eqref{(2.1.6)} are calculated in Appendix B. The results agree with also presented in these Appendices {\it independent} calculations of these condensates using roots of the Seiberg-Witten curve. And these last calculations with roots are valid {\it only for BPS particles}. This agreement confirms in a non-trivial way a self-consistency of the whole approach and BPS properties of dyons and original quarks $Q^i, {\ov Q}_j$.\\

According to reasonings given above in this section, the quarks with all $N_F$ flavors, $SU(2N_c-N_F)$ colors, and masses $\sim\lm$ are too heavy and too short ranged and cannot form a coherent condensate for this reason. In other words, they are {\it not higgsed}, i.e. $\langle Q^i_a\rangle=\langle{\ov Q}^{\,a}_j\rangle=0,\,\, a=\nd+1...N_c$. Indeed, in the range from high energies down to
$\mu^{\rm low}_{\rm cut}=(\mx\lm)^{1/2}$, {\it for independently moving} heavy quarks ${\ov Q}^{\,a}_j$ and $Q_a^i$  with $SU(\bb)$ colors, all  their physical (i.e. path-dependent) phases induced by interactions with the {\it light} $\,U^{(1)}(1)\times U^{\bb-1}(1)$ photons still fluctuate freely, so that the mean values of quark fields integrated over this energy range, as well as the mean values of factorizable parts of local quarks bilinears, look as (the non-factorizable parts are also zero in this energy range, see below)
\bq
\langle{\ov Q}^{\,a}_j\rangle_{\mu^{\rm low}_{\rm cut}}=\langle Q_a^i\rangle_{\mu^{\rm low}_{\rm cut}}=\langle{\ov Q}^{\,a}_j Q_a^i\rangle^{\rm factor}_{\mu^{\rm low}_{\rm cut}}=0\,,\quad a=\nd+1...N_c\,,\quad \mu^{\rm low}_{\rm cut}=(\mx\lm)^{1/2}\,.\,\,\label{(2.1.16)}
\eq

For this reason, at the scale $\mu_{\rm cut}=\lm/(\rm several)$ where all particles with masses $\sim\lm$ {\it decoupled already as heavy, there are no any contributions to particle masses in the Lagrangian \eqref{(2.1.2)} from mean values of heavy quark fields}, $\langle{\ov Q}^{\,a}_j\rangle_{\mu_{\rm cut}}=\langle Q_a^i\rangle_{\mu_{\rm cut}}=0,\,\,\mu_{\rm cut}=\lm/(\rm several)$. And after all heavy particles decoupled at $\mu<\mu_{\rm cut}$, resulting in the Lagrangian \eqref{(2.1.2)}, {\it they themselves do not affect then the behavior of lower energy theory}. (Really, because $\nt>N_c,\, \no<\nd$ in these br2 vacua, it is clear beforehand that the quarks $Q^2, {\ov Q}_2$ in $\nt>N_c$ equal condensates $\langle\Qt\rangle_{N_c}$ are definitely not higgsed at all due to the rank restrictions, because otherwise this will result in a wrong pattern of flavor symmetry breaking. Higgsed are the light quarks $Q^i_a, {\ov Q}^{\,a}_i,\, \, i,a=1...\no$ only, see \eqref{(2.1.14)} and the end of section 2.4).\\

It is worth emphasizing also the following. The equality $\langle\,\partial \w/\partial \phi\rangle=0$ {\it is valid only when all nonperturbative contributions are accounted for in the superpotential} $\w$. If nonperturbative contributions are smaller than the perturbative ones, they can be neglected and then the equality $\langle\,\partial \w_{\rm pert}/\partial \phi\rangle\approx 0$  will be approximately right. If nonperturbative contributions are of the same order as perturbative ones, then ignoring them will give at best only the order of magnitude estimate. Remind that, by definition for any operator $O=A-B$, the equality $\langle O\rangle=0$ denotes the {\it total mean value} integrated from the appropriate high energy down to $\mu^{\rm lowest}_{\rm cut}=0$. And following from it equality of {\it total mean values} of two operators, $\langle A\rangle=\langle B\rangle$, {\it does not mean in general that the numerical values of $\langle A\rangle$ and $\langle B\rangle$ originate from the same energy regions}. The equality of {\it such total mean values} $\langle A\rangle=\langle B\rangle$ means only that, when integrated from the appropriate high energies down to zero, {\it their numerical values} will be equal. In general, {\it the numerical values of $\langle A\rangle$ and $\langle B\rangle$ can originate from different energy regions}. This is the case with e.g. \eqref{(2.1.23)}, see below. And similarly in many cases in the text below.\\

As it is seen from \eqref{(2.1)},\eqref{(2.4.1)}, the $SU(\bb)$ singlet parts of heavy quark {\it total mean values}, $\langle{\ov Q}_j Q^i\rangle_{\bb}=\sum_{a=\nd+1}^{N_c}\langle{\ov Q}^{\,a}_j Q^i_{a}\rangle\sim\,\delta^i_j\,\mx m\ll\mx\lm$ are of the same size as condensates of light higgsed quarks $\langle\Qo\rangle_{\nd}\approx\langle\Qo\rangle_{\no}=\langle{\ov Q}^1_1\rangle\langle Q^1_1\rangle\sim\mx m$ (the total mean values $\langle{\ov Q}_i Q^i\rangle_{\nd}$ of all flavors $i=1...N_F$ of quarks are holomorphic in $\mx$ and their values in \eqref{(2.4.1)},\eqref{(2.4.5)},\eqref{(2.4.6)} are valid at smaller $\mx\ll m$ as well),
\bq
\langle\Qt\rangle_{\bb}=\langle\Qt\rangle_{N_c}-\langle\Qt \rangle_{\nd}\approx\mx m_1\Biggl [1+\frac{\bb}{\nt-N_c}\Bigl (\frac{m_1}{\lm}\Bigr )^{\frac{\bb}{\nt-N_c}}\Biggr ) \Biggr ]\,,\label{(2.1.17)}
\eq
\bbq
\langle\Qo\rangle_{\bb}=\langle\Qo\rangle_{N_c}-\langle\Qo\rangle_{\nd}\approx - \mx m_1\Biggl [1+\frac{\bb}{\nt-N_c}\Bigl (\frac{m_1}{\lm}\Bigr )^{\frac{\bb}{\nt-N_c}}\Biggr ) \Biggr ]\,,
\eeq
\bbq
\langle{\rm Tr\,}\qq\rangle_{\bb}=\nt\langle\Qt\rangle_{2N_c-N_F}+\no\langle\Qo\rangle_{2N_c-N_F}\approx  (\nt-\no)\mx m_1\Biggl [1+\frac{\bb}{\nt-N_c}\Bigl (\frac{m_1}{\lm}\Bigr )^{\frac{\bb}{\nt-N_c}}\Biggr ) \Biggr].
\eeq

Besides, the total mean value of the flavor singlet part $\langle{\rm Tr\,}\qq\rangle_{\bb}$ in \eqref{(2.1.17)} can be connected with the $\langle\Sigma_D\rangle$ part of the dyon condensate, $\Sigma_D =\sum_{j=1}^{\bb}{\ov D}_j D_j$. Considering 'm' as the background field, from \eqref{(1.1)},\eqref{(2.1.2)} and using $\delta_3=0$ from \eqref{(B.4)},
\bq
\langle\,\frac{\partial}{\partial m}\w_{\rm matter}\rangle=\langle\,\frac{\partial}{\partial m}
{\widehat\w}_{\rm matter}\rangle \ra\langle{\rm Tr\,}\qq\rangle_{\bb}=\langle\Sigma_D\rangle
\equiv\sum_{j=1}^{\bb}\langle{\ov D}_j D_j\rangle=\sum_{j=1}^{\bb}\langle{\ov D}_j\rangle\langle D_j\rangle.   \quad\,\,\,\,\label{(2.1.18)}
\eq

As for the value of $\langle\Sigma_D\rangle$ in br2 vacua of the $SU(N_c)$ theory, it is obtained {\it independently} in Appendix B using \eqref{(2.2.10)} and the values of roots of the curve \eqref{(1.2)}, see \eqref{(B.7)},\eqref{(B.15)}
\bq
\langle\Sigma_D\rangle=\sum_{j=1}^{\bb}\langle{\ov D}_j\rangle\langle D_j\rangle\approx (\nt-\no)\mx m_1 \Biggl [1+\frac{\bb}{\nt-N_c}\Bigl (\frac{m_1}{\lm}\Bigr )^{\frac{\bb}{\nt-N_c}}\Biggr ) \Biggr]\,,\quad m_1=\frac{N_c}{N_c-\nt}\,,\quad \label{(2.1.19)}
\eq
this agrees with \eqref{(2.1.17)},\eqref{(2.1.18)}. \\

As an example, the {\it operator} ${\rm Tr\,}(\qq)_{\bb}$ of heavy quarks with masses $\sim\lm$ in \eqref{(2.1.17)}, when integrated over the interval $(\rm several) m < \mu<(\rm several)\lm$, transforms in general into a linear combination of three {\it operators} of light fields, see \eqref{(2.1.1)}
\bq
\hspace*{-2mm} [\,{\rm Tr\,}(\qq)_{\bb}\,]^{(\rm several)\lm}_{(\rm several) m}=\Biggl [f_1 N_c\mx (m-c_1 a_1)+f_2 \Sigma_D+ f_3 {\rm Tr\,}(\qq)_{\nd}\Bigr ]_{,}^{(\rm several) m}\,\,\,\, f_{1,2,3}=O(1).\quad \label{(2.1.20)}
\eq

This {\it operator expansion} relates then {\it the numerical total mean values} of the left and right hand sides of \eqref{(2.1.20)}. But the total mean values of three operators in the right hand side of \eqref{(2.1.20)} originate and saturate at parametrically different energies. $\langle a_1\rangle$ originates and saturates at $\mu\sim m$, as otherwise all dyons would have masses $\sim m\gg (\mx\lm)^{1/2}$. They will not be higgsed then but will decouple as heavy at $\mu<m$. $\bb$ unequal double roots of the curve \eqref{(1.2)} will be absent, while e.g. all $U^{(1)}(1)\times U^{\bb}(1)$ photons will remain massless, etc.  $\langle\Sigma_D\rangle$ originates and saturates at $\mu\sim (\mx\lm)^{1/2}\ll\langle\Lambda^{SU(\nd-\no)}_{{\cal N}=2\,\, SYM}\rangle\ll m$. And $\langle{\rm Tr\,}(\qq)_{\nd}\rangle$ originates and saturates from the sum of three regions, $\mu\sim m\ll\lm\,,\,\,  \mu\sim\langle\Lambda^{SU(\nd-\no)}_{{\cal N}=2\,\, SYM}\rangle\ll m$, and $\mu\sim (\mx m)^{1/2}\ll \langle\Lambda^{SU(\nd-\no)}_{{\cal N}=2\,\, SYM}\rangle$, see below in this section. Besides, the {\it numerical total mean value} of the right hand side in \eqref{(2.1.20)} equals $\langle\Sigma\rangle$, see \eqref{(2.1.18)}. And because $\langle m-c_1 a_1\rangle=0$ from \eqref{(2.1.1)}, we obtain: $f_2=1,\, f_3=0$.

The contributions to the heavy quark total mean values $\langle (\qq)_{1,2}\rangle_{\bb}$ also originate on the whole: a) from the quantum loop effects at the scale $\mu\sim\lm$ in the strong coupling (and nonperturbative) regime, transforming on the first "preliminary" stage such bilinear {\it operators} of heavy quark fields with masses $\sim\lm$ into the appropriate {\it operators} of light fields, i.e. $\mx a_1$ and bilinear {\it operators} of light $SU(\nd)$ quarks and dyons, see \eqref{(2.1.20)} and \eqref{(2.1.22)}-\eqref{(2.1.25)} below; \, b) finally, from formed at lower energies $\mu\ll\lm$ the genuine condensates of these light higgsed fields.

In connection with \eqref{(2.1.17)}-\eqref{(2.1.19)}, it is worth asking the following question. Because the difference between $\w_{\rm matter}$ in \eqref{(1.1)} and $\widehat\w_{\rm matter}$ in \eqref{(2.1.2)} and formation of dyons originate from the color symmetry breaking $SU(N_c)\ra SU(\nd)\times U^{\bb}(1)$ in the high energy region $\mu\sim\lm$, how e.g. the condensate $\langle\Sigma_D\rangle$ of these dyons knows about the number $\no$ which originates only at much lower energies $\sim m$ ? (Put attention that even the leading term $\sim\mx m$ of $\langle\Sigma_D\rangle$ in \eqref{(2.1.20)} is much smaller than the separate terms $\langle{\ov D}_j\rangle\langle D_j\rangle\sim\mx\lm$, but these largest terms $\sim\mx\lm$ cancel in the sum over j). And similarly, because the color breaking $\langle X^{adj}_{SU(\bb)}\rangle\sim\lm$ does not break {\it by itself} the global flavor symmetry, why the condensates of heavy quarks $\langle\Qt\rangle_{\bb}$ and $\langle\Qo\rangle_{\bb}$ are different ?

First, about $\langle\Sigma_D\rangle$. The answer is that, indeed, both {\it operators} ${\rm Tr\,}(\qq)_{2N_c-N_F}$ and $\Sigma_D$ in \eqref{(2.1.18)} are by themselves $U(N_F)_{\rm flavor}\times U(\nd)_{\rm color}$ singlets. The explicit dependence on $\no$ appears in $\langle\Sigma_D\rangle$ in \eqref{(2.1.19)} (and in left hand sides of \eqref{(2.1.17)},\eqref{(2.1.25)},\eqref{(2.1.26)}) {\it only from the region of much lower energies, after taking the corresponding total vacuum averages $\langle...\rangle$ in br2 vacua}. Recall that in these vacua. -

1) $\langle a_1\rangle\sim m$ is higgsed and operates at the scale $\sim m$. It "eats" the masses $"m"$ of all dyons in \eqref{(2.1.5)}, as otherwise all these dyons would not be higgsed but would decouple as heavy at $\mu<m$. $\bb$ unequal double roots of the curve \eqref{(2.1)} will be absent, while $\bb$ photons will remain massless.

2) $SU(\nd)$ is broken also at the scale $\sim m$ to $SU(\no)\times U^{(2)}(1)\times SU(\nd-\no)$ by higgsed $\langle X^{adj}_{SU(\nd)}\rangle\sim\langle a_2\rangle\sim m$. Together with $\langle a_1\rangle$, this $\langle a_2\rangle$ "eats" masses "m" of all $SU(\no)$ quarks, $\langle(m-a_1-a_2)\rangle=0$. Otherwise all these $SU(\no)$ quarks will decouple as heavy at $\mu<m$, there will remain $SU(\no)\,\,{\cal N}=2$ SYM at $\mu<m$, the multiplicity will be wrong and $U(N_F)$ will remain unbroken. At the same time, all quarks in the $SU(\nd-\no)$ color sector remain with masses $\sim m$.

3) To avoid $g^2(\mu<\langle\Lambda^{SU(\nd-\no)}_{{\cal N}=2\,\, SYM}\rangle )< 0$ in ${\cal N}=2\,\, SU(\nd-\no)$ SYM, $\langle X^{adj}_{SU(\nd-\no)}\rangle$ is higgsed at the scale $\sim\langle\Lambda^{SU(\nd-\no)}_{{\cal N}=2\,\, SYM}\rangle,\,\, SU(\nd-\no)\ra U^{\nd-\no-1}(1)$.

The (non-factorizable) mean values $\langle (\qq)_{1,2}\rangle_{N_c-\no}$ of quarks with masses $\sim m$ in the $SU(N_c-\no)$ SYM sector are determined by the one-loop Konishi anomaly for these massive non-higgsed quarks. I.e., on the first "preliminary" {\it operator stage} at the scale $\mu\sim m$ in the weak coupling regime, the one-loop diagrams with scalar quarks and their fermionic superpartners with masses $\sim m$ inside, transform  the quark {\it operators} $[\,(\qq)_{1,2}\,]_{N_c-\no}$ into the {\it operator} $S_{N_c-\no}/m$ of lighter $SU(N_c-\no)$ gluino (plus operators with total superderivatives). And the mean value of $\langle S\rangle_{\nd-\no}=\wmu\langle\Lambda^{SU(\nd-\no)}_{{\cal N}=2\,\,SYM}\rangle^2$, originates
only at much lower energies $\mu\lesssim\langle\Lambda^{SU(\nd-\no)}_{{\cal N}=2\,\,SYM}\rangle$, see \eqref{(2.1.7)}.

4) The global $U(N_F)$ symmetry is broken to $U(\no)\times U(\nt)$ by higgsed quarks in the $SU(\no)$ color sector at the scale $\sim (\mx m)^{1/2}\ll\langle\Lambda^{SU(\nd-\no)}_{{\cal N}=2\,\, SYM}\rangle\ll m\ll\lm$.

5) The {\it total mean value} of $\langle\Sigma_{D}\rangle\sim \mx m\ll\mx\lm$ in \eqref{(2.1.19)} originates and saturates at energies $\sim (\mx\lm)^{1/2}$. The reason is that $D_j$ and ${\ov D}_j$ {\it move independently} in the whole energy interval $[\mu^{\rm low}_{\rm cut}=(\mx\lm)^{1/2}]< \mu <\lm/(\rm several)$, and their physical (i.e. path-dependent) phases induced by interactions with the light $U^{\bb-1}(1)\times U^{(1)}(1)$ photons {\it still fluctuate freely}, so that $\langle{\ov D}_j\rangle_{\mu^{\rm low}_{\rm cut}}=\langle D_j\rangle_{\mu^{\rm low}_{\rm cut}}=\langle\Sigma_D\rangle_{\mu^{\rm low}_{\rm cut}}=0$. The nonzero mean values of all light dyon fields are formed only at the lower scale $\sim (\mx\lm)^{1/2}\ll\lm$, in the weak coupling regime, from {\it coherent condensates} of these {\it higgsed light dyons} (resulting in the appearance of $\sim (\mx\lm)^{1/2}$ masses of all $U^{\bb-1}(1)\times U^{(1)}(1)$ $\,{\cal N}=2$ multiplets): $\langle{\ov D}_j D_j\rangle =\langle{\ov D}_j\rangle\langle D_j\rangle= - \mx\lm\,\omega^{j-1}+(\bb)^{-1}\langle\Sigma_D\rangle$, see \eqref{(2.1.6)},\eqref{(B.8)},\eqref{(B.15)}, and \eqref{(2.1.19)} for $\langle\Sigma_{D}\rangle$. (Clearly, the leading terms $\langle D_j\rangle=\langle {\ov D}_j\rangle\sim (\mx\lm)^{1/2},\, j=1...\bb$ cancel in $\langle\Sigma_D\rangle$).

{\it All particles} in this dyonic sector acquire masses $\sim (\mx\lm)^{1/2}$ and decouple as heavy at lower energies. No particles at all remain in this sector at lower energies.\\

The number $\no$ penetrates then into $\langle\Sigma_D\rangle$ from the {\it numerical} relation between the {\it total vacuum averages} in br2 vacua following from \eqref{(2.1.5)},\eqref{(2.1.12)} (see also \eqref{(2.4.6)},\eqref{(A.23)},\eqref{(B.4)})
\bq
\langle\frac{\partial\w^{\,\rm low}_{\rm tot}}{\partial a_{1}}\rangle=0\quad\ra\quad \frac{\nd}{\bb}\langle\Sigma_D\rangle= \langle{\rm Tr\,}\qq\rangle_{\nd}
-\Bigl [\langle\frac{\partial\w_{\rm a_{1}}}{\partial a_{1}}\rangle= -\mx\frac{\nd N_F N_c}{(\bb)^2}\langle a_1\rangle\,\Bigr ]\,,\label{(2.1.21)}
\eq
\bbq
\langle{\rm Tr\,}\qq\rangle_{\nd}=\langle{\rm Tr\,}\qq\rangle_{\no}+\langle{\rm Tr\,}\qq\rangle_{\nd-\no}\,,\quad \langle{\rm Tr\,}\qq\rangle_{\nd-\no}=
-\langle\frac{\partial\w^{(SYM)}_{SU(\nd-\no)}}{\partial a_{1}}\rangle\,,
\eeq
\bbq
\langle\Sigma_D\rangle\approx (\nt-\no)\mx m_1 \Biggl [1+\frac{\bb}{\nt-N_c}\Bigl (\frac{m_1}{\lm}\Bigr )^{\frac{\bb}{\nt-N_c}}\Biggr ) \Biggr]\,,
\eeq
because all $N_F$ flavors of quarks from the color $SU(\nd)$ and all dyons interact with $a_1$. \eqref{(2.1.21)} agrees with \eqref{(2.1.17)}-\eqref{(2.1.19)}.
Remind that, by definition, $\langle O\rangle$ means the total mean value of any operator $O$ integrated from very high energies down to $\mu^{\rm lowest}_{\rm cut}=0$. The equation \eqref{(2.1.21)} is the example of {\it numerical} relations between total mean values where each term originates and saturates at different energies.

Now, about $\langle\Qo\rangle_{\bb}$ and $\langle\Qt\rangle_{\bb}$. The heavy quark operators in \eqref{(2.1.17)} include both the $SU(N_F)$ adjoint and singlet in flavor parts. As for the singlet part, its mean value is given in \eqref{(2.1.18)},\eqref{(2.1.19)}. As for the adjoint parts, they look as, see \eqref{(2.1.18)}
\bbq
\langle{\ov Q}_j Q^i -\delta^i_j\frac{1}{N_F}{\rm Tr\,}(\qq)\rangle_{\bb}=A\langle{\ov Q}_j Q^i -\delta^i_j\frac{1}{N_F}{\rm Tr\,}(\qq)\rangle_{\nd}\,,
\eeq
\bq
\langle\Qt-\frac{1}{N_F}{\rm Tr\,}(\qq)\rangle_{\bb}=A\langle\Qt-\frac{1}{N_F}{\rm Tr\,}(\qq)\rangle_{\nd},  \label{(2.1.22)}
\eq
\bq
\langle\Qt\rangle_{\bb}=A\langle\Qt-\frac{1}{N_F}{\rm Tr\,}(\qq)\rangle_{\nd}+
\frac{1}{N_F}\langle\Sigma_D\rangle.\label{(2.1.23)}
\eq
\bq
\langle\Qo-\frac{1}{N_F}{\rm Tr\,}(\qq)\rangle_{\bb}=A\langle\Qo-\frac{1}{N_F}{\rm Tr\,}(\qq)\rangle_{\nd}  \label{(2.1.24)}
\eq
\bq
\langle\Qo\rangle_{\bb}=A\langle\Qo-\frac{1}{N_F}{\rm Tr\,}(\qq)\rangle_{\nd}+ \frac{1}{N_F}\langle\Sigma_D\rangle\,,\label{(2.1.25)}
\eq
where $A=O(1)$ is some constant originating from integrating out the loop (and nonperturbative) effects from the Lagrangian \eqref{(1.1)} over the energy region $\mu\sim\lm$. It is seen from \eqref{(2.1.22)},\eqref{(2.1.24)},\eqref{(2.4.1)},\eqref{(2.4.6)} that nonzero contributions to the "non-genuine" (i.e. non-factorizable) adjoint in $U(N_F)$ bilinear condensates of heavy non-higgsed quarks are induced in two stages. a) On the first "preliminary" operator stage from the energy range $\lm/(\rm several)<\mu<(\rm several)\lm$, transforming the heavy quarks {\it operators} in left parts of \eqref{(2.1.22)},
\eqref{(2.1.24)} into {\it operators} of light $SU(\nd)$ quarks, this is the operator expansion. b) Finally, at the second numerical stage at lower energies $\sim (\mx m)^{1/2}$, by the "genuine" (i.e. coherent) factorizable condensate $\langle\Qo\rangle_{\no}=\langle{\ov Q}^1_1\rangle\langle Q^1_1\rangle\approx\mx m_1$ \eqref{(2.1.14)} of light higgsed quarks which break spontaneously $U(N_F)\ra U(\no)\times U(\nt)$.

Unfortunately, we cannot calculate directly the coefficient 'A' originating from the scale $\mu\sim\lm$, but we can claim that it does not know the number $\no$ appearing only at much lower scale $\mu\sim m\ll\lm$. Besides, we can find 'A' e.g. from \eqref{(2.1.23)} using \eqref{(2.4.1)},\eqref{(2.4.6)},\eqref{(2.1.19)} for the right hand part, and then predict $\langle\Qo\rangle_{\bb}$ in \eqref{(2.1.25)}. As a result,
\bq
A= -2\,,\quad \langle\Qo\rangle_{\bb}\approx - \mx m_1\Biggl [1+\frac{\bb}{\nt-N_c}\Bigl (\frac{m_1}{\lm}\Bigr )^{\frac{\bb}{\nt-N_c}} \Biggr]\,, \label{(2.1.26)}
\eq
this agrees with \eqref{(2.1.17)}, and $A=\,-2$ in independent of $\no$ as it should be.\\

On the whole, the total decomposition of quark condensates $\langle (\qq)_{1,2}\rangle_{N_c}$ over their separate color parts look in these br2 vacua as follows.\\
I) The condensate $\langle\Qo\rangle_{N_c}$.\\
a) From \eqref{(2.1.14)}, the (factorizable) condensate of higgsed quarks in the $SU(\no)$ part
\bq
\langle\Qo\rangle_{\no}=\langle{\ov Q}^1_1\rangle\langle Q^1_1\rangle\approx\mx m_1 \Bigl [\, 1+\frac{\bb}{\nt-N_c}\Bigl (\frac{m_1}{\lm}\Bigr )^{\frac{2N_c-N_F}{\nt-N_c}}\,\Bigr ]\,.\label{(2.1.27)}
\eq
b) The (non-factorizable) condensate $\langle\Qo\rangle_{\nd-\no}$ is determined by the one-loop Konishi anomaly for the heavy non-higgsed quarks with the mass $m_2$ in the $SU(\nd-\no)$ SYM sector, see \eqref{(2.1.4)},\eqref{(2.1.15)}
\bq
\langle\Qo\rangle_{\nd-\no}=\frac{\langle S\rangle_{\nd-\no}}{m_2}\approx
\frac{\wmu\langle\Lambda^{SU(\nd-\no)}_{{\cal N}=2\,\, SYM}\rangle^2}{m_2}\approx\mx m_1\Bigl (\frac{m_1}{\lm}\Bigr )^{\frac{\bb}{\nt-N_c}}\,.\label{(2.1.28)}
\eq
c) From \eqref{(2.1.17)}, the (non-factorizable) condensate of heavy non-higgsed quarks with masses $\sim\lm$
\bq
\langle\Qo\rangle_{\bb}\approx - \mx m_1\Biggl [1+\frac{\bb}{\nt-N_c}\Bigl (\frac{m_1}{\lm}\Bigr )^{\frac{\bb}{\nt-N_c}} \Biggr ]\,.\label{(2.1.29)}
\eq
Therefore, on the whole
\bq
\langle\Qo\rangle_{N_c}=\langle\Qo\rangle_{\no}+\langle\Qo\rangle_{\nd-\no}+\langle\Qo\rangle_{\bb}\approx
\mx m_1\Bigl (\frac{m_1}{\lm}\Bigr )^{\frac{\bb}{\nt-N_c}}\,,\,\,\quad \label{(2.1.30)}
\eq
as it should be, see \eqref{(2.1)}.\\
II) The condensate $\langle\Qt\rangle_{N_c}$.\\
a) From \eqref{(2.1.6)} the (factorizable) condensate of the non-higgsed massless Nambu-Goldstone particles in the $SU(\no)$ part
\bq
\langle\Qt\rangle_{\no}=\sum_{a=1}^{\no}\langle{\ov Q}^{\,a}_2\rangle\langle Q^2_{a}\rangle=0\,.\label{(2.1.31)}
\eq
b) The (non-factorizable) condensate $\langle\Qt\rangle_{\nd-\no}$ is determined by the same one-loop Konishi anomaly for the heavy non-higgsed quarks with the mass $m_2$ in the $SU(\nd-\no)$ SYM sector,
\bq
\langle\Qt\rangle_{\nd-\no}=\frac{\langle S\rangle_{\nd-\no}}{m_2}\approx\mx m_1\Bigl (\frac{m_1}{\lm}\Bigr )^{\frac{\bb}{\nt-N_c}}\,.\label{(2.1.32)}
\eq
c) From \eqref{(2.1.19)}, the (non-factorizable) condensate of heavy non-higgsed quarks with masses $\sim\lm$
\bq
\langle\Qt\rangle_{\bb}\approx \mx m_1\Biggl [1+\frac{\bb}{\nt-N_c}\Bigl (\frac{m_1}{\lm}\Bigr )^{\frac{\bb}{\nt-N_c}} \Biggr ]\,.\label{(2.1.33)}
\eq
Therefore, on the whole
\bq
\langle\Qt\rangle_{N_c}=\langle\Qt\rangle_{\no}+\langle\Qt\rangle_{\nd-\no}+\langle\Qt\rangle_{\bb}\approx
\mx m_1 \Biggl [1+\frac{N_c-\no}{\nt-N_c}\Bigl (\frac{m_1}{\lm}\Bigr )^{\frac{\bb}{\nt-N_c}} \Biggr ],\,\,\,\quad \label{(2.1.34)}
\eq
as it should be, see \eqref{(2.1)}.\\

And finally, remind that the mass spectra in these br2 vacua depend essentially on the value of $m/\lm$ and all these vacua with $\nt>N_c$ evolve at $m\gg\lm$ to br1 vacua of sections 4.1 or 4.3 below, see section 3 in \cite{ch4} or section 4 in \cite{ch5}.

\numberwithin{equation}{subsection}
\subsection{$U(N_c)$\,,\,\,smallest $\mx$}

$U(N_c)$ theory is obtained by adding one $SU(N_c)$ singlet $U^{(0)}(1)$ ${\cal N}=2$ multiplet, with its scalar field $\sqrt{2} X^{(0)}=a_0 I$, where $I$ is the unit $N_c\times N_c$ matrix. Instead of \eqref{(1.1)}, the superpotential looks now as
\bq
{\cal W}_{\rm matter}=\frac{\mu_{0}}{2} N_c a^2_0+\mx{\rm Tr}\,(X^{\rm adj}_{SU(N_c)})^2 +{\rm Tr}\,\Bigl [ (m-a_0)\,{\ov Q} Q-{\ov Q}\sqrt{2} X^{\rm adj}_{SU(N_c)} Q \Bigr ]_{N_c}\,,\quad \mu_0=\mx\,,\label{(2.2.1)}
\eq
and we consider in this paper only the case with $\mu_0=\mx$. The only change in the Konishi anomalies \eqref{(1.3)} and in the value of $\langle\Qt\rangle_{N_c}$ in \eqref{(2.1)} will be that now
\bq
\langle \Qo+\Qt\rangle_{N_c}=\mx m\,,\quad  \langle\Qt\rangle_{N_c}\approx \mx m\gg \langle\Qo\rangle_{N_c}\approx\mx m\Bigl (\frac{m}{\lm}
\Bigr)^{\frac{\bb}{{\rm n}_2-N_c}},  \label{(2.2.2)}
\eq
\bbq
\langle S\rangle_{N_c}=\frac{\langle\Qo\rangle_{N_c}\langle\Qt\rangle_{N_c}}{\mx}\approx\mx m^2\Bigl (\frac{m}{\lm}\Bigr)^{\frac{\bb}{{\rm n}_2-N_c}},
\eeq
while in br2 vacua the small ratio $\langle \Qo\rangle_{N_c}/\langle\Qt\rangle_{N_c}\ll 1$ and small $\langle S\rangle_{N_c}\ll\mx m^2$ in \eqref{(2.1)} remain parametrically the same. Besides, from \eqref{(2.2.1)},\eqref{(2.2.2)} (neglecting power corrections)
\bq
\langle a_0\rangle=\frac{\langle\,{\rm Tr}\,(\qq)\rangle_{N_c}}{N_c\mx}=\frac{\no \langle\Qo\rangle_{N_c}+\nt \langle\Qt\rangle_{N_c}}{N_c \mx}\approx\frac{\nt}{N_c}\,m \,,\quad \langle m-a_0\rangle\approx -\,\frac{\nd-\no}{N_c}\,m\,.\label{(2.2.3)}
\eq

The changes in \eqref{(2.1.2)} and \eqref{(2.1.5)} are also very simple\,: a)\, the term "$\mx N_c a^2_{0}/2$"\, is added,\,\, b) "$m$"\, is replaced by "$m-a_0$" in all other terms. So, instead of \eqref{(2.1.5)},\eqref{(2.1.6)}, we have now
\bq
\w^{\,\rm low}_{\rm tot}=\w^{\,(SYM)}_{SU(\nd-\no)}+\w^{\,\rm low}_{\rm matter}+\dots\,,\quad
\w^{\,\rm low}_{\rm matter}=\w_{SU(\no)}+\w_{D}+\w_{a}\,,
\label{(2.2.4)}
\eq
\bbq
\w^{\,(SYM)}_{SU(\nd-\no)}=(\nd-\no)\,\wmu\Bigl (\Lambda^{SU(\nd-\no)}_{{\cal N}=2\,\, SYM}\Bigr )^2+\w^{\,(M)}_{SU(\nd-\no)}\,,\quad\langle\Lambda^{SU(\nd-\no)}_{{\cal N}=2\,\,SYM}\rangle^2
\approx m^2\Bigl (\frac{m}{\lm}\Bigr )^{\frac{2N_c-N_F}{\nt-N_c}}\,,
\eeq
\bbq
\w_{SU(\no)}=(m-a_0-a_1-a_2)\,{\rm Tr}\,({\ov Q} Q)_{\no}-\,{\rm Tr}\,({\ov Q}\sqrt{2} X^{\rm adj}_{SU(\no )} Q)_{\no}+\wmu{\rm Tr}
\,(X^{\rm adj}_{SU(\no)})^2\,,
\eeq
\bbq
\w_{D}=(m-a_0-c_1 a_1)\sum_{j=1}^{\bb}{\ov D}_j D_j-\sum_{j=1}^{\bb} a_{D,j}\,{\ov D}_j D_j\,-
\,\mx\lm\sum_{j=1}^{\bb}\omega^{j-1}\,a_{D,j}+\mx L \sum_{j=1}^{\bb} a_{D,j}\,,
\eeq
\bbq
\w_{a}=\frac{\mx}{2}\Biggl [N_c a^2_0+(1+\delta_1)\frac{\nd N_c}{\bt} a_1^2+(1+\delta_2)\frac{\no \nd}{\nd-\no} a_2^2+2 N_c \delta_3 a_1(m-a_0-c_1 a_1)+2 N_c \delta_4(m-a_0-c_1 a_1)^2\Biggr ].
\eeq

From \eqref{(2.2.3)},\eqref{(2.2.4)} (the leading terms only)
\bq
\langle a_1\rangle= -\,\frac{\bt}{\nd}\langle m-a_0\rangle\approx \frac{\bt(\nd-\no)}{\nd N_c}\,m\,,\quad \langle a_2\rangle=\langle m-a_0-a_1\rangle\approx -\,\frac{\nd-\no}{\nd}\,m\,, \label{(2.2.5)}
\eq
while, according to reasonings given in section 2.1, all $\delta_i$ should remain the same. Of course, this can be checked directly. Instead of \eqref{(2.1.8)} we have now for the leading terms
\bq
\langle\w^{\,\rm low}_{\rm tot}\rangle\approx\langle\w^{\,\rm low}_{\rm matter}\rangle\approx\frac{\mx}{2}\Biggl [\,N_c\langle a_0\rangle^2+\frac{\nd N_c}{\bt}(1+\delta_1)\langle a_1\rangle^2+\frac{{\rm n}_1\nd}{\nd-{\rm n}_1}(1+\delta_2)\langle a_2\rangle^2\,\Biggr ]\,, \label{(2.2.6)}
\eq
where $\langle a_i\rangle$ are given in \eqref{(2.2.3)},\eqref{(2.2.5)}. Instead of \eqref{(2.1.9)},\eqref{(2.1.10)} we have now (with the same accuracy)
\bq
\w^{\,\rm eff}_{\rm tot}(\Pi)=m\,{\rm Tr}\,({\ov Q} Q)_{N_c}-\frac{1}{2\mx}\Biggl [ \,\sum_{i,j=1}^{N_F} ({\ov Q}_j Q^i)_{N_c}({\ov Q}_{\,i} Q^j)_{N_c}\Biggr ]-\nd S_{N_c}\,\,\ra\,\, \langle\w^{\,\rm eff}_{\rm tot}\rangle\approx\frac{1}{2}\,\nt\,\mx m^2\,.\label{(2.2.7)}
\eq
It is not difficult to check that values of $\delta_{1,2}$ obtained from \eqref{(2.2.6)}=\eqref{(2.2.7)} are the same as in ~\eqref{(2.1.12)}.~
\footnote{\, Besides, one can check that \eqref{(2.2.4)} with all $\delta_{i}=0,\,i=1...4$, will result in the wrong value: $\langle a_0\rangle\approx [\,m\,(2 N_c-\nt)/N_c\,]$, this differs from \eqref{(2.2.3)}.
\label{(f6)}
}

Therefore, instead of \eqref{(2.1.13)}, we obtain now  finally for the leading term, see \eqref{(2.2.4)},\eqref{(2.2.5)},
\bbq
\langle{\rm Tr}\,({\ov Q} Q)\rangle_{\no}=\no\langle\Qo\rangle_{\no}+\nt\langle\Qt\rangle_{\no}= \no\langle\Qo\rangle_{\no}\approx \wmu \,\frac{{\rm n}_1\nd}{\nd-\no}\langle a_2\rangle\approx\no \mx m\,,
\eeq
\bq
\langle\Qo\rangle^{U(N_c)}_{\no}=\sum_{a=1}^{\no}\langle{\ov Q}_1^{\,a}Q^1_a\rangle=\langle{\ov Q}^1_1\rangle\langle Q^1_1\rangle\approx\mx m\approx \langle\Qt\rangle^{U(N_c)}_{N_c}\,,\,\, \langle\Qt\rangle_{\no}=0\,,\,\, \langle S\rangle_{\no}=0, \label{(2.2.8)}
\eq
while as before in $SU(N_c)$
\bq
\langle{\ov D}_j\rangle\langle D_j\rangle= -\mx\lm\omega^{j-1}+O(\mx m)\,.\label{(2.2.9)}
\eq

It is worth to remind that the number of double roots of the curve \eqref{(1.2)} in this case is still $N_c-1$, as two single roots with $e^{+}=-e^{-}\approx 2\langle\Lambda^{SU(\nd-{\rm n}_1)}_{{\cal N}=2\,\, SYM}\rangle$ still originate here from $SU(\nd-\no)\,\, {\cal N}=2$ SYM. In other words, the number of charged particles massless at $\mx\ra 0$ (and higgsed at small $\mx\neq 0$) remains $N_c-1$ as in the $SU(N_c)$ theory. But now, in comparison with the $SU(N_c)$ theory, one extra $U^{(0)}(1)$ gauge multiplet is added. As a result, in addition to $2\no\nt$ massless Nambu-Goldstone multiplets, there remains now one exactly massless ${\cal N}=1$ photon multiplet, while the corresponding scalar ${\cal N}=1$ multiplet has now smallest nonzero mass $\sim\mx$ due to breaking ${\cal N}=2\ra {\cal N}=1$.

In so far as we know from \eqref{(2.1.5)},\eqref{(2.2.4)} the charges and numbers of all particles massless at $\mx\ra 0$, we can establish the definite correspondence with the roots of the curve \eqref{(1.2)}: $2N_c-N_F$ unequal double roots $e_j\sim\lm,\, j=1...2N_c-N_F$ correspond to our dyons $D_j$,\, $\nd-\no-1$ unequal double roots $e_i\sim\langle\Lambda^{SU(\nd-{\rm n}_1)}_{{\cal N}=2\,\, SYM}\rangle,\, i=1...N_c-\no-1$ correspond to pure magnetic monopoles from $SU(\nd-\no)$ SYM, and $\no$ equal double roots $e_k\sim m,\, k=1...\no$ correspond to $\no$ higgsed original pure electric quarks from $SU(\no)$. Besides, we know from \eqref{(2.2.8)},\eqref{(2.2.9)} the leading terms of their condensates (see \eqref{(2.1.6)} for the monopole condensates). Then, as a check, we can compare the values of condensates in \eqref{(2.2.8)},\eqref{(2.2.9)} with the formulas proposed in \cite{SY5} where these condensates are expressed through the roots of the $U(N_c)$ curve \eqref{(1.2)}.
\footnote{\,
Remind that these formulas were derived in \cite{SY5} from and checked then on a few simplest examples only, and proposed after this to be universal.

Besides, in general, the main problem with the use of the expressions like \eqref{(2.2.10)} from \cite{SY5} for finding the condensates of definite light BPS particles (massless at $\mx\ra 0$) is that one needs to find first the values of all roots of the curve \eqref{(1.2)}. But for this, in practice, one has to understand, at least, the main properties of the color and flavor symmetry breaking in different vacua, the corresponding mass hierarchies, and multiplicities of each type roots. \label{(f7)}
}

Specifically, these roots look in these $U(N_c)$ br2 vacua as: 1) the quark double roots $e^{(Q)}_k= - m,\,\, k=1...\no$,\,\, 2) the dyon double roots $e^{(D)}_j
\approx \omega^{j-1}\lm,\,\,j=1...(2N_c-N_F)$,\,\, 3) two single roots \cite{DS,CIV,CSW} which we know originate from $SU(\nd-\no)\,\,{\cal N}=2$ SYM,\,\, $e^{+}=-e^{-}\approx 2\langle\Lambda^{SU(\nd-{\rm n}_1)}_{{\cal N}=2\,\, SYM}\rangle$. Therefore, with this knowledge, the formulas from \cite{SY5} look as
\bq
\langle{\ov Q}_k Q^k\rangle^{U(N_c)}_{\no}= - \mx\sqrt{(e^{(Q)}_k-e^+)(e^{(Q)}_k-e^-)}\approx -\mx e^{(Q)}_k\approx\mx m\,,\quad k=1...\no, \label{(2.2.10)}
\eq
\bbq
\langle{\ov D}_j D_j\rangle= - \mx\sqrt{(e^{(D)}_j-e^+)(e^{(D)}_j-e^-)}\approx -\mx e^{(D)}_j\approx -\mx\omega^{j-1}\lm\,,\quad j=1...2N_c-N_F\,.
\eeq
( Both $e^{\pm}$ are neglected here because they are non-leading). It is seen that the leading terms of \eqref{(2.2.8)},\eqref{(2.2.9)} and \eqref{(2.2.10)} agree, this is non-trivial as the results were obtained very different methods. The calculations of power corrections to \eqref{(2.2.8)},\eqref{(2.2.9)}, and \eqref{(2.2.10)} are presented in important Appendices A and~ B.

\subsection{$SU(N_c)$\,,\,\,larger $\mx\,,\,\,\,\langle\Lambda^{SU(\nd-\no)}_{{\cal N}=2\,\, SYM}\rangle\ll\mx\ll\, m$}

Significant changes for such values of $\mx$ occurs only in the ${\cal N}=2\,\, SU(\nd-\no)$ SYM sector. All $SU(\nd-\no)$ adjoint scalars $X^{adj}_{SU(\nd-\no)}$ are {\it too heavy and too short ranged} now, their physical (i.e. path dependent) $SU(\nd-\no)$ phases induced by interactions with the lighter $SU(\nd-\no)$ gluons fluctuate freely at all scales $\mu\gtrsim\langle\Lambda^{SU(\nd-\no)}_{{\cal N}=1\,\,SYM}\rangle$ in this case, and they are not higgsed, i.e. $\langle X^{adj}_{SU(\nd-\no)}\rangle_{\mx^{\rm pole}/(\rm several)}=0$. Instead, they all decouple as heavy already at the scale $\langle\Lambda^{SU(\nd-\no)}_{{\cal N}=1\,\,SYM}\rangle\ll\mu<[\mx^{\rm pole}={\it g}^2\mx]/(\rm several)\ll m$ {\it in the weak coupling region} and do not affect then by themselves the lower energy dynamics. There remains ${\cal N}=1\,\,SU(\nd-\no)$ SYM with the scale factor of its gauge coupling $\langle\Lambda^{SU(\nd-\no)}_{{\cal N}=1\,\, SYM}\rangle=[\,\wmu\langle\Lambda^{SU(\nd-\no)}_{{\cal N}=2\,\,SYM}\rangle^2\,]^{1/3},\\\langle\Lambda^{SU(\nd-\no)}_{{\cal N}=2\,\, SYM}\rangle\ll
\langle\Lambda^{SU(\nd-\no)}_{{\cal N}=1\,\, SYM}\rangle\ll\mx\ll~ m$, see \eqref{(2.1.4)},\eqref{(2.1.15)}. The small nonzero (non-factorizable) value $\langle {\,\rm Tr\,}(X^{adj}_{SU(\nd-\no)})^2\rangle=(\nd-\no)\langle S\rangle_{\nd-\no}/\wmu=(\nd-\no)\langle\Lambda^{SU(\nd-\no)}_{{\cal N}=2\,\, SYM}\rangle^2 \ll
\langle\Lambda^{SU(\nd-\no)}_{{\cal N}=1\,\, SYM}\rangle^2$ arises here not because $X^{adj}_{SU(\nd-\no)}$ are higgsed, but only due to the Konishi anomaly. I.e., on the first "preliminary" operator stage, from one-loop diagrams with heavy scalars $X^{adj}_{SU(\nd-\no)}$ and their fermionic superpartners with masses $\sim\mx$ inside, transforming (in the weak coupling regime at the scale $\sim\mx$) this {\it operator} ${\,\rm Tr\,}(X^{adj}_{SU(\nd-\no)})^2$ of heavy scalars into the bilinear {\it operator} $\sim ({\rm Tr\,}\lambda\lambda)_{\nd-\no}/\mx$ of lighter gluinos of ${\cal N}=1\,\, SU(\nd-\no)$ SYM. But the mean vacuum value of this latter originates and saturates only in the strong coupling (and non-perturbative) region at much smaller energies $\mu\sim\langle\Lambda^{SU(\nd-\no)}_{{\cal N}=1\,\, SYM}\rangle\ll\mx$.

The multiplicity of vacua of this ${\cal N}=1 \,\, SU(\nd-\no)$ SYM is also $\nd-\no$, as it should be.. There appears now a large number of strongly coupled ${\cal N}=1$ gluonia with the mass scale $\sim\langle\Lambda^{SU(\nd-\no)}_{{\cal N}=1\,\,SYM}\rangle$. All heavier $SU(\nd-\no)$ charged original pure electric quarks and hybrids with masses $\sim m$ and scalars $X^{adj}_{SU(\nd-\no)}$ with masses $\sim\mx\gg\langle\Lambda^{SU(\nd-\no)}_{{\cal N}=1\,\,SYM}\rangle$ are still weakly confined, but the tension of the confining string is larger now, $\sigma^{1/2}_{SU(\nd-\no)}\sim\langle\Lambda^{SU(\nd-\no)}_{{\cal N}=1\,\,SYM}\rangle\ll\mx\ll m$.

At the same time, in this range $\langle\Lambda^{SU(\nd-\no)}_{{\cal N}=1\,\,SYM}\rangle\ll\mx\ll m$, nothing has happened yet with $\langle a_2\rangle\sim m$ {\it still higgsed} at the scale $\mu\sim m\gg\mx$, resulting in $SU(\nd)\ra SU(\no)\times U^{(2)}(1)\times SU(\nd-\no)$, and in the $SU(\no)$ sector with its quarks higgsed at $\mu\sim (\mx m)^{1/2}\gg\mx$.

\subsection{$SU(N_c)$\,,\,\, even larger $\mx\,,\,\,\,m\ll\mx\ll\lm$}

\hspace*{4mm} The case with $m\ll\mx\ll\lm$ and $N_c+1<N_F<3N_c/2$ is described in section 8.1 of  \cite{ch6}. Therefore, we consider here in addition the region $3N_c/2<N_F<2N_c-1$.

Remind, that the whole dyonic $\bt=2N_c-N_F$ sector, including the $U^{(1)}(1)\,\, {\cal N}=2$ photon multiplet with its scalar partner $a_1$, acquires masses $\sim (\mx\lm)^{1/2}\gg\mx$ and decouples at lower energies. But finally, the trace of the heavier scalar $a_1$ remains in the lower energy ${\cal W}_{SU(\nd)}$ superpotential at scales $\mu<(\mx\lm)^{1/2},\,\, \mx\ll(\mx\lm)^{1/2}\ll\lm$, in the form: $\,m\ra{\wt m}= (m-\langle a_1\rangle)=m N_c/\nd\,$. Remind also that, see \eqref{(2.1.2)},\eqref{(2.1.12)}, $\mx\ra\wmu=(1+\delta_2)\mx= - \mx$.

Therefore, after integrating out all particles with masses $\sim (\mx\lm)^{1/2}$, the lower energy superpotential at the scale $\mu= (\mx\lm)^{1/2}/{(\rm several)}\gg\mx$ looks as, see also \eqref{(2.4.5)},\eqref{(2.4.6)},
\bq
\w^{({\cal N}=2,\,\nd)}_{\rm matter}={\rm Tr}\,\Bigl [\,{\ov Q}\,(\tm-\sqrt{2}X^{\rm adj}_{SU(\nd)})\,Q\,
\Bigr ]_{\nd}+\wmu{\rm Tr}\,(X^{\rm adj}_{SU(\nd)})^2\,,\quad  \tm=\frac{N_c}{\nd}m\,,\quad \wmu=-\mx\,,\label{(2.4.1)}
\eq
\bbq
\langle\Qo\rangle_{\nd}= [\,\frac{\nd\,\wmu \tm}{\nd-\no}=\wmu m_2=\mx m_1\,]+\frac{N_c-\no}{\nd-\no}\langle\Qt\rangle_{\nd}\approx\mx m_1\approx\langle\Qt\rangle_{N_c}\,,
\eeq
\bbq
\langle\Qt\rangle_{\nd}\approx\mx m_1\Bigl (\frac{m_1}{\lm}\Bigr)^{\frac{\bb}{{\rm n}_2-N_c}}\approx\langle\Qo\rangle_{N_c}\,,\quad m_2= - m_1=\frac{N_c}{\nt-N_c}\,m\,,
\eeq
\bbq
\langle{\rm Tr}(\sqrt{2}X^{\rm adj}_{SU(\nd)})^2\rangle=\frac{1}{\wmu}\Bigl [(2\nd-N_F)\langle S\rangle_{\nd}+\tm\langle{\rm Tr}({\ov Q} Q)\rangle_{\nd}\Bigr ]\approx \frac{\no\nd}{\nd-\no}\tm^2\,.
\eeq

All $\nd^{\,2}-1\,\,X^{\rm adj}_{SU(\nd)}$ {\it are not higgsed and decouple as heavy} at scales $\mu<\mx^{\rm pole}/(\rm several)={\it g}^2(\mu=\mx^{\rm pole})\mx/(\rm several)$ {\it in the weak coupling region}, and can be integrated out. Therefore, the ${\cal N}=1\,\, SU(\nd)$ SQCD with $N_F$ flavors of light quarks with masses $\ll\mx$ emerges already at the scale $\mu=\mx^{\rm pole}/(\rm several)$. The scale factor $\lt$ of its gauge coupling is, see \eqref{(2.1.15)},
\bq
(\,\lt\,)^{3\nd-N_F}=\Bigl [\Lambda_{SU(\nd)}= -\lm\Bigr ]^{2\nd-N_F}\,\wmu^{\,\nd}\,,\quad  \frac{\lt}{\wmu}=\Bigl (\frac{\mx}{\lm}\Bigr )^{\frac{2N_c-N_F}{2N_F-3N_c}}\ll 1 \,.\label{(2.4.2)}
\eq

This ${\cal N}=1$ theory is not IR free at $3N_c/2<N_F<2N_c-1$ and so its logarithmically small coupling $g^2(\mu\sim\mx)$ begins to grow logarithmically at $\mu<\mx$. There is a number of variants of the mass spectrum. To deal with them it will be convenient to introduce colorless but flavored auxiliary (i.e. sufficiently heavy, with masses $\sim\mx$, and dynamically irrelevant at $\mu<\mx$) fields $\Phi_i^j$ \cite{ch5}, so that the Lagrangian at the scale $\mu=\mx^{\rm pole}/(\rm several)$ instead of
\bq
K={\rm Tr}\,(Q^\dagger Q+{\ov Q}^\dagger {\,\ov Q}),\,\, \w^{\,({\cal N}=1,\,\nd)}_{\rm matter}=\tm\, {\rm Tr}\,({\ov Q} Q)_{\nd}-\frac{1}{2\wmu}\Biggl ( {\rm Tr}\,({\ov Q} Q)_{\nd}^2-\frac{1}{\nd}\Bigl ({\rm Tr}({\ov Q} Q)_{\nd}\Bigr )^2  \Biggr ) \label{(2.4.3)}
\eq
can be rewritten as
\footnote{\,
all factors of the ${\cal N}=1$ RG evolution in Kahler terms are ignored below for simplicity if they are logarithmic only, as well as factors of $g$ if $g$ is either constant or logarithmically small \label{(f8)}
}
\bq
K={\rm Tr}\,(\Phi^\dagger \Phi)+{\rm Tr}\,(Q^\dagger Q+{\ov Q}^\dagger {\,\ov Q})\,,\quad
\w^{\,({\cal N}=1,\,\nd)}_{\rm matter}=\w_{\Phi}+{\rm Tr}\,\Bigl ({\ov Q}\,\tm^{\rm tot} Q \Bigr )\,,\label{(2.4.4)}
\eq
\bbq
\w_{\Phi}=\frac{\wmu}{2}\Bigl ( {\rm Tr}\,(\Phi^2)-\frac{1}{N_c}({\rm Tr}\,\Phi)^2 \Bigr )\,,\quad {\tm}
=\frac{N_c}{\nd}\,m\,,\quad (\tm^{\rm tot})^j_i=\tm\,\delta^j_i -\Phi^j_i\,,\quad \wmu= - \mx\,,
\eeq
and {\it its further evolution at lower energies is determined now by the dynamics of the genuinely ${\cal N}=1$ theory \eqref{(2.4.4)}}.

The Konishi anomalies for \eqref{(2.4.4)} look as, see \eqref{(2.4.2)}, compare with \eqref{(1.3)},\eqref{(1.4)},
\bq
\langle\Qo+\Qt-\frac{1}{\nd}{\rm Tr}\,\qq\rangle_{\nd}=\wmu \tm\,,\label{(2.4.5)}
\eq
\bbq
\langle S\rangle_{\nd}=\frac{\langle\Qo\rangle_{\nd}\langle\Qt\rangle_{\nd}}{\wmu}=\Biggl(\frac{\langle
\det\qq\rangle_{\nd}}{\lt^{3\nd-N_F}}\Biggr)^{1/N_c}=\Biggl(\,\frac{[\langle\Qo\rangle_
{\,\nd}]^{\no}[\langle\Qt\rangle_{\,\nd}]^{\nt}}{\Lambda_{SU(\nd)}^{2\nd-N_F}\,\wmu^{\,\nd}}\,\Biggr )^{1/N_c}\,,
\eeq
\bbq
\langle\tm^{\rm tot}_1=\tm-\Phi^1_1\rangle=\frac{\langle\Qt\rangle_{\nd}}{\wmu}\,,\quad \langle\tm^{\rm tot}_2=\tm-\Phi^2_2\rangle=\frac{\langle\Qo\rangle_{\nd}}{\wmu}\,.
\eeq
\bbq
\langle\Phi^i_j\rangle=\frac{1}{\wmu}\Biggl [\langle{\ov Q}_{j} Q^i\rangle_{\nd}-\delta^i_j\,\frac{1}{\nd}\langle{\rm Tr\,}\qq\rangle_{\nd} \,\Biggr ]\,.
\eeq

From \eqref{(2.4.5)}, see \eqref{(2.4.1)},\eqref{(2.1.15)} and section 8.1.1 in \cite{ch6} (neglecting all power corrections for simplicity),
\bq
\langle\Qo\rangle_{\nd}= [\,\frac{\nd\,\wmu \tm}{\nd-\no}=\wmu m_2=\mx m_1\,]+\frac{N_c-\no}{\nd-\no}
\langle\Qt\rangle_{\nd}\approx\mx m_1\approx\langle\Qt\rangle_{N_c}\,,\label{(2.4.6)}
\eq
\bbq
\langle\Qt\rangle_{\nd}\approx\wmu m_2\Bigl (\frac{m_2}{\Lambda_{SU(\nd)}}\Bigr)^{\frac{\bb}{\nt-N_c}}
\approx\mx m_1\Bigl (\frac{m_1}{\lm}\Bigr)^{\frac{\bb}{{\rm n}_2-N_c}}\approx\langle\Qo\rangle_{N_c}\,,
\quad \Lambda_{SU(\nd)}= -\lm\,,
\eeq
\bbq
\langle S\rangle_{\nd}=\langle S\rangle_{\,SU(\nd-\no)}^{\,(SYM)}= -\langle S\rangle_{N_c}\,,\quad
m_2=\frac{N_c}{\nt-N_c}\,m= -m_1\,.
\eeq

Besides, it is not difficult to check that accounting for both the leading terms $\sim \mx m^2$ and the leading power corrections $\sim \langle S\rangle_{N_c}= -\langle S\rangle_{\nd}$, see
\eqref{(2.4.3)}-\eqref{(2.4.6)},\, \eqref{(2.1)},\eqref{(2.1.9)},\eqref{(2.1.12)}
\bq
\Bigl [\langle\w^{\,({\cal N}=1,\,\nd)}_{\rm tot}\rangle=\langle\w^{\,({\cal N}=1,\,\nd)}_{\rm matter}\rangle-N_c\langle S\rangle_{\nd}\Bigr ]+\Bigl [\langle\w_{a_1}\rangle=\frac{\mx}{2}(1+\delta_1)\frac{\nd N_c}{\bt}\langle a_1\rangle^2\Bigr ]=\langle \w^{\,\rm eff}_{\rm tot}\rangle\,,\label{(2.4.7)}
\eq
as it should be.\\

{\bf A)} If $m$ is not too small so that $(\mx m)^{1/2}\gg\lt$, see \eqref{(2.4.2)}. Then $\no<\nd$ quarks are higgsed still {\it in the weak coupling regime} at the scale $\mu\sim(\mx m)^{1/2}\gg\lt,\,\, SU(\nd)\ra SU(\nd-\no)$, and $\no(2\nd-\no)$ gluons acquire masses $\mu_{\rm gl,1}\sim (\mx m)^{1/2}$. There remains at lower energies ${\cal N}=1$ SQCD with the unbroken $SU(\nd-\no)$ gauge group, $\nt>N_c$ flavors of still active lighter quarks $Q^2,\,{\ov Q}_2$ with $SU(\nd-\no)$ colors, $\no^2$ pions $\Pi^1_{1^\prime}\ra \sum_{a=1}^{\nd}({\ov Q}^{\,a}_{i^\prime} Q^i_a),\,\,i^\prime,i=1,\,...,\,\no,\,\, \langle\Pi^1_1\rangle=\langle \Qo\rangle_{\nd}\sim \mx m$, and $2\no\nt$ hybrids $\Pi^2_1,\,\Pi^1_2$ (these are in essence the quarks $Q^2,\,{\ov Q}_2$ with broken colors). The scale factor $\widehat\Lambda$ of the gauge coupling is, see \eqref{(2.4.2)},
\bq
{\wh\Lambda}^{\,\wh{\rm b}_{\rm o}}=\frac{\lt^{3\nd-N_F}}{\det \Pi^1_1}\,,\quad
\Bigl (\frac{{\langle\widehat\Lambda}\rangle^{2}}{\mx m}\Bigr )^{\wh{\rm b}_{\rm o}}\sim\Bigl (\frac{\lt^{2}}{\mx m} \Bigr )^{2N_F-3N_c}\ll 1\,.\,\quad \widehat{\rm b}_{\rm o}= 3(\nd-\no)-\nt\,.\label{(2.4.8)}
\eq

{\bf a1)} If $\wh{\rm b}_{\rm o}<0$. The $SU(\nd-\no)$ theory at $\mu<\mu_{\rm gl,1}$ is then IR free, its coupling is small and still decreases logarithmically with diminishing energy. The quarks $Q^2,\,{\ov Q}_2$ with $\nd-\no$ still active colors and $\nt>N_c$ flavors have masses $m^{\rm pole}_{Q,2}\sim m$ and decouple as heavy at $\mu<\,m$. Notice that {\it all this occurs in the weak coupling regime} in this case, so that {\it there is no need to use the assumed dynamical scenario from} \cite{ch3}.

There remains at lower energies ${\cal N}=1\,\, SU(\nd-\no)$ SYM with the scale factor of its coupling
\bq
\Bigl (\Lambda^{SU(\nd-\no)}_{{\cal N}=1\,\,SYM}\Bigr )^{3(\nd-\no)} = \frac{{\lt}^{\,3\nd-N_F}\,\det\tm^{\rm tot}_2}{\det \Pi^1_1},\quad (\tm^{\rm tot}_2)^{2^\prime}_2=\tm\,\delta^{2^\prime}_2-\Phi^{2^\prime}_2\,, \,\, \frac{\langle\Lambda^{SU(\nd-\no)}_{{\cal N}=1\,\,SYM}\rangle}{m}\ll 1.\label{(2.4.9)}
\eq

After integrating this ${\cal N}=1$ SYM via the VY-procedure \cite{VY}, the Lagrangian looks as, see \eqref{(2.1.15)},\eqref{(2.4.9)},\\ \eqref{(2.4.2)},
\bbq
K={\rm Tr}\,(\Phi^\dagger\Phi)+  {\rm Tr}\,\Biggl [2\sqrt{(\Pi^1_1)^{\dagger}\Pi^1_1}+\Pi_2^1\frac{1}
{\sqrt{(\Pi^1_1)^{\dagger}\Pi^1_1}}\,(\Pi_2^1)^{\dagger}+(\Pi_1^2)^{\dagger}\frac{1}{\sqrt{
(\Pi^1_1)^{\dagger}\Pi^1_1}}\,\Pi_1^2 \Biggr ]\,,
\eeq
\bq
\w^{\,(\nd)}=\w_{\rm non-pert}+\w_{\Phi}+{\rm Tr}\,\Bigl (\tm^{\rm tot}_1\,\Pi^1_1\Bigr )+\w_{\rm hybr}\,, \label{(2.4.10)}
\eq
\bbq
\w_{\rm non-pert}=(\nd-\no)\Bigl (\Lambda^{SU(\nd-\no)}_{{\cal N}=1\,\,SYM}\Bigr )^3= (\nd-\no)\Biggl [\frac{(\Lambda_{SU(\nd)}=-\lm)^{2\nd-N_F}\wmu^{\,\nd}\,\det\tm^{\rm tot}_2}{\det \Pi^1_1}\Biggr ]^{1/(\nd-\no)}\,,
\eeq
\bbq
\w_{\rm hybr}={\rm Tr}\,\Bigl (\tm^{\rm tot}_2\,\Pi_2^1\frac{1}{\Pi^1_1}\Pi_1^2-\Phi_1^2\Pi_2^1-
\Phi_2^1\Pi_1^2\Bigr )\,.
\eeq

The masses of $\no^{\,2}$ pions $\Pi^1_1$ and hybrids from \eqref{(2.4.10)} are, see \eqref{(2.4.4)} for $\w_{\Phi}$,
\bq
\mu(\Pi^1_1)\sim\frac{\langle\Pi^1_1\rangle=\langle\Qo\rangle_{\nd}}{\mx}\sim m\sim m^{\rm pole}_{Q,2}\,,\quad\langle\Pi^1_{1^\prime}\rangle=\langle({\ov Q}_{1^\prime} Q^1)\rangle_{\nd}\approx - \delta^1_{1^\prime} \frac{N_c}{\nd-\no}\,\mx m\,,\label{(2.4.11)}
\eq
\bbq
\mu(\Pi^1_2)=\mu(\Pi^2_1)=0\,,
\eeq
(the main contribution to $\mu(\Pi^1_1)$ gives the term $\sim (\Pi^1_1)^2/\mx$ originating from  ${\rm Tr}\, (\tm^{\rm tot}_1\,\Pi^1_1 )$ in \eqref{(2.4.10)} after integrating out heavier $\Phi^1_1$ with masses $\sim\mx\gg m$ ).\\

On the whole, the mass spectrum of this ${\cal N}=1\,\,SU(\nd)$ theory look in this case as follows.\\
1) There are $\no(2\nd-\no)\,\,{\cal N}=1$ multiplets of massive gluons, $\mu_{\rm gl,1}\sim (\mx m)^{1/2}$.\\
2) A large number of hadrons made from non-relativistic and weakly confined ${\ov Q}_2, Q^2$ quarks with $\nt>N_c$ flavors and $\nd-\no$ colors, their mass scale is $\mu_H\sim m,\,\, \Lambda^{SU(\nd-\no)}_{{\cal N}=1\,\,SYM}\ll \mu_H\ll\mu_{\rm gl,1}$ (the tension of the confining string is $\sigma^{1/2}\sim\Lambda^{SU(\nd-\no)}_{{\cal N}=1\,\,SYM}\ll m$).\\
3) $\no^2$ pions $\Pi^1_1$ with masses $\sim m$.\\
4) A large number of ${\cal N}=1\,\,SU(\nd-\no)$ SYM strongly coupled gluonia with the mass scale $\sim\Lambda^{SU(\nd-\no)}_{{\cal N}=1\,\,SYM}$.\\
5) $2\no\nt$ massless (complex) Nambu-Goldstone ${\cal N}=1$ multiplets $\Pi^1_2,\,\Pi^2_1$.\\
6) All $N_F^2$ fields $\Phi$ have masses $\mu(\Phi)\sim\mx\gg m$. They are dynamically irrelevant and not observable as real particles at scales $\mu<\mx$.\\
Comparing with the mass spectrum at $N_c+1<N_F<3N_c/2$ in "$\bf A$" of section 8.1.1 in \cite{ch6} it is seen that it is the same (up to different logarithmic factors).

The multiplicity of these vacua is $N_{\rm br2}=(\nd-\no)C^{\,\no}_{N_F}$, as it should be. The factor $\nd-\no$ originates from ${\cal N}=1\,\,SU(\nd-\no)$ SYM and the factor $C^{\,\no}_{N_F}$ - from spontaneous flavor symmetry breaking $U(N_F)\ra U(\no)\times U(\nt)$ due to higgsing of $\no$ quarks $\,Q^1, {\ov Q}_1$.\\

{\bf a2)} If $\wh{\rm b}_{\rm o}>0$. Then, because $\wh\Lambda\ll m$, the quarks $Q^2,\,{\ov Q}_2$ decouple as heavy at $\mu\sim m$, still in the weak coupling region. As a result, the mass spectrum will be the same as in "{\bf a1}" above (up to different logarithmic factors).\\

{\bf B)} If $(\mx m)^{1/2}\ll\lt$. Then ${\cal N}=1$ SQCD with $SU(\nd)$ colors and $3N_c/2<N_F<2N_c-1$ quark flavors enters first at $\mu<\lt$ the strongly coupled conformal regime with the frozen gauge coupling $a_*=\nd g^2_*/2\pi=O(1)$.

{\it And only now we use for the first time the dynamical scenario proposed in} \cite{ch3} {\it to calculate the mass spectrum}. Really, what is only assumed in this scenario in the case of the standard ${\cal N}=1$ SQCD conformal regime is the following.

In this ${\cal N}=1$ theory without colored adjoint scalars, unlike the very special ${\cal N}=2$ theory with its additional colored scalar fields $X^{\rm adj}$ and enhanced supersymmetry, {\it no parametrically lighter solitons} (in addition to the ordinary mass spectrum described below) {\it are formed at those scales where the standard ${\cal N}=1$ conformal regime is broken explicitly by nonzero particle masses} (see also the footnote \ref{(f3)}).

The potentially competing masses look then as
\bbq
{\wt m}_{Q,2}^{\rm pole}\sim\frac{m}{z_Q(\lt,m_{Q,2}^{\rm pole})}\sim\lt\Bigl (\frac{m}{\lt}\Bigr )^{N_F/3\nd},\,\, z_Q(\lt,\mu\ll\lt)=\Bigl (\frac{\mu}{|\lt|}\Bigr )^{\gamma_Q}\ll 1,\,\, 0<\gamma_Q=\frac{3\nd-N_F}{N_F}<\frac{1}{2}\,,
\eeq
\bq
\mu^2_{\rm gl,1}\sim z_Q(\lt,\mu_{\rm gl,1})\Bigl [\langle\Qo\rangle_{\nd}\sim\mx m\Bigr ]\,,\quad \mu_{\rm gl,1}\sim\lt\Bigl (\frac{\mx m}{\lt^2}\Bigr )^{N_F/3N_c}\ll\lt\,, \label{(2.4.12)}
\eq
\bbq
m\ll\mu_{\rm gl,1}\ll\lt\ll\mx\ll\lm\,,\quad \frac{{\wt m}_{Q,2}^{\rm pole}}{\mu_{\rm gl,1}}\ll 1\,.
\eeq
Therefore, the quarks $Q^i, {\ov Q}_j,\,\, i,j=1...\no,$ are higgsed at $\mu=\mu_{\rm gl,1}$, the flavor symmetry is broken spontaneously, $U(N_F)\ra U(\no)\times U(\nt)$, and there remains at lower energy ${\cal N}=1$ SQCD with $SU(\nd-\no)$ colors and $\nt$ flavors of still active quarks $Q^2, {\ov Q}_2\,$ with $SU(\nd-\no)$ colors.

{\bf b1)} If $\wh{\rm b}_{\rm o}<0$. Then this theory is IR free and the gauge coupling becomes logarithmically small at $\Lambda^{SU(\nd-\no)}_{{\cal N}=1\,\,SYM}\ll\mu\ll\mu_{\rm gl,1}$. The RG evolution at $\Lambda^{SU(\nd-\no)}_{{\cal N}=1\,\,SYM}<\mu<\mu_{\rm gl,1}$ is logarithmic only (and neglected for simplicity). The pole mass of $Q^2, {\ov Q}_2$ quarks with still active unbroken $\nd-\no$ colors looks really as
\bq
m_{Q,2}^{\rm pole}\sim \frac{m}{z_Q(\lt,\mu_{\rm gl,1})}\sim\lm\Bigl (\frac{\mx}{\lm}\Bigr )^{1/3}\Bigl (\frac{m}{\lm} \Bigr )^{\frac{2(3N_c-N_F)}{3N_c}}\ll\mu_{\rm gl,1}\,.\label{(2.4.13)}
\eq
They decouple then as heavy at $\mu<m_{Q,2}^{\rm pole}$ {\it in the weak coupling region} and there remains ${\cal N}=1\,\, SU(\nd-\no)$ SYM with $\Lambda^{SU(\nd-\no)}_{{\cal N}=1\,\,SYM}\ll m_{Q,2}^{\rm pole}$, see \eqref{(2.4.9)}. Integrating it via the VY procedure \cite{VY}, the lower energy Kahler terms looks now as \cite{ch5},
\bbq
K=z_{\Phi}(\lt,\mu_{\rm gl,1}){\rm Tr} (\Phi^\dagger\Phi )+z_Q(\lt,\mu_{\rm gl,1})
{\rm Tr}\,\Biggl [\,2\,\sqrt{(\Pi^1_1)^{\dagger}\Pi^1_1}+\Pi_2^1\frac{1}
{\sqrt{(\Pi^1_1)^{\dagger}\Pi^1_1}}\,(\Pi_2^1)^{\dagger}+(\Pi_1^2)^{\dagger}\frac{1}{\sqrt{
(\Pi^1_1)^{\dagger}\Pi^1_1}}\,\Pi_1^2 \Biggr ]\,,
\eeq
\bq
z_{\Phi}(\lt,\mu_{\rm gl,1})= \Bigl (\frac{\mu_{\rm gl,1}}{|\lt|}\Bigr )^{\gamma_{\Phi}}\,,\quad
-1<\gamma_{\Phi}=-2\gamma_{Q}<0\,,\quad z_{\Phi}(\lt,\mu_{\rm gl,1})=1/z^2_{Q}(\lt,\mu_{\rm gl,1})\gg 1\,,\label{(2.4.14)}
\eq
while the superpotential is as in \eqref{(2.4.10)}. Therefore, in comparison with \eqref{(2.4.11)}, only the pion masses have changed and are now
\bq
\mu(\Pi^1_1)\sim\frac{\langle\Pi^1_1\rangle=\langle\Qo\rangle_{\nd}}{\mx}\frac{1}{z_Q(\lt,\mu_{\rm gl,1})}\sim\lm\Bigl (\frac{\mx}{\lm}\Bigr )^{1/3}\Bigl (\frac{m}{\lm} \Bigr )^{\frac{2(3N_c-N_F)}{3N_c}}\sim m_{Q,2}^{\rm pole}\,, \label{(2.4.15)}
\eq
while, because $0<\gamma_Q<1/2$, all $N_F^2$ fions $\Phi$ remain too heavy, dynamically irrelevant and not observable as real particles at $\mu<\mx^{\rm pole}$.\\

{\bf b2)} If $\wh{\rm b}_{\rm o}>0$. Then the ${\cal N}=1\,\, SU(\nd-\no)$ theory remains in the conformal window at $\mu<\mu_{\rm gl,1}$ with the frozen gauge coupling $a^{*}=O(1)$.

The anomalous dimensions at $\mu<\mu_{\rm gl,1}$ look now as, see \eqref{(2.4.8)},
\bq
0<{\wh\gamma}_Q=\frac{{\rm {\wh b}}_{\rm o}}{\nt}<\gamma_Q<\frac{1}{2}\,,\quad {\wh\gamma}_{\Phi}=-2{\wh\gamma}_Q\,,\quad  {\wh z}_Q(\mu_{\rm gl,1},\mu\ll\mu_{\rm gl,1})=\Bigl (\frac{\mu}{\mu_{\rm gl,1}}\Bigr )^{{\wh\gamma}_Q}\ll 1\,.\label{(2.4.16)}
\eq
The pole mass of $Q^2, {\ov Q}_2$ quarks with unbroken colors looks now as, see \eqref{(2.4.9)},
\eqref{(2.4.12)},
\bq
{\wh m}_{Q,2}^{\rm pole}\sim\frac{m}{z_Q(\lt,\mu_{\rm gl,1})}\frac{1}{{\wh z}_Q(\mu_{\rm gl,1},m_{Q,2}^{\rm pole})}\sim\lm\Bigl (\frac{\mx}{\lm}\Bigr )^{1/3}\Bigl (\frac{m}{\lm}\Bigr )^{\frac{\nt-\no}{3(\nd-\no)}}\sim (\rm several)\,\Lambda^{SU(\nd-\no)}_{{\cal N}=1\,\,SYM}\,,\label{(2.4.17)}
\eq
while the masses of $\no^2$ pions $\Pi^1_1$ remain the same as in \eqref{(2.4.15)} and all $N_F^2\,\, \Phi^j
_i$ also remain dynamically irrelevant. The overall hierarchies of nonzero masses look in this case as
\bq
\mu(\Pi^1_1)\ll\Lambda^{SU(\nd-\no)}_{{\cal N}=1\,\,SYM}\sim {\wh m}_{Q,2}^{\rm pole}\ll\mu_{\rm gl,1}\ll\lt\ll\mx\ll\lm\,.\label{(2.4.18)}
\eq

On the whole for this section 2.4 with $m\ll\mx\ll\lm$. -

In all cases considered the overall phase $\rm\mathbf{Higgs_1-HQ_2}$ (HQ=heavy quark) remains the same in the non-Abelian $SU(\nd)$ sector. At scales $\mx\ll\mu\ll\lm$ it behaves as the effectively massless IR free ${\cal N}=2$ theory. At the scale $\mu\sim \mx^{\rm pole}=g^2\mx$ all $X^{adj}_{SU(\nd)}$ decouple as heavy and its dynamics becomes those of the ${\cal N}=1$ theory. $2\no^2$ quarks $Q^1, {\ov Q}_1$ with $SU(\no)$ colors and $\no$ flavors are higgsed while $2\nt (\nd-\no)$ quarks $Q^2, {\ov Q}_2$ with $SU(\nd-\no)$ colors and $\nt$ flavors are in the HQ phase and confined by the ${\cal N}=1\, SU(\nd-\no)$ SYM. $2\no\nt$ massless Nambu-Goldstone particles in all cases are in essence the hybrid quarks with $\nt$ flavors and broken $SU(\no)$ colors. And the overall qualitative picture is also the same in all variants considered, see e.g. the text after \eqref{(2.4.11)}. Changes only the character of the ${\cal N}=1$ RG evolution, i.e. it is logarithmic or power-like, and this influences the values of nonzero masses and mass hierarchies.\\

Finally, we think it will be useful to make the following short additional comments.\\
In this section, the largest masses $\mx^{\rm pole}=g^2(\mu=\mx^{\rm pole})\mx\gg (\mx m)^{1/2}\gg m$ in the whole $SU(\nd)$ sector have  $\nd^{\,2}-1$ adjoint scalars $X^{\rm adj}_{SU(\nd)}$. They all are {\it too heavy and too short ranged} now, and in the whole energy region $\mu > (\rm several) (\mx m)^{1/2}$ at least
their physical (i.e. path dependent) $SU(\nd)$ phases induced by interactions with the {\it lighter} (and effectively massless at such energies) $SU(\nd)$ gluons fluctuate freely. Therefore, the mean value of $X^{\rm adj}_{SU(\nd)}$, integrated not only over the interval from the high energy down to $\mx^{\rm pole}/({\rm several})$ as in the Lagrangian \eqref{(2.4.3)},\eqref{(2.4.4)}, but even down to much lower energies $\mu_{\rm cut}^{\rm low}= (\rm several) (\mx m)^{1/2}\ll\mx$ is zero,
\bbq
\langle\sqrt{2} X^{\rm adj}_{SU(\nd)}\rangle_{\mu^{\rm low}_{\rm cut}}=2\sum_{i=1}^{N_F}\sum_{A=1}^{\nd^{\,2}-1}T^{A}\langle{\rm Tr\,}{\ov Q}_i
T^{A} Q^i\rangle_{\mu^{\rm low}_{\rm cut}}/\wmu=
\eeq
\bq
= 2\sum_{i=1}^{\no}\sum_{A=1}^{\nd^{\,2}-1}T^{A} {\rm Tr}\Biggl [\langle{\ov Q}_i\rangle_{\mu^{\rm low}_{\rm cut}} T^{A} \langle Q^i\rangle_{\mu^{\rm low}_{\rm cut}}\Biggr ]/\wmu=0\,,\label{(2.4.19)}
\eq
because all light quark fields $Q^i_a$ and ${\ov Q}^{\,a}_i,\, a=1...\nd,\, i=1...N_F$ (and all light $SU(\nd)$ gluons) {\it still fluctuate independently and freely} in this higher energy region.

But its formal {\it total mean value} $\langle X^{\rm adj}_{SU(\nd)}\rangle$, integrated by definition down to $\mu^{\rm lowest}_{\rm cut}=0$, contains the small but nonzero, $\sim m\ll (\mx m)^{1/2}$, contribution originating only in the lower energy region $\mu\sim (\mx m)^{1/2}\ll\mx$ from the classical coupling of heavy $X^{\rm adj}_{SU(\nd)}$ with {\it lighter higgsed quarks} ${\ov Q}^{\,a}_i, Q^i_a,\, a,i=1...\no$, see e.g. \eqref{(2.4.1)} or \eqref{(2.1.14)},

\bbq
\langle\sqrt{2}X^{\rm adj}_{SU(\nd)}\rangle=\langle\sqrt{2}\sum_{A=1}^{\nd^{\,2}-1} X^{A} T^{A}\rangle\approx 2\sum_{i=1}^{N_F}\sum_{A=1}^{\nd^{\,2}-1}T^{A}\langle{\rm Tr\,}{\ov Q}_i T^{A} Q^i\rangle/\wmu \approx
\eeq
\bq
\approx 2\sum_{i=1}^{\no}\sum_{A=1}^{\nd^{\,2}-1}T^{A}{\rm Tr\,}[\,\langle{\ov Q}\rangle_i T^{A} \langle Q^i\rangle\,]/\wmu\approx m_2\,{\rm diag}\,(\,\underbrace{\,1-\frac{\no}{\nd}}_{\no}\,;\, \underbrace{\,-\frac{\no}{\nd}}_{\nd-\no}\,)\,.\label{(2.4.20)}
\eq
This leads to the main contribution $\sim m^2$ to ${\rm Tr\,}[\langle (\sqrt{2} X^{\rm adj}_{SU(\nd)})^2\rangle ]$, which in the case considered looks as
${\rm Tr\,}[\langle (\sqrt{2} X^{\rm adj}_{SU(\nd)})^2\rangle ]\approx{\rm Tr\,}[\langle \sqrt{2} X^{\rm adj}_{SU(\nd)}\rangle\langle \sqrt{2} X^{\rm adj}_{SU(\nd)}\rangle]$, see \eqref{(2.4.20)},
\bq
{\rm Tr\,} [\langle (\sqrt{2} X^{\rm adj}_{SU(\nd)})^2\rangle ]\approx \Bigl [\frac{\no (\nd-\no)}{\nd}\, m_2^2=\frac{\no\nd}{\nd-\no}\,\tm^2\,\Bigr ], \,\, \tm=\frac{N_c}{\nd} m,\,\, m_2=\frac{N_c}{\nd-\no} m\,, \,\,\,\,\label{(2.4.21)}
\eq
this agrees with the exact relation in the last line of \eqref{(2.4.1)} following from the Konishi anomaly (up to small nonperturbative power corrections $\sim\langle S\rangle_{\nd}/\mx\sim m^2_1  (m_1/\lm)^{(\bb)/(\nt-N_c)}$ originating finally at even lower energies from the SYM part), and explains the origin of a smooth holomorphic behavior of $\langle{\rm Tr\,}(\sqrt{2}X^{\rm adj}_{SU(\nd)})^2\rangle$ when $\mx$ is increased from $\mx\ll m$ in section 2.1 to $\mx\gg m$ in this section.

But, in the case considered, this small formal nonzero total mean value of $\langle X^{\rm adj}_{SU(\nd)}
\rangle\sim m\ll (\mx m)^{1/2}$ does not mean that these heavy adjoint scalars are {\it higgsed}, in the sense that they give by themselves {\it the additional contributions} $\sim m$ {\it to particle masses} (those of hybrid gluons and $SU(\nd)$ quarks in this case), as it really occurs at $\mx\ll m$ in section 2.1. (Finally, the physical reason that the fields $X^{\rm adj}_{SU(\nd)}$ do not give by themselves originating from the scale $\mu\sim\mx$ contributions to particle masses is that they are too heavy now and so too short ranged and cannot form the coherent condensate).

Clearly, {\it there are no any additional contributions to particle masses in the Lagrangian \eqref{(2.4.3)},\eqref{(2.4.4)} at the scale $\mu=\mu_{\rm x}^{\rm pole}/({\rm several})$ from $\langle X^{\rm adj}_{SU(\nd)}\rangle_{\mu_{\rm x}^{\rm pole}/(\rm several)}=0$}, see \eqref{(2.4.19)}, because all $\nd^{\,2}-1$ fields $X^{\rm adj}_{SU(\nd)}$ with masses $\mx^{\rm pole}$ {\it decoupled already as heavy} at $\mu=\mx^{\rm pole}/(\rm several)$ where \eqref{(2.4.19)} is definitely valid, and they {\it do not affect by themselves the dynamics of the lower energy ${\cal N}=1$ theory at $\mu < \mx^{\rm pole}/(\rm several)$}. This is the reason for a qualitative difference in patterns of color symmetry breaking and mass spectra at $\mx\ll m$ in section 2.1 and at $\mx\gg m$ in this section. Remind that the color breaking (and corresponding mass splitting) occurs at the scale $\mu\sim m\gg (\mx m)^{1/2}\gg\mx$ and the unbroken gauge symmetry looks as $SU(\nd)\ra SU(\no)\times U^{(2)}(1)\times SU(\nd-\no)$ in section 2.1, while it occurs at the scale $m\ll\mu\sim (\mx m)^{1/2}\ll\mx$ and the unbroken gauge symmetry looks as $SU(\nd)\ra SU(\nd-\no)$ in this section.

As it is seen from \eqref{(2.4.3)}-\eqref{(2.4.5)} and the whole content of this section, the whole contributions $\sim (\mx m)^{1/2}$ to the masses of $SU(\no)$ and hybrid gluons (see the footnote \ref{(f8)}) originate only at lower energies $\mu\sim (\mx m)^{1/2}\ll\mx$ in D-terms of the genuine ${\cal N}=1$ SQCD \eqref{(2.4.3)},\eqref{(2.4.4)} from higgsed quarks only. And the whole additional F-term contributions $\sim m$ to the quark masses originate really from the quark self-interaction term $\sim{\rm Tr\,}(\qq)^2/\mx$ with higgsed quarks $\langle\Qo\rangle_{\nd}=\langle{\ov Q}^1_1\rangle\langle Q^1_1\rangle$ in \eqref{(2.4.3)} or, what is the same, from $\langle\Phi_{1,2}\rangle\sim m$ in \eqref{(2.4.4)}, this last also originating finally from these higgsed quarks, forming the coherent condensate and giving masses to corresponding gluons, see the last line of \eqref{(2.4.5)}. And the crucial difference between the section 2.1 with higgsed light $\langle X^{adj}_{SU(\nd)}\rangle\sim\langle a_2\rangle\sim m$ at $\mx\ll m$ in\eqref{(2.1.3)},
\eqref{(2.1.6)} contributing to quark masses, and replacing it $\langle\Phi^j_i\rangle\sim m$ with the mass $\sim\mx\gg m$ in this section, see \eqref{(2.4.4)},\eqref{(2.4.5)}, is that $ X^{adj}_{SU(\nd)}$ acts in the color space, while $\Phi^j_i$ acts in the flavor space.\\

\section{Unbroken flavor symmetry, \,\,S-vacua,\,\,$SU(N_c)$}
\numberwithin{equation}{section}

As in the br2 vacua in section 2.1, the non-trivial discrete  $Z_{\bb\geq 2}$ symmetry is also unbroken in these vacua, i.e. they also belong to the baryonic branch in the language \cite{APS}, this case corresponds to $\no=0$. From \eqref{(1.3)},\eqref{(2.1.9)}, the quark condensates $\langle\qq\rangle_{N_c}$ in these vacua  with the multiplicity $\nd$ look as (neglecting smaller power corrections), see section 3 in \cite{ch4} or section 4 in \cite{ch5},
\bq
\langle\qq\rangle_{N_c}\approx -\,\frac{N_c}{\nd}\,\mx m = -\mx\tm\,,\quad \langle S\rangle_{N_c}=\Bigl ( \frac{\det\langle\qq\rangle_{N_c}}{\lm^{\bb}\mx^{N_c}}\Bigr )^{1/\nd}\approx\mx \tm^2\Bigl (\frac{-\tm}{\lm}\Bigr )^{(\bb)/\nd}\ll\mx m^2,\,\,\, \label{(3.1)}
\eq
\bbq
\langle{\rm Tr\,}(\sqrt{2}\,X^{adj}_{SU(N_c)})^2\rangle=\Bigl [(2N_c-N_F)\langle S\rangle_{N_c}+m\langle{\rm Tr\,}\qq\rangle_{N_c}\Bigr ]\approx m\,(-N_F\mx\tm)\,,\quad \tm=\frac{N_c}{\nd} m\,.
\eeq

As before, the scalar $X^{\rm adj}_{SU(N_c)}$ higgses the $SU(N_c)$ group at the scale $\mu\sim\lm$ as $SU(N_c)\ra SU(\nd)\times U^{(1)}(1)\times U^{\bt-1}(1)$, see \eqref{(2.1.1)},\eqref{(2.1.16)},
\bq
\langle X \rangle=\langle\,X^{adj}_{SU(\nd)}+ X^{(1)}_{U(1)}+ X^{adj}_{\bt}\,\rangle\,,\quad \nd=N_F-N_c\,,\quad \bt=2N_c-N_F\,,\label{(3.2)}
\eq
\bbq
\langle\sqrt{2}\,X^{adj}_{\bt}\,\rangle=C_{\bt}\lm\,{\rm diag}(\,\underbrace{\,0}_{\nd}\,; \underbrace
{\,\omega^0,\,\,\omega^1,\,...\,,\,\omega^{\bt-1}}_{\bt}\,)\,,\quad \omega=\exp\{\frac{2\pi i}{\,\bt}\,\}\,,
\eeq
\bbq
\sqrt{2}\,X^{(1)}_{U(1)}=a_1\,{\rm diag}(\,\underbrace{\,1}_{\nd}\,;\, \underbrace{\,c_1}_{\bt}\,),\quad c_1=-\,\frac{\nd}{\bt}\,,\quad \langle a_1\rangle=\frac{1}{c_1}m=\,-\frac{\bt}{\nd} m\,.
\eeq

I.e., as in br2 vacua, {\it the same} $\bt=2N_c-N_F\geq 2$ dyons $D_j$, massless in the limit $\mx\ra 0$ (this number is required by the unbroken $Z_{2N_c-N_F\geq 2}$ discrete symmetry) are formed at the scale $\mu\sim\lm$ and at sufficiently small $\mx$ the superpotential $\wh\w$ at $m\ll\mu\ll\lm$ is as in \eqref{(2.1.2)}.

The difference is in the behavior of the $SU(\nd)$ part. In these vacua with $\no=0$, there is no analog of the additional gauge symmetry breaking $SU(\nd)\ra SU(\no)\times U^{(2)}(1)\times SU(\nd-\no)$ in br2 vacua at the scale $\mu\sim m$.

{\bf A)}\,\, At $\mx\ll\Lambda^{SU(\nd)}_{{\cal N}=2\,\,SYM}\ll m$ and $N_c<N_F<2N_c-1$, {\it all quarks} in the $SU(\nd)$ part in these S vacua have now masses $\tm=\langle m-a_1\rangle=m N_c/\nd$ and decouple as heavy {\it in the weak coupling regime} at scales $\mu\,<\, \tm/(\rm several)$, there remains ${\cal N}=2\,\,\,SU(\nd)$ SYM with the scale factor $\Lambda^{SU(\nd)}_{{\cal N}=2\,\, SYM}$ of its gauge coupling, see \eqref{(2.1.15)},
\bq
\langle\Lambda^{SU(\nd)}_{{\cal N}=2\,\, SYM}\rangle^{2\nd}=\Bigl (\Lambda_{SU(\nd)}= -\lm\Bigr )^{2\nd-N_F}\,\tm^{N_F}\,,\quad \tm=m\,\frac{N_c}{\nd}\,,\quad \langle\Lambda^{SU(\nd)}_{{\cal N}=2\,\, SYM}\rangle\ll \,m\,.\label{(3.3)}
\eq

And the field $X^{adj}_{SU(\nd)}$ higgses this UV free ${\cal N}=2\,\,\, SU(\nd)$ SYM at the scale $\sim\langle\Lambda^{SU(\nd)}_{{\cal N}=2\,\, SYM}\rangle$ in a standard way, $SU(\nd)\ra U^{\nd-1}(1)$ \cite{DS}. This results in the right multiplicity $N_s=\nd=N_F-N_c$ of these S vacua. The two single roots with $(e^{+}-e^{-})\sim \Lambda^{SU(\nd)}_{{\cal N}=2\,\,SYM}$ of the curve \eqref{(1.2)} originate here from this ${\cal N}=2\,\,\, SU(\nd)$ SYM. Other $\nd-1$ unequal double roots originating from this $SU(\nd)$ sector correspond to massless at $\mx\ra 0\,\,\nd-1$ pure magnetic monopoles $M_{\rm n}$ with the $SU(\nd)$ adjoint charges.

As a result of all described above, the low energy superpotential at $\mx\ll (\langle\Lambda^{SU(\nd)}_{{\cal N}=2\,\, SYM}\rangle)^2/\lm$ and at the scale
$\mu<\langle\Lambda^{SU(\nd)}_{{\cal N}=2\,\, SYM}\rangle$
can be written in these S-vacua as (the coefficients $f_n=O(1)$ in \eqref{(3.4)} are known from \cite{DS})
\bq
\w^{\,\rm low}_{\rm tot}=\w^{\,(SYM)}_{SU(\nd)}+\w^{\,\rm low}_{\rm matter}+\dots\,,\quad
\w^{\,\rm low}_{\rm matter}=\w_{D}+\w_{a_1}\,,\label{(3.4)}
\eq
\bbq
\w^{\,(SYM)}_{SU(\nd)}=\nd \mx(1+\delta_2)\Bigl (\Lambda^{SU(\nd)}_{{\cal N}=2\,\, SYM}\Bigr )^2+\w^{\,(M)}_{SU(\nd)}\,,
\eeq
\bbq
\w^{\,(M)}_{SU(\nd)}= - \sum_{n=1}^{\nd-1} {\tilde a}_{M,\rm n}\Biggl [\, {\ov M}_{\rm n} M_{\rm n}+\mx(1+\delta_2)\Lambda^{SU(\nd)}_{{\cal N}=2\,\,SYM}\Biggl (1+O\Bigl (\frac{\langle\Lambda^{SU(\nd)}_{{\cal N}=2\,\,SYM}\rangle}{m}\Bigr )\Biggr )\,f_{\rm n}\,\Biggr ]\,,
\eeq
\bbq
\w_{D}=(m-c_1 a_1)\sum_{j=1}^{\bb}{\ov D}_j D_j-\sum_{j=1}^{\bb} a_{D,j}\,{\ov D}_j D_j\,-
\,\mx\lm\sum_{j=1}^{\bb}\omega^{j-1}\,a_{D,j}+\mx \, L_S \sum_{j=1}^{\bb} a_{D,j}\,,
\eeq
\bbq
\w_{a_1}=\frac{\mx}{2}(1+\delta_1)\frac{\nd N_c}{\bt} a_1^2+\mx N_c \delta_3 a_1(m-c_1 a_1)+\mx N_c\delta_4 (m-c_1 a_1)^2\,,
\eeq
where $\delta_{1,2}$ are given in \eqref{(2.1.12)}, while $\delta_3=0$, see \eqref{(B.4)}.

From \eqref{(3.4)}:
\bq
\langle a_1\rangle=\frac{1}{c_1} m= - \frac{2N_c-N_F}{\nd} m\,,\quad \langle a_{D,j}\rangle=\langle {\tilde a}_{M,\rm n}\rangle=0\,,\label{(3.5)}
\eq
\bbq
\langle{\ov M}_{\rm n} M_{\rm n}\rangle=\langle{\ov M}_{\rm n}\rangle\langle M_{\rm n}\rangle\approx \mx\langle\Lambda^{SU(\nd}_{{\cal N}=2\,\,SYM}\rangle f_n\,,\quad (2N_c-N_F)\mx\langle L_S\rangle=\langle\Sigma_D\rangle =\sum_{i=1}^{N_F}\langle{\ov Q}_i Q^i\rangle_{N_c}\,,
\eeq
\bbq
\langle{\ov D}_j D_j\rangle=\langle{\ov D}_j\rangle\langle D_j\rangle\approx \Bigl [\,-\mx\lm\omega^{j-1}- \frac{N_c N_F}{\nd (\bb)}\mx m\,\Bigr ]\,,\quad \Sigma_D=\sum_{j=1}^{\bb}\langle{\ov D}_j\rangle\langle D_j\rangle\,.
\eeq

And the qualitative situation with the condensates of $N_F$ flavors of heavy non-higgsed quarks with $SU(2N_c-N_F)$ colors and masses $\sim\lm$ in these S-vacua with $\no=0$ is the same as those in br2 vacua described at the end of section 2.1, see \eqref{(2.1.16)}-\eqref{(2.1.26)}. The mean vacuum values of all lighter non-higgsed quarks with $N_F$ flavors, $\nd$ colors and masses $\tm$ in these S-vacua are power suppressed, see \eqref{(2.4.1)} with $\no=0,\, \nt=N_F$ or  (8.2.2) in \cite {ch6}, their small nonzero values originate in this case only from the Konishi anomaly for quarks in the $SU(\nd)$ sector, see \eqref{(3.3)}, $\,\langle\qq\rangle_{\nd}=\langle S\rangle_{\nd}/{\tm}= - \mx\langle\Lambda^{SU(\nd)}_{{\cal N}=2\,\, SYM}\rangle^{2}/{\tm}\sim\mx m\Bigl (m/\lm\Bigr)^{(\bb)/\nd}\ll\mx m\,$.

Besides, from \eqref{(3.1)},\eqref{(3.5)} (the leading terms only for simplicity, compare with \eqref{(2.1.18)}\,)
\bq
\langle\Sigma_D\rangle=\langle{\rm Tr\,}\qq\rangle_{N_c}\approx\langle{\rm Tr\,}\qq\rangle_{\bb}\approx - \frac{N_c N_F}{\nd}\mx m\,. \label{(3.6)}
\eq

On the whole for the mass spectrum in these $\nd$ S-vacua with the unbroken flavor symmetry at $\mx\ll(\langle\Lambda^{SU(\nd)}_{{\cal N}=2\,\,SYM}\rangle)^2/
\lm$. - \\
1) All original electric quarks $Q, {\ov Q}$ are not higgsed but confined. The masses of all original electric particles charged under $SU(\bb)$ are the largest ones, $\sim\lm\,$, and they all are weakly confined because all $\bb$ dyons $D_j$ are higgsed, $\langle{\ov D}_j\rangle=\langle D_j\rangle\sim (\mx\lm)^{1/2}$, the string tension in this sector is $\sigma^{1/2}_{SU(\bb)}\sim \langle D_j\rangle\sim (\mx\lm)^{1/2}\ll\lm\,$. These confined particles form a large number of hadrons with masses $\sim\lm$.\\
2) The next mass scale is $\tm$, these are masses of original electric quarks with $SU(\nd)$ colors and $N_F$ flavors, they are also weakly confined due to higgsing of $\nd-1$ pure magnetic monopoles $M_{\rm n}$ of ${\cal N}=2\,\,SU(\nd)$\,\,SYM, the string tension here is much smaller, $\sigma^{1/2}_{SU(\nd)}\sim \langle M_{\rm n}\rangle_{SU(\nd)}\sim (\mx\langle\Lambda^{SU(\nd)}_{{\cal N}=2\,\,SYM}\rangle)^{1/2}
\ll (\mx\lm)^{1/2}$. They form hadrons with masses $\sim m$.\\
3) The next mass scale is $\langle\Lambda^{SU(\nd)}_{{\cal N}=2\,\,SYM}\rangle\ll m$, see \eqref{(3.3)}, these are masses of charged $SU(\nd)$ gluons and scalars. They are also weakly confined due to higgsing of $\nd-1$ magnetic monopoles $M_{\rm n}$, the tension of the confining string is the same, $\sigma^{1/2}_{SU
(\nd)}\sim \langle M_{\rm n}\rangle_{SU(\nd)}\sim (\mx\langle\Lambda^{SU(\nd)}_{{\cal N}=2\,\,SYM}\rangle)^{1/2}$. But the mass scale of these hadrons is $\sim\langle\Lambda^{SU(\nd)}_{{\cal N}=2\,\,SYM}\rangle\,$.\\
4) The other mass scale is $\sim (\mx\lm)^{1/2}$ due to higgsing of $\bt$ dyons $D_j$. As a result, $\bt=2N_c-N_F$ long ${\cal N}=2$ multiplets of massive photons with masses $\sim (\mx\lm)^{1/2}$ are formed (including $U^{(1)}(1)$ with its scalar $a_1$).\\
5) The lightest are $\nd-1$ long ${\cal N}=2$ multiplets of massive dual photons with masses $\sim (\mx\langle\Lambda^{SU(\nd)}_{{\cal N}=2\,\,SYM}\rangle)^{1/2}$.\\
6) Clearly, there are no massless Nambu-Goldstone particles because the global flavor symmetry remains unbroken. And there are no massless particles at all at $\mx\neq 0,\,\, m\neq 0$.

The corresponding scalar multiplets have additional small contributions $\sim\mx$ to their masses, these small corrections break ${\cal N}=2$ down to ${\cal N}=1$.\\

{\bf B)}\,\, At larger $\langle\Lambda^{SU(\nd)}_{{\cal N}=2\,\,SYM}\rangle\ll\mx\ll\, m$, the difference is that all $SU(\nd)$ adjoint scalars $X^{adj}_{SU(\nd)}$ are now {\it too heavy}, their physical $SU(\nd)$ phases induced by interactions with light $SU(\nd)$ gluons fluctuate freely at all scales $\mu\gtrsim\langle\Lambda^{SU(\nd)}_{{\cal N}=1\,\,SYM}\rangle$, and they are not higgsed. Instead, they decouple as heavy much before, already at the scale $\mu<[\,\mx^{\,\rm pole}=|g^2\mx|\,]/{\rm several)}\ll m$, and there remains ${\cal N}=1 \,\, SU(\nd)$ SYM with the scale factor $\langle\Lambda^{SU(\nd)}_{{\cal
N}=1\,\,SYM}\rangle=[\,\wmu \langle\Lambda^{SU(\nd)}_{{\cal N}=2\,\,SYM}\rangle^2\,]^{1/3}$ of its gauge coupling, $\,\,\langle\Lambda^{SU(\nd)}_{{\cal N}=2\,\,SYM}\rangle\ll\langle\Lambda^{SU(\nd)}_{{\cal N}=1\,\,SYM}\rangle\ll\mx\ll\, m$. As in section 2.3, the small nonzero value $\langle {\rm Tr\,} (X^{adj}_{SU(\nd)})^2\rangle=\nd\langle S\rangle_{\nd}/\wmu=\nd\langle\Lambda^{SU(\nd)}_{{\cal N}=2\,\, SYM}\rangle^2\ll\langle\Lambda^{SU(\nd)}_{{\cal N}=1\,\, SYM}\rangle^2$, see \eqref{(3.3)}, arises here only due to the Konishi anomaly, i.e. from one-loop Feynman diagrams with heavy scalars $X^{adj}_{SU(\nd)}$ and their fermionic superpartners with masses $\sim\mx$ inside, not because $X^{adj}_{SU(\nd)}$ are higgsed.

The multiplicity of vacua of this ${\cal N}=1 \,\, SU(\nd)$ SYM is also $\nd$ as it should be. There appears now a large number of strongly coupled gluonia with the mass scale $\sim\langle\Lambda^{SU(\nd)}_{{\cal N}=1\,\,SYM}\rangle$. All $SU(\nd)$ electrically  charged particles are still confined, the tension of the confining string is larger now, $\sigma^{1/2}_{SU(\nd)}\sim\langle\Lambda^{SU(\nd)}_{{\cal N}=1\,\,SYM}\rangle\ll\mx\ll\, m$.

The case with $m\ll\mx\ll\lm$ is described in section 8.2 of  \cite{ch6}.

\addcontentsline{toc}{section}
{\hspace*{4cm} \bf {Part II.\,\,Large quark masses}, $\Large\mathbf{m\gg\lm}$}
\vspace*{3mm}

\begin{center}{\hspace*{1cm}\bf\Large {Part II.\,\,Large quark masses}, $\Large\mathbf{m\gg\lm}$} \end{center}

\section{Broken flavor symmetry}
\numberwithin{equation}{section}

\hspace*{4mm} We present below in this section the mass spectra at $m\gg\lm$ of the direct (electric) $SU(N_c)$ theory \eqref{(1.1)} in vacua  with spontaneously broken flavor symmetry, $U(N_F)\ra U(\no)\times U(\nt)$,\, $N_c+1<N_F<2N_c-1$. There are the br1-vacua (br=breaking) with $1\leq\no< N_F/2\,,\,\, N_F=\no+\nt\,,$ in which $\langle\Qo\rangle_{N_c}\gg\langle\Qt\rangle_{N_c}$, see section 3 in \cite{ch4} and \eqref{(1.3)},\eqref{(1.4)}. The quark and gluino condensates look in these br1 vacua of $SU(N_c)$ as (the leading terms only, see the footnote \ref{(f4)})
\bq
\langle\Qo\rangle_{N_c}\approx\mx m_3\,,\quad \langle\Qt\rangle_{N_c}\approx\mx m_3\Bigl (\frac{\lm}{m_3}\Bigr )^{\frac{2N_c-N_F}{N_c-\no}} \,,\quad m_3=\frac{N_c}{N_c-\no}\, m\,, \label{(4.1)}
\eq
\bbq
\langle S\rangle_{N_c}=\frac{\langle\Qo\rangle_{N_c}\langle\Qt\rangle_{N_c}}{\mx}\approx\mx m_3^2\Bigl (\frac{\lm}{m_3}\Bigr )^{\frac{2N_c-N_F}{N_c-\no}}\,,\quad\frac{\langle\Qt\rangle_{N_c}}
{\langle\Qo\rangle_{N_c}}\approx\Bigl (\frac{\lm}{m_3}\Bigr)^{\frac{2N_c-N_F}{N_c-\no}}\ll 1\,,
\eeq
while $\no\leftrightarrow\nt$ in \eqref{(4.1)} in br2-vacua, but with $N_F/2<\nt<N_c$ in this case. The multiplicity of these br1 and br2 vacua is correspondingly $N_{\rm br1}=(N_c-\no)C_{N_F}^{\,\no}$ and $N_{\rm br2}=(N_c-\nt)C_{N_F}^{\,\nt},\,\, C_{N_F}^{\,\no}=C_{N_F}^{\,\nt}=[\,N_{F} !/\no!\,\nt!\,],\,\,\, 1\leq \no< N_F/2,\,\, \nt> N_F/2$. It is seen from \eqref{(4.1)} that the discrete $Z_{\bb}$ symmetry is unbroken at $m\gg\lm$ in all these br1 and br2 vacua with $\no\neq\nd$. In contrast, $Z_{\bb}$ symmetry is (formally) broken spontaneously in special vacua with $\no=\nd,\,\,\nt=N_c$ and the multiplicity $N_{spec}=(\bb)  C_{N_F}^{\,\nd}$ (as it is really broken spontaneously in these special vacua at $m\ll\lm$).

Out of them, see section 3 in \cite{ch4} and/or section 4 in ~\cite{ch5}, $(\nd-\no)C_{N_F}^{\,\no}$ part of br1-vacua with $1\leq\no <\nd$ evolves at $m\ll\lm$ into the br2-vacua of section 2.1 with $\langle\Qt\rangle^{(\rm br2)}_{N_c}\sim\mx m\gg\langle\Qo\rangle^{(\rm br2)}_{N_c}$, while the other part of br1-vacua and br2-vacua evolve at $m\ll\lm$ into $Lt$ vacua with $\langle\Qo\rangle^{(\rm Lt)}_{N_c}\sim\langle\Qt\rangle^{(\rm Lt)}_{N_c}\sim\mx\lm,\, ,\, \langle S\rangle_{Lt}\sim\mx\lm^2$.

\subsection{$SU(N_c),$  br1 vacua, smaller $\mx\,,\,\,\,\mx\ll\Lambda^{SU(N_{c}-\no)}_{{\cal N}=2\,\,SYM}$}
\numberwithin{equation}{subsection}

In the case considered, $X^{\rm adj}_{SU(N_c)}$ breaks the $SU(N_c)$ group {\it in the weak coupling regime} at the largest scale $\mu\sim m\gg\lm$ as\,: $\,SU(N_c)\ra SU(\no)\times U(1)\times SU(N_c-\no)$, see \eqref{(4.1.4)} below,
\bq
\langle X^{\rm adj}_{SU(N_c)}\rangle= \langle\, X^{adj}_{SU(\no)}+ X_{U(1)}+ X^{adj}_{SU(N_c-\no)}\,\rangle\,,\label{(4.1.1)}
\eq
\bbq
\sqrt{2}\, X_{U(1)}=a \,{\rm diag}\Bigl(\underbrace{\,1}_{\no}\,;\,\underbrace{\,c}_{N_c-\no}\, \Bigr )\,,\quad c=-\frac{\no}{N_c-\no}\,,\quad a\equiv \langle a\rangle+{\hat a}\,,\quad \langle a \rangle=m\,,
\quad \langle{\hat a}\rangle\equiv 0\,.
\eeq
As a result, all quarks $Q^i_a,\,{\ov Q}_j^{\,a}$ with flavors $i, j=1...N_F$ and colors $a=(\no+1)...N_c$ have large masses $m_3=m-c\langle a\rangle=m N_c/(N_c-\no)$, the hybrid gluons and hybrid $X$ also have masses $m_3\gg\lm$, and they all decouple at scales $\mu\lesssim\, m_3$. The lower energy theory at $\mu<\, m_3$ consists of ${\cal N}=2\,\, SU(N_c-\no)$\,\, SYM, then ${\cal N}=2\,\, SU(\no)$ SQCD with $N_F$ flavors of massless at $\mx\ra 0$ quarks $Q^{i}_{\rm b},\, {\ov Q}^{\,\rm b}_j$ with $\no$ colors, ${\rm b}=1\,...\,\no$, and finally one ${\cal N}=2\,\, U(1)$ photon multiplet with its scalar superpartner $X_{U(1)}$. The scale factor of the ${\cal N}=2\,\,\, SU(N_{c}-\no)$ SYM gauge coupling is $\langle\Lambda^{SU(N_{c}-{\rm n}_1)}_{{\cal N}=2\,\,SYM}\rangle\ll m$. The field $X^{\rm adj}_{SU(N_c-\no)}$ in \eqref{(4.1.1)}) breaks this ${\cal N}=2\,\, SU(N_{c}-\no)$ SYM at the scale $\mu\sim\langle\Lambda^{SU(N_{c}-{\rm n}_1)}_{{\cal N}=2\,\,SYM}\rangle$ in a standard way, $SU(N_c-\no)\ra U(1)^{N_c-\no-1}$ \cite{DS}, see also \eqref{(4.1)},
\bq
\langle\sqrt{2}\, X^{adj}_{SU(N_c-\no)}\rangle\sim\langle\Lambda^{SU(N_{c}-\no)}_{{\cal N}=2\,\,SYM}\rangle
\,{\rm diag}\Bigl (\,\underbrace{\,0}_{\no}\,; k_1,k_2,\,...\,, k_{N_c-\no}\Bigr )\,,\quad k_i=O(1)\,,\label{(4.1.2)}
\eq
\bbq
\langle\Lambda^{SU(N_{c}-\no)}_{{\cal N}=2\,\,SYM}\rangle^2\approx\Bigl (\frac{\lm^{\bb}{(m_3)}^{N_F}}{{(m_3)}^{2\no}}\Bigr )^{\frac{1}{N_c-\no}}\approx
m^2_3\Bigl (\frac{\lm}{m_3}\Bigr )^{\frac{2N_c-N_F}{N_c-\no}}\ll m^2\,,
\eeq
\bbq
\langle S\rangle_{N_c-\no}=\mx\langle\Lambda^{SU(N_{c}-\no)}_{{\cal N}=2\,\,SYM}\rangle^2=\langle S\rangle_{N_c}\,,
\eeq
where $k_i$ are known numbers \cite{DS}. We note that $\Lambda^{SU(N_{c}-\no)}_{{\cal N}=2\,\,SYM}$ in \eqref{(4.1.2)} and $\langle a\rangle=m$ in \eqref{(4.1.1)} have the same charge 2 under the discrete $Z_{2N_c-N_F}$ transformations as $X$ itself, this is due to unbroken $Z_{\bb}$ symmetry. There are $N_c-\no$ physically equivalent vacua and $N_c-\no-1$ light pure magnetic monopoles ${\ov M}_n,\, M_n,\,\, n=1... N_c-\no-1$ with the $SU(N_{c}-\no)$ adjoint charges (massless at $\mx\ra 0$) in each of these ${\cal N}=2$ SYM vacua. The relevant part of the low energy superpotential of this SYM part looks as
\bq
{\cal W}^{\,\rm low}_{\rm tot}={\cal W}^{(M)}_{SYM}+{\cal W}_{\no}+\dots\,,\label{(4.1.3)}
\eq
\bbq
{\cal W}^{(M)}_{SYM}= - \sum_{n=1}^{N_c-\no-1} {\tilde a}_{M,\,\rm n}\Biggl [\, {\ov M}_{\rm n} M_{\rm n}+\mx\langle\Lambda^{SU(N_{c}-{\rm n}_1)}_{{\cal N}=2\,\,SYM}\rangle\Biggl (1+O\Bigl (\frac{\langle\Lambda^{SU(N_{c}-{\rm n}_1)}_{{\cal N}=2\,\,SYM}\rangle}{m}\Bigr )\Biggr )\,z_{\rm n}\, \,\Biggr ]\,,\quad z_{\rm n}=O(1),
\eeq
where $z_{\rm n}$ are known numbers \cite{DS} and dots denote as always smaller power corrections.

All monopoles in \eqref{(4.1.3)} are higgsed at $\mx\ne 0$ with $\langle M_{\rm n}\rangle=\langle{\ov M}_{\rm n} \rangle\sim [\,\mx\langle\Lambda^{SU(N_{c}-{\rm n}_1)}_{{\cal N}=2\,\,SYM}\rangle\,]^{1/2}$, so that all light particles of this ${\cal N}=2\,\, U(1)^{N_c-\no-1}$ Abelian magnetic part form $N_c-\no-1$ long ${\cal N}=2$ multiplets of massive dual photons with masses $\sim (\mx\langle\Lambda^{SU(N_{c}-\no}_{{\cal N}=2\,\,SYM})^{1/2}\rangle$ .

Besides, for this reason, all heavier original $SU(N_{c}-\no)$ electrically charged particles  with masses $\sim m$ or $\sim\langle\Lambda^{SU(N_{c}-{\rm n}_1)}_{{\cal N}=2\,\,SYM}\rangle$ are weakly confined (the confinement is weak in the sense that the tension of the confining string is much smaller than their masses, $\sigma^{1/2}_{SYM}\sim (\mx\langle\Lambda^{SU(N_{c}-{\rm n}_1)}_{{\cal N}=2\,\,SYM}\rangle)^{1/2}\ll\langle\Lambda^{SU(N_{c}-{\rm n}_1)}_{{\cal N}=2\,\,SYM}\rangle\ll m$).\\

And finally, as for the (independent) ${\cal N}=2\,\,SU(\no)\times U(1)$ part with $N_F$ flavors of original electric quarks ${\ov Q}^b_j,  Q^{\,i}_b,\,\, i,j=1...N_F,\,\, b=1...\no$. Its superpotential at the low scale looks as
\bq
{\cal W}_{\no}=(\, m-a\,)\,{\rm Tr}\,({\ov Q} Q)_{\no} - {\rm Tr}\,({\ov Q}\sqrt{2} X^{adj}_{SU(\no)} Q) +\mx {\rm Tr}\,( X^{adj}_{SU(\no)})^2+\, \label{(4.1.4)}
\eq
\bbq
+ \,\frac{\mx}{2}\,\frac{\no N_c}{N_c-\no}\,a^2+(N_c-\no)\mx\Biggl [\,\Bigl (\Lambda^{SU(N_{c}-{\rm n}_1)}_{{\cal N}=2\,\,SYM}\Bigr )^2\approx\Bigl (\frac{\lm^{2N_c-N_F}(m_3-c\,{\hat a})^{N_F}}{[\,m_3+(1-c){\hat a}\,]^{2\no}}\Bigr )^{\frac{1}{N_c-\no}}\,\Biggr ]\,,\quad a\equiv \langle a\rangle+{\hat a}\,.
\eeq

Because the type of color breaking $SU(N_c)\ra SU(\no)\times U(1)\times SU(N_c-\no)$ in \eqref{(4.1.1)} is qualitatively different from $SU(N_c)\ra SU(N_F-N_c)\times U^{(1)}(1)\times U(1)^{\bb-1}$ in section 2.1 and occurs here {\it in the weak coupling regime} $a(\mu\sim m)=N_c g^2(\mu\sim m)/8\pi^2\ll 1$, the nonperturbative instanton contributions do not operate in this high energy region. They originate and operate in these vacua only at much lower energies $\sim\langle\Lambda^{SU(N_{c}-{\rm n}_1)}_{{\cal N}=2\,\,SYM}\rangle$ from the SYM part, see \eqref{(4.1.3)}. For this reason, there are no in \eqref{(4.1.4)} analogs of terms $\sim\delta_i$ in \eqref{(2.1.2)}.

From \eqref{(4.1.4)} (neglecting here small power corrections in the quark condensate $\langle\Qo\rangle
_{\no}$ originating from the SYM part, see \eqref{(A.4)}\,), compare also with \eqref{(4.1)},
\bq
\langle a\rangle=m\,,\quad \langle X^{adj}_{SU(\no)}\rangle=0\,,\quad\langle{\rm Tr}\,({\ov Q} Q)_{\no}\rangle=\no\langle\Qo\rangle_{\no}+\nt\langle\Qt\rangle_{\no}\approx\no \mx m_3\,,\label{(4.1.5)}
\eq
\bbq
\langle\Qo\rangle^{SU(N_c)}_{\no}=\langle{\ov Q}^{1}_{1}\rangle\langle Q^{1}_{1}\rangle\approx\mx m_3\approx\langle\Qo\rangle^{SU(N_c)}_{N_c}\,,\quad \langle\Qt\rangle_{\no}=\sum_{a=1}^{\no}\langle{\ov Q}^{\,a}_2\rangle\langle Q^2_a\rangle=0\,,\quad \langle S\rangle_{\no}=0\,,
\eeq
\bbq
\langle Q^i_b\rangle=\langle {\ov Q}_i^{\,b}\rangle\approx\delta^i_b\,(\mx m_3)^{1/2},\,\, b=1... \no,\,\,i=1...N_F\,,\,\, m_3=\frac{N_c}{N_c-\no}\,.
\eeq

On the whole, all this results in the right multiplicity of these br1 vacua, $N_{\rm br1}=(N_c-\no) C_{N_F}^{\,\no}$, the factor $N_c-\no$ originates from ${\cal N}=2\,\,\, SU(N_{c}-\no)$ SYM, while $C_{N_F}^{\,\no}$ is due to the spontaneous breaking $U(N_F)\ra U(\no)\times U(\nt)$ by $\no$ higgsed quarks in the $SU(\no)\times U(1)$ sector.

It is worth emphasizing that the value $\langle a\rangle=m$ in \eqref{(4.1.5)} is {\it exact} (i.e. there is no any small correction). This is seen from the following. At very small $\mx\ra 0$, the possible correction $\delta m\neq 0$ is independent of $\mx$, and so $\delta m\gg\mx$. Then quarks $Q^i_b, {\ov Q}_i^{\,b},\,\, i=1...N_F,\,\, b=1...\no$ will acquire masses $\delta m$, much larger than the scale of their potentially possible coherent condensate, $\delta m\gg (\mx m)^{1/2},\,\, \mx\ra 0$. In this case, these quarks will be not higgsed but will decouple as heavy ones at scales $\mu<\delta m$ and the flavor symmetry will remain unbroken. There will remain ${\cal N}=2\,\, SU(\no)\times U(1)$ SYM  at scales $\mu<\delta m$. This lowest energy ${\cal N}=2\,\, SU(\no)$ SYM will give then its own multiplicity factor $\no$, so that the overall multiplicity of vacua will be $\no (N_c-\no)$, while the right multiplicity is $(N_c-
\no)C^{\,\no}_{N_F}$, see \eqref{(4.1)} and section 3 in \cite{ch4}. To have this right multiplicity and spontaneous flavor symmetry breaking all $\no$ quarks have to be higgsed at arbitrary small $\mx$, and this is only possible when they are {\it exactly massless} in the limit $\mx\ra 0$, i.e. at $\delta m=0$.~
~\footnote{\, These considerations clearly concern also all other similar cases.
}

As a result of higgsing of $\no$ out of $N_F$ quark flavors in $SU(\no)\times U(1)$ in \eqref{(4.1.4)}, the flavor symmetry is broken spontaneously as $U(N_F)\ra U(\no)\times U(\nt)$, the quarks $Q^{\,l}_b, {\ov Q}^{\,b}_l$ with flavors $l=(\no+1)...N_F$ and colors $b=1...\no$ will be the massless Nambu-Goldstone particles ($\,2\no\nt$ complex degrees of freedom), while all other particles in \eqref{(4.1.4)} will acquire masses $\sim (\mx m)^{1/2}$ and will form $\no^2$ long ${\cal N}=2$ multiplets of massive gluons.

On the whole, the hierarchies of nonzero masses look in this case as
\bq
\langle M_{\rm n}\rangle\sim (\mx\langle\Lambda^{SU(N_{c}-{\rm n}_1)}_{{\cal N}=2\,\,SYM}\rangle)^{1/2}\ll\langle Q^i_{b=i}\rangle\sim (\mx m)^{1/2}\,,\quad \lm\ll\langle\Lambda^{SU(N_{c}-\no)}_{{\cal N}=2\,\,SYM}\rangle\ll m\,,\label{(4.1.6)}
\eq
where ${\rm n}=1...N_{c}-\no-1\,,\,\,\, i=1...\no$\,. There are no massless particles, except for the $\,2\no\nt$ Nambu-Goldstone multiplets.

The curve \eqref{(1.2)} has $N_c-1$ double roots in these vacua. $N_c-\no-1$ unequal roots correspond to pure magnetic monopoles $M_{\rm n}$ in \eqref{(4.1.3)} and $\no$ equal ones correspond to higgsed quarks from $SU(\no)$. The two single roots with $(e^{+}-e^{-})\sim\langle\Lambda^{SU(N_{c}-\no)}_{{\cal N}=2\,\,SYM}\rangle$ originate from SYM.

The calculation of leading power corrections to $\langle{\cal W}^{\,\rm low}_{\rm tot}\rangle$ is presented in Appendix A, and those to $\langle\Qo\rangle^{SU(N_c)}_{\no}$ are presented in Appendices A and B. These latter are also compared therein with the {\it independent} calculations of $\langle\Qo\rangle^{SU(N_c)}_{\no}$ using the roots of the curve \eqref{(1.2)}.\\

Now, let us recall some additional explanations about notations in \eqref{(4.1.4)},\eqref{(4.1.5)}. As for {\it higgsed} $\langle a\rangle$, it looks as $\langle a\rangle=\langle a\rangle^{(\rm several) m}_{m/(\rm several)}=m$. In other words, its total mean value originates and saturates at $\mu\sim m$. If not,  all quarks will decouple as heavy at $\mu=m/(\rm several)$, the multiplicity of these br1 vacua will be wrong and $U(N_F)$ will remain unbroken. As for $\langle Q_a\rangle,\,\, a=1...\no$, of light quarks with $SU(\no)$ colors. - At scales $[\mu^{\rm low}_{\rm cut}=(\mx m_3)^{1/2}]<\mu<m/(\rm several)$, all $SU(\no)$ gluons are effectively massless and all light quarks with $SU(\no)$ colors move freely and {\it independently}. Therefore, the physical (i.e. path dependent) $SU(\no)$ phases of these quarks induced by their interactions with $SU(\no)$ gluons fluctuate {\it freely} in this higher energy region, so that
\bq
\langle{\ov Q}^{\,b}_j Q^i_a\rangle_{\mu^{\rm low}_{\rm cut}}=\langle{\ov Q}^{\,b}_j\rangle_{\mu^{\rm low}_{\rm cut}}\langle Q^i_a\rangle_{\mu^{\rm low}_{\rm cut}}=0\,,\quad a,b=1...\no\,,\quad i,j=1...N_F\,\quad \mu^{\rm low}_{\rm cut}=(\mx m_3)^{1/2}\,.\label{(4.1.7)}
\eq
Therefore, in the whole region $0\leq \mu<m/(\rm several)$, the nonzero quark mean values in \eqref{(4.1.5)}, i.e. $[\langle\Qo\rangle_{\no}=\langle{\ov Q}^{1}_{1}\rangle\langle Q^{1}_{1}\rangle\,]
/\mx\approx [\langle a\rangle=m]N_c/(N_c-\no)]$ (the approximation here is controllable and consists in neglecting small power corrections originating from the SYM part, see \eqref{(4.1.3)},\eqref{(4.1.4)} and Appendix A) originate and saturate {\it only} in the region $g_Q (\mx m_3)^{1/2}/(\rm several)<\mu< (\rm several){\it g}_Q (\mx m_3)^{1/2}$, after $\no$ out of $N_F$ quarks are higgsed (i.e. form the coherent condensate), {\it giving masses} $g_Q (\mx m)^{1/2}$ to themselves and all $SU(\no)\times U(1)$ gluons. And there remain only $2\no\no$ massless Nambu-Goldstone multiplets in this sector at lower energies.

And there is no contradiction in the relation following from \eqref{(4.1.4)}
\bbq
\langle\Qo\rangle_{\no}=\Bigl (\langle{\ov Q}^1_1\rangle\langle Q^1_1\rangle\Bigr )^{(\rm several){\it g}_Q (\mx m_3)^{1/2}}_{g_Q (\mx m_3)^{1/2}/(\rm several)}\approx \Bigl [\mx\frac{N_c}{N_c-\no}\Bigl (\langle a\rangle^{(\rm several) m}_{m/(\rm several)}=m\,\Bigr )=\mx m_3\,\Bigr ],
\eeq
in that the {\it total mean values} of $\langle a\rangle$ and $\langle Q^1_1\rangle$ originate and saturate in different energy regions, because this is {\it not} the equality between mean values of two equal to each other operators, but only the numerical equality between two total mean values of two different operators.\\

On the whole, the total decomposition of quark condensates $\langle (\qq)_{1,2}\rangle_{N_c}$ over their separate color parts look in these br1 vacua as follows.\\
I) The condensate $\langle\Qo\rangle_{N_c}$.\\
a) From \eqref{(4.1.5)} and accounting for the leading power correction, see \eqref{(A.5)}, the (factorizable) condensate of higgsed quarks in the $SU(\no)$ part
\bq
\langle\Qo\rangle_{\no}\approx\mx m_3\Biggl [1-\frac{2N_c-N_F}{N_c-\no}\Bigl (\frac{\lm}{m_3}\Bigr )^{\frac{2N_c-N_F}{N_c-\no}}\Biggr ]\,,\quad m_3=\frac{N_c}{N_c-\no} m\gg
\langle\Lambda^{SU(N_{c}-\no)}_{{\cal N}=2\,\,SYM}\rangle\gg\lm\,. \label{(4.1.8)}
\eq
b) The (non-factorizable) condensate $\langle\Qo\rangle_{N_c-\no}$ is determined by the one-loop Konishi anomaly for the heavy non-higgsed quarks with the mass $m_3$ in the $SU(N_c-\no)$ SYM sector. I.e., on the first "preliminary" stage, from the one-loop diagrams with heavy scalar quarks and their fermionic superpartners inside, transforming in the weak coupling regime at the scale $\mu\sim m$ the heavy quark {\it operator} $(\Qo)_{N_c-\no}$ into the {\it operator} $\sim (\lambda\lambda)_{N_c-\no}/m_3$ of lighter $SU(N_c-\no)$ gluino. And the mean value of this latter originates only at much lower energies $\mu\lesssim\langle\Lambda^{SU(N_{c}-\no)}_{{\cal N}=2\,\,SYM}\rangle$, see \eqref{(4.1.2)},
\bq
\langle\Qo\rangle_{N_c-\no}=\frac{\langle S\rangle_{N_c-\no}}{m_3}=\frac{\mx\langle\Lambda^{SU(N_{c}-\no)}
_{{\cal N}=2\,\,SYM}\rangle^2}{m_3}\approx\mx m_3\Bigl (\frac{\lm}{m_3}\Bigr )^{\frac{\bb}{N_c-\no}}\,.\label{(4.1.9)}
\eq

Therefore, on the whole
\bq
\langle\Qo\rangle_{N_c}=\langle\Qo\rangle_{\no}+\langle\Qo\rangle_{N_c-\no}\approx
\mx m_3\Biggl [ 1+\frac{\nt-N_c}{N_c-\no}\Bigl (\frac{\lm}{m_3}\Bigr )^{\frac{\bb}{\nt-N_c}}\Biggr ]\,,\,\,\quad \label{(4.1.10)}
\eq
as it should be, see \eqref{(A.1)}.\\
II) The condensate $\langle\Qt\rangle_{N_c}$.\\
a) From \eqref{(4.1.5)}, the (factorizable) condensate of non-higgsed massless Nambu-Goldstone particles in the $SU(\no)$ part
\bq
\langle\Qt\rangle_{\no}=\sum_{a=1}^{\no}\langle{\ov Q}^{\,a}_2\rangle\langle
Q^2_{a}\rangle=0\,.\label{(4.1.11)}
\eq
b) The (non-factorizable) condensate $\langle\Qt\rangle_{N_c-\no}$ is determined by the same one-loop Konishi anomaly for the heavy non-higgsed quarks with the mass $m_3$ in the $SU(\nd-\no)$ SYM sector,
\bq
\langle\Qt\rangle_{N_c-\no}=\frac{\langle S\rangle_{N_c-\no}}{m_3}\approx\mx m_3\Bigl (\frac{\lm}{m_3}\Bigr )^{\frac{\bb}{N_c-\no}}\,.\label{(4.1.12)}
\eq

Therefore, on the whole
\bq
\langle\Qt\rangle_{N_c}=\langle\Qt\rangle_{\no}+\langle\Qt\rangle_{N_c-\no}\approx
\mx m_3\Bigl (\frac{\lm}{m_3}\Bigr )^{\frac{\bb}{N_c-\no}}\,, \label{(4.1.13)}
\eq
as it should be, see \eqref{(A.1)}.\\

\subsection{$U(N_c)$, br1 vacua, smaller $\mx\,,\,\, \mx\ll\Lambda^{SU(N_{c}-\no)}_{{\cal N}=2\,\,SYM}$}

When the additional $U^{(0)}(1)$ is introduced, the Konishi anomalies look as in \eqref{(2.2.2)} but $\langle\Qo\rangle_{N_c}$ is dominant in br1 vacua, i.e. (see the footnote \ref{(f4)}, the leading terms only)
\bq
\langle\Qo+\Qt\rangle_{N_c}=\mx m\,,\quad \langle\Qo\rangle_{N_c}=\mx m-\langle\Qt\rangle_{N_c}\approx\mx m\,,\label{(4.2.1)}
\eq
\bbq
\langle\Qt\rangle_{N_c}\approx\mx m\Bigl (\frac{\lm}{m}\Bigr )^{\frac{2N_c-N_F}{N_c-\no}}\approx
\mx m\,\frac{\langle\Lambda^{SU(N_{c}-\no)}_{{\cal N}=2\,\,SYM}\rangle^2}{m^2}\,,
\eeq
while
\bq
\frac{\langle a_0\rangle}{m}=\frac{1}{\mx m\, N_c}\langle{\rm Tr}\,\qq\rangle_{N_c}\approx\Bigl [\,\frac{\no}{N_c} +\frac{\nt-\no}{N_c}\Bigl (\frac{\lm}{m}\Bigr )^{\frac{2N_c-N_F}{N_c-\no}}\,\Bigr ]
\approx\frac{\no}{N_c}\,.\label{(4.2.2)}
\eq
(Put attention that $\langle\Lambda^{SU(N_{c}-\no)}_{{\cal N}=2\,\,SYM}\rangle$ in this section is different from those in section 4.1, see \eqref{(4.1.2)}).

Instead of \eqref{(4.1.4)} the superpotential of the $SU(\no)\times U^{(0)}(1)\times U(1)$ part looks now as
\bbq
{\cal W}^{\,\rm low}_{\rm matter}=\w_{\no}+\w_{a_0,a}+\dots\,,
\eeq
\bq
{\cal W}_{\no}=( m-a_0-a){\rm Tr}({\ov Q} Q)_{\no}-{\rm Tr}({\ov Q}\sqrt{2} X^{adj}_{SU(\no)} Q)+\mx {\rm Tr}( X^{adj}_{SU(\no)})^2,\,\,\,\label{(4.2.3)}
\eq
\bbq
\w_{a_0,a}=\frac{\mx}{2}N_c a^2_0+\frac{\mx}{2}\frac{\no N_c}{N_c-\no} a^2\,,
\eeq
where dots denote smaller power corrections.

From \eqref{(4.2.3)} (for the leading terms)
\bbq
\langle a_0\rangle\approx\frac{\no}{N_c} m\,,\,\,\,\langle a\rangle=\langle m-a_0\rangle\approx\frac{N_c-
\no}{N_c} m\,,\,\,\, \langle{\rm Tr}\,({\ov Q} Q)\rangle_{\no}\approx\mx N_c\langle a_0\rangle
\approx\mx\frac{\no N_c}{N_c-\no}\langle a\rangle\approx\no \mx m\,,
\eeq
\bq
\langle X^{adj}_{SU(\no)}\rangle=0,\,\,\langle\Qo\rangle^{U(N_c)}_{\no}=\sum_{a=1}^{\no}\langle{\ov Q}^{\,a}_1 Q^1_a\rangle=\langle{\ov Q}^{\,a}_1\rangle\langle Q^1_a\rangle\approx\mx m\approx\langle\Qo\rangle^{U(N_c)}_{N_c},\,\, \langle\Qt\rangle_{\no}=0.\quad\label{(4.2.4)}
\eq

The qualitative difference with section 4.1 is that one extra ${\cal N}=1$ photon multiplet remains massless now in this $U(N_c)$ theory while corresponding scalar multiplet has the smallest mass $\sim\mx$ because ${\cal N}=2$ is broken down to ${\cal N}=1$ at the level $O(\mx)$\,.

The calculation of leading power corrections to $\langle{\cal W}^{\,\rm low}_{\rm tot}\rangle$ in these $U(N_c)$ br1 vacua is presented in Appendix A, and those to $\langle\Qo\rangle^{U(N_c)}_{\no}$ are presented in Appendices A and B. These latter are also compared therein with the {\it independent} calculations of $\langle\Qo\rangle^{U(N_c)}_{\no}$ using the roots of the curve \eqref{(1.2)}.\\

\subsection{$SU(N_c)$, br1 vacua, larger $\mx\,,\, \Lambda^{SU(N_{c}-\no)}_{{\cal N}=2\,\,SYM}\ll\mx\ll m$}

Nothing changes significantly in this case with the $SU(\no)\times U(1)$ part in \eqref{(4.1.4)}. The mean value $\langle a\rangle=m\gg\mx$ of the field $\langle\sqrt{2}\,X_{U(1)}\rangle$ in \eqref{(4.1.4)} stays intact as far as $\mx\ll m$, and so the contributions $\sim (\mx m)^{1/2}$ to the masses of $\no^2-1$ long ${\cal N}=2\,\,\,SU(\no)$ multiplets of massive gluons and to the mass of the long ${\cal N}=2\,\,\,U(1)$ multiplet of the massive photon due to higgsing of $\no$ quarks remain dominant. At this level the ${\cal N}=2$ SUSY remains unbroken in this sector. It breaks down here to ${\cal N}=1$ only at the level $\sim\mx$ by the additional smaller contributions $\sim\mx\ll (\mx m)^{1/2}$ to the masses of corresponding scalar multiplets.

The situation with $SU(N_c-\no)$ SYM in \eqref{(4.1.2)},\eqref{(4.1.3)} is different. At $\langle\Lambda^
{SU(N_c-\no)}_{{\cal N}=2\,\,SYM}\rangle\ll\mx\ll m$, the adjoint scalars $X^{adj}_{SU(N_c-\no)}$ of the $SU(N_c-\no)$ subgroup in \eqref{(4.1.2)} become too heavy and too short ranged. Their light $SU(N_c-\no)$ physical (i.e. path dependent) phases induced by interactions with effectively massless at all scales $\mu\gtrsim\langle\Lambda^{SU(N_c-\no)}_{{\cal N}=1\,\,SYM}\rangle\,\, SU(N_c-\no)$ gluons fluctuate freely at the scale $\mu>\mx^{\rm pole}/(\rm several)={\it g}^2\mx/(\rm several)$, and they are {\it not higgsed}, i.e. $\langle X^{adj}_{SU(N_c-\no)}\rangle_{\mx^{\rm pole}/(\rm several)}=0$. Instead, they decouple as heavy already at the scale $\mu<\mx^{\rm pole}/(\rm several)\ll m$ in the weak coupling region, can all be integrated out and do not affect by themselves the lower energy dynamics in this $SU(N_c-\no)$ sector. There remains ${\cal N}=1 \,\, SU(N_c-\no)$ SYM with the scale factor $\langle\Lambda^{SU(N_c-\no)}_{{\cal N}=1\,\,SYM}\rangle=[\,\mx \langle\Lambda^{SU(N_c-\no)}_{{\cal N}=2\,\,SYM}\rangle^2\,]^{1/3}$ of its gauge coupling, $\,\,\langle\Lambda^{SU(N_c-\no)}_{{\cal N}=2\,\,SYM}\rangle\ll\langle\Lambda^{SU(N_c-\no)}_{{\cal N}=1\,\,SYM}\rangle\ll\mx\ll\, m$,
\bq
\langle S\rangle_{N_c-\no}=\langle\Lambda^{SU(N_c-\no)}_{{\cal N}= 1\,\,SYM}\rangle^3=\mx m_3^2\Bigl (\frac{\lm}{m_3}\Bigr )^{\frac{2N_c-N_F}{N_c-\no}},\quad \frac{\langle\Lambda^{SU(N_c-\no)}_{{\cal N}=1\,\,SYM}\rangle}{\langle\Lambda^{SU(N_{c}-{\rm n}_1)}_{{\cal N}=2\,\,SYM}\rangle}\sim\Biggl (\frac{\mx}{\langle\Lambda^{SU(N_{c}-{\rm n}_1)}_{{\cal N}=2\,\,SYM}\rangle}\Biggr )^{1/3}\gg 1\,. \label{(4.3.1)}
\eq
The small (non-factorizable) value $\langle{\rm Tr\,} (X^{adj}_{SU(N_c-\no)})^2\rangle=
(N_c-\no)\langle S\rangle_{N_c-\no}/\mx=(N_c-\no)\langle\Lambda^{SU(N_c-\no)}_{{\cal N}=2\,\, SYM}\rangle^2 \\ \ll \langle\Lambda^{SU(N_c-\no)}_{{\cal N}=1\,\, SYM}\rangle^2$ arises here not because $X^{adj}_{SU(N_c-\no)}$ are higgsed, but only due to the Konishi anomaly. I.e. from one-loop diagrams with heavy scalars $X^{adj}_{SU(N_c-\no)}$ and their fermionic superpartners with masses $\sim\mx$ inside. On the first "preliminary" stage these loop effects transform, in the weak coupling regime at the scale $\sim \mx$, the heavy field {\it operator} ${\rm Tr\,} (X^{adj}_{SU(N_c-\no)})^2$ into the {\it operator} $\sim {\rm Tr\,}(\lambda\lambda)_{SU(N_c-\no)}/\mx$ of lighter ${\cal N}=1\,\, SU(N_c-\no)$ SYM gluinos. And the nonzero mean value of this latter originates and saturates only in the strong coupling and non-perturbative regime at much lower energies $\sim\langle\Lambda^{SU(N_c-\no)}_{{\cal N}=1\,\, SYM}\rangle\ll\mx$ .

There is now a large number of strongly coupled gluonia with the mass scale $\sim\langle\Lambda^
{SU(N_c-\no)}_{{\cal N}=1\,\,SYM}\rangle$ in this ${\cal N}=1$ SYM. The multiplicity of vacua in this sector remains equal $N_c-\no$ as it should be. All heavier original electric particles with masses $\sim m$ or $\sim\mx$ charged with respect to $SU(N_c-\no)$ remain weakly confined, but the string tension is larger now, $\sigma^{1/2}_{{\cal N}=2}\ll\sigma^{1/2}_{{\cal N}=1}\sim\langle\Lambda^{SU(N_c-\no)}_{{\cal N}=1\,\,SYM}\rangle\ll\mx\ll m$.

\subsection{$SU(N_c)$\,, \, special vacua}

The values of quark condensates in these special vacua with $\no=\nd,\, \nt=N_c$, see \cite{ch4,ch6} and \eqref{(4.1.2)},
\bq
\langle\Qo+\Qt-\frac{1}{N_c}{\rm Tr}\,(\qq)\rangle_{N_c}=(1-\frac{\no=\nd}{N_c})\langle\Qo\rangle_{N_c}=\mx m\,,\label{(4.4.1)}
\eq
\bbq
\langle\Qo\rangle_{N_c}=\frac{N_c}{\bb}\mx m=\mx m_3\,,\quad \langle\Qt\rangle_{N_c}=\mx\lm\,,\quad
\langle S\rangle_{N_c}=\frac{\langle\Qo\rangle_{N_c}\langle\Qt\rangle_{N_c}}{\mx}=
\eeq
\bbq
=\Bigl (\frac{\det \langle\qq\rangle_{N_c}}{\lm^{2N_c-N_F}\mx^{N_c}} \Bigr )^{1/\nd}=\langle S\rangle_{2N_c-N_F}=\mx\langle\Lambda^{SU(\bb)}_{{\cal N}=2\,\,SYM}\rangle^2=\mx m_3\lm\,,
\eeq
are exact and valid at any values $m\gtrless\lm$. Therefore, the discrete $Z_{\bb}$ symmetry is formally broken spontaneously therein at large $m\gg\lm$ also, as it is really broken at $m\ll\lm$. But practically they behave at $m\gg\lm$ as the br1 vacua described above in section (4.1) with $\no=\nd\,$. I.e., the factor $\bb$ in their multiplicity $N_{\rm spec}=(\bb) C^{\,\nd}_{N_F}$ originates at $m\gg\lm$ from the multiplicity $\bb$ of the $SU(\bb)\,\,{\cal N}=2$ SYM at $\mx\ll\langle\Lambda^{SU(\bb)}_{{\cal N}=2\,\,SYM}\rangle$ or ${\cal N}=1$ SYM at $\langle\Lambda^{SU(\bb)}_{{\cal N}=1\,\,SYM}\rangle\ll\mx\ll m$, after the breaking $SU(N_c)\ra SU(\nd)\times U(1)\times SU(\bb)$ by $\langle X_{U(1)}\rangle$ at the scale $\mu= m_3\gg\lm$, see \eqref{(4.1.1)}. The factor $C^{\,\nd}_{N_F}$ originates due to higgsing of $\no=\nd$ original electric quarks in the $SU(\nd)$ sector, $\langle\Qo\rangle_{\nd}=\langle{\ov Q}^{\,1}_1
\rangle\langle Q^1_1\rangle=\mx m_3$, see \eqref{(4.1.4)}, and spontaneous breaking $U(N_F)\ra U(\nd)\times U(N_c)$. The two single roots with $( e^{+}-e^{-})\sim\langle\Lambda^{SU(\bb)}_{{\cal N}=2\,\,SYM}\rangle$ of the curve \eqref{(1.2)} at small $\mx$ originate from the ${\cal N}=2\,\,SU(\bb)$ SYM. There are no massless particles at $\mx\neq 0$, except for $2\no\nt$ Nambu-Goldstone multiplets.

\section{Unbroken flavor symmetry,\,\, SYM vacua,\,\, $SU(N_c)$}

These SYM vacua of the $SU(N_c)$ theory have the multiplicity $N_c\,$. $\langle X^{\rm adj}\rangle_{N_c}\sim
\langle\Lambda^{SU(N_c)}_{{\cal N}=2\, SYM}\rangle\ll m$ in these vacua, so that all quarks have here large masses $m\gg\langle\Lambda^{SU(N_c)}_{{\cal N}=2\, SYM}\rangle$ and decouple in the weak coupling regime
at $\mu<m/(\rm several)$. There remains at this scale the ${\cal N}=2\,\, SU(N_c)$ SYM with the scale factor of its gauge coupling: $\langle\Lambda^{SU(N_c)}_{{\cal N}=2\, SYM}\rangle^2=(\lm^{\bb} m^
{N_F})^{1/N_c}=m^2\, (\lm/m)^{(\bb)/N_c},\,\,\,\lm\ll\langle\Lambda^{SU(N_c)}_{{\cal N}=2\, SYM}\rangle\ll m$.

If $\mx\ll\langle\Lambda^{SU(N_c)}_{{\cal N}=2\, SYM}\rangle$, then the field $X^{\rm adj}_{SU(N_c)}$ is higgsed at the scale $\mu\sim\langle\Lambda^{SU(N_c)}_{{\cal N}=2\, SYM}\rangle$ in a standard way for the ${\cal N}=2$ SYM, $\langle\sqrt{2} X^{\rm adj}_{SU(N_c)}\rangle\sim\langle\Lambda^{SU(N_c)}_{{\cal N}=2\, SYM}\rangle\,{\rm diag}\,(k_1,k_2,\,...\,k_{N_c}),\,\, k_i=O(1)$ \cite{DS}, $SU(N_c)\ra U^{N_c-1}(1)$. Note that the value $\sim\langle\Lambda^{SU(N_c)}_{{\cal N}=2\, SYM}\rangle$ of $\langle X^{\rm adj}_{SU(N_c)}\rangle$ is consistent with the unbroken $Z_{\bb}$ discrete symmetry.

All original electrically charged gluons and scalars acquire masses $\sim\langle\Lambda^{SU(N_c)}_{{\cal N}=2\, SYM}\rangle$ and $N_c-1$ lighter Abelian pure magnetic monopoles $M_k$ are formed. These are higgsed at $\mx\neq 0$ and $N_c-1$ long ${\cal N}=2$ multiplets of massive dual photons are formed, with masses $\sim (\mx\langle\Lambda^{SU(N_c)}_{{\cal N}=2\, SYM}\rangle)^{1/2}\ll\langle\Lambda^{SU(N_c)}_{{\cal N}=2\, SYM}\rangle$ (there are non-leading contributions $\sim\mx\ll (\mx\langle\Lambda^{SU(N_c)}_{{\cal N}=2\, SYM}\rangle)^{1/2}$ to the masses of corresponding scalars, these break slightly ${\cal N}=2$ down to ${\cal N}=1$ ). As a result, all original electrically charged quarks, gluons and scalars $X$ are weakly confined (i.e. the tension of the confining string, $\sigma^{1/2}_2\sim (\mx\langle\Lambda^{SU(N_c)}_{{\cal N}=2\, SYM}\rangle)^{1/2}$, is much smaller than their masses $\sim m$ or $\sim\langle\Lambda^{SU(N_c)}_{{\cal N}=2\, SYM}\rangle\,$).

All $N_c-1$ double roots of the curve \eqref{(1.2)} correspond in this case to $N_c-1$ pure magnetic monopoles (massless at $\mx\ra 0$), while two single roots with $(e^+ - e^-)\sim\langle\Lambda^{SU(N_c)}_{{\cal N}=2\, SYM}\rangle$ also originate from this ${\cal N}=2 \,\,SU(N_c)$ SYM.\\

The mass spectrum is different if $\langle\Lambda^{SU(N_c)}_{{\cal N}=2\, SYM}\ll\mx\ll m$. All scalar fields $X^{\rm adj}_{SU(N_c)}$ are then too heavy and not higgsed. Instead, they decouple as heavy at scales $\mu<\mx^{\rm pole}/(\rm several)$, still in the weak coupling regime, and there remains at lower energies ${\cal N}=1\,\, SU(N_c)$ SYM with its $N_c$ vacua and with the scale factor $\langle\Lambda^{SU(N_c)}_{{\cal N}=1\, SYM}\rangle$ of its gauge coupling, $\langle\Lambda^{SU(N_c)}_{{\cal N}=1\, SYM}\rangle=[\,\mx(\langle\Lambda^{SU(N_c)}_{{\cal N}=2\, SYM}\rangle)^{\,2}\,]^{1/3},\,\, \langle\Lambda^{SU(N_c)}_{{\cal N}=2\, SYM}\rangle\ll\langle\Lambda^{SU(N_c)}_{{\cal N}=1\, SYM}\rangle\ll\mx\ll m$. A large number of strongly coupled gluonia with the mass scale $\sim\langle\Lambda^{SU(N_c)}_{{\cal N}=1\, SYM}\rangle$ is formed in this ${\cal N}=1$ SYM theory, while all heavier original charged particles are weakly confined (the tension of the confining string, $\sigma^{1/2}_1\sim\langle\Lambda^{SU(N_c)}_{{\cal N}=1\, SYM}\rangle$, is much smaller than the quark masses $\sim m$ or the scalar masses $\sim\mx$).

\section{Very special vacua with $\langle S\rangle_{N_c}=0$ in $U(N_c)$ gauge theory}
\numberwithin{equation}{section}

\hspace{4mm} We consider in this section the vs (very special) vacua with $\no=\nd,\,\nt=N_c,\, N_c < N_F<2 N_c-1$\,, $\langle S\rangle_{N_c}=0$ (\,as always, $\langle S\rangle_{N_c}$ is the bilinear gluino condensate summed over all its colors) and the multiplicity $N_{vs}=C^{\nd}_{N_F}=C^{N_c}_{N_F}$ in the $U(N_c)=SU(N_c)\times U^{(0)}(1)$ theory, when the additional Abelian $U^{(0)}(1)$ with $\mu_0=\mx$ is added to the $SU(N_c)$ theory (see e.g. \cite{SY2} and references therein, these are called $r=N_c$ vacua in \cite{SY2}). Note that vacua with $\langle S\rangle_{N_c}=0$ are absent in the $SU(N_c)$ theory with $m\neq 0,\,\, \mx\neq 0$, see Appendix B in \cite{ch5}.

The superpotential of this $U(N_c)$ theory looks at high energies as
\bq
{\cal W}_{\rm matter}=\mx\Bigl [{\rm Tr}\,(X_0)^2=\frac{1}{2}N_c a_0^2\Bigr ]+\mx{\rm Tr}\,(X^{\rm adj}_{SU(N_c)})^2+{\rm Tr}\,\Biggl [\,(m-a_0)\,{\ov Q} Q-{\ov Q}\sqrt{2} X^{adj}_{SU(N_c)} Q \Biggr ]_{N_c}\,,\label{(6.1)}
\eq
\bbq
\sqrt{2} X_{0}=a_0\, {\rm diag}(\,\underbrace{\,1}_{N_c}\,)\,,\quad
X^{\rm adj}_{SU(N_c)}=X^A T^A\,,\,\, {\rm Tr}\,(T^A T^B)=\frac{1}{2}\,\delta^{AB}\,, \quad A, B=1,\,...,\,N_c^2-1\,.
\eeq

From \eqref{(6.1)} and Konishi  anomalies
\bq
\langle a_0\rangle=\frac{\langle {\rm Tr}({\ov Q}\, Q)\rangle_{N_c}=\nd\langle\Qo\rangle_{N_c}
+N_c\langle\Qt\rangle_{N_c}}{\mx N_c}\,,\label{(6.2)}
\eq
\bbq
\langle m-a_0\rangle\sum_{a=1}^{N_c}\langle{\ov Q}_i^a Q^i_a\rangle-\sum_{A=1}^{N^2_c-1}\sum_{a,b=1}^{N_c}\langle{\ov Q}_i^b\sqrt{2} X^A (T^{A})^a_b Q^i_a\rangle=\langle S\rangle_{N_c},\,\, i=1...N_F,\,\, \rm {no\,\,summation\,\,over\,\, flavor},
\eeq
\bbq
\mx\langle{\rm Tr}\,(\sqrt{2} X^{\rm adj}_{SU(N_c)})^2\rangle=(2N_c-N_F)\langle S\rangle_{N_c}+\langle m-a_0\rangle\langle{\rm Tr}\,(\qq)\rangle_{N_c}\,.
\eeq

Note that, as a consequence of a unique property $\langle S\rangle_{N_c}=0$ of these $U(N_c)$ vs -vacua, the curve \eqref{(1.2)} at $\mx\ra 0$ for theory \eqref{(6.1)} has not $N_c-1$ but $N_c$ double roots: $\,\nd$ equal double roots $e_k= - m$ corresponding to low energy non-Abelian $SU(\nd)$, and $2N_c-N_F$ unequal double roots $e_j= -m+\omega^{j-1}\lm,\, j=1...\bb$, see e.g. \cite{Bo}. There are no single roots of the curve \eqref{(1.2)} in these vacua.

As previously in section 1, taking $\mx$ sufficiently large
\footnote{\,
The scale factor $\Lambda_0$ of the Abelian coupling $g_0(\Lambda_0/\mu)$ is taken sufficiently large, the theory (6.1) is considered only at scales $\mu\ll\Lambda_0$ where $g_0$ is small, the large $\mx$ means here  $\lm\ll m\ll\mx\ll\Lambda_0$ if $m\gg\lm$ (and only here, while $\mx\ll\lm$ in all other cases).
}
and integrating out all $N^2_c$ scalars $X$ as heavy, the superpotential of the $U(N_c)\,\,\, {\cal N}=1$ SQCD looks then as
\bq
{\cal W}_{\rm matter}=m\,{\rm Tr}\, (\,{\ov Q} Q)_{N_c}-\frac{1}{2\mx}\,\sum_{i,j=1}^{N_F}\,({\ov Q}_j Q^i)_{N_c}({\ov Q}_i Q^j)_{N_c}\,,\quad ({\ov Q}_j Q^i)_{N_c}=\sum_{a=1}^{N_c}({\ov Q}_j^{\,a} Q^i_a)\,,\label{(6.3)}
\eq
while, instead of \eqref{(1.3)}, $\w^{\,\rm eff}_{\rm tot}$ looks now as
\bq
\w^{\,\rm eff}_{\rm tot}={\cal W}_{\rm matter}-\nd S_{N_c}\,,\quad S_{N_c}=\Bigl(\frac{\det (\qq)_{N_c}}{\lm^{2N_c-N_F}\mx^{N_c}}\Bigr )^{1/\nd}\,,\quad
\langle S\rangle_{N_c}=\Bigl (\frac{\langle\Qo\rangle^{\nd}\langle\Qt\rangle^{N_c}}
{\lm^{2N_c-N_F}\mx^{N_c}}\Bigr )^{1/\nd}_{N_c}.\,\,\label{(6.4)}
\eq
From \eqref{(6.4)}, in these specific vs-vacua with $\langle S\rangle_{N_c}=0$, the equations for determining quark condensates look as
\bq
\langle{\ov Q}_k\frac{\partial\w^{\rm eff}_{\rm tot}}{\partial{\ov Q}_k}\rangle= m\langle (\qq)_k\rangle_
{N_c}-\frac{\langle (\qq)_k\rangle^2_{N_c}}{\mx}-\langle S\rangle_{N_c}=0\,,\quad k=1,\,2\,,\label{(6.5)}
\eq
\bbq
\langle\Qo+\Qt\rangle_{N_c}=\mx m\,,\quad
\langle S\rangle_{N_c}=\langle\Qo\rangle_{N_c}\langle\Qt\rangle_{N_c}/\mx\,.
\eeq
The only self-consistent solution for these vs-vacua with $\no=\nd,\, \nt=N_c,\,\, \langle S\rangle_{N_c}=0$ looks as, see \eqref{(6.1)}-\eqref{(6.3)},
\bq
\langle\Qt\rangle_{N_c}=\mx m,\,\, \langle\Qo\rangle_{N_c}=0,\,\,
\langle S\rangle_{N_c}=\frac{\langle\Qo\rangle_{N_c}\langle\Qt\rangle_{N_c}}{\mx}=0\,,\label{(6.6)}
\eq
\bbq
\sum_{A=1}^{N^2_c-1}\sum_{a,b=1}^{N_c}\langle{\ov Q}_i^b\sqrt{2} X^A (T^{A})^a_b Q^i_a\rangle=0\,, \quad i=1...N_F\,,\quad \rm{no\,\, summation\,\, over\,\, flavor},
\eeq
\bbq
\langle a_0\rangle=\frac{\langle\Qt\rangle_{N_c}}{\mx} = m\,,\quad \langle{\rm Tr}\,(\sqrt{2} X^{\rm adj}_{SU(N_c)})^2\rangle=0\,, \quad\langle\w^{\,\rm eff}_{\rm tot}\rangle=
\langle{\cal W}_{\rm matter}\rangle=\frac{\mx}{2}N_c \langle a_0\rangle^2=\frac{\mx}{2}N_c m^2\,.
\eeq
(The variant with $\langle\Qo\rangle_{N_c}=\mx m,\,\,\langle\Qt\rangle_{N_c}=0$ and $\langle a_0\rangle=\nd m/N_c$ from \eqref{(6.2)}, will result at $m\gg\lm$ in all heavy non-higgsed quarks with masses ${\hat m}=\langle m-a_0\rangle=(\bb) m/N_c\gg\lm$, the lower energy $SU(N_c)$ \,\,${\cal N}=2$ SYM, the multiplicity of vacua equal $N_c$, $\,\langle S\rangle_{N_c}=\mx (\lm^{\bb}{\hat m}^{N_F})^{1/N_c}\neq 0$, the unbroken flavor symmetry, and the massless photon. There will be not $N_c$ but $N_c-1$ unequal ${\cal N}=2$ SYM double roots at $\mx\ra 0$. This is incompatible with $\langle S\rangle_{N_c}=\langle\Qo\rangle_{N_c}\langle\Qt
\rangle_{N_c}/\mx\ra 0$ from \eqref{(6.5)}, and the multiplicity is wrong. While at $m\ll\lm$ this variant with $\langle a_0\rangle=\nd m/N_c$ from \eqref{(6.2)} will be at least incompatible with \eqref{(6.2.2)},\eqref{(6.2.3)}).

\subsection {$m\gg\lm\,,\,\, \mx\gg\lm^2/m \,,\,\, N_c < N_F < 2 N_c-1$}
\numberwithin{equation}{subsection}

\hspace*{4mm} Consider first this simplest case with $0<\mx\ll\lm,\,\,m\gg\lm$ and $\mx m\gg\lm^2$. Note that $\langle a_0\rangle=m$ "eats" all quark masses $m$ in \eqref{(6.1)} in the range of scales $m/(\rm several)<\mu<(\rm several) m$, see \eqref{(6.6)} (as otherwise all quarks will decouple as heavy at $\mu<m/(\rm several)$ resulting in the wrong lower energy theory). In this region of parameters, $\nt=N_c$ quarks are higgsed at $\mu\sim (\mx m)^{1/2}\gg\lm$ {\it in the weak coupling regime}, e.g. (compare with \eqref{(2.4.19)},\eqref{(2.4.20)}),
\bq
\langle{\ov Q}_{i=b+\nd}^{\, b}\rangle=\langle Q^{i=b+\nd}_b\rangle= (\mx m)^{1/2}\gg\lm\,,\quad  b=1...N_c\,,\quad i=\nd+1...N_F\,, \label{(6.1.1)}
\eq
\bbq
\langle X^{adj,A}_{SU(N_c)}\rangle={\sqrt 2}\,{\rm Tr\,}\Bigl [\langle{\ov Q}\rangle T^{A}\langle Q\rangle\Bigr ]/\mx =0\,,\quad A=1...N_c^2-1\,,\quad \langle a_0\rangle=m\,,\quad \langle S\rangle_{N_c}=0\,.
\eeq

Higgsed quarks with $N_c$ colors and $\nt=N_c$ flavors give in this case large masses $\sim (\mx m)^{1/2}\gg\lm$ to themselves and to all gluons and scalars, and simultaneously {\it prevent} all adjoint scalars (which are light by themselves, in the sense $\mx\ll\lm$) from higgsing, see second line in \eqref{(6.2)}, $\langle a_0\rangle=m,\, \langle S\rangle_{N_c}=0$.

From \eqref{(6.3)},\eqref{(6.1.1)}, the multiplicity of these vs-vacua is $N_{vs}=C^{N_c}_{N_F}$, the factor $C^{N_c}_{N_F}$ originates in this phase from the spontaneous flavor symmetry breaking, $U(N_F)\ra U(N_c)\times U(\nd)$, due to higgsing of $\nt=N_c$ out of $N_F$ quarks. This multiplicity shows that the non-trivial at $2N_c-N_F\geq 2$ discrete $Z_{\bb}$ symmetry is {\it unbroken}.

As far as $\mx$ remains sufficiently larger than $\lm^2/m$, the mass spectrum includes in this phase: $N_c^2$ long ${\cal N}=2$ multiplets of massive gluons and their ${\cal N}=2$ superpartners with masses $\mu_{\rm gl}\sim (\mx m)^{1/2}\gg\lm$ and $2\no\nt=2\nd N_c$ massless Nambu-Goldstone multiplets (these are remained original electric quarks with $\nd$ flavors and $N_c$ colors). There are no heavier particles with masses $~\sim m$.

\subsection{$m\gg\lm\,,\,\, \mx\ll\lm^2/m\,,\,\, N_c < N_F < 2 N_c-1$}
\numberwithin{equation}{subsection}

\hspace*{4mm} Consider now the case of smaller $\mx$, such that $(\mx m)^{1/2}\ll\lm$, while $m$ remains large, $m\gg\lm$. The quark condensates still look as in \eqref{(6.6)}, and $\,\langle a_0\rangle=m$ {\it stays intact} and still "eats"  all quark masses $"m"$ in \eqref{(6.1)}. Therefore, because the $SU(N_c)$ theory is UV free, its gauge coupling grows logarithmically with diminished energy and, if nothing prevents, it will become $g^2(\mu)<0$ at $\mu<\lm$. Clearly, even if quarks were higgsed as above in section 6.1, i.e. $\langle Q^2\rangle=\langle{\ov Q}_2\rangle\sim (\mx m)^{1/2}$ (while $\langle X^{adj}_{SU(N_c)}\rangle
\ll\lm$), see \eqref{(6.1.1)}, this would result only in appearance of small particle masses $\sim (\mx m)^{1/2}\ll\lm$, so that {\it all particles would remain effectively massless at the scale $\mu\sim\lm\gg (\mx m)^{1/2}$} (this is especially clear in the unbroken ${\cal N}=2$ theory at $\mx\ra 0$). Therefore, this will not help and the problem with $g^2(\mu<\lm)<0$ cannot be solved in this way (these quarks ${\ov Q}_2, Q^2$ are really not higgsed now at all, i.e. $\langle Q^i_b\rangle=\langle{\ov Q}_i^{\,b}\rangle=0,\, i=\nd+1...N_F,\, b=1...N_c$, see below).

As explained in Introduction, to really avoid the unphysical $g^2(\mu<\lm)<0$, the field $X^{adj}_{SU(N_c)}$ is higgsed necessarily in this case at $\mu\sim\lm$. Because the non-trivial $Z_{\bb\geq 2}$ discrete symmetry is {\it unbroken} in these vs-vacua, this results, as in section 2.1, in $SU(N_c)\ra SU(\nd)\times U^{(1)}(1)\times U^{\bb-1}(1)$, and main contributions $\sim\lm$ to particle masses originate now from this higgsing of $X^{adj}_{SU(N_c)}$, see \eqref{(2.1.1)},\eqref{(2.1.16)},
\bq
\langle\, X^{adj}_{SU(N_c)}\rangle=\langle  X^{adj}_{SU(\nd)}+X^{(1)}_{U(1)}+ X^{adj}_{SU(\bb)}\rangle,\label{(6.2.1)}
\eq
\bbq
\langle \sqrt{2} X^{adj}_{SU(\bb)}\rangle=C_{\bb} \lm\,{\rm diag}\,\Bigl (\,\underbrace{0}_{\nd}\,;\,\underbrace{\omega^0,\,
\omega^1,\,...\,\omega^{\bb-1}}_{\bb} \,\Bigr )\,,\quad \omega=\exp\{\frac{2\pi i}{\bb}\}\,,
\eeq
\bbq
\sqrt{2}\, X^{(1)}_{U(1)}=a_{1}\,{\rm diag}(\,\underbrace{\,1}_{\nd}\,;\,\underbrace{\,c_1}
_{\bb}),\, c_1=-\,\frac{\nd}{\bb},\,\, \langle Q^i_k\rangle=\langle{\ov Q}_i^{\,k}\rangle=0,\,\,k=\nd+1...\,N_c,\, i=1...\,N_F.
\eeq

Now, vice versa, higgsed $\langle X^{A}_{SU(\bb)}\rangle\sim\lm$ {\it prevent} quarks $Q^i_k,\, {\ov Q}^{\,k}_i,\, k=\nd+1...\,N_c,\, i=1...\,N_F$ from higgsing, i.e. $\langle Q^i_k\rangle=\langle{\ov Q}_i^{\,k}\rangle=0$, see second line in \eqref{(6.2)}, $\langle a_0\rangle=m,\, \langle S\rangle_{N_c}=0$. The physical reason is that these quarks acquire now masses $\sim\lm$, much larger than a potentially possible scale $\sim (\mx m)^{1/2}$ of their coherent condensate, see \eqref{(6.1.1)}. These quarks are too  heavy and too short ranged now and it looks unrealistic that they can form the coherent condensate with  $\langle Q^i_k\rangle\neq 0$ as before in section 6.1.

Besides, for independently moving heavy ${\ov Q}$ and $Q$, their physical (i.e. path dependent) phases induced by interactions with the {\it light} $U^{2N_c-N_F}(1)$ photons fluctuate {\it independently and freely} at least at all scales $\mu > \mu^{\rm low}_{\rm cut}=(\rm several){\rm max}\{(\mx m)^{1/2},\, (\mx\lm)^{1/2} \}$. I.e., in any case, $\langle Q^i_k\rangle_{\mu^{\rm low}_{\rm cut}}=\langle{\ov Q}_i^{\,k}\rangle_{\mu^{\rm low}_{\rm cut}}=0$. Clearly, this concerns also all other heavy charged particles with masses $\sim\lm$. Therefore, after integrating out the theory \eqref{(6.1)} over the interval from the high energy down to the scale $\mu^{\rm low}_{\rm cut}$, all particles with masses $\sim\lm$ {\it decouple} as heavy already at the scale $\mu<\lm/(\rm several)$, i.e. much above $\mu^{\rm low}_{\rm cut}$, and definitely {\it the heavy quarks by themselves do not give any contributions to masses of remaining lighter degrees of freedom in the Lagrangian \eqref{(6.2.2)}} because $\langle Q^i_k\rangle_{\mu^{\rm low}_{\rm cut}}=\langle{\ov Q}_i^{\,k}\rangle_{\mu^{\rm low}_{\rm cut}}=0$.
And after the heavy particles decoupled at $\mu<\lm/(\rm several)$, they themselves do not affect the lower energy theory. This is exactly the same situation as with heavy quarks with masses $\sim\lm$ discussed in detail in section 2.1 (and qualitatively the same as with $X^{adj}_{SU(\nd)}$ with masses $\sim\mx\gg (\mx m)^{1/2}$ in section 2.4), and only the parameter $m\gg\lm$ in this section while $m\ll\lm$ in sections 2.1 and 2.4, and always $\mx\ll\lm$.

The lower energy original electric ${\cal N}=2\,\,SU(\nd)$ theory at the scale $\mu=\mu^{\rm low}_{\rm cut}\ll\lm$, with $2\nd<N_F<2N_c-1$ flavors of remained light quarks $Q^i_a, {\ov Q}_j^{\,a},\,\, a=1...\nd,\,\, i,j=1...N_F$, is IR free and weakly coupled. The whole matter superpotential has the same form as \eqref{(2.1.5)}, but with  omitted $a_2$ and $SU(\nd-\no)$ SYM part, $\no=\nd$, $m\ra (m-a_0)$, and with the addition of $\sim\mx a_0^2\,$,
\bq
\w_{N_c}=\w_{\,\nd}+\w_{D}+\w_{a_0,\,a_1}\,,\label{(6.2.2)}
\eq
\bbq
\w_{\,\nd} =(m-a_0-a_1){\rm Tr}\,({\ov Q} Q)_{\nd}-{\rm Tr}\,\Bigl ({\ov Q}\sqrt{2}X_{SU(\nd)}^{\rm adj} Q\Bigr )_{\nd}+\mx (1+\delta_2){\rm Tr}\,(X^{\rm adj}_{SU(\nd)})^2\,,
\eeq
\bbq
\w_{D}=\Bigl ( m-a_0-c_1 a_1 \Bigr )\sum_{j=1}^{\bb}{\ov D}_j D_j\,-\sum_{j=1}^{\bb} a_{D,j}{\ov D}_j D_j
-\,\mx\lm \sum_{j=1}^{\bb}\omega^{j-1} a_{D,j}+\mx  L\,\Bigl (\,\sum_{j=1}^{\bb} a_{D,j}\Bigr )\,,
\eeq
\bbq
\w_{a_0,\,a_1}=\frac{\mx}{2}N_c\,a^2_0+\frac{\mx}{2}\frac{\nd N_c}{\bt}\,(1+\delta_1)\,a^2_1+
\mx N_c\delta_3\, a_1(m-a_0-c_1 a_1)+\mx N_c\delta_4\, (m-a_0-c_1 a_1)^2\,,
\eeq
where $\delta_1$ and $\delta_2$ are the same as in \eqref{(2.1.12)} (and $\delta_3=0$, see \eqref{(B.4)}).

Proceeding now similarly to section 2.1, we obtain from \eqref{(6.2.2)}
\bq
\langle a_0\rangle=m\,,\quad \langle a_1\rangle=0\,,\quad \langle a_{D,j}\rangle=0\,,
\quad \langle X_{SU(\nd)}^{\rm adj}\rangle=0\,,\label{(6.2.3)}
\eq
\bbq
\langle{\ov D}_j D_j\rangle=\langle{\ov D}_j\rangle\langle D_j \rangle=\,-\mx\lm\,\omega^{j-1}+\mx \langle L\rangle,\,\, \langle L\rangle=m,\,\, \langle\Sigma_D\rangle=\sum_{j=1}^{\bb}\langle{\ov D}_j\rangle\langle D_j \rangle=(\bb)\mx m,
\eeq
\bbq
\langle\Qo\rangle_{\nd}=\langle{\ov Q}^1_1\rangle\langle Q^1_1\rangle=\mx m=\langle\Qt\rangle_{N_c},\,\, \langle\Qt\rangle_{\nd}=\sum_{a=1}^{\nd}\langle{\ov Q}^{\,a}_2\rangle\langle Q^2_a\rangle=0=\langle\Qo\rangle_{N_c}\,,
\eeq
\bbq
\langle{\rm Tr} ({\ov Q} Q)\rangle_{\nd}=\nd\langle\Qo\rangle_{\nd}+N_c\langle\Qt\rangle_{\nd}=\nd\mx m\,, \quad \wmu\langle S\rangle_{\nd}=\langle\Qo\rangle_{\nd}\langle\Qt\rangle_{\nd}=0.\,\,\,
\eeq

From \eqref{(6.2.3)}, the multiplicity of these vs -vacua in this phase is $N_{vs}=C^{\nd}_{N_F}=C^{N_c}_{N_F}$ as it should be, the factor $C^{\nd}_{N_F}$ originates now from the spontaneous flavor symmetry breaking, $U(N_F)\ra U(\nd)\times U(N_c)$, due to higgsing of $\nd$ quarks in the $SU(\nd)$ color sector.

On the whole, the mass spectrum looks now as follows.\\
1) Due to higgsing of $X^{adj}_{SU(\bb)}$ at the scale $\sim\lm$, $\,\,SU(N_c)\ra SU(\nd)\times U^{(1)}(1)\times U^{\bb-1}(1)$, there is a large number of original pure electrically $SU(\bb)$ charged particles with the largest masses $\sim\lm$. They all are weakly confined due to higgsing at the scale $\mu\sim (\mx m)^{1/2}$ of mutually non-local with them $\bb$ BPS dyons $D_j, \ov D_j$ with the nonzero $SU(\bb)$ adjoint magnetic charges, the string tension is $\sigma^{1/2}_D\sim \langle{\ov D}_j D_j\rangle^{1/2}\sim (\mx m)^{1/2}\ll\lm$. These original electrically charged particles form hadrons with the mass scale $\sim\lm\,$. Clearly, higgsed $\langle X^{adj}_{SU(\bb)}\rangle\sim\lm$ does not break {\it by itself} the global flavor symmetry. All heavy quarks with $SU(\bb)$ colors have equal masses $|C_{\bb}\lm|$ and are in the (anti)fundamental representation of the non-Abelian $SU(N_F)$, they form e.g. a number of $SU(N_F)$ adjoint hadrons with masses $\sim\lm$. Original $SU(\bb)$ adjoint heavy gluons and scalars are flavorless and have different masses $|\Lambda_j-\Lambda_k|,\, \Lambda_j=C_{\bb}\lm\omega^{j-1},\, \omega=\exp\{2\pi i/(\bb)\}$. {\it There is no color-flavor locking in this heaviest sector}\,: the global $U(N_F)$ is unbroken while color $SU(\bb)\ra U^{\bb-1}(1)$. \\
2) Due to higgsing of these dyons $D_j, {\ov D}_j$, there are $\bb$ long ${\cal N}=2\,\, U(1)$ multiplets of massive photons, all with masses $\sim \langle{\ov D}_j D_j\rangle^{1/2}\sim (\mx m)^{1/2}$.\\
3) There are $\nd^{\,2}$ long ${\cal N}=2$ multiplets of massive gluons with masses $\sim \langle\Qo\rangle^{1/2}_{\nd}\sim (\mx m)^{1/2}\ll\lm$ due to higgsing of $\nd$ flavors of original electric quarks with $SU(\nd)$ colors.\\
4) $2 \no\nt=2\nd N_c$ (complex) Nambu-Goldstone multiplets are massless (in essence, these are remained original electric quarks with $\nt=N_c=N_F-\nd$ flavors and $SU(\nd)$ colors). There are no other massless particles at $\mx\neq 0,\, m\neq 0$.\\
5) There are no particles with masses $\sim m\gg\lm$.

The $U(N_c)$ curve \eqref{(1.2)} has in this phase at $\mx\ra 0$\,\, $\nd$ equal double roots $e_k= - m$ corresponding to $\nd$ out of $N_F$ higgsed original electric quarks with $SU(\nd)$ colors, and $\bb$ unequal double roots $e_j= -m+\omega^{j-1}\lm$ corresponding to dyons $D_j$.

It is worth recalling once more that higgsed $\langle\, X^{adj}_{SU(\bb)}\rangle\sim\lm$ in \eqref{(6.2.1)} {\it does not break by itself the global flavor symmetry, it is really broken only in the $SU(\nd)$ color sector by higgsed $\nd$ quarks with $N_F$ flavors and $SU(\nd)$ colors}.\\

In essence, the situation here with the heavy non-higgsed quarks $Q^i_k, {\ov Q}^{\,k}_i,\, i=1...N_F,\, k=\nd+1...N_c$ with masses $\sim\lm$ and with their chiral bilinear condensates (really, it is more adequate to speak about mean values, using the word "condensate" only for "genuine", i.e. coherent, condensates of higgsed fields) {\it is the same} as in br2 vacua of section 2.1, see \eqref{(2.1.16)}-\eqref{(2.1.26)} and accompanying text. In the lower energy phase of this section at $\mx\ll\lm^2/m$, the nonzero mean values of {\it non-factorizable} local bilinears $\langle(\qq)_{1.2}\rangle_{\bb}$ of these heavy quarks with masses $\sim\lm$ originate from: a) the quantum loop effects in the strong coupling (and non-perturbative) regime at the scale $\mu\sim\lm$, transforming these bilinears {\it operators} of heavy quark fields into bilinears {\it operators} of light $SU(\nd)$ quark fields and the {\it operator} $\Sigma_D$; b) and finally, from formed at much lower scales $\mu\sim (\mx m)^{1/2}\ll\lm$ the genuine (i.e. coherent) condensates of light higgsed quarks and dyons. I.e., the equations \eqref{(2.1.17)}-\eqref{(2.1.19)} and \eqref{(2.1.22)}-\eqref{(2.1.26)} (with $A=-2$) remain the same and only the values of condensates are different in br2 and vs vacua, see \eqref{(6.2.3)}.

In this section, the corresponding values of condensates of heavy quarks with $SU(\bb)$ colors, $N_F$ flavors, and masses $\sim\lm$ in these vs-vacua with $\nt=N_c,\, \no=\nd$, and $\langle S\rangle_{N_c}=0$ differ from those in the br2 vacua with $\nt>N_c,\, \no<\nd$ and $\langle S\rangle_{N_c}\neq 0$ in section 2.1 only by: 1) $m_1\ra m$ due to $SU(N_c)\ra U(N_c)$,\,\, 2) the absence of non-leading power corrections originating in br2 vacua from the SYM part, see \eqref{(6.3)},\eqref{(6.2.3)}, compare with \eqref{(6.1.1)}, \eqref{(2.1.17)},\eqref{(2.1.19)},
\bq
\langle\Qt\rangle_{N_c}=[\langle\Qt\rangle_{\bb}=\mx m\,]+[\langle\Qt\rangle_{\nd}=\sum_{a=1}^{\nd}\langle
{\ov Q}^{\,a}_2\rangle\langle Q^2_a\rangle=0\,]=\mx m\,, \label{(6.2.4)}
\eq
\bbq
\langle\Qo\rangle_{N_c}=[\langle\Qo\rangle_{\nd}=\langle{\ov Q}^{\,1}_1\rangle\langle Q^1_1\rangle=\mx m\,]+[\langle\Qo\rangle_{\bb}= - \mx m\,]=0\,,
\eeq
\bbq
\langle{\rm Tr\,}\qq\rangle_{\bb}=N_c\langle\Qt\rangle_{\bb}+\nd\langle\Qo\rangle_{\bb}=\langle
\Sigma_D\rangle=(2N_c-N_F)\mx m\,,
\eeq
while $\langle \sqrt{2} X^{adj}_{SU(\bb)}\rangle$ in \eqref{(6.2.1)} is the same as \eqref{(2.1.17)} and differs from \eqref{(6.1.1)}.\\

There is one important qualitative difference between the vs vacua and br2 vacua. Because \eqref{(6.3)},
\eqref{(6.4)},\eqref{(6.2.2)},\eqref{(6.2.3)} remain the same at $m\lessgtr\lm$, and $\langle a_0\rangle=m$ still "eats" all quark masses $"m"$ in \eqref{(6.1)} at all $m\lessgtr\lm$, the whole physics in vs vacua is qualitatively independent separately of the ratio $m/\lm$ and {\it depends only on a competition between $(\mx m)^{1/2}$ and $\lm$}. For this reason, the lower energy phase with $m\gg\lm,\, (\mx m)^{1/2}\ll\lm$ in vs vacua of this section continues smoothly into the region $\mx\ll m\ll\lm$. This is in contrast with the br2 vacua with $\mx\ll m\ll\lm$ in section 2.1. In the latter, all properties depend essentially on the ratio $m/\lm$ and they evolve into the qualitatively different br1 vacua of section 4 at $m\gg\lm$.\\

Comparing the mass spectra at $(\mx m)^{1/2}\gg\lm$ in section 6.1 with those in this section at $(\mx m)^{1/2}\ll\lm$ one can see the qualitative difference between these two cases. {\it The lower energy phase with $\mx\ll\lm^2/m$ in this section is, in essence, the phase of the unbroken ${\cal N}=2$ theory because it continues smoothly down to $\mx\ra 0$. While the higher energy phase of section 6.1 with $\mx\gg\lm^2/m$, because it cannot be continued as it is down to $\mx\ra 0$, is, in essence, the phase of the strongly broken ${\cal N}=2$ theory (i.e. already the genuine ${\cal N}=1$ theory in this sense)}. Remind that the applicability of the whole machinery with the use of roots of the curve \eqref{(1.2)} is guaranteed only at $\mx\ra 0$. I.e., it is applicable only to the lower energy phase of this section, and clearly not applicable to the higher energy phase of section 6.1, in spite of that $\mx\ll\lm$ therein also.

And these two phases at $\mx\gtrless\lm^2/m$ are qualitatively different, compare e.g. \eqref{(6.1.1)} and \eqref{(6.2.1)}. (In the higher energy phase with
$\langle{\ov Q}^{\,b}_{i=b+\nd}\rangle=\langle Q^{i=b+\nd}_b\rangle=(\mx m)^{1/2},\,\,b=1...N_c,\,\,i=\nd+1...N_F$ and $\langle m - a_0\rangle=0,\,\langle S\rangle_{N_c}=0$, the mean value $\langle X^{\rm adj}_{SU(N_c)}\rangle\neq 0$ {\it is incompatible with the Konishi anomaly in} \eqref{(6.2)},\eqref{(6.6)}). Besides, in this higher energy phase in section 6.1 all charged and neutral gluons from $SU(\bb)$ have {\it exactly equal masses $| g (\mx m)^{1/2}|$ in the whole wide region from $(\mx m)^{1/2}\gg\lm$ down to $(\mx m)^{1/2} >(\rm several)\lm$}, see  \eqref{(6.1.1)}. On the other hand, in the lower energy phase of this section, the masses of charged $SU(\bb)$ gluons differ greatly at $\mx\ra 0:\, \mu_{\rm gl}^{ij}\sim |(\omega^{i}-\omega^{j})\lm |\,,\,\, i\neq j,\,\,i,j=0...\bb-1$, see \eqref{(6.2.1)}, and differ from masses of neutral gluons. This shows that the dependence of masses of these gluons on $\mx m$ is non-analytic. The mean values of $\langle X^{adj}_{SU(N_c)}\rangle$ and $\langle Q^i_k\rangle=\langle{\ov Q}_i^{\,k}\rangle,\, k=\nd+1...\,N_c,\, i=1...\,N_F$ also behave non-analytically, compare \eqref{(6.1.1)} and \eqref{(6.2.1)}. In this sense, we can say that there is a phase transition somewhere at $(\mx m)^{1/2}\sim\lm$. This is in contradiction with a widespread opinion that there are no phase transitions in the supersymmetric theories because the gauge invariant mean values of chiral fields ( e.g. $\langle (\qq)_{1,2}\rangle_{N_c},\,\langle S\rangle_{N_c},\, \,\langle{\rm Tr\,}(X^{adj}_{SU(N_c)})^2\rangle$ in \eqref{(6.2)}-\eqref{(6.4)}) depend holomorphically on the superpotential parameters.

This phenomenon with the phase transition of the type described above occurring at very small $\mx\ll\lm$ from the nearly unbroken ${\cal N}=2$ at $\mx\ll\lm^2/m$ to the strongly broken ${\cal N}=2$ at $\mx\gg\lm^2/m$, i.e. ${\cal N}=2\ra {\cal N}=1$, where the use of roots of the curve \eqref{(1.2)} becomes not legitimate, appears clearly {\it only} at $m\gg\lm$. Let us start e.g. with the nearly unbroken ${\cal N}=2$ theory at {\it sufficiently small} $\mx$ where, in particular, the whole machinery with the use of roots of the curve \eqref{(1.2)} is still legitimate (for independent of $\mx$ quantities). In this section, when $\mx$ begins to increase, the phase transition occurs at $\mx\sim (\lm^2/m)\ll\lm$. But in other cases, see sections 7 and 8 below, it occurs even at parametrically smaller values of $\mx$ than in this section. In br2 vacua of section 7 it occurs at $\mx\sim \langle\Lambda_{SU(\nt)}\rangle^2/m,\, \langle\Lambda_
{SU(\nt)}\rangle\ll\lm$, see \eqref{(7.1.6)}, and in S- vacua of section 8 it occurs at $\mx\sim \langle\Lambda_{SU(N_F)}\rangle^2/m,\, \langle\Lambda_{SU(N_F)}\rangle\ll\lm$, see \eqref{(8.4)}. \\

Let us compare now the corresponding parts of unbroken non-Abelian global symmetries of two phases, those in section 6.1 and here. It is $SU(N_c)_{C+F}\times SU(\nd)_F$ of original pure electric particles in section 6.1, where "C+F" denotes the color-flavor locking. The higgsed massive electric quarks $Q^{i=b+\nd}_b, {\ov Q}^{\,b}_{i=b+\nd},\,\, b=1...N_c,\, i=\nd+1...N_F$, with masses $m_Q\sim (\mx m)^{1/2}\gg\lm$, together with massive electric $SU(N_c)$ gluons and $X^{adj}_{SU(N_c)}$ scalars form the long ${\cal N}=2$ multiplet in the adjoint representation of  $SU(N_c)_{C+F}$. The $2\nd N_c$ massless Nambu-Goldstone multiplets are bifundamental, these are original non-higgsed quarks $Q^i_a$ and ${\ov Q}^a_i$ with $\nd$ flavors and $N_c$ colors.

The corresponding non-Abelian global symmetry in this section is $SU(\nd)_{C+F}\times SU(N_c)_F$ in the lighter sector, and higgsed massive electric quarks $Q^{i=a}_a, {\ov Q}^{\,a}_{i=a},\,\, a,i=1...\nd$, with masses $m_Q\sim (\mx m)^{1/2}\ll\lm$, together with $SU(\nd)$ gluons and $X^{adj}_{SU(\nd)}$ scalars form the long ${\cal N}=2$ multiplet in the adjoint representation of $SU(\nd)_{C+F}$, see the point "3" just above. The $2\nd N_c$ massless Nambu-Goldstone multiplets are also bifundamental. But, in addition, {\it all $N_F$ flavors of original electric quarks with $SU(\bb)$ colors have equal large masses $\sim\lm$, so that the global $SU(N_F)$ is unbroken in this heavy sector. In the limit $\mx\ra 0$ these heavy quarks are not confined and form the (anti)fundamental representation of $SU(N_F)$}. All original unconfined at $\mx\ra 0$ $\,SU(\bb)$-adjoint and hybrid gluons and scalars with the large masses $\sim\lm$ are clearly flavor singlets. At small $\mx\neq 0$ all these heavy pure electrically charged particles are weakly confined and form hadrons with masses $\sim\lm$, but this confinement does not break {\it by itself} $SU(N_F)$. In particular, the heavy quarks form a number of $SU(N_F)$ adjoint hadrons.

Besides, it is worth noting the following.  It is erroneous to imagine that, because $\langle\Qt\rangle_{N_c}=\mx m$ still in the $\mx m\ll\lm^2$ phase of this section, then all quarks with $n_2 = N_c$ flavors are still higgsed in this lower energy phase as previously in section 6.1 at $\mx m\gg\lm^2$, i.e. $\langle{\ov Q}^{\,b}_{i=b+\nd}\rangle=\langle Q^{i=b+\nd}_b\rangle=(\mx m)^{1/2},\,\,b=1...N_c,\,\,i=\nd+1...N_F$, and so give corresponding contributions to particle masses, but because $(\mx m)^{1/2}\ll\lm$ now, this will be of little importance. It is sufficient to notice that, with $\langle X^{\rm adj}_{SU(\bb)}\rangle\sim\lm\neq 0$ now, see \eqref{(6.2.1)}, this variant with still higgsed quarks with all $N_c$ colors is incompatible with the Konishi anomalies in \eqref{(6.2)},\eqref{(6.6)}.\\

The difference of adjoint representations $SU(N_c)_{C+F}\leftrightarrow SU(\nd)_{C+F}$ at $\mx m\gtrless\lm^2$ was considered e.g. in \cite{SY4,SY7,SY2} as an evidence that all $SU(\nd)$ particles cannot be the same original pure electric particles and, by analogy with \cite{SY3}, it was finally claimed that they all are dyons. E.e., the dyonic quarks ${\ov d}^{\,b}_i,\, d^i_b$ with $N_F$ flavors of this $SU(\nd)$ are composites of original flavored quarks with $SU(\nd)$ colors and corresponding flavorless magnetic monopoles.

This line of reasonings is in a clear contradiction with the picture described above in this section. The reason for different adjoint representations is that the phase and mass spectrum are determined by $N_c$ out of $N_F$ higgsed quarks of $SU(N_c)$ at $\mx m\gg\lm^2$ (and $X^{adj}_{SU(N_c)}$ is not higgsed,
$\langle X^{adj}_{SU(N_c)}\rangle=0$, see \eqref{(6.1.1)}), while they are determined by higgsed both $X^{\rm adj}_{SU(2N_c-N_F)}$ and $\nd$ out of $N_F$ quarks of $SU(\nd)$ at $\mx m\ll\lm^2$, see \eqref{(6.2.1)},\eqref{(6.2.3)}, (both phases in section 6.1 and here at $m\gg\lm$, and the case $\mx\ll m\ll\lm$ clearly corresponds to the regime $\mx m <\lm^2$, see section 6.3 below).

Besides, the above described in this section  picture with the {\it phase transition} at $\mx m\gtrless\lm^2$ (to avoid entering the unphysical regime with $g^2(\mu<\lm)<0$ without higgsed $\langle X^{adj}_{SU(\bb)}\rangle\sim\lm$), is in a clear contradiction with the proposed by M.Shifman and A.Yung  {\it a smooth analytic crossover} from the region $\mx m\gg\lm^2$ to the "instead-of-confinement" regime in the region $\mx m\ll\lm^2$, with {\it still unbroken the non-Abelian}  $SU(N_c)_{C+F}$ global symmetry as it was in section 6.1 above, see e.g. \cite{SY4,SY7,SY2}. And the smooth analytic crossover implies that
$N_c$ of $N_F$ original $SU(N_c)$ quarks are {\it still higgsed}, see \eqref{(6.1.1)}, and only the value of their condensate is smaller now,
\bq
\langle{\ov Q}_{i=b+\nd}^{\, b} Q^{i=b+\nd}_b\rangle=\langle{\ov Q}_{i=b+\nd}^{\, b}\rangle\langle Q^{i=b+\nd}_b\rangle= (\mx m)^{1/2}\ll\lm\,,\quad  b=1...N_c\,,\quad i=\nd+1...N_F\,.\label{(6.2.5)}
\eq

\vspace*{2mm}

At $\mx m\ll\lm^2$, according to M.Shifman and A.Yung, see e.g. \cite{SY6} and refs therein to their previous related papers. -

a) There is a large number of light particles with masses $\ll\lm$ which all are dyonic solitons. The first part of this set is the $SU(\nd)$ gauge theory with $N_F$ flavors of above dyonic quarks $d^i_b,\,{\ov d}^{\,b}_i,\, i=1...N_F,\, b=1...\nd$. $\,\nd$ them of $N_F$ are higgsed at small $\mx\neq 0$ and, together with dyonic $SU(\nd)$ gluons and scalars, form the adjoint representation of the unbroken $SU(\nd)_{C+F}$ global symmetry. The second part of this set includes $2 N_c\nd$ massless Nambu-Goldstone particles which are remained non-higgsed dyonic quarks $d^i_b,\,{\ov d}^{\,b}_i,\, i=\nd+1...N_F,\, b=1...\nd$. The third part includes $U^{\bb}(1)$ photon multiplets and $\bb$ dyons $\ov{\textsf{D}}_n,\, {\textsf{D}}_n$ which are composites of diagonal original quarks ${\ov Q}^{\,n}_n,\, Q^n_n,\, n=1 ... \bb$ and corresponding $SU(\bb)$ magnetic monopoles\,: ${\textsf{D}}_n=( Q^n_n+M_n)$. All quark-like dyons (except for Nambu-Goldstone particles) are higgsed  at $\mx\neq 0$.

As was emphasized in section 2.1, as a result of this picture, {\it the dyons} ${\textsf{D}}^i_n,\, n=1...\bb,\, i=1...N_F$ {\it will really realize the fundamental representation $N_F$ of the flavor} $SU(N_F)$ and this will result in a wrong pattern of the flavor symmetry breaking at $\mx\neq 0$. Besides, the number of such dyons ${\textsf{D}}^i_n$ will be too large.

b) At the same time, in the region $\mx m\ll\lm^2$ of this section {\it all original electric particles from the unbroken adjoint representation of $SU(N_c)_{C+F}$ in section 6.1 mysteriously acquire large equal masses} $\sim\lm$ (in addition to equal contributions $\sim (\mx m)^{1/2}$ to their masses from still higgsed quarks, see \eqref{(6.2.5)}).

It is not even attempted to recognize the dynamical mechanism responsible for the origin of these large equal masses $\sim\lm$ of all original particles forming the long ${\cal N}=2$ adjoint representation of $SU(N_c)_{C+F}$.

Let us recall that in the whole weak coupling region $\mx m\gg\lm^2$, with quarks higgsed as in \eqref{(6.1.1)}, it follows from the Konishi anomalies \eqref{(6.2)},\eqref{(6.6)} (valid for each flavor separately) that all $\langle X^A\rangle=0,\, A=1...N_c^2-1$. If, according to M.Shifman and A.Yung, the transition from $\mx m\gg\lm^2$ to $\mx m\ll\lm^2$ is a smooth analytic crossover, then all $\langle X^A\rangle=0$ at $\mx m\ll\lm^2$ also and quarks are higgsed still as in \eqref{(6.1.1)}. From where then appear large masses $\sim\lm$ of all original particles forming the adjoint representation of $SU(N_c)_{C+F}$ ? This contradicts the BPS properties of original particles.

And it is not attempted to answer the question: are the properties of this mysterious mechanism responsible for the appearance of these large masses $\sim\lm$  compatible with the smooth analytic crossover and with the unbroken global $SU(N_c)_{C+F}$ symmetry ? Recall that the described above in this section concrete dynamical mechanism responsible for the appearance of large masses $\sim\lm$ of all original {\it charged} particles with $SU(\bb)$ colors, i.e. higgsed $\langle X^{adj}_{SU(\bb)}\rangle\sim\lm$ \eqref{(6.2.1)}, is incompatible with the unbroken global $SU(N_c)_{C+F}$ requiring equal masses of all $N_c^2-1$  $\,SU(N_c)_{C+F}$ adjoint particles. E.g., charged $SU(\bb)$ adjoint gluons have different masses.

Besides, as explained in detail in section 2.1, e.g. the quarks $Q^i_a,\,{\ov Q}^a_i,\, a=1...\bb,\, i=1...N_F$ with masses $\sim\lm$ at $(\mx m)^{1/2}\ll\lm$ are too heavy and too short ranged and cannot form the coherent condensate with $\langle Q\rangle_{\lm/(\rm several)}=\langle{\ov Q}\rangle_{\lm/(\rm several)}\sim (\mx m)^{1/2}$, because they decouple already at the scale $\sim\lm/(\rm several)\gg (\mx m)^{1/2}$ where $\langle Q\rangle_{\lm/(\rm several)}=\langle{\ov Q}\rangle_{\lm/(\rm several)}=0$ and do not affect by themselves further lower energy evolution of remained lighter fields. In other words, as opposed to \eqref{(6.2.5)}, the nonzero total mean values of heavy quarks bilinears $\langle\Qt\rangle_{\bb}$ and $\langle\Qo\rangle_{\bb}$ in \eqref{(6.2.4)} are non-factorizable. As explained in detail in section 2.1, they are induced by nonzero genuine condensates of light higgsed quarks with $SU(\nd)$ colors and light dyons ${\ov D}_j, D_j$.

Moreover, e.g. the appearance at $(\mx m)^{1/2}\ll\lm$ of the quark mass terms $\sim\lm \sum_{a,i=1}^{N_c}{\ov Q}^a_i Q^i_a$ in the superpotential contradicts the unbroken in these vs-vacua $Z_{\bb}$ symmetry, and also the mass terms $\sim\lm{\rm Tr} (X^{adj}_{SU(N_c)})^2$ of all $N_c^2-1$ scalars in the superpotential contradict both the unbroken $U_{R}(1)$ and $Z_{\bb}$ symmetries. The very idea that original quarks acquired their masses $\sim\lm$ at $\mx\ra 0$ not from higgsed $\langle X^{adj}_{SU(\bb)}\rangle\sim\lm$ but from some (unrecognized) outside sources contradicts their BPS properties.  Recall that in similar br2 vacua in section 2.1, with higgsed $\langle X^{\rm adj}_{SU(\bb)}\rangle\sim\lm$ in \eqref{(2.1.1)}, the condensates of light higgsed original quarks $\langle\Qo\rangle=\langle{\ov Q}^1_1\rangle\langle Q^1_1\rangle$ in \eqref{(2.1.14)} agree with {\it independently} calculated their values from roots of the curve \eqref{(1.2)} (and the same in vs-vacua considered here with $\no=\nd$). And because the values calculated from roots are valid only for BPS particles,
{\it this confirms the BPS properties of original quarks}.

c) In addition, in this variant with the smooth analytic crossover, all dyons $d^k_b,\, k,b=1...\nd$, ${\textsf{D}}_n,\, n=1...\bb$ {\it and still quarks} $Q^i_a,\, a=1...N_c,\, i=\nd+1...N_F$ are {\it higgsed simultaneously} with $\langle d\rangle\sim \langle Q\rangle\sim (\mx m)^{1/2}\,,\, \langle{\textsf{D}}\rangle\sim \max [(\mx\lm)^{1/2},\, (\mx m)^{1/2}]$, see e.g. \eqref{(6.2.5)}. But, in any case, these quarks $Q^i_a$ are definitely mutually non-local with respect to these dyons. How then can it be?

Comparing properties of proposed in \cite{SY4,SY7,SY2,SY6} the smooth analytic crossover between regions $\mx m\gtrless\lm^2$ (with the "instead-of-confinement" regime in the region $\mx m\ll\lm^2$), and those with the phase transition described above in this section, it is seen that these properties are, so to say, "orthogonal" to each other. And, on account of critical remarks presented above, we consider the proposal of the smooth analytic crossover and the "instead-of-confinement" regime as not self-consistent. \\

In addition, we would like to especially emphasize (and this concerns the whole content of this paper, not only those of this section) that, for {\it higgsed fields giving nonzero contributions to particle masses}, {\it the mean values of fields themselves should be nonzero}, i.e. $\langle Q\rangle=\langle{\ov Q}\rangle\neq 0$ or $\langle X\rangle\neq 0$, {\it not only} something like $\langle Q^{\dagger} Q\rangle^{1/2}\neq 0$ or $\langle X^{\dagger} X\rangle^{1/2}\neq 0$ for the D-terms, or $\langle{\ov Q} Q\rangle^{1/2}\neq 0,\, \langle X^2\rangle^{1/2}\neq 0$ for the F-terms (which are, as a rule, nonzero even for those fields which are really not higgsed).

This is especially clearly seen  e.g. from the mass terms of fermionic superpartners. For the D-terms of {\it higgsed} quarks or $X$ these look as:
\footnote{\,
Really $\langle ...\rangle$ in \eqref{(6.2.6)},\eqref{(6.2.7)} have to be understood as $\langle ...\rangle_{\mu^{low}_{cut}=M/(\rm several)}$, where $M$ is the pole mass of $Q,\, {\ov Q}$ or $X$.
}
\bq
{\rm Tr\,}\Bigl [\,\langle Q^{\,\dagger}\rangle\lambda \chi+{\ov\chi}\lambda\langle{\ov Q}^{\,\dagger}\rangle\,\,\Bigr ]\quad {\rm or}\quad {\rm Tr\,}\Bigl (\,\langle X^{\dagger}\rangle\,[\psi, \lambda]\,\Bigr )\,,\label{(6.2.6)}
\eq
and for the F-terms they look as
\bq
{\rm Tr\,}\Bigl [\,{\ov\chi}\psi\langle Q\rangle+\langle{\ov Q}\rangle\psi\chi\,\Bigr ]\quad {\rm or}\quad {\rm Tr\,}\Bigl [\,{\ov \chi}\langle X\rangle \chi\,\Bigr ]\,.\label{(6.2.7)}
\eq

\subsection{$m\ll\lm$}

\hspace*{4mm} Comparing with the case $\mx\ll\lm,\,\, m\gg\lm$ in sections 6.1-6.2, the case $\mx\ll m\ll\lm$ corresponds clearly the regime $\mx\ll\lm^2/m$. The quark condensates remain the same as in \eqref{(6.2.3)}, and $\langle a_0\rangle=m$ remains the same as in \eqref{(6.2.3)} and still "eats" all quark masses $"m"$. {\it The phase remains the same as in section} 6.2, and at $m\ll\lm$ the difference with the case $m\gg\lm,\,\, \mx\ll\lm^2/m$ in section 6.2 is only quantitative, not qualitative, see \eqref{(6.2.3)}. The main difference is that $\langle D_j\rangle\sim (\mx m)^{1/2}\gg (\mx\lm)^{1/2}$ at $m\gg\lm$, while $\langle D_j\rangle\sim (\mx\lm)^{1/2}\gg (\mx m)^{1/2}$ at $m\ll\lm$. Still, there are no particles with masses $\sim m$, the only particle masses independent of $\mx$ are $\sim\lm$.

\subsection{$N_F=N_c$}

Now, in short about the vs vacuum with $N=N_F=N_c=\nt,\, \nd=0=\no$ at $m\lessgtr\lm$ and $\mx\ll\lm$. There is only one such vacuum. And really, there is nothing extraordinary with this case.

\subsubsection{Higher energy phase at $(\mx m)^{1/2}\gg\lm$}

As always in all vs vacua, see \eqref{(6.3)},\eqref{(6.4)}, $\langle\qq\rangle_{N_c}=\mx m,\, \langle S\rangle_{N_c}=0$ in this vacuum, and $\langle a_0\rangle=m$ "eats" all quark masses. The whole color group $U(N_c)$ is broken in this higher energy phase as before in vs-vacua of section 6.1 by higgsed quarks at the scale $\mu\sim (\mx m)^{1/2}\gg\lm$ in the weak coupling regime, while $\langle X^{adj}_{SU(N_c)}\rangle=0$ is not higgsed, see \eqref{(6.1.1)}. {\it All $SU(N_c)$ gluons have equal masses $\sim (\mx m)^{1/2}$ in this phase in the whole wide region from $(\mx m)^{1/2}\gg\lm$ down to $(\mx m)^{1/2} >(\rm several)\lm$}. There is the residual global non-Abelian $SU(N)_{C+F}$ symmetry (i.e. color-flavor locking). $N^2-1$ long ${\cal N}=2$ multiplets of massive gluons are formed in the adjoint representation of this remained unbroken global symmetry $SU(N)_{C+F}$. There are no particles with masses $\sim m$.

The global $SU(N)_{F}$ is substituted by global $SU(N)_{C+F}$, so that there are no massless Nambu-Goldstone particles in this vacuum. And there are no massless particles at all in this phase.

\subsubsection{Lower energy phase at $(\mx m)^{1/2}\ll\lm$}

As in all vs-vacua, the total mean value of higgsed $a_0=\langle a_0\rangle^{(\rm several) m}_{m/(\rm several)}+[\,{\hat a}_0^{\rm soft}\,]^{m/(\rm several)}$ originates and saturates at $\mu\sim m$ and "eats" masses "m" of all quarks
\bq
\langle a_0\rangle=\frac{\langle{\,\rm Tr\,}\qq\rangle_{N_c}}{N_c\mx }=m\,,\quad \langle a_0\rangle=[\,\langle a_0\rangle^{(\rm several) m}
_{m/(\rm several)}=m\,]+[\,\langle {\hat a}_0^{\rm soft}\rangle^{m/(\rm several)}_{\mu_{\rm cut}^{\rm lowest}=0}=0 \,]\,. \label{(6.4.1)}
\eq
In this phase, at all $m\gtrless\lm$, to avoid entering the unphysical regime $g^2(\mu<\lm)<0$ in this UV free theory, the non-Abelian color $SU(N_c)$ is also broken in the strong coupling and non-perturbative regime at the scale $\mu\sim \lm$ by higgsed $\langle X^{adj}_{SU(N_c)}\rangle\sim\lm,\, SU(N_c)\ra U^{N_c-1}(1)$, see \eqref{(2.1.1)},\eqref{(6.2.1)},
\bq
X^{adj}_{SU(N_c)}=\langle X^{adj}_{SU(N_c)}\rangle^{(\rm several)\lm}_{\lm/(\rm several)}+[\, {\hat X}^{adj,\rm soft}_{SU(N_c)} \,]^{\lm/(\rm several)}\,, \label{(6.4.2)}
\eq
\bbq
\langle\sqrt{2} X^{adj}_{SU(N_c)}\rangle^{(\rm several)\lm}_{\lm/(\rm several)}=C_N\lm\,{\rm diag}\,\Bigl(\,\rho^0,\,\rho^1,\,...,\,\rho^{N_c-1}\,\Bigr ),\quad \rho=\exp\{\,\frac{2\pi i}{N_c}\,\}\,,\quad \langle{\hat X}^{adj,\rm soft}_{SU(N_c)}\rangle^{\lm/(\rm several)}_{\mu_{\rm cut}^{\rm lowest}=0}=0\,. \eeq

The unbroken non-trivial discrete symmetry is $Z_{2N_c-N_F}=Z_{N\geq 2}$ in this vacuum, and it determines the standard pattern of this color symmetry breaking by higgsed $\langle X^{adj}_{SU(N_c)}\rangle$ giving the largest contributions $\sim\lm$ to particle masses. All original charged particles acquire masses $\sim\lm$, while $N_c=N$ light flavorless BPS dyons $D_j$ (massless at $\mx\ra 0$) are formed at the scale $\mu\sim\lm$.

Clearly, higgsed flavorless $X^{adj}_{SU(N_c)}$ does not break by itself the flavor symmetry $SU(N_F)$, i.e. {\it there is no color-flavor locking}. The global non-Abelian flavor symmetry $SU(N_F)=SU(N)_F$ remains unbroken, while the non-Abelian color $SU(N_c)=SU(N)_C$ is broken down to Abelian one.

As before in section 6.2 for heavy quarks with $SU(\bb)$ colors, {\it all quarks have equal masses}, but they are too heavy and too short ranged now and cannot form the coherent condensate. I.e., they are {\it not higgsed} in this phase. This is in accord with the Konishi anomaly \eqref{(6.6)} (valid for each flavor separately) which requires  $\langle Q^i_i\rangle=\langle{\ov Q}^{\,i}_i\rangle=0,\, i=1...N$ when $X^{adj}_{SU(N_c)}$ is higgsed as in \eqref{(6.4.2)}.

All equal mass quarks are clearly {\it in the (anti)fundamental representation} of unbroken global $SU(N)_F$, while {\it the non-Abelian color $SU(N)_C$ is broken at the scale $\mu\sim\lm$ down to Abelian $U^{N-1}(1)$, and masses of all charged flavorless gluons (and scalars) differ greatly in this lower energy phase at $\mx\ra 0:\, \mu_{\rm gl}^{ij}\sim |(\rho^{i}-\rho^{j})\lm |\,,\,\, i\neq j,\,\,i,j=1...N$}, see \eqref{(6.4.2)}. There is the phase transition at $(\mx m)^{1/2}\sim\lm$.

After all heavy particles with masses $\sim\lm$ decoupled at $\mu=\lm/(\rm several)$, the low energy superpotential looks as
\bq
\w_{N_F}=\w_{D}+\w_{a}\,, \label{(6.4.3)}
\eq
\bbq
\w_{D}= (m-a_0)\sum_{j=1}^{N}{\ov D}_j D_j -\sum_{j=1}^{N} a_{D,\,j}{\ov D}_j D_j -\mx\lm\sum_{j=1}^{N}\rho^{\,{j-1}} a_{D,\,j}+\mx L\,\Bigl (\sum_{j=1}^{N} a_{D,\,j}\Bigr )\,
\eeq
\bbq
\w_{a}=\frac{\mx}{2} N \,a_0^2+\mx N\,{\hat\delta}_4\, (m-a_0)^2\,.
\eeq

We would like to emphasize that all dyons in \eqref{(2.1.5)},\eqref{(2.2.4)},\eqref{(3.4)},\eqref{(6.2.2)} and \eqref{(6.4.3)} (and in \eqref{(7.2.2)},
\eqref{(8.6)} below) are {\it the same} particles, and only corresponding values of $2N^\prime_c-N^\prime_F$ in different vacua may be different.

From \eqref{(6.4.3)}
\bbq
\langle a_0\rangle=m\,,\quad \langle a_{D,\,j}\rangle=0\,,\quad \langle{\ov D}_j\rangle\langle D_j \rangle= -\mx\lm\rho^{\,{j-1}}+\mx m\,,
\eeq
\bq
\langle\Sigma_D\rangle=\sum_{j=1}^{N}\langle{\ov D}_j\rangle\langle D_j\rangle=N\mx m=\langle{\rm Tr\,}\qq\rangle_{N_c}\,.\label{(6.4.4)}
\eq

All $N$ dyons in \eqref{(6.4.3)} are higgsed at small $\mx$ at the scale $\mu_D\sim\langle{\ov D}_j D_j\rangle^{1/2},\, \mu_D\sim\max [ (\mx m)^{1/2},\\(\mx\lm)^{1/2}]$ (at $m\gg\lm$ below in this section for definiteness), in the weak coupling regime $g_D(\mu_D)\ll 1$, and $N$ long ${\cal N}=2$ multiplets of massive photons are formed with masses $\sim \mu_D\ll\lm$. All original heavy charged particles with masses $\sim\lm$ are weakly confined, the string tension is $\sigma^{1/2}\sim\langle{\ov D}_j D_j\rangle^{1/2}\ll\lm$. In particular, the heavy flavored quarks form a number of $SU(N_F)$ adjoint hadrons with masses $\sim\lm$. There are no massless Nambu-Goldstone particles because the global flavor symmetry $U(N_F)$ is unbroken.\\

As for non-factorizable condensates of heavy non-higgsed in this lower energy phase quarks with masses $\sim\lm$, \eqref{(2.1.16)} looks similarly here, with $(\bb)\ra N_c$, as well as \eqref{(2.1.18)}. Instead of \eqref{(2.1.20)} we have here
\bq
[\,{\rm Tr\,}(\qq)_{N_c}\,]^{(\rm several)\lm}_{(\rm several)(\mx\lm)^{1/2}}=\Bigl [\,{\hat d}_1 N_c
\mx {\hat a}^{\rm soft}_0+\Sigma_D\,\Bigr ]^{(\rm several)(\mx\lm)^{1/2}},\quad
{\hat d}_1=O(1)\,.\,\label{(6.4.5)}
\eq
This operator expansion leads to the numerical equality
\bq
\langle {\rm Tr\,}(\qq)_{N_c}\rangle^{(\rm several)\lm}_{\mu_{\rm cut}^{\rm lowest}=0}=\langle\Sigma_D\rangle=\langle\Sigma_D\rangle^{(\rm several) (\mx m)^{1/2}}_{\mu_{\rm cut}^{\rm lowest}=0}\,,\quad
\langle {\hat a}^{\rm soft}_0\rangle^{\lm/(\rm several)}_{\mu_{\rm cut}^{\rm lowest}=0}=0\,,\quad m\gg\lm \label{(6.4.6)}
\eq
as it should be, see \eqref{(6.4.1)},\eqref{(6.4.4)}. As before in section 6.2 for $\langle{\rm Tr\,}\qq\rangle_{\bb}$ at $\nd\geq 1$, the nonzero total mean value of the heavy quark non-factorizable bilinear operator originates and saturates not at the scale $\mu\sim\lm$, but only at much lower energies
$\mu_D\sim (\mx m)^{1/2}$.\\

Moreover, at $\mx\ra 0$, because $\langle a_0\rangle=m$ "eats" all quark masses, all charged solitons will have masses either $\sim\lm$ or zero. But the $U(N_c)$ curve \eqref{(1.2)} has only $N_c$ unequal double roots $e^{D}_j= - m +\rho^{j-1}\lm,\, j=1...N_c$ in this vacuum corresponding to $N_c=N$ massless BPS dyons $D_j$, and shows that {\it there are no {\it additional} charged ${\cal N}=2$ BPS solitons massless at $\mx\ra 0$}. In particular, there are no massless pure magnetic monopoles or other additional dyons, they all have masses $\sim\lm$. And there are no massless particles at all at small $\mx\neq 0$. Besides, there are no particles with masses $\sim m\gg\lm$, the largest masses are $\sim\lm$.\\

According to e.g. \cite{SY6} (and refs therein to their previous related papers), the transition from $(\mx m)^{1/2}\gg\lm$ to $(\mx m)^{1/2}\ll\lm$ is a smooth analytical crossover with {\it still higgsed quarks}, $\langle Q^i_a\rangle=\delta^i_a (\mx m)^{1/2},\, i,a=1...N$, {\it and still unbroken $SU(N)_{C+F}$ global symmetry}. And all original particles from the long ${\cal N}=2\,\, SU(N)_{C+F}$ adjoint multiplet mysteriously acquire equal masses $\sim\lm$ at $(\mx m)^{1/2}<\lm$. Besides, e.g. each heavy original quark $Q^i$ with the mysteriously appeared mass $\sim\lm$, decays into the (flavored dyon-flavorless antimonopole) pair $({\cal D}^i+{\ov{\cal M}})$ (both particles with masses $\sim\lm$) in the "instead-of-confinement" regime at $(\mx m)^{1/2} < \lm$. Due to higgsed at small $\mx\neq 0$ {\it light} dyons $\ov{\textsf{D}}_n,\, \textsf{D}_n,\, n=1...N$ (composed, according to \cite{SY6}, from diagonal quarks $Q^n_n$ and flavorless magnetic monopoles $M_n:\, \textsf{D}_n=(Q^n_n+M_n)$\,), these heavy ${\cal D}^i$ and ${\ov{\cal M}}$ are confined in this "instead-of-confinement" regime, and this pair to which quark decays forms a "stringy meson" with the mass $\sim\lm$. This is, so to say, "internal decay", so that, as a whole, each pair $({\cal D}^i+{\ov{\cal M}})$ remains the same quark $Q^i$.

Let us recall that, in any case, the considered theory is not at the arbitrary point on the formal moduli space existing at the mathematical point $\mx\equiv 0$ only, but stays in definite isolated vacua - those at $\mx\neq 0$, even in the formal limit $\mx\ra 0$. And e.g. massive BPS quarks are at least marginally stable at $\mx\ra 0$ in vacua considered in this paper and can in this case only formally dissolve "emitting" massless quanta.

And, in any case, similarly to section 6.2, it remains completely unclear in the variant with the smooth crossover transition from $(\mx m)^{1/2}\gg\lm$ to $(\mx m)^{1/2}\ll\lm$ proposed in \cite{SY6}\,: a) from where mysteriously appeared large equal masses $\sim\lm$ of all particles from still unbroken long ${\cal N}=2\,\, SU(N)_{C+F}$ adjoint multiplet (because, with still higgsed quarks $\langle Q^i_a\rangle=\delta^i_a  (\mx m)^{1/2},\, i,a=1...N$, the Konishi anomaly \eqref{(6.6)} requires $\langle X^{adj}_{SU(N)}\rangle=0,\, A=1...N^2-1$);\, b) because all $N^2-1\,\, SU(N)_{C+F}$ adjoint gluons acquired masses $\sim\lm$, from where then mysteriously appeared $U^{N-1}(1)$ massless at $\mx\ra 0$ photon multiplets;\, c) how their dyons $\textsf{D}_n=(Q^n_n+M_n)$ and mutually non-local with them quarks $Q^n_n$ can be higgsed simultaneously at small $\mx\neq 0$.

\subsection{$m\gg\lm\,,\,\, 2N_c-N_F=1$}
\numberwithin{equation}{subsection}

At $\mx m\gg\lm^2$ the behavior of all vs-vacua is the same and is described in section 6.1. Therefore, we consider below only the case $\mx m\ll\lm^2$ when the vs-vacua with $\bb=1$ and $\bb\geq 2$ behave differently.

The pattern of flavor symmetry breaking remains fixed, $\no=\nd=N_c-1\,,\,\, \nt=N_c\,,\,\, U(N_F)\ra U(\no)\times U(\nt)$, as well as the multiplicity $N_{vs}=C^{\,\nt=N_c}_{N_F}=C^{\,\no=\nd}_{N_F}$. And $\langle a_0\rangle=m$ also remains the same in all vs -vacua (really, at all $m\gtrless\lm$) and still "eats" all quark masses $"m"$. But the discrete symmetry $Z_{2N_c-N_F}$ becomes trivial at $\bb=1$ and gives no restrictions on the form of $\langle X^{\rm adj}_{SU(N_c)}\rangle\sim\lm$ (remind that $\langle X^{\rm adj}_{SU(N_c)}\rangle\sim\lm$ is higgsed necessarily at $\mx m\ll\lm^2$ to avoid $g^2(\mu<\lm)<0$ ). As above in section 6.2, the right flavor symmetry breaking implies the unbroken $SU(\nd)$ lower energy group and a presence of $N_F$ flavors of quark-like particles massless at $\mx\ra 0$, then $\no=\nd$ of them will be higgsed at small $\mx\neq 0$ and there will remain $2\nd N_c$ Nambu-Goldstone multiplets.

However, it is not difficult to see that a literal picture with {\it light all BPS original electric particles} of remained unbroken at the scale $\mu\sim\lm$ electric subgroup $SU(\nd)$ cannot be right at $\bb=1$. Indeed, $X^{\rm adj}_{SU(N_c)}$ is higgsed now as $\langle X^{\rm adj}_{SU(N_c)}\rangle\sim\lm\, {\rm diag}(\,1... 1\,;\, -\nd\,),\,\,\nd=N_c-1$. And all original quarks $Q^i,\,{\ov Q}_i$ with $SU(N_c)$ colors acquire masses $\sim\lm$ and decouple at $\mu\lesssim\lm$.
\footnote{\,
Clearly, at $m\ll\lm,\, \mx\ra 0$ and $\bb=1$, this argument concerns also $SU(N_c)$ br2 vacua with $0<\no<\nd$ and the multiplicity $(\nd-\no)C^{\no}_{N_F}$, and S vacua with the multiplicity $\nd$, both with $\langle S\rangle_{N_c}\neq 0$ and with the $SU(\nd=N_c-1)$ lower energy gauge group at energy $\mu<\lm/(\rm several)$.
}

Therefore, the vs-vacua with $2N_c-N_F=1$ are not typical vacua, they are really exceptional. The absence of the non-trivial unbroken $Z_{2N_c-N_F}$ symmetry is crucial.
\footnote{\,
Considered in \cite{SY3} example of $U(N_c)$ vs-vacua ($r=N_c$ vacua in the language of \cite{SY3}) with $N_c=3,\, N_F=5,\, \no=\nd=N_c-1=2$ and $\langle S\rangle_{N_c}=0$ belongs just to this exceptional type of vs-vacua with $2N_c-N_F=1$.
}
The possible exceptional properties of vacua with $\bb=1$ at $\mx\ra 0$ are indicated also by the curve \eqref{(1.2)}, its form is changed at $\bb=1$\,: $m\ra m+(\lm/N_c)$ \cite{APS}.

For this reason we do not deal with such vacua in this paper.

\section{$\mathbf{SU(N_c),\,\,m\gg\lm,\,\, U(N_F)\ra U(\no)\times U(\nt)},$\,\,  br2 vacua}

\subsection{Larger $\mx$}

\hspace*{4mm} The quark condensates in these br2 vacua with $\no\geq 1,\, \nt<N_c$ are obtained from \eqref{(4.1)},\eqref{(4.1.1)} by the replacement $\no\leftrightarrow\nt$, i.e.\,: $\langle\Qt\rangle_{N_c}\approx\mx m_1\gg\langle\Qo\rangle_{N_c},\, m_1=m N_c/(N_c-\nt)$, see sections 3 and 11 in \cite{ch4}. The discrete $Z_{\bb\geq 2}$ symmetry is also unbroken in these vacua and the multiplicity is $N_{\rm br2}=(N_c-\nt) C^{\,\nt}_{N_F}$. They evolve at $m\ll\lm$ to Lt-vacua with $\langle\Qo\rangle_{N_c}\sim\langle\Qt\rangle_{N_c}\sim\mx\lm$, see section 3 in \cite{ch4}.

Similarly to br1 vacua, the scalars $X^{adj}$ are also higgsed at the largest scale $\mu\sim m\gg\lm$ in the weak coupling region, now as $SU(N_c)\ra SU(\nt)\times U(1)\times SU(N_c-\nt),\,\, N_F/2<\nt<N_c$, so that (the leading terms only)
\bq
\langle\Qt\rangle_{N_c}\approx\mx m_1\,,\quad \langle\Qo\rangle_{N_c}\approx\mx m_1\Bigl (\frac{\lm}{m_1}\Bigr )^{\frac{2N_c-N_F}{N_c-\nt}} \,,\quad m_1=\frac{N_c}{N_c-\nt}\, m\,, \label{(7.1.1)}
\eq
\bbq
\langle S\rangle_{N_c}=\frac{\langle\Qo\rangle_{N_c}\langle\Qt\rangle_{N_c}}{\mx}\approx\mx m_1^2\Bigl (\frac{\lm}{m_1}\Bigr )^{\frac{2N_c-N_F}{N_c-\nt}}\ll\mx m^2\,,
\quad\frac{\langle\Qo\rangle_{N_c}}{\langle\Qt\rangle_{N_c}}\approx\Bigl (\frac{\lm}{m_1}\Bigr)^{\frac{2N_c-N_F}{N_c-\nt}}\ll 1\,,
\eeq
\bq
\langle X^{\rm adj}_{SU(N_c)}\rangle\equiv \langle\, X^{adj}_{SU(\nt)}+ X_{U(1)}+ X^{adj}_{SU(N_c-\nt)}\,\rangle\,,\label{(7.1.2)}
\eq
\bbq
\sqrt{2}\, X_{U(1)}=a \,{\rm diag}\Bigl (\underbrace{\,1}_{\nt}\,;\,\underbrace{\,\wh{c}}_{N_c-\nt}\, \Bigr )\,,\quad \wh{c}=-\frac{\nt}{N_c-\nt}\,,\quad \langle a \rangle= m\,,
\eeq
\bbq
\mx\langle{\rm Tr}\,(\sqrt{2} X^{\rm adj}_{SU(N_c)})^2\rangle=(2N_c-N_F)\langle S\rangle_{N_c}+ m \langle{\rm Tr}\,(\qq)\rangle_{N_c}\approx m (\nt\mx m_1)\,.
\eeq
As a result, all quarks charged under $SU(N_c-\nt)$ and hybrids $SU(N_c)/[\,SU(\nt)\times SU(N_c-\nt)\times U(1)\,]$ acquire large masses $m_1=m-{\wh c}\,\langle a\rangle=m N_c/(N_c-\nt)$ and decouple at scales $\mu<\,m_1$, there remains ${\cal N}=2\,\, SU(N_c-\nt)$\, SYM with the scale factor $\Lambda^{SU(N_c-\nt)}_
{{\cal N}=2\, SYM}$ of its gauge coupling, see also \eqref{(7.1.1)},
\bq
\langle\Lambda^{SU(N_c-\nt)}_{{\cal N}=2\, SYM}\rangle^2=\Bigl (\frac{\lm^{\bb}{(m_1)}^{N_F}}{{(m_1)}^{2\nt}
}\Bigr )^{\frac{1}{(N_c-\nt)}}={m^2_1}\Bigl (\frac{\lm}{{m_1}}\Bigr )^{\frac{\bb}{(N_c-\nt)}}\ll\lm^2\,,\quad m_1=\frac{N_c}{N_c-\nt} m\,,\label{(7.1.3)}
\eq
\bbq
\langle S\rangle_{N_c-\nt}=\mx\langle\Lambda^{SU(N_c-\nt)}_{{\cal N}=2\, SYM}\rangle^2\approx\langle S\rangle_{N_c}\,.
\eeq

At $\mx\ll\langle\Lambda^{SU(N_c-\nt)}_{{\cal N}=2\, SYM}\rangle$ it is higgsed in a standard way \cite{DS}, $SU(N_c-\nt)\ra U^{N_c-\nt-1}(1)$, due to higgsing of $X^{adj}_{SU(N_c-\nt)}$
\bq
\langle {\sqrt 2} X^{adj}_{SU(N_c-\nt)}\rangle\sim\langle\Lambda^{SU(N_c-\nt)}_{{\cal N}=2\, SYM}\rangle\, {\rm diag} \Bigl (\underbrace{0}_{\nt}\,;\,\underbrace{k_1,\,...\,k_{N_c-\nt}}_{N_c-\nt}\Bigr )\,,\quad k_i=O(1)\,.\label{(7.1.4)}
\eq
Note that the value $\langle\Lambda^{SU(N_c)}_{{\cal N}=2\, SYM}\rangle$ of $\langle X^{adj}_{SU(N_c-\nt)}\rangle$ in \eqref{(7.1.3)},\eqref{(7.1.4)} is consistent with the unbroken $Z_{\bb}$ discrete symmetry.

As a result, $\,N_c-\nt-1$ pure magnetic monopoles $M_{\rm n}$ (massless at $\mx\ra 0$) with the $SU(N_c-\nt)$ adjoint charges are formed at the scale $\mu\sim\langle\Lambda^{SU(N_c-\nt)}_{{\cal N}=2\, SYM}\rangle$ in this SYM sector. These correspond to $N_c-\nt-1$ unequal double roots of the curve \eqref{(1.2)} connected with this SYM sector. Note also that two single roots with $(e^{+} -e^{-})
\sim\langle\Lambda^{SU(N_c)}_{{\cal N}=2\, SYM}\rangle$ of the curve \eqref{(1.2)} also originate from this SYM sector. Other $\nt$ double roots of the curve at $\mx\ra 0$ originate in these vacua from the $SU(\nt)$ sector, see below.

These $N_c-\nt-1$ monopoles are all higgsed with $\langle M_{\rm n}\rangle=\langle {\ov M}_{\rm n}\rangle\sim (\mx\langle\Lambda^{SU(N_c-\nt)}_{{\cal N}=2\, SYM}\rangle)^{1/2}$, so that $N_c-\nt-1$ long ${\cal N}=2$ multiplets of massive photons appear, with masses $\sim (\mx\langle\Lambda^{SU(N_c-\nt)}_{{\cal N}=2\, SYM}\rangle)^{1/2}$. Besides, this leads to a weak confinement of all heavier original $SU(N_c-\nt)$ electrically charged particles, i.e. quarks and all hybrids with masses $\sim m$, and all ${\cal N}=2\,\, SU(N_c-\nt)$ SYM adjoint charged particles with masses $\sim\langle\Lambda^{SU(N_c-\nt)}_{{\cal N}=2\, SYM}\rangle$, the tension of the confining string is $\sigma^{1/2}_2\sim (\mx\langle\Lambda^{SU(N_c-\nt)}_{{\cal N}=2\, SYM}\rangle)^{1/2}\ll\langle\Lambda^
{SU(N_c-\nt)}_{{\cal N}=2\, SYM}\rangle\ll\lm\ll m$. The factor $N_c-\nt$ in the multiplicity of these br2 vacua arises just from this ${\cal N}=2\,\, SU(N_c-\nt)$\, SYM part.

At larger $\mx$ in the range $\langle\Lambda^{SU(N_c-\nt)}_{{\cal N}=2\, SYM}\rangle\ll\mx\ll m$, the scalars $X^{adj}_{SU(N_c-\nt)}$ become too heavy, their light  $SU(N_c-\nt)$ physical phases fluctuate then freely at the scale $\sim\mx^{\rm pole}=g^{2}(\mu=\mx^{\rm pole})\mx$ and they are not higgsed. Instead, they decouple as heavy in the weak coupling region at scales $\mu<\mx^{\rm pole}/(\rm several)$. There remains then ${\cal N}=1\,\, SU(N_c-\nt)$ SYM with the scale factor $\Lambda^{SU(N_c-\nt)}_{{\cal N}=1\, SYM}$ of its gauge coupling
\bq
\langle S\rangle_{{\cal N}=1\, SYM}=\langle\Lambda^{SU(N_c-\nt)}_{{\cal N}=1\, SYM}\rangle^3
=\mx\langle\Lambda^{SU(N_c-\nt)}_{{\cal N}=2\, SYM}\rangle ^2,\,\,
\,\, \langle\Lambda^{SU(N_c-\nt)}_{{\cal N}=2\, SYM}\rangle\ll\langle\Lambda^{SU(N_c-\nt)}_{{\cal N}=1\, SYM}\rangle\ll\mx\ll m\,.\quad\label{(7.1.5)}
\eq
Therefore, there will be in this case a large number of strongly coupled $SU(N_c-\nt)$ gluonia with the mass scale $\sim\langle\Lambda^{SU(N_c-\nt)}_{{\cal N}=1\, SYM}\rangle$, while all original heavier charged particles with $SU(N_c-\nt)$ electric charges and masses either $\sim m\gg\langle\Lambda^{SU(N_c-\nt)}_{{\cal N}=1\, SYM}\rangle$ or $\sim g^{2}\mx\gg\langle\Lambda^{SU(N_c-\nt)}_{{\cal N}=1\, SYM}\rangle$ still will be weakly confined, the tension of the confining string is larger in this case, $\sigma^{1/2}_1\sim\langle\Lambda^{SU(N_c-\nt)}_{{\cal N}=1\, SYM}\rangle$. This ${\cal N}=1\,\,SU(N_c-\nt)\,$ SYM gives the same factor $N_c-\nt$ in the multiplicity of these vacua.\\

Now, about the electric $SU(\nt)$ part decoupled from $SU(N_c-\nt)$ SYM. The scale factor of its gauge coupling is
\bq
\langle\Lambda_{SU(\nt)}\rangle=\Bigl (\,\frac{\lm^{\bb}}{(m_1)^{2(N_c-\nt)}}\,\Bigr )^{\frac{1}{2\nt-N_F}}=m_1\Bigl (\frac{\lm}{m_1} \Bigr )^{\frac{\bb}{2\nt-N_F}}\ll\lm\,,\quad m_1=\frac{N_c}{N_c-\nt} m\,.\label{(7.1.6)}
\eq
Note that the value of $\langle\Lambda_{SU(\nt)}\rangle$ in \eqref{(7.1.6)} is also consistent with the unbroken $Z_{\bb}$ discrete symmetry.

What {\it qualitatively differs} this $SU(\nt)$ part in br2 vacua at scales $\mu<\, m$ from its analog $SU(\no)$ in br1 vacua of section 4.1 is that $SU(\no)$ with $N_F>2\no$ is IR free, while $SU(\nt)$ with $N_F<2\nt$ is UV free, and its small ${\cal N}=2$ gauge coupling at the scale $\mu\sim m\gg\lm$ begins to grow logarithmically with diminished energy at $\mu<m$. Therefore, {\it if nothing prevents}, $\,\,X^{adj}_{SU(\nt)}$ will be higgsed necessarily at the scale $\sim\langle\Lambda_{SU(\nt)}\rangle$, with $\langle X^{adj}_{SU(\nt)}\rangle\sim\langle\Lambda_{SU(\nt)}\rangle$, to avoid unphysical $g^2(\mu<\langle\Lambda_{SU(\nt)}\rangle)<0$ (see section 7.2 below).

But if $(\mx m)^{1/2}\gg\langle\Lambda_{SU(\nt)}\rangle$, {\it the phase will be different}. The leading effect in this case will be the breaking of the whole $SU(\nt)$ group due to higgsing of $\nt<N_c$ out of $N_F$ quarks $Q^i,\,{\ov Q}_i$ in the weak coupling region, with e.g. $\langle Q^{i=b+\no}_b\rangle\sim \delta^{i}_b (\mx m)^{1/2},\,\, b=1...\nt,\,\, i=\no+1... N_F$. This will give the additional factor $C^{\,\nt}_{N_F}$ in the multiplicity of these br2 vacua due to spontaneous flavor symmetry breaking $U(N_F)\ra U(\no)\times U(\nt)$, so that the overall multiplicity will be $(N_c-\nt)C^{\,\nt}_{N_F}$, as it should be. The corresponding parts of the superpotential will be as in \eqref{(4.1.3)},\eqref{(4.1.4)}, with a replacement $\no\ra\nt$,
\bq
{\cal W}_{\nt}=(\, m-a\,)\,{\rm Tr}\,({\ov Q} Q)_{\nt} - {\rm Tr}\,({\ov Q}\sqrt{2} X^{\rm adj}_{SU(\nt)} Q)_{\nt} +\mx {\rm Tr}\,( X^{\rm adj}_{SU(\nt)})^2+\frac{\mx}{2}\,\frac{\nt N_c}{N_c-\nt}\,a^2\,. \label{(7.1.7)}
\eq

From \eqref{(7.1.7)} in the case considered (no summation over $j$ in \eqref{(7.1.8)}\,)
\bbq
\langle a\rangle=m,\,\, \langle  X^{adj}_{SU(\nt)}\rangle=0,\,\,\langle \Qt\rangle_{\nt}=
\sum_{b=1}^{\nt}\langle{\ov Q}_2^{\,b} Q^2_b\rangle=\langle{\ov Q}_2^{\,2}\rangle\langle Q^2_2\rangle\approx\frac{N_c}{N_c-\nt}\mx m=\mx m_1\,,
\eeq
\bq
\langle \Qo\rangle_{\nt}=\sum_{b=1}^{\nt}\langle{\ov Q}_j^{\,b}\rangle\langle Q^j_b\rangle=0\,,\,\, j=1...\no,\,\,\langle { S}\rangle_{\nt}=\frac{1}{\mx}\langle \Qt\rangle_{\nt}\langle \Qo\rangle_{\nt}=0\,.\quad\label{(7.1.8)}
\eq

$2\no\nt$ massless Nambu-Goldstone multiplets are formed as a result of the spontaneous flavor symmetry breaking, $\Pi_j^i= ({\ov Q}_{j} Q^i)_{\nt}\,\ra\,({\ov Q}_{j}\langle Q^i\rangle)_{\nt},\,\,\Pi_i^j= ({\ov Q}_{i} Q^j)_{\nt}\,\ra\,(\langle{\ov Q}_{i}\rangle Q^j)_{\nt},\,\, i=\no+1...N_F,\,\, j=1...\no,\,\,
\langle\Pi^i_j\rangle=\langle \Pi^j_i\rangle=0$ (in essence, these are non-higgsed quarks with $\no$ flavors and $SU(\nt)$ colors). Besides, there are $\nt^2$ long ${\cal N}=2$ multiplets of massive gluons with masses $\sim (\mx m)^{1/2}$.

We emphasize that higgsing of $\nt$ quarks at the higher scale $(\mx m)^{1/2}\gg\langle\Lambda_{SU(\nt)}
\rangle$, see \eqref{(7.1.7)}, \eqref{(7.1.8)}, {\it prevents $X^{adj}_{SU(\nt)}$ from higgsing}, i.e. $\langle  X^{adj}_{SU(\nt)}\rangle_{\nt}$ is not simply smaller, but exactly zero (see also section 6.1 for a similar regime).

On the whole for these br2 vacua in the case considered, i.e. at $(\mx m)^{1/2}\gg\langle\Lambda_{SU(\nt)}
\rangle$, all qualitative properties are similar to those of br1 vacua in section 4.1.

\subsection{Smaller $\mx$}

\hspace*{4mm} Consider now the behavior of the $SU(\nt)\times U(1)$ part in the opposite case of smaller $\mx,\, \mx\ll\langle\Lambda_{SU(\nt)}\rangle^2/m,\, \langle\Lambda_{SU(\nt)}\rangle\ll\lm\ll m$. As will be seen below, in this case even all qualitative properties of mass spectra in this part will be quite different (the behavior of the $SU(N_c-\nt)$ SYM part decoupled from $SU(\nt)\times U(1)$ was described above in section 7.1).

Now, at smaller $\mx$, instead of quarks, the scalar field $X^{adj}_{SU(\nt)}$ is higgsed first at the scale $\sim\langle\Lambda_{SU(\nt)}\rangle$ to avoid unphysical $g^2(\mu<\langle\Lambda_{SU(\nt)}\rangle)<0$ of UV free ${\cal N}=2\,\, SU(\nt)$. Although $\langle\Lambda_{SU(\nt)}\rangle$ respects the original $Z_{\bb}$ discrete symmetry of the UV free $SU(N_c)$ theory, see \eqref{(7.1.6)}, because the $SU(N_c-\nt)$ SYM part is decoupled, there is now the analog of $Z_{\bb}$ in the $SU(\nt)$ group with $N_F$ flavors of quarks, this is the unbroken non-trivial $Z_{2\nt-N_F}=Z_{\nt-\no}$ discrete symmetry with $\nt-\no\geq 2$, \,${\rm b}_2=(\bb)\ra {\rm b}^{\prime}_2=(2\nt-N_F)=(\nt-\no)$. Therefore, the field $X^{adj}_{SU(\nt)}$ is higgsed at $\nt-\no\geq 2$ at the scale $\mu\sim\langle\Lambda_{SU(\nt)}\rangle$ (in the strong coupling and nonperturbative regime), qualitatively similarly to \eqref{(2.1.1)},
\eqref{(2.1.2)} in section 2.1\,: $SU(\nt)\ra SU(\no)\times U^{(1)}(1)\times U^{\nt-\no-1}(1)$. There will be similar dyons $D_j$ (massless at $\mx\ra 0$) etc., and a whole qualitative picture will be similar to those in section 2.1, with evident replacements of parameters. The only qualitative difference is that the additional SYM part is absent here because $\no$ quark flavors are higgsed now in $SU(\no)$, see \eqref{(2.1.1)},\eqref{(2.1.16)},
\bq
\langle X^{adj}_{SU(\nt)}\rangle=\langle  X^{adj}_{SU(\no)}+X^{(1)}_{U(1)}+ X^{adj}_{SU(\nt-\no)}\rangle,\label{(7.2.1)}
\eq
\bbq
\langle\sqrt{2} X^{adj}_{SU(\nt-\no)}\rangle=C_{\nt-\no}\langle\Lambda_{SU(\nt)}\rangle\,{\rm diag}\,\Bigl(\,\underbrace{0}_{\no}\,;\,\underbrace{\tau^0,\,\tau^1,\,...,\,\tau^{\nt-\no-1}}_{\nt-\no}\,;
\underbrace{0}_{N_c-\nt}\,\Bigr )\,,\quad \tau=\exp\{\frac{2\pi i}{\nt-\no}\}\,,
\eeq
\bbq
\sqrt{2}\, X^{(1)}_{U(1)}=a_{1}\,{\rm diag}\,(\,\underbrace{\,1}_{\no}\,;\, \underbrace{\,\wh{c}_1}
_{\nt-\no}\,;\underbrace{0}_{N_c-\nt}\,),\quad \wh{c}_1=-\,\frac{\no}{\nt-\no}\,.\quad
\eeq
Therefore, similarly to \eqref{(2.1.2)},\eqref{(6.2.2)}, in this case the low energy superpotential of this $SU(\nt)\times U(1)$ sector can be written as
\footnote{\,
In other words, in this case the $SU(N^\prime_c=\nt)\times U(1)$ part of $SU(N_c)$ with $m\gg\lm$ and with decoupled $SU(N_c-\nt)$ SYM part, {\it is in its own vs-vacuum} of section 6.2, with $\langle\,{S}\,\rangle_
{N^\prime_c}=0$ and with $a$ of $U(1)$ in \eqref{(7.2.2)} playing a role of $a_0$ in \eqref{(6.2.2)}, etc.
}
\bq
\w_{\nt}=\w_{\no}+\w_{D}+\w_{a,\,a_1}+\dots\,,\label{(7.2.2)}
\eq
\bbq
\w_{\no} =(m-a-a_1){\rm Tr}\,({\ov Q} Q)_{\no}-{\rm Tr}\,\Bigl ({\ov Q}\sqrt{2}X_{SU(\no)}^{\rm adj} Q\Bigr )_{\no}+\mx (1+\wh{\delta}_2){\rm Tr}\,(X^{\rm adj}_{SU(\no)})^2\,,
\eeq
\bbq
\w_{D}=\Bigl ( m-a-\wh{c}_1 a_1 \Bigr )\sum_{j=1}^{\nt-\no}{\ov D}_j D_j -\sum_{j=1}^{\nt-\no} a_{D,j}{\ov D}_j D_j -\mx\Lambda_{SU(\nt)}\sum_{j=1}^{\nt-\no}\tau^{j-1} a_{D,j}+\mx{\wh L}\,\Bigl (\sum_{j=1}^{\nt-\no} a_{D,j}\Bigr ),
\eeq
\bbq
\w_{a,\,a_1}=\frac{\mx}{2}\frac{\nt N_c}{N_c-\nt}\,a^2+\frac{\mx}{2}
\frac{\no\nt}{\nt-\no}(1+\wh{\delta}_1)\,a^2_1+\mx\nt\wh{\delta}_3\,a_1(m-a-\wh{c}_1 a_1)+\mx\nt\wh{\delta}_4\, (m-a-\wh{c}_1 a_1)^2\,,
\eeq
where $\wh{L}$ is the Lagrange multiplier field, $\langle\wh{L}\rangle=O(m)$ and dots denote small power corrections. The additional terms with $\wh{\delta}_{1,2,3,4}$, as previously in br2 vacua in section 2.1 and in vs -vacua in section 6.2, originate from integrating out all heavier fields with masses $\sim\Lambda_{SU(\nt)}$ in the soft background. Here these soft background fields are $(m-a-\wh{c}_1 a_1)$ and $[(1-\wh{c}_1) a_1+\sqrt{2}X^{\rm adj}_{SU(\no)}]$ (\,while $a_1$ plays here a role of $a_1$ in \eqref{(2.1.1)}\,). These constants $\wh\delta_i$ differ from \eqref{(2.1.12)} only because the number of colors is different, $N_c\ra N^\prime_c=\nt$, i.e. $\wh\delta_1=[\,- 2N^\prime_c/(2N^\prime_c-N_F)]
=[\,-2\nt/(\nt-\no)\,],\,\, \wh\delta_2=-2$ (while $\wh\delta_3=0$ as in br2 vacua, see \eqref{(B.4)}).

The charges of fields and parameters entering \eqref{(7.2.2)} under $Z_{\nt-\no}=\exp\{i\pi/(\nt-\no)\}$ transformation are\,: $q_{\lambda}=q_{\rm\theta}=1,\,\, q_{X_{SU(\no)}^{\rm adj}}=q_{a}=q_{a_1}=q_{a_{D,j}}=q_{\rm m}=q_{\wh{L}}=2,\,\,q_{Q}=q_{\ov Q}=q_{D_j}=q_{{\ov D}_j}=q_{\Lambda_{SU(\nt)}}=0,\,\, q_{\mx}=-2$. The non-trivial $Z_{\nt-\no\geq 2}$ transformations change only numerations of dual scalars $a_{D,j}$ and dyons in \eqref{(7.2.2)}, so that $\int d^2\theta\,\w_{\nt}$ is $Z_{\nt-\no}$-invariant.

From \eqref{(7.2.2)}:
\bq
\langle a\rangle=m\,,\quad\langle a_1\rangle=0\,,\quad \langle a_{D,j}\rangle=0\,,
\quad \langle X_{SU(\no)}^{\rm adj}\rangle=0\,,\quad \langle S\rangle_{\no}=0\,,\label{(7.2.3)}
\eq
\bbq
\langle{\ov D}_j D_j\rangle=\langle{\ov D}_j\rangle\langle D_j \rangle\approx -\mx\langle\Lambda_{SU(\nt)}\rangle\tau^{j-1}+\mx \langle{\wh L}\rangle\,,\quad \langle\Sigma_D\rangle=\sum_{j=1}^{\nt-\no}\langle{\ov D}_j\rangle\langle D_j \rangle=(\nt-\no)\mx \langle{\wh L}\rangle,
\eeq
\bbq
\langle{\wh L}\rangle\approx\frac{N_c}{N_c-\nt}\, m=m_1\,, \quad \langle{\rm Tr}\,({\ov Q} Q)\rangle_{\no}=\no\langle\Qo\rangle_{\no}+\nt\langle\Qt\rangle_{\no}\approx\no\mx m_1\,,
\eeq
\bbq
\langle\Qo\rangle_{\no}=\langle{\ov Q}^{\,1}_1\rangle\langle Q^1_1\rangle\approx\frac{N_c}{N_c-\nt}\,\mx m\approx\mx m_1\approx\langle\Qt\rangle_{N_c}\,,\quad \langle\Qt\rangle_{\no}=\sum_{a=1}^{\no}\langle{\ov Q}^{\,a}_2\rangle\langle Q^2_a\rangle=0\,.
\eeq

The multiplicity of these br2 vacua in the case considered is $N_{\rm br2}=(N_c-\nt)C^{\,\no}_{N_F}=(N_c-\nt)C^{\,\nt}_{N_F}$, as it should be. The factor $(N_c-\nt)$ originates from the $SU(N_c-\nt)$ SYM. The factor $C^{\,\no}_{N_F}$ is finally a consequence of the color breaking $SU(\nt)\ra SU(\no)\times U(1)^{\nt-\no}$ with the unbroken $Z_{2\nt-N_F}=Z_{\nt-\no}$ discrete symmetry, and spontaneous breaking of flavor symmetry due to higgsing of $\no$ quarks flavors (massless at $\mx\ra 0$) in the $SU(\no)$ color subgroup of $SU(\nt)$, $\langle Q^i_a\rangle=\langle{\ov Q}^{\,a}_i\rangle\approx \delta_a^i (\mx m_1)^{1/2}$,  $a=1...\no,\,\,i=1...N_F$.

Note also that if the non-trivial discrete symmetry $Z_{\nt-\no\geq 2}$ were broken spontaneously in the $SU(\nt)$ subgroup at the scale $\mu\sim\langle\Lambda_{SU(\nt)}\rangle\gg(\mx m)^{1/2}$, this would lead then
to the factor $\nt-\no$ in the multiplicity of these vacua, and this is a wrong factor.

Due to higgsing of $\nt-\no$ dyons with the nonzero $SU(\nt-\no)$ adjoint magnetic charges, $\langle{\ov D_j}\rangle\langle D_j\rangle\sim \mx m\gg\mx\lm\gg\mx\langle\Lambda_{SU(\nt)}\rangle$, all original pure electrically charged particles with $SU(\nt-\no)$ colors and masses either $\sim m$ or $\sim\langle\Lambda_{SU(\nt)}\rangle$ are weakly confined, the string tension is $\sigma^{1/2}\sim (\mx m)^{1/2}\ll\langle\Lambda_{SU(\nt)}\rangle\ll\lm\ll m$. Besides, $\nt-\no$ long ${\cal N}=2$ multiplets of massive photons with masses $\sim (\mx m)^{1/2}$ are formed. Remind that, due to higgsing of $N_c-\nt-1$ magnetic monopoles $M_{\rm n}$ from $SU(N_c-\nt)$ SYM at $\mu\sim (\mx\langle\Lambda^{SU(N_c-\nt)}_{{\cal N}=2\, SYM}\rangle)^{1/2}$, all original pure electrically charged particles with $SU(N_c-\nt)$ charges and masses either $\sim m$ or $\sim\langle\Lambda^{SU(N_c-\nt)}_{{\cal N}=2\, SYM}\rangle$ are also weakly confined, but this string tension $\sigma^{1/2}_2\sim (\mx\langle\Lambda^{SU(N_c-\nt)}_{{\cal N}=2\, SYM}\rangle)^{1/2}\ll (\mx m)^{1/2}$ is much smaller.

Due to higgsing of $\no$ flavors of original electric quarks in the IR free color $SU(\no)$ with $N_F>2\no$ flavors, there are $\no^2$ long ${\cal N}=2$ multiplets of massive gluons with masses $\sim (\mx m)^{1/2}\ll\langle\Lambda_{SU(\nt)}\rangle$, while the original non-higgsed quarks with $\nt$ flavors and $SU(\no)$ colors form $2\no\nt$ massless Nambu-Goldstone multiplets. There are no other massless particles at $\mx\neq 0$. And there are no particles with masses $\sim m\gg\lm$ in this $SU(\nt)\times U(1)$ sector.

Remind that the curve \eqref{(1.2)} has $N_c-1$ double roots in these br2 vacua at $\mx\ra 0$. Of them: $\nt-\no$ unequal roots correspond to $\nt-\no$ dyons $D_j,\,\, \no$ equal roots correspond to $\no$ higgsed quarks $Q^i$ of $SU(\no)$, the remaining unequal $N_c-\nt-1$ double roots correspond to $N_c-\nt-1$ pure magnetic monopoles $M_{\rm n}$ from $SU(N_c-\nt)\,\, {\cal N}=2$ SYM. Two single roots with $(e^{+}-e^
{-})\sim\langle\Lambda^{SU(N_c-\nt)}_{{\cal N}=2\, SYM}\rangle$ originate from this ${\cal N}=2$ SYM.

Besides, similarly to vs-vacua in section 6, there is a phase transition in these br2 vacua in the region $\mx\sim\langle\Lambda_{SU(\nt)}\rangle^2/m$. E.g., all $SU(\nt-\no)$ charged gluons have equal masses in the whole wide range $(\rm several)\langle\Lambda_{SU(\nt)}\rangle < (\mx m)^{1/2}\ll m$ of the higher energy phase of section 7.1, while masses of these gluons differ greatly at $\mx\ra 0$ in the lower energy phase of this section, $\mu^{ij}_{gl}\sim |(\tau^i-\tau^j)\langle\Lambda_{SU(\nt)}\rangle|,\,i\neq j,\, i,j=0...\nt-\no-1$. And $\langle X^{adj}_{SU(\nt-\no)}\rangle$, $\,\langle\Qt\rangle_{\nt}$ also behave non-analytically at $\mx\gtrless\langle\Lambda_{SU(\nt)}\rangle^2/m$. But, unlike the phase transition at $\mx\sim\lm^2/m$ in
vs-vacua of section 6, the phase transition in these br2 vacua is at much smaller $\mx\sim\langle\Lambda_
{SU(\nt)}\rangle^2/m\,,\,\,\langle\Lambda_{SU(\nt)}\rangle\ll\lm$, see \eqref{(7.1.6)}.\\

And finally, remind that the mass spectra in these br2 vacua with $\nt<N_c$ and $m\gg\lm$ depend essentially on the value of $m/\lm$, and all these vacua  evolve at $m\ll\lm$ to Lt-vacua with the spontaneously broken discrete $Z_{\bb}$ symmetry, see section 3 in \cite{ch4} or section 4 in \cite{ch5}.

\section{\bf Large quark masses $m\gg\lm,\,\,\, SU(N_c),\,\,\, N_F<N_c$}
\numberwithin{equation}{section}

\hspace*{4mm} The vacua with the unbroken $U(N_F)$ flavor symmetry (here and in what follows at not too small $N_c$ and $N_F$) are in this case the SYM-vacua with the multiplicity $N_c$ and S-vacua with the multiplicity $N_c-N_F$, see section 2 in \cite{ch4} and section 3 in \cite{ch5}. The vacua with the broken $U(N_F)\ra U(\no)\times U(\nt)$ symmetry are the br1 and br2 vacua with multiplicities $(N_c-{\rm n}_i)C^{\,{\rm n}_i}_{N_F},\, i=1,\,2$.

Except for S-vacua, all formulae for this case $N_F<N_c,\,\, m\gg\lm$ are the same as described above for br1 vacua in section 4, for SYM vacua in section 5 and for br2 vacua in section 7 (and only $N_F<N_c$ now). Therefore, we describe below only the mass spectra in S-vacua with $m\gg\lm,\,\mx\ll\lm\,$.

From
\bbq
\w^{\,\rm eff}_{\rm tot}(\Pi)=m\,{\rm Tr}\,({\ov Q} Q)_{N_c}-\frac{1}{2\mx}\Biggl [ \,\sum_{i,j=1}^{N_F} ({\ov Q}_j Q^i)_{N_c}({\ov Q}_{\,i} Q^j)_{N_c}-\frac{1}{N_c}\Bigl ({\rm Tr}\,({\ov Q} Q)_{N_c}\Bigr )^2\Biggr ]+(N_c-N_F) S_{N_c} \,,
\eeq
\bbq
S_{N_c}=\Bigl(\frac{\lm^{2N_c-N_F}\mx^{N_c}}{\det (\qq)_{N_c}}\Bigr )^{\frac{1}{N_c-N_F}}\,,
\eeq
the quark and gluino condensates look in these $SU(N_c)$ S-vacua with $\no=0,\, \nt=N_F$ as,
\bq
\langle\qq\rangle_{N_c}\approx\mx m_1\Bigl [1-\frac{N_c}{N_c-N_F}\Bigl (\frac{\lm}{m_1}\Bigr )^{\frac{2N_c-N_F}{N_c-N_F}} \,\Bigr ],\,\,  m_1=\frac{N_c}{N_c-N_F}\,m,\,\,\,
\langle S\rangle_{N_c}\approx\mx m_1^2\Bigl (\frac{\lm} {m_1}\Bigr )^{\frac{2N_c-N_F}{N_c-N_F}}.\,\,\, \label{(8.1)}
\eq
It is seen from \eqref{(8.1)} that the non-trivial discrete symmetry $Z_{2N_c-N_F\geq 2}$ is not broken.

The scalars $ X^{adj}_{SU(N_c)}$ are higgsed at the highest scale $\sim m\gg\lm$ {\it in the weak coupling regime}, $SU(N_c)\ra SU(N_F)\times U(1)\times SU(N_c-N_f)$, see \eqref{(8.7)} below,
\bq
\langle X^{adj}_{SU(N_c)}\rangle=\langle  X^{adj}_{SU(N_F)}+X_{U(1)}+ X^{adj}_{SU(N_c-N_F)}\rangle,\label{(8.2)}
\eq
\bbq
\sqrt{2}\, X_{U(1)}=a\,{\rm diag}\,(\,\underbrace{\,1}_{N_F}\,;\, \underbrace{\,\wh{c}}
_{N_c-N_F}\,),\quad \wh{c}=-\,\frac{N_F}{N_c-N_F}\,,\quad \langle a\rangle= m\,.
\eeq

The quarks in the $SU(N_c-N_F)$ sector and $SU(N_c)/[SU(N_F)\times SU(N_c-N_F)\times U(1)]$ hybrids have large masses $m_Q=\langle m-{\wh c}\, a\rangle=N_c m/(N_c-N_F)=m_1$. After they are integrated out, the scale factor of the $SU(N_c-N_F)$ SYM gauge coupling is, see also \eqref{(8.1)},
\bq
\langle\Lambda^{SU(N_c-N_F)}_{{\cal N}=2\, SYM}\rangle^2=\Bigl (\frac{\lm^{\bb}{(m_1)}^{N_F}}{(m_1)^{2 N_F}}\Bigr )^{\frac{1}{N_c-N_F}}={m^2_1}\Bigl (\frac{\lm}{{m_1}}\Bigr )^{\frac{\bb}{N_c-N_F}}\ll\lm^2\,,\quad m_1=\frac{N_c}{N_c-N_F} m\,,\label{(8.3)}
\eq
\bbq
\langle S\rangle_{N_c-N_F}=\mx\langle\Lambda^{SU(N_c-N_F)}_{{\cal N}=2\, SYM}\rangle^2=\langle S\rangle_{N_c}\,,
\eeq
while those of $SU(N_F)$ SQCD is
\bq
\langle\Lambda_{SU(N_F)}\rangle=\Bigl (\,\frac{\lm^{\bb}}{(m_1)^{2(N_c-N_F)}}\,\Bigr )^{\frac{1}{N_F}}=m_1\Bigl (\frac{\lm}{m_1} \Bigr )^{\frac{\bb}{N_F}}\ll\lm\,.\label{(8.4)}
\eq

If $\mx\ll\langle\Lambda^{SU(N_c-N_F)}_{{\cal N}=2\, SYM}\rangle$, the scalars $X^{adj}_{SU(N_c-N_F)}$ of ${\cal N}=2$ SYM are higgsed at $\mu\sim\langle\Lambda^{SU(N_c-N_F)}_{{\cal N}=2\, SYM}\rangle$ \eqref{(8.3)} in a known way, $SU(N_c-N_F)\ra U^{N_c-N_F-1}(1)$ \cite{DS}, giving $N_c-N_F$ vacua and $N_c-N_F-1$ massless at $\mx\ra 0$ magnetic monopoles ${\ov M}_{\rm n}, M_{\rm n}$. These all are higgsed at $\mx\neq 0$ giving $N_c-N_F-1$ $\,{\cal N}=2$ multiplets of dual massive photons with masses $\sim (\mx\langle\Lambda^{SU(N_c-N_F)}_{{\cal N}=2\, SYM}\rangle)^{1/2}$. All heavy original particles with masses $\sim m$ and all charged $SU(N_c-N_F)$ adjoints with masses $\sim\langle\Lambda^{SU(N_c-N_F)}_{{\cal N}=2\, SYM}\rangle$ are weakly confined, the string tension in this sector is $\sigma^{1/2}_2\sim (\mx\langle\Lambda^{SU(N_c-N_F)}_
{{\cal N}=2\, SYM}\rangle)^{1/2}\ll\langle\Lambda^{SU(N_c-N_F)}_{{\cal N}=2\, SYM}\rangle$.

On the other hand, if e.g. $\langle\Lambda^{SU(N_c-N_F)}_{{\cal N}=2\, SYM}\rangle\ll\mx\ll m$, see \eqref{(8.3)}, all scalars  $X^{adj}_{SU(N_c-N_F)}$ are too heavy and not higgsed, they all decouple in the weak coupling regime and can be integrated out at $\mu<g^2\mx/(\rm several)$, there remains then ${\cal N}=1\,\, SU(N_c-N_F)$ SYM with $\langle\Lambda^{SU(N_c-N_F)}_{{\cal N}=1\, SYM}\rangle=[\,\mx\langle
\Lambda^{SU(N_c-N_F)}_{{\cal N}=2\, SYM}\rangle^2\,]^{1/3}\gg\langle\Lambda^{SU(N_c-N_F)}_{{\cal N}=2\, SYM}\rangle$. There will be a large number of strongly coupled $SU(N_c-N_F)$ gluonia with the mass scale $\sim\langle\Lambda^{SU(N_c-N_F)}_{{\cal N}=1\, SYM}\rangle$. All original heavier $SU(N_c-N_F)$ charged particles with masses $\sim m$ or $g^2\mx$ are still weakly confined, but the string tension in this sector is larger now, $\langle\Lambda^{SU(N_c-N_F)}_{{\cal N}=2\, SYM}\rangle\ll\sigma^{1/2}_1
\sim\langle\Lambda^{SU(N_c-N_F)}_{{\cal N}=1\, SYM}\rangle\ll\mx$.\\

Now, as for the $SU(N_F)\times U(1)$ sector. At the scale $\mu= m/(\rm several)$ this is the UV free ${\cal N}=2$ SQCD with $N_F$ flavors of massless quarks ${\ov Q}^{\,a}_j, Q^i_a,\,\,i,a=1...N_F,\,\,
\langle m-a\rangle=0$, its gauge coupling grows logarithmically with diminished energy. The dynamics of this part is the same as in vs-vacua of section 6.4. When $\langle\Lambda_{SU(N_F)}\rangle^2/m\ll\mx\ll m$, see \eqref{(8.4)}, then the low energy superpotential in this sector will be as e.g. in \eqref{(7.1.7)} with $\nt=N_F$. I.e., all quarks in this $SU(N_F)$ sector will be higgsed in the weak coupling regime in this higher energy phase (while $\langle X^{adj}_{SU(N_F)}\rangle=0$), and $[(N_F^2-1)+1]$ long ${\cal N}=2$ multiplets of massive gluons (including $U(1)$ with its scalar "a") with masses $\sim \langle\qq\rangle^{1/2}_{N_F}\sim (\mx m)^{1/2}$ will be formed, see \eqref{(7.1.8)}. The global symmetry $SU(N_F)_{C+F}$ is unbroken in this higher energy phase, and $(N_F^2-1)\,\, {\cal N}=2$ long multiplets form the $SU(N_F)_{C+F}$ adjoint representation.
There are no particles with mass $\sim m$ and there are no massless particles. The multiplicity of vacua in this phase is determined by the multiplicity $N_c-N_F$ of $SU(N_c-N_F)$ SYM.

But at $(\mx m)^{1/2}\ll\langle\Lambda_{SU(N_F)}\rangle$ the lower energy phase is different. To avoid $g^2(\mu<\langle\Lambda_{SU(N_F)}\rangle)<0$, the main contributions to masses originate now in this UV free $SU(N_F)$ sector from higgsed $\langle X^{adj}_{SU(N_F)}\rangle\sim\langle\Lambda_{SU(N_F)}\rangle$, in the strong coupling and nonperturbative regime. The situation in this phase is also qualitatively similar to those in section 6.4 (or in 7.2 with $\no\geq 1$, the difference is that $\no=0,\, \nt=N_F$ now as in section 6.4, so that the whole $SU(N_F)$ color group will be broken at the scale $\sim\langle\Lambda_{SU(\nt=N_F)}\rangle$, see below). Because $SU(N_c-N_F)$ sector is decoupled, there is the residual unbroken discrete symmetry $Z_{2N_F-N_F\geq\, 2}=Z_{N_F\geq\, 2}$. Therefore, $X^{adj}_{SU(N_F)}$ are higgsed as, see \eqref{(2.1.1)},\eqref{(2.1.16)},
\bq
\langle\sqrt{2} X^{adj}_{SU(N_F)}\rangle=C_{N_F}\langle\Lambda_{SU(N_F)}\rangle\,{\rm diag}\,\Bigl(\,
\underbrace{{\tilde\rho}^{\,0},\,{\tilde\rho}^{\, 1},\,...,\,{\tilde\rho}^{\, N_F-1}}_{N_F}\,;\underbrace{0}
_{N_c-N_F}\,\Bigr ),\quad \tilde\rho=\exp\{\,\frac{2\pi i}{N_F}\,\}\,,\label{(8.5)}
\eq
and the whole $SU(N_F)_C$ color group is broken, $SU(N_F)_C\ra U^{N_F-1}(1)$. All quarks with $SU(N_F)_C$ colors acquire now masses $m_Q\sim\langle\Lambda_{SU(N_F)}\rangle$ which are large in comparison with the potentially possible scale of their coherent condensate, $m_Q\gg (\mx m)^{1/2}$, so that {\it they are not higgsed but decouple as heavy at $\mu<\langle\Lambda_{SU(N_F)}\rangle/(\rm several)$}, as well as all charged $SU(N_F)_C$ adjoints. Instead, $N_F$ massless at $\mx\ra 0$ dyons ${\ov D}_j, D_j$ are formed at the scale $\sim\langle\Lambda_{SU(N_F)}\rangle$.

The relevant part of the low energy superpotential of this sector is in this case qualitatively the same as in vs-vacua of section 6.4, see \eqref{(6.4.3)}, and only $\w_{SYM}$ is added
\bq
\w^{\,(N_F)}_{\rm low}=\w_{D}+\w_{a}+\w_{SYM}\,, \label{(8.6)}
\eq
\bbq
\w_{D}= (m-a)\sum_{j=1}^{N_F}{\ov D}_j D_j -\sum_{j=1}^{N_F} a_{D,\,j}{\ov D}_j D_j -\mx\Lambda_{SU(N_F)}\sum_{j=1}^{N_F}{\tilde\rho}^{\,{j-1}} a_{D,\,j}+\mx{\,\tilde L}\,
\Bigl (\sum_{j=1}^{N_F} a_{D,\,j}\Bigr )\,,
\eeq
\bbq
\w_{a}=\frac{\mx}{2}\frac{N_c N_F}{N_c-N_F}\,a^2+\mx N_F\,{\tilde\delta}_4\, (m-a)^2\,,
\eeq
\bbq
\w_{SYM}= (N_c-N_F)\mx \Bigl (\Lambda^{SU(N_{c}-N_F)}_{{\cal N}= 2\,\,SYM}\Bigr )^2= (N_c-N_F)\mx\Biggl (\frac{\lm^{2N_c-N_F}(m-{\hat c}\,a)^{N_F}}{[\,(1-{\hat c}) a\,]^{2 N_F}}\Biggr )^{\frac{1}{N_c-N_F}}\approx
\eeq
\bbq
\approx \,\mx\langle\Lambda^{SU(N_c-N_F)}_{{\cal N}=2\,\,SYM}\rangle^2\Biggl [(N_c-N_F) -N_F \frac{(2N_c-N_F)}{N_c-N_F}\,\frac{\hat a}{m_1}\,\Biggr ]\,,\quad m_1=\frac{N_c}{N_c-N_F} m\,,\quad a=\langle a\rangle+{\hat a}\,.
\eeq

From \eqref{(8.6)}
\bbq
\langle a\rangle=m\,,\quad \langle a_{D,\,j}\rangle=0\,,\quad \langle{\ov D}_j\rangle\langle D_j \rangle= -\mx\langle\Lambda_{SU(N_F)}\rangle{\tilde\rho}^{\,{j-1}}+\mx \langle{\tilde L}\rangle\,,\quad \langle\Sigma_D\rangle=\sum_{j=1}^{N_F}\langle{\ov D}_j\rangle\langle D_j\rangle=N_F\mx \langle{\tilde L}\rangle,
\eeq
\bq
\langle\Sigma_D\rangle\approx N_F\mx m_1\Bigl [\,1-\frac{2N_c-N_F}{N_c-N_F}\frac{\langle
\Lambda^{SU(N_c-N_F)}_{{\cal N}=2\,\,SYM}\rangle^2}{m^2_1}\,\Bigr ]\approx N_F\mx m_1\Bigl [\,1-\frac{2N_c-N_F}{N_c-N_F}\Bigl (\frac{\lm}{{m_1}}\Bigr )^{\frac{\bb}{N_c-N_F}}\Bigr ]\,,\label{(8.7)}
\eq
\bbq
\langle S\rangle_{N_F}=0,\quad\langle{\ov D}_j\rangle\langle D_j\rangle\approx \mx m_1\Bigl [\,1-\frac{2N_c-N_F}{N_c-N_F}\frac{\langle\Lambda^{SU(N_c-N_F)}_{{\cal N}=2\,\,SYM}\rangle^2}{m^2_1}\,\Bigr ]-\mx\langle\Lambda_{SU(N_F)}\rangle{\tilde\rho}^{\,{j-1}}\,,\quad j=1...N_F\,.
\eeq

As an example, we present also the analog of \eqref{(2.1.20)} in these S-vacua. From \eqref{(1.1)},\eqref{(8.6)},\eqref{(8.7)}
\bq
\langle\,\frac{\partial}{\partial m}\w_{\rm matter}\rangle=\langle\,\frac{\partial}{\partial m}
{\w}^{\,(N_F)}_{\rm low}\rangle\ra\langle{\rm Tr\,}\qq\rangle_{N_c}=\langle\Sigma_D\rangle+
\langle\,\frac{\partial}{\partial m}{\w_{SYM}}\rangle\approx \label {(8.8)}
\eq
\bbq
\approx  N_F\mx m_1\Bigl [\,1-\frac{N_c}{N_c-N_F}\Bigl (\frac{\lm}{{m_1}}\Bigr )^{\frac{\bb}{N_c-N_F}}\Bigr ]\,,
\eeq
this agrees with \eqref{(8.1)}.\\

The curve \eqref{(1.2)} has $N_c-1$ double roots in these $N_c-N_F$ S-vacua. From these, $N_c-N_F-1$ unequal double roots correspond to massless at $\mx\ra 0$ magnetic monopoles $M_{\rm n}$ from $SU(N_c-N_F)\,\, {\cal N}=2$ SYM, and remaining $N_F$ unequal double roots correspond to massless at $\mx\ra 0$ dyons $D_j$. Two single roots with $(e^{+}-e^{-})\sim \langle\Lambda^{SU(N_c-N_F)}_{{\cal N}=2\, SYM}\rangle$ \eqref{(8.3)} originate at $\mx\ra 0$ from the $SU(N_c-N_F)\,\,{\cal N}=2$ SYM sector. The global flavor symmetry $U(N_F)$ is unbroken and there are no massless Nambu-Goldstone particles. And there are no massless particles at all at $\mx\neq 0$. At the same time, as in section 6.4.2, the color is broken in this sector, $SU(N_F)_C\ra U^{N_F-1}(1)$, and masses of charged $SU(N_F)_C$ adjoints are different, see \eqref{(8.5)}. There is no color-flavor locking in this lower energy phase.

All original charged electric particles are confined in the lower energy phase at $\mx<\langle\Lambda_
{SU(N_F)}\rangle^2/m_1$, see \eqref{(8.4)}. But the phase changes in the $SU(N_F)$ sector already at very small $\mx :\,\langle\Lambda_{SU(N_F)}\rangle^2/m_1<\mx\ll\langle\Lambda_{SU(N_F)}\rangle\ll\lm\ll m$. Instead of higgsed $\langle X^{adj}_{SU(N_F)}\rangle\sim\langle\Lambda_{SU(N_F)}\rangle,\,\, SU(N_F)\ra U^{N_F-1}(1)$ and subsequent condensation of $N_F$ dyons $D_j$ \eqref{(8.7)}, all $N_F$ flavors of original electric quarks ${\ov Q}^{\,a}_i,\,Q^i_a,\,\,a,i=1...N_F$ are then higgsed as $\langle{\ov Q}^{\,a}_i
\rangle=\langle Q_a^i\rangle\approx\delta^i_a (\mx m_1)^{1/2}>\langle\Lambda_{SU(N_F)}\rangle$ (and clearly not confined but screened), forming $N_F^2$ long ${\cal N}=2$ multiplets of massive gluons.

Remind that, at $N_F<N_c$ and $m\gg\lm$, the mass spectra in these S vacua with the multiplicity $N_c-N_F$, as well as in the SYM vacua with the multiplicity $N_c$ in section 5, depend essentially on the value of $m/\lm$, and all these vacua with the unbroken global flavor symmetry $U(N_F)$ evolve at $\mx\ll m\ll\lm$ to L-vacua with the multiplicity $2N_c-N_F$ and with the spontaneously broken discrete $Z_{\bb}$ symmetry, see sections 3.1 and 12 in \cite{ch5}.

\section{Conclusions}

\hspace*{4mm} We presented above in this paper the calculations of mass spectra of softly broken ${\cal N}=2\,\, SU(N_c)$ or $U(N_c)$ gauge theories in a large number of their various vacua with the unbroken non-trivial $Z_{\bb\geq 2}$ discrete symmetry. The quantum numbers of light particles in each vacuum were determined, as well as forms of corresponding low energy Lagrangians and mass spectra at different hierarchies between the Lagrangian parameters $m,\,\mx,\,\lm$. A crucial role in obtaining these results was played by the use\,: a) the unbroken $Z_{\bb\geq 2}$ symmetry,\, b) the pattern of spontaneous flavor symmetry breaking, $U(N_F)\ra U(\no)\times U(\nt),\, 0\leq \no<N_F/2,\,$ c) the knowledge of multiplicities of various vacua. Besides, the knowledge of the quark and gluino condensates $\langle\Qo\rangle_{N_c},\,\langle\Qt\rangle_{N_c}$, and $\langle S\rangle_{N_c}$ (obtained from the effective superpotential $\w_{\rm tot}^{\rm eff}$) was also of great importance. In addition, two dynamical {\it assumptions} "A" and "B\," of general character formulated in Introduction were used. They concern the BPS properties of original particles and the absence of extra fields massless at $\mx\neq 0,\, m\neq 0$. All this appeared to be sufficient to calculate mass spectra in vacua considered at small but nonzero values of $\mx$, in particular at $0<\mx<M_{\rm min}$, where $M_{\rm min}$ is the independent of $\mx$ mass scale appropriate for each given vacuum, such that the phase and mass spectrum continue smoothly down to $\mx\ra 0$. In other words, the theory stays at $\mx<M_{\rm min}$ in the same regime and all hierarchies in the mass spectrum are the same as in the ${\cal N}=2$ theory at $\mx\ra 0$.

Within this framework we considered first in detail in section 2.1 the br2 vacua of $SU(N_c)$ theory with $N_c+1<N_F<2N_c-1$ (these are vacua of the baryonic branch in \cite{APS} or zero vacua in \cite{SY1,SY2}), at $m\ll\lm$ and $0<\mx\ll M^{(\rm br2)}_{\rm min}=\langle\Lambda^{SU(\nd-{\rm n}_1)}_{{\cal N}=2\,\,SYM}
\rangle^2/\lm\ll\langle\Lambda^{SU(\nd-{\rm n}_1)}_{{\cal N}=2\,\,SYM}\rangle\ll m\ll\lm$. The original color symmetry is broken spontaneously in these vacua by higgsed $\langle X^{adj}_{SU(N_c)}\rangle\neq 0$ in three stages. The first stage is at the highest scale $\mu\sim\lm$, $\,\langle X^{adj}_{SU(2N_c-N_F)}\rangle\sim\lm$, $\,\,SU(N_c)\ra SU(N_F-N_c)\times U^{(1)}(1)\times U^{2N_c-N_F-1}(1)$, this pattern is required by the unbroken discrete $Z_{2N_c-N_F\geq 2}$ symmetry and is necessary to avoid $g^2(\mu<\lm)$ in this effectively massless at the scale $\mu\sim\lm$ $\,\,{\cal N}=2$ UV free theory. The second stage is in the $SU(N_F-N_c)$ color sector at the scale $\mu\sim m$, $\,\langle X^{adj}_{SU(N_F-N_c)}\rangle\sim m\ll\lm$, $\,SU(N_F-N_c)\ra SU(\no)\times U^{(2)}(1)\times SU(N_F-N_c-\no)$. And the third stage is in the $SU(N_F-N_c-\no)$ color sector at the scale $\mu\sim\langle\Lambda^
{SU(\nd-{\rm n}_1)}_{{\cal N}=2\,\,SYM}\rangle\ll m$, $\,SU(N_F-N_c-\no)\ra U^{N_F-N_c-\no-1}(1)$.

The global flavor symmetry $U(N_F)$ is broken spontaneously at $\mx\neq 0$ in these br2-vacua at the scale $\mu\sim (\mx m)^{1/2}$ in the $SU(\no)$ color sector by higgsed original pure electric quarks, $\,\langle{\ov Q}^{\,a}_i\rangle=\langle Q^i_a\rangle\sim \delta_a^i\,(\mx m)^{1/2}\,,\,\, a=1...\no,\,\, i=1...N_F$, as\,: $U(N_F)\ra U(\no)\times U(\nt),\, 1\leq\no< N_F-N_c$.

It was shown that the lightest charged BPS particles (massless at $\mx\ra 0$) are the following.\,-\\
1) $\bb$ flavorless mutually local dyons $D_j, {\ov D}_j$ (this number $\bb$ is required by the unbroken $Z_{\bb\geq 2}$ discrete symmetry which operates interchanging them among each other), with the two nonzero pure electric charges, the $SU(N_c)$ baryon charge and $U^{(1)}(1)$ one. The nonzero magnetic charges of all these dyons are corresponding $SU(\bb)$ adjoints. The whole $\bb$ set of these dyons is coupled with $\bb$ Abelian ${\cal N}=2$ photon multiplets, $U^{\bb-1}(1)$ magnetic and $U^{(1)}(1)$ electric. These dyons are formed at the scale $\mu\sim\lm$, at the first stage of the color breaking, and are mutually local with respect to all particles of the $SU(N_F-N_c)$ sector and between themselves.\\
2) All original electric particles of remaining unbroken at $\mx\ra 0$ $\,SU(\no)$ subgroup of original $SU(N_c)$.\\
3) $\nd-\no-1$ pure magnetic monopoles with the $SU(\nd-\no)$ adjoint charges. These last are formed at the low scale $\mu\sim\langle\Lambda^{SU(\nd-{\rm n}_1)}_{{\cal N}=2\,\,SYM}\rangle\ll m\ll\lm$ in the ${\cal N}=2\,\,SU(\nd-\no)$ SYM sector, with the scale factor $\langle\Lambda^{SU(\nd-{\rm n}_1)}_{{\cal N}=2\,\,SYM}\rangle$ of its gauge coupling. These magnetic monopoles are coupled with $U^{\nd-\no-1}(1)$
${\cal N}=2$ Abelian magnetic photon multiplets.\\
4) In addition to original electric ${\cal N}=2\,\, SU(\no)$ adjoints, there are magnetic $U^{\nd-\no-1}(1)$ and  $U^{\bb-1}(1)$, and one $U^{(1)}(1)$ electric ${\cal N}=2$ Abelian photon multiplets, all massless at $\mx\ra 0$.

Massless at $\mx\ra 0$ $\,2N_c-N_F$ dyons $D_j,\,\no$ original pure electric quarks $Q^i_a, {\ov Q}^a_i,\, a=1...\no,\, i=1...N_F$ from $SU(\no)$, and $\nd-\no-1$ pure magnetic monopoles are all higgsed at small $\mx\neq 0$, and no massless particles remains (except for $2\no\nt$ Nambu-Goldstone multiplets due to spontaneous breaking of global flavor symmetry, $U(N_F)\ra U(\no)\times U(\nt)$) in the $SU(\no)$ color sector due to higgsed original pure electric quarks with $N_F$ flavors). The mass spectrum was described in these br2-vacua at smallest $0<\mx\ll\langle\Lambda^{SU(\nd-{\rm n}_1)}_{{\cal N}=2\,\,SYM}\rangle^2/\lm$.

The material of this section 2.1 served then as a basis for similar regimes in sections 3,\,6,\,7, and 8.

The mass parameter $\mx$ was increased then in two different stages\,:\, 1) $\langle\Lambda^{SU(\nd-{\rm n}_1)}_{{\cal N}=2\,\,SYM}\rangle\ll\mx\ll m$ in section 2.3\,;\,\,2) $m\ll\mx\ll\lm$ in section 2.4, and changes in the mass spectra were described. In all those cases when the corresponding ${\cal N}=1$ SQCD lower energy theories were weakly coupled, we needed no additional dynamical assumptions.  And only in those few cases in section 2.4, when the corresponding ${\cal N}=1$ SQCD theories were in the strongly coupled conformal regime at $3N_c/2<N_F<2N_c$, we used additionally the assumption of the dynamical scenario introduced in \cite{ch3}. In essence, this assumption from \cite{ch3} looks here as follows\,: {\it no additional parametrically light composite solitons are formed in this ${\cal N}=1\,\, SU(\nd)$ SQCD without decoupled  colored adjoint scalars $X^{\rm adj}_{SU(\nd)}$ at those even lower scales where this ${\cal N}=1$ conformal regime is broken by nonzero particle masses} (see also the footnote \ref{(f3)}). As was shown in \cite{ch3,ch5,ch6}, this dynamical scenario does not contradict to any proven properties of ${\cal N}=1$ SQCD, satisfies all those checks of Seiberg's duality hypothesis for ${\cal N}=1$ SQCD which were used in \cite{S2}, and allows to calculate the corresponding mass spectra.

Many other vacua with the broken or unbroken flavor symmetry and at different hierarchies between the Lagrangian parameters $m,\, \mx,\, \lm$ were considered in sections 3-8 and corresponding mass spectra were calculated within the above described framework. Besides, calculations of power corrections to the leading terms of the low energy condensates of original quarks $Q,\,{\ov Q}$ and dyons in a number of vacua are presented in two important Appendices. The results agree with also presented in these Appendices {\it independent} calculations of these quark and dyon condensates using roots of the Seiberg-Witten spectral curve. And because these last calculations with roots are valid {\it only for BPS particles}, these agreements serve as an independent confirmation of the main assumption "A" from Introduction about BPS properties of original quarks and as numerous checks of a self-consistency of the whole approach.

We consider that there is no need to repeat in detail in these conclusions all results obtained in this paper (see the table of contents). But at least two points have to be emphasized. 1) As shown e.g. in section 6.2 (and similarly in sections 6.4, 7 and 8), the widespread opinion that the holomorphic dependence of gauge invariant chiral condensates (e.g. $\langle (\qq)_{1,2}\rangle_{N_c},\, \langle S\rangle_{N_c},\, \langle {\rm Tr\,}(X^{adj}_{SU(N_c)})^2\rangle\,)$ on parameters of the superpotential implies the absence of phase transitions in supersymmetric theories is not right. 2) In contradiction with the smooth analytical crossover transition from the higher energy region to lower energy one (in sections 6,7 and 8) and the emergence of the "instead-of-confinement" regime proposed in a number of papers by M.Shifman and A.Yung, see e.g. the latest paper \cite{SY6} and references therein to their previous papers with this crossover, we have found that this transition is not a crossover but a phase transition. See the critique of the M.Shifman and A.Yung proposal in sections 6.2, 6.4.

Clearly, this article does not pretend on strict proofs as the two dynamical assumptions of general character presented in Introduction were used (i.e., the assumption "A" about the BPS nature of original electric particles, and "B" about the absence of extra massless particles). But, to the best of our knowledge, the results obtained look self-consistent, satisfy a large number of independent checks, and do not contradict to any proven results.

In comparison with corresponding results from recent related papers \cite{SY1,SY2,SY6} of M.Shifman and A.Yung (and from a number of their previous numerous papers on this subject), our results are essentially different. In addition to critical remarks given in Introduction and in sections 2.1, 6.2 and 6.4, an extended criticism of a number of results from \cite{SY1,SY2} is given in section 8 of \cite{ch6}\,.\\

\appendix{\Large \bf Quark and dyon condensates}\\

\hspace*{4mm} As examples, we present below in two Appendices the calculation of leading power corrections to the low energy quark condensates $\langle\Qo\rangle_{\no}$ in br1 vacua of $SU(N_c)$ and $U(N_c)$ in sections 4.1 and 4.2, and in br2 vacua of $SU(N_c)$ and $U(N_c)$ in sections 2.1 and 2.2, and to the dyon condensates in sections 2.1 and 2.2. The results agree with also presented in these Appendices {\it independent} calculations of these condensates using roots of the Seiberg-Witten spectral curve. These agreements serve as {\it the independent numerous checks} of a self-consistency of the whole approach.

\section{Calculations of quark condensates}

{\bf 1) br1 vacua of $SU(N_c),\,\, m\gg\lm$ in section 4.1}\\

From \eqref{(1.3)},\eqref{(1.4)}, in these vacua (up to even smaller power corrections)
\bq
\langle\Qo\rangle_{N_c}=\mx m_3-\frac{N_c-\nt}{N_c-\no}\langle\Qt\rangle_{N_c}\,,\quad
\langle\Qt\rangle_{N_c}\approx\mx m_3\Bigl (\frac{\lm}{m_3}\Bigr )^{\frac{2N_c-N_F}{N_c-\no}}\,,\label{(A.1)}
\eq
\bbq
\langle S\rangle_{N_c}=\frac{\langle\Qo\rangle_{N_c}\langle\Qt\rangle_{N_c}}{\mx}\approx m_3\langle\Qt\rangle_{N_c}\,,\quad m_3=\frac{N_c}{N_c-\no} m\,.
\eeq
From \eqref{(2.1.9)},\eqref{(A.1)}, the leading power correction to $\langle\w^{\,\rm eff}_{\rm tot}\rangle$ is
\bq
\langle\delta \w^{\,\rm eff}_{\rm tot}\rangle\approx (N_c-\no)\mx m^2_3\Bigl (\frac{\lm}{m_3}\Bigr )^{\frac{2N_c-N_F}{N_c-\no}}\approx (N_c-\no)\langle S\rangle_{N_c}\,.\label{(A.2)}
\eq
On the other hand, this leading power correction to $\langle\w^{\,\rm low}_{\rm tot}\rangle$ originates from the $SU(N_c-\no)$ SYM part, see \eqref{(4.1.2)},\eqref{(4.1.4)},
\bbq
\langle\delta \w^{\,\rm low}_{\rm tot}\rangle\approx\mx\langle\,{\rm Tr\,}\Bigl (X^{adj}_{SU(N_c-\no)}\Bigr )^2\,\rangle\approx(N_c-\no)\langle S\rangle_{N_{c}-\no}
\approx (N_c-\no)\mx\langle\Lambda^{SU(N_{c}-\no)}_{{\cal N}=2\,\,SYM}\rangle^2\approx
\eeq
\bq
\approx (N_c-\no)\mx  m_3^2\Bigl (\frac{\lm}{m_3}\Bigr )^{\frac{2N_c-N_F}{(N_c-\no)}}\,,\quad \ra\quad \langle\delta \w^{\,\rm low}_{\rm tot}\rangle=\langle\delta \w^{\,\rm eff}_{\rm tot}\rangle\,,\label{(A.3)}
\eq
as it should be.

To calculate the leading power correction $\delta\langle\Qo\rangle_{\no}\sim\mx\langle\Lambda^
{SU(N_{c}-\no)}_{{\cal N}=2\,\,SYM}\rangle^2/m$ to the low energy quark condensate $\langle\Qo\rangle_{\no}$, we have to account for the terms with the first power of quantum fluctuation ${\hat a}$ of the field $a=\langle a\rangle+{\hat a},\,\,\langle{\hat a}\rangle=0$, in the $SU(N_c-\no)$ SYM contribution to $\delta \w^{\,\rm low}_{\rm tot}$, see \eqref{(4.1.2)}-\eqref{(4.1.4)},
\bbq
\delta \w^{\,\rm low}_{\rm tot}\approx (N_c-\no)\mx\, \delta\, \Bigl (\Lambda^{SU(N_{c}-\no)}_{{\cal N}=2\,\,SYM}\Bigr )^2\approx (N_c-\no)\mx\,\delta\,\Bigl (\frac{\lm^{2N_c-N_F}(m_3-c\,{\hat a})^{N_F}}{[\,m_3+(1-c){\hat a}\,]^{2\no}}\Bigr )^{\frac{1}{N_c-\no}}\approx
\eeq
\bq
\approx \,\mx\langle\Lambda^{SU(N_{c}-\no)}_{{\cal N}=2\,\,SYM}\rangle^2\,\Bigl [\,-\no \frac{(2N_c-N_F)}{N_c-\no}\,\Bigr ]\,\frac{{\hat a}}{m_3}\,,\quad m_3=\frac{N_c}{N_c-\no} m\,.\label{(A.4)}
\eq
As a result, from \eqref{(4.1.4)},\eqref{(A.4)},\eqref{(A.1)},
\bbq
\delta\langle\Qo\rangle_{\no}\approx\frac{1}{\no}\langle\frac{\partial}{\partial {\hat a}}\delta \w^{\,\rm low}_{\rm tot}\rangle\approx - \frac{(2N_c-N_F)}{N_c-\no}\,\mx\frac{\langle\Lambda^{SU(N_{c}-\no)}_{{\cal N}=2\,\,SYM} \rangle^2}{m_3}\,,
\eeq
\bq
\langle\Qo\rangle^{SU(N_c)}_{\no}=\langle{\ov Q}^1_1\rangle\langle Q^1_1\rangle\approx \mx m_3\Biggl [1-\frac{2N_c-N_F}{N_c-\no} \,\frac{\langle\Lambda^{SU(N_{c}-\no)}_{{\cal N}=2\,\,SYM} \rangle^2}{m^2_3}\Biggr ]\approx \label{(A.5)}
\eq
\bbq
\approx\mx m_3\Biggl [1-\frac{2N_c-N_F}{N_c-\no}\Bigl (\frac{\lm}{m_3}\Bigr )^{\frac{2N_c-N_F}{N_c-\no}} \Biggr ]\,,\quad\langle\Qo\rangle^{SU(N_c)}
_{N_c}\approx\mx m_3\Biggl [1-\frac{N_c-\nt}{N_c-\no}\Bigl (\frac{\lm}{m_3}\Bigr )^{\frac{2N_c-N_F}{N_c-\no}} \Biggr ]\,.
\eeq
It is seen from \eqref{(A.5)} that, although the leading terms are the same in $\langle\Qo\rangle_{\no}$ and $\langle\Qo\rangle_{N_c}$, the leading power corrections are different.\\

{\bf 2) br1 vacua of $U(N_c),\,\, m\gg\lm$ in section 4.2}

From \eqref{(1.4)},\eqref{(4.2.1)},\eqref{(4.2.2)},\eqref{(4.2.3)} (keeping now leading power corrections)
\bq
\langle\Qo\rangle_{N_c}=\mx m-\langle\Qt\rangle_{N_c}\,,\label{(A.6)}
\eq
\bbq
\langle\Qt\rangle_{N_c}=\mx\lm^{\frac{\bb}{N_c-\no}}\Bigl (\frac{\langle\Qo\rangle_{N_c}}{\mx}\Bigr )^{\frac{\nd-\no}{N_c-\no}}\approx\mx m\Bigl (\frac{\lm}{m}\Bigr )^{\frac{2N_c-N_F}{N_c-\no}}
\Biggl (1+O\Bigl (\frac{\langle\Lambda^{SU(N_{c}-\no)}_{{\cal N}=2\,\,SYM}\rangle^2}{m^2} \Bigr )\Biggr )\,,
\eeq
\bbq
\langle S\rangle_{N_c}=\frac{\langle\Qo\rangle_{N_c}\langle\Qt\rangle_{N_c}}{\mx}\approx \mx m^2\Bigl (\frac{\lm}{m}\Bigr )^{\frac{2N_c-N_F}{N_c-\no}}\Biggl (1+O\Bigl (\frac{\langle\Lambda^{SU(N_{c}-\no)}_{{\cal N}=2\,\,SYM}\rangle^2}{m^2} \Bigr )\Biggr )\,,
\eeq
\bq
\frac{\langle a_0\rangle}{m}=\frac{1}{N_c\mx m}\langle{\rm Tr}\,\qq\rangle_{N_c}\approx\frac{\no}{N_c}  +\frac{\nt-\no}{N_c}\Bigl (\frac{\lm}{m}\Bigr )^{\frac{2N_c-N_F}{N_c-\no}}\,,\quad \langle a\rangle=\langle m-a_0\rangle\,, \label{(A.7)}
\eq
\bq
\langle\w^{\,\rm low}_{\rm tot}\rangle\approx \Bigl [\langle\w_{\no}\rangle=0\,\Bigr ]+\Biggl [
\langle\w_{a_0,a}\rangle=\frac{\mx}{2}\Bigl (\,N_c\langle a_0\rangle^2+\frac{\no N_c}{N_c-\no}\langle a\rangle^2\Bigr )\Biggr ]+\langle\w_{SU(N_c-\no)}^{\,SYM}\rangle\,,\label{(A.8)}
\eq
\bbq
\langle\w_{SU(N_c-\no)}^{\,SYM}\rangle=\mx\langle{\,\rm Tr\,}\Bigl ( X^{adj}_{SU(N_c-\no)}\Bigr )^2\rangle=
(N_c-\no)\langle S\rangle_{SU(N_c-\no)}=(N_c-\no)\mx\langle\Lambda^{SU(N_{c}-\no)}_{{\cal N}=2\,\,SYM}\rangle^2\approx
\eeq
\bbq
\approx (N_c-\no)\mx\langle\Biggl (\frac{\lm^{2N_c-N_F}(m-a_0-c \,a)^{N_F}}{[\,(1-c) a \,]^{2\no}} \Biggr )^{\frac{1}{N_c-\no}}\rangle\approx (N_c-\no)\mx m^2\Bigl (\frac{\lm}{m}\Bigr )^{\frac{2N_c-N_F}{N_c-\no}}\,.
\eeq
From \eqref{(A.6)}-\eqref{(A.8)}, the leading power correction $\langle\,\delta\,\w^{\,\rm low}_{\rm tot}\rangle\sim\mx\langle\Lambda^{SU(N_{c}-\no)}_{{\cal N}=2\,\,SYM}\rangle^2$ to $\langle\w^{\,\rm low}
_{\rm tot}\rangle$ looks as
\bbq
\langle\,\delta\, \w^{\,\rm low}_{\rm tot}\rangle=\Bigl (\langle\,\delta\,\w_{a_0,a}\rangle=0\Bigr )+
\langle\,\delta\,\w_{SU(N_c-\no)}^{\,SYM}\rangle\approx
\eeq
\bq
\approx (N_c-\no)\mx\langle\Lambda^{SU(N_{c}-\no)}_{{\cal N}=2\,\,SYM}\rangle^2\approx (N_c-\no)\,m\langle\Qt\rangle_{N_c}\approx (N_c-\no)\mx m^2\Bigl (\frac{\lm}{m}\Bigr )^{\frac{2N_c-N_F}{N_c-\no}}\,.\label{(A.9)}
\eq

Instead of \eqref{(2.1.9)},\eqref{(2.1.10)} we have now, see \eqref{(2.2.7)},
\bq
\w^{\,\rm eff}_{\rm tot}(\Pi)=m\,{\rm Tr}\,({\ov Q} Q)_{N_c}-\frac{1}{2\mx}\Biggl [ \,\sum_{i,j=1}^{N_F} ({\ov Q}_j Q^i)_{N_c}({\ov Q}_{\,i} Q^j)_{N_c}\Biggr ]-\nd \Biggl [\,S_{N_c}=\frac{\langle\Qo\rangle_{N_c}\langle\Qt\rangle_{N_c}}{\mx}\,\Biggr ],\,\,\,\,\,\label{(A.10)}
\eq
and using \eqref{(A.6)} (with the same accuracy)
\bbq
\langle\delta\w^{\,\rm eff}_{\rm tot}(\Pi)\rangle\approx (N_c-\no)m\langle\Qt\rangle_{N_c}\approx (N_c-\no)\mx m^2\Bigl (\frac{\lm}{m}\Bigr )^{\frac{2N_c-N_F}{N_c-\no}}\approx\langle\delta\w^{\,\rm low}
_{\rm tot}\rangle\,,
\eeq
as it should be.

As for the leading power correction $\delta\langle\Qo\rangle_{\no}\sim \mx m\Bigl (\lm/m\Bigr )^
{\frac{2N_c-N_F}{N_c-\no}}$ to the quark condensate, we have now instead of \eqref{(A.4)}
\bbq
\delta\w^{\,\rm low}_{\rm tot}=\delta \w_{a_0,a}+\delta \w_{SU(N_c-\no)}^{\,SYM}\,,\quad a_0=\langle a_0\rangle +{\hat a}_0\,,\quad a=\langle a\rangle +{\hat a}\,,
\eeq
\bbq
\delta \w_{SU(N_c-\no)}^{\,SYM}=(N_c-\no)\mx\, \delta\, \Bigl (\Lambda^{SU(N_{c}-\no)}_{{\cal N}=2\,\,SYM}\Bigr )^2\approx (N_c-\no)\mx\,\delta\,\Bigl (\frac{\lm^{2N_c-N_F}(m-{\hat a}_0-c\,{\hat a})^{N_F}}{[\,m+(1-c){\hat a}\,]^{2\no}}\Bigr )^{\frac{1}{N_c-\no}}\approx
\eeq
\bq
\approx \Bigl [\,-N_F{\hat a_0}-\frac{\no(2N_c-N_F)}{N_c-\no}\,{\hat a}\, \Bigr ) \mx\frac{\langle\Lambda^{SU(N_{c}-\no)}_{{\cal N}=2\,\,SYM}\rangle^2}{m}\,,
\label{(A.11)}
\eq
while from \eqref{(4.2.3)},\eqref{(A.7)}
\bq
\delta \w_{a_0,a}\approx (\nt-\no)\Bigl [\,{\hat a}_0-\frac{\no}{N_c-\no}\,{\hat a}\,\Bigr ]\mx\frac{\langle\Lambda^{SU(N_{c}-\no)}_{{\cal N}=2\,\,SYM}\rangle^2}{m}\,.\label{(A.12)}
\eq

Instead of \eqref{(A.5)} (with the same accuracy), on the one hand, see \eqref{(4.2.3)},\eqref{(4.2.4)},\eqref{(A.6)},
\bbq
\langle{\,\rm Tr\,}\qq\rangle_{\no}=\no\mx m+\delta\langle{\,\rm Tr\,}\qq\rangle_{\no}=\no\Bigl (\mx m+\delta\,\langle\Qo\rangle_{\no}\Bigr )\,,
\eeq
\bq
\delta\langle{\,\rm Tr\,}\Qo\rangle_{\no}=\frac{1}{\no}\langle\frac{\partial}{\partial {\hat a}_0} \Bigl (\delta \w_{SU(N_c-\no)}^{\,SYM}+\delta \w_{a_0,a}\Bigr )\rangle\approx -2 \mx\frac{\langle\Lambda^{SU(N_{c}-\no)}_{{\cal N}=2\,\,SYM}\rangle^2}{m}\,,\label{(A.13)}
\eq
while on the other hand,
\bbq
\delta\langle{\,\rm Tr\,}\Qo\rangle_{\no}=\frac{1}{\no}\langle\,\frac{\partial}{\partial {\hat a}}\Bigl (\delta \w_{SU(N_c-\no)}^{\,SYM}+\delta \w_{a_0,a}\Bigr )\approx -2 \mx\frac{\langle\Lambda^{SU(N_{c}-\no)}_{{\cal N}=2\,\,SYM}\rangle^2}{m} \,,
\eeq
as it should be.

Therefore, on the whole for these $U(N_c)$ br1 vacua, see \eqref{(A.6)} for $\langle\Qo\rangle^{U(N_c)}_{N_c}$,
\bq
\langle\Qo\rangle^{U(N_c)}_{\no}=\langle{\ov Q}^1_1\rangle\langle Q^1_1\rangle\approx\mx m\Bigl [\,1-2\frac{\langle\Lambda^{SU(N_{c}-\no)}_{{\cal N}=2\,SYM}\rangle^2}{m^2}\,\Bigr ]\approx\mx m \Bigl [\, 1-2\Bigl (\frac{\lm}{m}\Bigr )^{\frac{2N_c-N_F}{N_c-\no}}\,\Bigr ]\,,\label{(A.14)}
\eq
\bbq
\langle\Qo\rangle^{U(N_c)}_{N_c}\approx\mx m\Bigl [\,1-\frac{\langle\Lambda^{SU(N_{c}-\no)}_{{\cal N}=2\,\,SYM}\rangle^2}{m^2}\,\Bigr ].
\eeq

We can compare also the result for $\langle\Qo\rangle^{U(N_c)}_{\no}$ in \eqref{(A.14)} with those from \eqref{(2.2.10)}.
\footnote{\,
In these simplest br1 vacua of $U(N_c)$ the charges and multiplicities of particles massless at $\mx\ra 0$ are evident, these are original electric particles from $SU(\no)\times U^{(0)}(1)$ and $N_c-\no-1$ magnetic monopoles from $SU(N_c-\no)$ SYM.
}
To obtain definite predictions from \eqref{(2.2.10)} one has to find first the values of roots entering \eqref{(2.2.10)}. In the case  considered, to obtain definite values of all roots of the curve \eqref{(1.2)} in the br1 vacua of $U(N_c)$ theory one needs only one additional relation \cite{CIV,CSW} for the two single roots $e^{\pm}$ of the curve \eqref{(1.2)}
\bq
e^{\pm}=\pm 2\sqrt{\langle S\rangle_{N_c}/\mx}\,,\quad e^{\pm}_c=\frac{1}{2} (e^{+}+e^{-})=0\,.\label{(A.15)}
\eq
One can obtain then from the $U(N_c)$ curve \eqref{(1.2)} the values of $\no$ equal double roots $e^{(Q)}_i$ of original electric quarks from $SU(\no)$ and $N_c-\no-1$ unequal double roots $e^{(M)}_k$ of magnetic monopoles from $SU(N_c-\no)$ SYM. We find from \eqref{(1.2)},\eqref{(A.15)} with our accuracy, see \eqref{(A.6)},\eqref{(A.8)},
\bbq
e^{+}=-e^{-}\approx 2 m\Bigl (\frac{\lm}{m}\Bigr )^{\frac{2N_c-N_F}{2(N_c-\no)}}\Biggl (1+O\Bigl (\frac{\langle\Lambda^{SU(N_{c}-\no)}_{{\cal N}=2\,\,SYM}\rangle^2}{m^2} \Bigr ) \Biggr )
\approx 2\langle\Lambda^{SU(N_{c}-\no)}_{{\cal N}=2\,\,SYM}\rangle,\quad e^{(Q)}_i= - m,\,\, i=1...\no\,,
\eeq
\bq
e^{(M)}_k\approx 2\cos (\frac{\pi k}{N_c-\no})\langle\Lambda^{SU(N_{c}-\no)}
_{{\cal N}=2\,\,SYM}\rangle-\frac{\nt-\no}{N_c-\no-1}\frac{\langle\Lambda^{SU(N_{c}-\no)}
_{{\cal N}=2\,\,SYM}\rangle^2}{m}\,,\quad k=1...(N_c-\no-1)\,.\label{(A.16)}
\eq
Let us note that these roots satisfy the sum rule, see \eqref{(A.7)},
\bq
\sum_{n=1}^{N_c}\phi_n=\frac{1}{2}\sum_{n=1}^{2 N_c}(-e_n)=N_c\langle a_0\rangle\,, \label{(A.17)}
\eq
as it should be.

Therefore, from \eqref{(2.2.10)},\eqref{(A.16)}
\bbq
\langle\Qo\rangle^{U(N_c)}_{\no}= - \mx\sqrt{(e^{(Q)}_i-e^+)(e^{(Q)}_i-e^-)}\approx
\eeq
\bq
\approx\mx m \Bigl (1-2\frac{\langle\Lambda^{SU(N_{c}-\no)}_{{\cal N}=2\,\,SYM}\rangle^2}{m^2} \Bigr)\approx
\mx m \Bigl [\, 1-2\Bigl (\frac{\lm}{m}\Bigr )^{\frac{2N_c-N_F}{N_c-\no}}\,\Bigr ]\,,\quad i=1...\no\,,\,\label{(A.18)}
\eq
this agrees with \eqref{(A.14)}.\\

Besides, we can use the knowledge of roots of the curve \eqref{(1.2)} for the $U(N_c)$ theory and of $\langle a_0\rangle$, see \eqref{(A.7)},\eqref{(A.16)}, to obtain the values of $\langle\Qo\rangle^{SU(N_c)}_{\no}$ condensates in the $SU(N_c)$ theory. For this, the curve \eqref{(1.2)} for the $U(N_c)$ theory
\bq
y^2=\prod_{i=1}^{N_c}(z+\phi_i)^2-4\lm^{\bb}(z+m)^{N_F}\,,\quad \sum_{i=1}^{N_c}\phi_i=\frac{1}{2}\sum_{n=1}^{2 N_c}(-e_n)=N_c\langle a_0\rangle{\rm\,\,\, in\,\,\, U(N_c)}\,,\label{(A.19)}
\eq
we rewrite in the form of the $SU(N_c)$ theory
\bq
y^2=\prod_{i=1}^{N_c}({\hat z}+{\hat \phi}_i)^2-4\lm^{\bb}({\hat z}+{\hat m})^{N_F}\,,\,\, \sum_{i=1}^{N_c}{\hat\phi}_i=\frac{1}{2}\sum_{n=1}^{2 N_c}(-{\hat e}_n)=0\,,
\,\, {\hat m}=m\, (1-m^{-1}\langle a_0\rangle)\,.\,\label{(A.20)}
\eq

But clearly, it remained the same $U(N_c)$ theory with the same condensates \eqref{(A.18)}. Therefore, for the $SU(N_c)$ curve \eqref{(1.2)} the quark condensates will be as in \eqref{(A.18)} with $m$ replaced by
${m^\prime}=m/(1-m^{-1}\langle a_0\rangle)$, both in the leading terms and in all power corrections, i.e., see \eqref{(A.7)},
\bbq
\frac{\langle a_0\rangle}{m}\,\,\ra\,\,\approx \Biggl (\,\frac{\no}{N_c}+\frac{\nt-\no}{N_c}\Bigl (\frac{\lm}{m_3}\Bigr )^{\frac{2N_c-N_F}{N_c-\no}}\,\Biggr )\,,
\eeq
\bq
m\ra {m^\prime}\approx m_3\Biggl (1+\frac{\nt-\no}{N_c-\no}\Bigl (\frac{\lm}{m_3}\Bigr )^{\frac{\bb}{N_c-\no}}\Biggr )\,,\quad m_3=\frac{N_c}{N_c-\no}\,m\,.\label{(A.21)}
\eq
Then, from \eqref{(A.18)},\eqref{(A.21)},
\bq
\langle\Qo\rangle^{SU(N_c)}_{\no}\approx \mx m_3 \Biggl (1+\frac{\nt-\no}{N_c-\no}\Bigl (\frac{\lm}{m_3}\Bigr )^{\frac{\bb}{N_c-\no}} \Biggr )\Bigl [\, 1-2\Bigl (\frac{\lm}{m_3}\Bigr )^{\frac{2N_c-N_F}{N_c-\no}}\,\Bigr ]\approx \label{(A.22)}
\eq
\bbq
\approx \mx m_3 \Bigl [\, 1-\frac{\bb}{N_c-\no}\Bigl (\frac{\lm}{m_3}\Bigr )^{\frac{2N_c-N_F}{N_c-\no}}\,\Bigr ]\,,
\eeq
this agrees with \eqref{(A.5)}.\\

{\bf 3) br2 vacua of $SU(N_c),\,\, m\ll\lm$ in section 2.1}

The calculations of leading power corrections to $\langle\,\delta \w_{\rm tot}^{\,\rm low}\rangle$ and $\langle\,\delta \w^{\,\rm eff}_{\rm tot}\rangle$ have been presented in \eqref{(2.1.15)}. Therefore, we present here the calculation of the leading power correction $\delta\,\langle\Qo\rangle_{\no}\sim \mx \langle\Lambda^{SU(N_{c}-\no)}_{{\cal N}=2\,\,SYM}\rangle^2/m$ to the quark condensate. For this, see \eqref{(2.1.4)},\eqref{(2.1.15)}, we write ($a_{1,2}=\langle a_{1,2}\rangle+{\hat a}_{1,2}$)
\bbq
\delta \w^{(SYM)}_{SU(\nd-\no)}\approx (\nd-\no)\wmu\, \delta\,\Bigl (\Lambda^{SU(\nd-\no)}_{{\cal N}=2\,\,SYM}\Bigr )^2\approx (\nd-\no)\wmu\,\delta\,\Biggl (\frac{\Lambda_{SU(\nd)}^{2\nd-N_F}(m_2-{\hat a}_1-c_2\,{\hat a}_2)^{N_F}}{[\,m_2+(1-c_2)\,{\hat a}_2\,]^{\,2\no}}\Biggr )^{\frac{1}{\nt-N_c}}\approx
\eeq
\bq
\approx \wmu\frac{\langle\Lambda^{SU(N_{c}-\no)}_{{\cal N}=2\,\,SYM}\rangle^2}{m_1}\,\Bigl [\, N_F{\hat a}_1+\frac{\no(2N_c-N_F)}{N_c-\nt}\,{\hat a}_2 \Bigr ],
\,\, m_1= \frac{N_c}{N_c-\nt}m= -\, m_2,\,\, \wmu=-\,\mx,\label{(A.23)}
\eq
\bbq
\langle{\,\rm Tr\,}\qq\rangle_{\no}=\no\mx m_1+\delta\langle{\,\rm Tr\,}\qq\rangle_{\no}=\no\Bigl (\mx m_1+\delta\,\langle\Qo\rangle_{\no}\Bigr )\,,\quad \Lambda_{SU(\nd)}=-\lm\,,
\eeq
From this, see \eqref{(2.1.5)}, and \eqref{(2.1)} for $\langle\Qt\rangle^{SU(N_c)}_{N_c}$,
\bbq
\delta\,\langle\Qo\rangle_{\no}=\frac{1}{\no}\langle\frac{\partial}{\partial {\hat
a}_2}\w^{(SYM)}_{SU(\nd-\no)}\rangle\approx\wmu m_1\Bigl [\,\frac{2N_c-N_F}{N_c-\nt}\frac{\langle\Lambda^
{SU(\nd-\no)}_{{\cal N}=2\,\,SYM}\rangle^2}{m^2_1}\,\Bigr ],\,\,\frac{\langle\Lambda^{SU(\nd-\no)}_{{\cal N}=2\,\,SYM}\rangle^2}{m^2_1}\approx\Bigl (\frac{m_1}{\lm}\Bigr )^{\frac{2N_c-N_F}{\nt-N_c}},
\eeq
\bq
\langle\Qo\rangle^{SU(N_c)}_{\no}=\langle{\ov Q}^1_1\rangle\langle Q^1_1\rangle\approx\mx m_1\Bigl [\,1+\frac{2N_c-N_F}{\nt-N_c}\,
\frac{\langle\Lambda^{SU(\nd-\no)}_{{\cal N}=2\,\,SYM}\rangle^2}{m^2_1}\,\Bigr ]\approx \label{(A.24)}
\eq
\bbq
\approx\mx m_1\Bigl [\,1+\frac{2N_c-N_F}{\nt-N_c}\,\Bigl (\frac{m_1}{\lm}\Bigr )^{\frac{2N_c-N_F}{\nt-N_c}}\Bigr ]\,,\quad m_1= \frac{N_c}{N_c-\nt}m\,,
\eeq
\bbq
\langle\Qt\rangle^{SU(N_c)}_{N_c}=\mx m_1-\frac{N_c-\no}{N_c-\nt}\,\langle\Qo\rangle_{N_c}
\approx\mx m_1\Bigl [\, 1+\frac{N_c-\no}{\nt-N_c}\frac{\langle\Lambda^{SU(\nd-\no)}_{{\cal N}=2\,\,SYM}\rangle^2}{m^2_1}\,\Bigr ]\,.
\eeq
It is seen that, as in \eqref{(A.5)}, the leading corrections to $\langle\Qo\rangle_{\no}$ and $\langle\Qt\rangle_{N_c}$ are different. \\

{\bf 4) br2 vacua of $U(N_c),\,\, m\ll\lm$ in section 2.2}

Keeping the leading order corrections $\sim\langle\Lambda^{SU(\nd-\no)}_{{\cal N}=2\,\, SYM}\rangle^2/m$, from \eqref{(2.2.2)},\eqref{(2.2.4)},\eqref{(2.1.15)} (put attention that the quark and gluino condensates, and $\Lambda^{SU(\nd-\no)}_{{\cal N}=2\,\, SYM}$ are different in br2 vacua of $SU(N_c)$ or $U(N_c)$ theories, see \eqref{(A.24)},\eqref{(A.25)},
\bq
\langle\Qt\rangle_{N_c}=\mx m-\langle\Qo\rangle_{N_c}\,,\quad \langle\Qo\rangle_{N_c}\approx\mx m\Bigl ( \frac{m}{\lm}\Bigr )^{\frac{2N_c-N_F}{\nt-N_c}}\approx \mx\frac{\langle\Lambda^{SU(\nd-\no)}_{{\cal N}=2\,\, SYM}\rangle^2}{m} \,,\label{(A.25)}
\eq
\bbq
\frac{\langle a_0\rangle}{m}=\frac{1}{N_c\mx m}\langle{\rm Tr\,}\qq\rangle_{N_c}\approx \frac{\nt}{N_c}+\frac{\no-\nt}{N_c}\,
\Bigl ( \frac{m}{\lm}\Bigr )^{\frac{2N_c-N_F}{\nt-N_c}}\,.
\eeq
\bq
\delta \w_{SU(\nd-\no)}^{\,SYM}=(\nd-\no)\wmu\,\delta\,\Biggl [\Bigl (\Lambda^{SU(\nd-\no)}_{{\cal N}=2\,\, SYM}\Bigr )^{2}\approx\Biggl (\frac{\Lambda_{SU(\nd)}^{2\nd-N_F}(-m-{\hat a}_0-{\hat a}_1-c_2\,{\hat a}_2)^{N_F}}
{[\,-m+(1-c_2)\,{\hat a}_2\,]^{\,2\no}}\Biggr )^{\frac{1}{\nt-N_c}}\Biggr ]\,\,\,\label{(A.26)}
\eq
\bbq
\approx\wmu\,\frac{\langle\Lambda^{SU(\nd-\no)}_{{\cal N}=2\,\, SYM}\rangle^2}{m}\, \Bigl [\,N_F({\hat a}_0+{\hat a}_1)-\no\frac{ 2N_c-N_F}{\nt-N_c}\,{\hat a}_2\,\Bigr ],\,\, \frac{\langle\Lambda^{SU(\nd-\no)}_{{\cal N}=2\,\,SYM}\rangle^2}{m^2}=
\Bigl (\frac{m}{\lm}\Bigr )^{\frac{2N_c-N_F}{\nt-N_c}},\,\, \Lambda_{SU(\nd)}=-\,\lm\,,
\eeq
\bbq
\delta \w_{a}\approx\mx\frac{\langle\Lambda^{SU(\nd-\no)}_{{\cal N}=2\,\, SYM}\rangle^2}{m}\,\Bigl [(\no-\nt) {\hat a}_0-\frac{N_F}{2N_c-N_F} {\hat a}_1-\no\frac{\nt-\no}{\nt-N_c}\,{\hat a}_2\,\Bigr ]\,,\quad a_i=\langle a_i\rangle+{\hat a}_i,\, i=0,\,1,\,2\,,\,
\eeq
\bbq
\delta\langle\Qo\rangle^{(U(N_c)}_{\no}=\frac{1}{\no}\langle\frac{\partial}{\partial {\hat a}_2}\Bigl (\delta\w_{SU(\nd-\no)}^{\,SYM}+\delta \w_{a}\Bigr )\rangle \approx - 2\mx m \frac{\langle\Lambda^{SU(\nd-\no)}_{{\cal N}=2\,\, SYM}\rangle^2}{m^2}\approx - 2\mx m\Bigl (\frac{m}{\lm}\Bigr )^{\frac{2N_c-N_F}{\nt-N_c}}\,.
\eeq
Therefore, on the whole for these $U(N_c)$ br2 vacua after accounting for the leading power corrections, see \eqref{(A.25)},\eqref{(A.26)},
\bq
\langle\Qo\rangle^{U(N_c)}_{\no}=\langle{\ov Q}^1_1\rangle\langle Q^1_1\rangle\approx\mx m\Bigl [1-2\Bigl (\frac{m}{\lm}\Bigr )^{\frac{2N_c-N_F}{\nt-N_c}}
\Bigr ],\,\, \langle\Qt\rangle^{U(N_c)}_{N_c}\approx\mx m\Bigl [1-\Bigl (\frac{m}{\lm}\Bigr )^{\frac{2N_c-N_F}{\nt-N_c}}\Bigr ].\,\,\,\,\label{(A.27)}
\eq

It is seen that, as above in $SU(N_c)$ br2 vacua \eqref{(A.24)}, the non-leading terms in $\langle\Qo\rangle_{\no}$ and $\langle\Qt\rangle_{N_c}$ are different.\\

We can compare now the value of $\langle\Qo\rangle^{U(N_c)}_{\no}$ from \eqref{(A.27)} with those from
\eqref{(2.2.10)} (see however the footnote \ref{(f7)}, in these br2 vacua the charges of massless at $\mx\ra 0$ particles are non-trivial and not obvious beforehand, as well as their multiplicities). But we know the charges and multiplicities of massless at $\mx\ra 0$ particles from sections 2.1 and 2.2. To obtain definite predictions from \eqref{(2.2.10)} for the quark condensates $\langle\Qo\rangle_{\no}$ we need the values of $\no$ double roots $e^{(Q)}_k$ corresponding to original electric quarks from $SU(\no)$ and the values of two single roots $e^{\pm}$. Similarly to \eqref{(A.15)},\eqref{(A.16)}, these look here as
\bq
e^{(Q)}_k= - m\,,\quad k=1...\no\,,\quad  e^{+}=-e^{-}\approx 2\langle\Lambda^{SU(\nd-\no)}_{{\cal N}=2\,\, SYM}\rangle\,,\quad \frac{\langle\Lambda^{SU(\nd-\no)}_{{\cal N}=2\,\, SYM}\rangle^2}{m^2}\approx\Bigl (\frac{m}{\lm}\Bigr )^{\frac{2N_c-N_F}{\nt-N_c}}\ll 1\,.\,\,\,\label{(A.28)}
\eq
From this
\bq
\langle\Qo\rangle^{U(N_c)}_{\no}= - \mx\sqrt{(e^{(Q)}_k-e^+)(e^{(Q)}_k-e^-)}\approx\mx m \Bigl [\,1-2\Bigl (\frac{m}{\lm}\Bigr )^{\frac{2N_c-N_F}{\nt-N_c}}\,\Bigr ]\,,\label{(A.29)}
\eq
this agrees with \eqref{(A.27)}.

Besides, proceeding similarly to br1 vacua in \eqref{(A.19)}-\eqref{(A.22)}, we obtain for $\langle\Qo\rangle^{SU(N_c)}_{\no}$ in br2 vacua of $SU(N_c)$, see \eqref{(A.25)},
\bbq
\frac{\langle a_0\rangle}{m}\,\,\ra\,\,\approx \Biggl (\,\frac{\nt}{N_c}+\frac{\no-\nt}{N_c}\Bigl (\frac{m_1}{\lm}\Bigr )^{\frac{2N_c-N_F}{\nt-N_c}}\,\Biggr )\,,
\eeq
\bq
m\ra {m^\prime}\approx m_1\Biggl (1+\frac{\no-\nt}{N_c-\nt}\Bigl (\frac{m_1}{\lm}\Bigr )^{\frac{\bb}{\nt-N_c}} \Biggr )\,,\quad m_1=\frac{N_c}{N_c-\nt}\,m\,,\label{(A.30)}
\eq
and from \eqref{(A.29)},\eqref{(A.30)}
\bq
\langle\Qo\rangle^{SU(N_c)}_{\no}\approx \mx m_1 \Biggl (1+\frac{\no-\nt}{N_c-\nt}\Bigl (\frac{m_1}{\lm}\Bigr )^{\frac{\bb}{\nt-N_c}}\Biggr )\Bigl [\, 1-2\Bigl (\frac{m_1}{\lm}\Bigr )^{\frac{2N_c-N_F}{\nt-N_c}}\,\Bigr ]\approx \label{(A.31)}
\eq
\bbq
\approx \mx m_1 \Bigl [\, 1+\frac{\bb}{\nt-N_c}\Bigl (\frac{m_1}{\lm}\Bigr )^{\frac{2N_c-N_F}{\nt-N_c}}\,\Bigr ]\approx \mx m_1\Biggl [\,1+\frac
{2N_c-N_F}{\nt-N_c}\frac{\langle\Lambda^{SU(\nd-\no)}_{{\cal N}=2\,\,SYM}\rangle^2}{m^2_1}\,\Biggr ]\,,
\eeq
this agrees with \eqref{(A.24)}.

\section{Calculations of dyon condensates}

Consider first the br2 vacua of $\mathbf{U(N_c)}$ with $m\ll\lm$ in section 2.2. Because, unlike the br1 vacua of $U(N_c)$ theory in section 4.2, there are now in addition massless dyons in br2 vacua, to obtain with our accuracy the definite values of roots from the $U(N_c)$ curve \eqref{(1.2)}, one needs to use not only \eqref{(A.15)}, but also the sum rule \eqref{(A.17)}. The sum rule \eqref{(A.17)} looks in these br2 vacua as, see \eqref{(A.25)},
\bq
N_c\langle a_0\rangle\approx \Bigl (\nt m+(\no-\nt)\frac{\langle\Lambda^{SU(\nd-\no)}_{{\cal N}=2\,\, SYM}\rangle^2}{m} \Bigr )\approx \Biggl (\nt m+(\no-\nt) m\Bigl (\frac{m}{\lm}\Bigr )^{\frac{\bb}{\nt-N_c}}\Biggr )\approx \label{(B.1)}
\eq
\bbq
\approx\Bigl (\no m+(\nd-\no-1)A^{(M)}_c\frac{\langle\Lambda^{SU(\nd-\no)}_{{\cal N}=2\,\, SYM}\rangle^2}{m}-\frac{1}{2}(e^{+}+e^{-}=0)+(2N_c-N_F)C^{(D)}_c \Bigr )\,,
\eeq
where $\no(-e^{(Q)}_i)=\no m$ is the contribution of $\no$ equal double roots of original electric quarks from $SU(\no)$, the term with $A^{(M)}_c$ is the contribution from the center of $\nd-\no-1$ unequal double roots of magnetic monopoles, and the term with $C^{(D)}_c$ is the contribution from the center of $2N_c-N_F$ unequal double roots of dyons. We obtain then from the curve \eqref{(1.2)} together with \eqref{(B.1)} (with our accuracy)
\bq
A^{(M)}_c=\frac{\nt-\no}{\nd-\no-1}\,,\quad C^{(D)}_c=\frac{\nt-\no}{2N_c-N_F} m \Bigl ( 1-2 \frac{\langle\Lambda^{SU(\nd-\no)}_{{\cal N}=2\,\, SYM}\rangle^2}{m^2} \Bigr )\,,\,\,\,\label{(B.2)}
\eq
\bbq
e^{(M)}_k\approx 2\cos (\frac{\pi k}{\nd-\no})\langle\Lambda^{SU(\nd-\no)}
_{{\cal N}=2\,\,SYM}\rangle-\frac{\nt-\no}{\nd-\no-1}\frac{\langle\Lambda^{SU(\nd-\no)}
_{{\cal N}=2\,\,SYM}\rangle^2}{m}\,,\quad k=1...(\nd-\no-1)\,,
\eeq
\bbq
e^{(D)}_j\approx \omega^{j-1}\lm\Biggl (1+O\Bigl (\frac{m}{\lm}\Bigr )^{\bb}\Biggr )-\frac{\nt-\no}{2N_c-N_F}\,m\Biggl (1-2\Bigl
(\frac{m}{\lm}\Bigr )^{\frac{\bb}{\nt-N_c}}\Biggr )\,,\,\, j=1...(2N_c-N_F)\,,
\eeq
while $e^{\pm}$ are given in \eqref{(A.28)}.

From \eqref{(2.2.10)},\eqref{(B.2)} we obtain now for the dyon condensates in these br2 vacua of $U(N_c)$ at $m\ll\lm$ (with the same accuracy)
\bq
\langle{\ov D}_j D_j\rangle_{U(N_c)}=\langle{\ov D}_j\rangle\langle D_j\rangle\approx \mx\Biggl [\,-\, \omega^{j-1}\lm+\frac{\nt-\no}{2N_c-N_F}\,m
\Biggl (1-2\Bigl (\frac{m}{\lm}\Bigr )^{\frac{\bb}{\nt-N_c}}\Biggr ) \Biggr ]\,, \label{(B.3)}
\eq
\bbq
\langle\Sigma^{U(N_c)}_D\rangle\equiv\sum_{j=1}^{2N_c-N_F}\langle{\ov D}_j D_j\rangle_{U(N_c)}=\sum_{j=1}^{2N_c-N_F}\langle{\ov D}_j\rangle\langle D_j\rangle
\approx \mx m\,(\nt-\no)\Biggl (1-2\Bigl (\frac{m}{\lm}\Bigr )^{\frac{\bb}{\nt-N_c}}\Biggr ), \,\,\, j=1...\bb\,.
\eeq

Now, using only the leading terms $\sim \mx m$ of \eqref{(B.3)},\eqref{(2.2.3)},\eqref{(2.2.5)},
\eqref{(2.2.8)}, we can find the value of $\delta_3=O(1)$ in \eqref{(2.1.2)},\eqref{(2.1.5)},\eqref{(2.2.4)},\eqref{(6.2.3)}. From \eqref{(2.2.4)}
\bq
\langle\frac{\partial}{\partial a_0}\w^{\,\rm low}_{\rm tot}\rangle=0= - \langle{\rm Tr\,}\qq\rangle_{\no}-\langle\Sigma^{U(N_c)}_D\rangle+\mx N_c\langle a_0\rangle-\mx N_c\delta_3\langle a_1\rangle= - \mx N_c\delta_3\langle a_1\rangle \ra {\boldsymbol\delta}{\mathbf{_3=0}}.\,\,\quad\label{(B.4)}
\eq
(The same result follows from $\langle\partial \w^{\,\rm low}_{\rm tot}/{\partial a_1}\rangle=0$ in \eqref{(2.2.4)}\,).

As for the non-leading terms $\delta\langle\Sigma^{U(N_c)}_D\rangle\sim \mx m (m/\lm)^{\frac{\bb}{\nt-N_c}}$ of $\langle\Sigma^{U(N_c)}_D\rangle$ in \eqref{(B.3)}, with $\delta_3=0$ from \eqref{(B.4)}, these can now be also calculated {\it independently} from \eqref{(2.2.4)},\eqref{(A.25)},\eqref{(A.26)}, i.e.
\bbq
\langle\frac{\partial}{\partial {\hat a}_0}\delta\w^{\,\rm low}_{\rm tot}\rangle=0\ra \delta\langle\Sigma^{U(N_c)}_D\rangle\approx\Bigl [-\delta\langle{\rm Tr}\qq\rangle_{\no}=2\no\mx m\Bigl (\frac{m}{\lm}\Bigr )^{\frac{\bb}{\nt-N_c}}\Bigr ]_Q+\Bigl [(\no-\nt)\mx m\Bigl (\frac{m}{\lm}\Bigr )^{\frac{\bb}{\nt-N_c}}\Bigr ]_{\delta\w_{a_0}}\,\,\,
\eeq
\bq
+\Bigl [\langle\frac{\partial}{\partial {\hat a}_0}\delta \w_{SU(\nd-\no)}^{\,SYM}\rangle= - N_F\mx m\Bigl (\frac{m}{\lm}\Bigr )^{\frac{\bb}{\nt-N_c}}\Bigr ]_{\delta\w_{SYM}}\approx -2\mx m\,(\nt-\no)
\Bigl (\frac{m}{\lm}\Bigr )^{\frac{\bb}{\nt-N_c}}\,,\label{(B.5)}
\eq
this agrees with \eqref{(B.3)}.\\

As for the value of $\langle\Sigma^{SU(N_c)}_D\rangle$ in br2 vacua of $\mathbf{SU(N_c)}$, with $\delta_3=0$ from \eqref{(B.4)}, it can be found now directly from \eqref{(2.1.5)},\eqref{(2.1.12)},\eqref{(A.23)},\eqref{(A.24)}
\bq
\langle\frac{\partial}{\partial {\hat a}_1}\w^{\,\rm low}_{\rm tot}\rangle=0\ra \langle\Sigma^{SU(N_c)}_D\rangle\approx \mx m_1(\nt-\no)\Biggl [1+\frac{\bb}
{\nt-N_c}\Bigl (\frac{m_1}{\lm}\Bigr )^{\frac{\bb}{\nt-N_c}} \Biggr ]\,,\,\, m_1=\frac{N_c}{N_c-\nt}\,m\,.\,\,\,\label{(B.6)}
\eq
(Really, the power suppressed term in \eqref{(B.6)} can be found {\it independently of the value of $\delta_3$} because it is parametrically different and, unlike \eqref{(B.5)} in $U(N_c)$ theory, there are no power corrections to $\langle a_1\rangle$ in $SU(N_c)$ theory, see \eqref{(2.1.6)}).

It can also be obtained from \eqref{(B.3)} in br2 vacua of $U(N_c)$ proceeding similarly to \eqref{(A.30)},\eqref{(A.31)} for the quark condensates. We obtain from \eqref{(A.30)} and \eqref{(B.3)}
\bbq
\langle\Sigma^{SU(N_c)}_D\rangle=\sum_{j=1}^{2N_c-N_F}\langle{\ov D}_j\rangle\langle D_j\rangle\approx \mx m_1\Biggl (1+\frac{\no-\nt}{N_c-\nt}\Bigl (\frac{m_1}{\lm}\Bigr )^{\frac{\bb}{\nt-N_c}}\Biggr )(\nt-\no)\Bigl [\, 1-2\Bigl (\frac{m_1}{\lm}\Bigr )^{\frac{2N_c-N_F}{\nt-N_c}}\,\Bigr ]\approx
\eeq
\bq
\approx\mx m_1(\nt-\no)\Biggl [1+\frac{\bb}{\nt-N_c}\Bigl (\frac{m_1}{\lm}\Bigr )^{\frac{\bb}{\nt-N_c}} \Biggr ]\,,\label{(B.7)}
\eq
this agrees with \eqref{(B.6)}.

On the whole, the dyon condensates in br2 vacua of $SU(N_c)$ look at $m\ll\lm$ as, $j=1,...,\bb$,
\bq
\hspace*{-4mm}\langle{\ov D}_j D_j\rangle^{SU(N_c)}\approx \mx\Biggl [-\omega^{j-1}\lm\Biggl (1+O\Bigl (\frac{m}{\lm}\Bigr )^{\bb}\Biggr )+\frac{\nt-\no}{2N_c-N_F}\,m_1\Biggl (1+\frac{\bb}{\nt-N_c}\Bigl (\frac{m_1}{\lm}\Bigr )^{\frac{\bb}{\nt-N_c}}\Biggr ) \Biggr ].\,\,\,\label{(B.8)}
\eq
\vspace*{2mm}

Finally, not going into details, we present below the values of roots of the $\mathbf{SU(N_c)}$ spectral curve \eqref{(1.2)} in br1 vacua of section 4.1 and br2 vacua of section 2.1.

{\bf 1) br1 vacua}. There are at $\mx\ra 0$ $\,\no$ equal double roots $e^{(Q)}_i$ of massless original electric quarks $Q_i$, then $N_c-\no-1$ unequal double roots $e^{(M)}_k$ of massless pure magnetic monopoles $M_k$ and two single roots $e^{\pm}$ from ${\cal N}=2\,\, SU(N_c-\no)$ SYM. They look (with our accuracy) as, see \eqref{(A.3)},
\bbq
e^{(Q)}_i= - m\,,\,\,i=1...\no\,,\quad e^{\pm}\approx e^{\pm}_{c}\pm 2\langle\Lambda^{SU(N_{c}-\no)}_{{\cal N}=2\,\,SYM}\rangle,\quad \langle\Lambda^{SU(N_{c}-\no)}_{{\cal N}=2\,\,SYM}\rangle\approx  m_3\Bigl (\frac{\lm}{m_3}\Bigr )^{\frac{2N_c-N_F}{2(N_c-\no)}},
\eeq
\bq
e^{\pm}_{c}=\frac{1}{2}(e^{+}+e^{-})\approx m_3\Biggl [\,\frac{\no}{N_c} +\frac{\nt-\no}{N_c-\no}\frac{\langle\Lambda^{SU(N_{c}-\no)}_{{\cal N}=2\,\,SYM}\rangle^2}{m^2_3} \,\Biggr ], \quad m_3=\frac{N_c}{N_c-\no} m\,,\label{(B.9)}
\eq
\bbq
e^{(M)}_k\approx \Biggl [\,e^{\pm}_{c}-\frac{\nt-\no}{N_c-\no-1}\frac{\langle\Lambda^{SU(N_{c}-\no)}_{{\cal N}=2\,\,SYM}\rangle^2}{m_3}\,\Biggr ]+ 2\cos (\frac{\pi k}{N_c-\no})\langle\Lambda^{SU(N_{c}-\no)}
_{{\cal N}=2\,\,SYM}\rangle,\quad \frac{1}{2}\sum_{n=1}^{2 N_c}(-e_n)=0\,.
\eeq
It should be emphasized that instead of $e^{\pm}_{c}=0$ in $\mathbf{U(N_c)}$ theory, see \eqref{(A.15)}, it looks in $\mathbf{SU(N_c)}$ theory as, see \eqref{(A.21)},\eqref{(B.9)},
\bbq
e^{\pm}_{c}\approx \Bigl [\frac{\langle a_0\rangle}{m}\Bigr ]\Bigl [\,m\,\Bigr ]\,\ra\,\Biggl [\,\frac{\langle a_0\rangle}{m}\,\,\ra\,\,\, \approx\Biggl (\,\frac{\no}{N_c}+\frac{\nt-\no}{N_c}\Bigl (\frac{\lm}{m_3}\Bigr )^{\frac{2N_c-N_F}{N_c-\no}}\,\Biggr )\,\Biggr ]\Biggl [m\,\ra\, {m^\prime}
\approx m_3\Biggl (\,1+\frac{\nt-\no}{N_c-\no}
\eeq
\bq
\cdot\Bigl (\frac{\lm}{m_3}\Bigr )^{\frac{\bb}{N_c-\no}}\Biggr )\,\Biggr ]\approx m_3\Biggl [\frac{\no}{N_c}+\frac{\nt-\no}{N_c-\no}\Bigl (\frac{\lm}{m_3}\Bigr )^{\frac{2N_c-N_F}{N_c-\no}} \Biggr ]\approx m_3\Biggl [\,\frac{\no}{N_c}+ \frac{\nt-\no}{N_c-\no}\frac{\langle\Lambda^{SU(N_{c}-\no)}_{{\cal N}=2\,\,SYM}\rangle^2}{m^2_3} \,\Biggr ]\,.\label{(B.10)}
\eq

{\bf 2) br2 vacua}. There are at $\mx\ra 0$ $\,\no$ equal double roots $e^{(Q)}_i$ of massless original electric quarks $Q_i$, then $\nd-\no-1$ unequal double roots $e^{(M)}_k$ of massless pure magnetic monopoles $M_k$ and two single roots $e^{\pm}$ from ${\cal N}=2\,\, SU(\nd-\no)$ SYM, and $2N_c-N_F$ unequal double roots of massless dyons $D_j$. They all look (with our accuracy) as, see \eqref{(A.24)},\eqref{(A.25)},\eqref{(A.30)},
\bbq
e^{(Q)}_i= - m\,,\,\,i=1...\no\,,\quad e^{\pm}\approx e^{\pm}_{c}\pm 2\langle\Lambda^{SU(\nd-\no)}_{{\cal N}=2\,\,SYM}\rangle,\quad \langle\Lambda^{SU(\nd-\no)}_{{\cal N}=2\,\,SYM}\rangle^2\approx  m^2_1\Bigl (\frac{m_1}{\lm}\Bigr )^{\frac{2N_c-N_F}{\nt-N_c}},
\eeq
\bbq
e^{\pm}_{c}=\frac{1}{2}(e^{+}+e^{-})\approx\Biggl [\,\frac{\langle a_0\rangle}{m}\,\ra\,\, \approx \Biggl (\,\frac{\nt}{N_c}+\frac{\no-\nt}{N_c}\Bigl (\frac{m_1}{\lm}\Bigr )^{\frac{2N_c-N_F}{\nt-N_c}}\,\Biggr )\,\Biggr ]\Biggl [\,m\,\ra\,\,{m^\prime}\approx m_1\Biggl (1+
\eeq
\bq
+\frac{\no-\nt}{N_c-\nt}\Bigl (\frac{m_1}{\lm}\Bigr )^{\frac{\bb}{\nt-N_c}}\Biggr )\,\Biggr ]\approx m_1\Biggl [\,\frac{\nt}{N_c}+\frac{\nt-\no}{\nt-N_c}
\frac{\langle\Lambda^{SU(\nd-\no)}_{{\cal N}=2\,\,SYM}\rangle^2}{m^2_1} \,\Biggr ], \quad m_1=\frac{N_c}{N_c-\nt} m\,,\label{(B.11)}
\eq
\bbq
e^{(M)}_k\approx \Biggl [\,e^{\pm}_{c}-\frac{\nt-\no}{\nd-\no-1}\frac{\langle\Lambda^{SU(\nd-\no)}_{{\cal N}=2\,\,SYM}\rangle^2}{m_1}\,\Biggr ]+ 2\cos (\frac{\pi k}{\nd-\no})\langle\Lambda^{SU(\nd-\no)}_{{\cal N}=2\,\,SYM}\rangle,\quad k=1...(\nd-\no-1)\,,
\eeq
\bq
e^{(D)}_j\approx \omega^{j-1}\lm+\frac{N_F}{2N_c-N_F}\,m\,,\quad j=1...2 N_c-N_F\,,
\quad \frac{1}{2}\sum_{n=1}^{2 N_c}(-e_n)=0\,.\label{(B.12)}
\eq

Now, with \eqref{(B.9)}-\eqref{(B.12)}, we can perform checks of possible applicability of \eqref{(2.2.10)} not only to softly broken $\mathbf{U(N_c)}$ $\,{\cal N}=2$ SQCD, but also directly to softly broken $\mathbf{SU(N_c)}$ $\,{\cal N}=2$ SQCD at appropriately small $\mx$ (clearly, this applicability is far not evident beforehand, see the footnote \ref{(f7)}).

{\bf a)} From \eqref{(B.9)} for the quark condensates in br1 vacua of section 4.1
\bq
\langle\Qo\rangle^{SU(N_c)}_{\no}= - \mx\sqrt{(e^{(Q)}_k-e^+)(e^{(Q)}_k-e^-)}\approx\mx m_3\Biggl [\,1-\frac{2N_c-N_F}{N_c-\no}
\frac{\langle\Lambda^{SU(N_{c}-\no)}_{{\cal N}=2\,\,SYM}\rangle^2}{m^2_3}\, \Biggr ]\,,\label{(B.13)}
\eq
this agrees with the {\it independent} calculation in \eqref{(A.5)}.\\

{\bf b)} From \eqref{(B.11)} for the quark condensates in br2 vacua of section 2.1
\bq
\langle\Qo\rangle^{SU(N_c)}_{\no}= - \mx\sqrt{(e^{(Q)}_k-e^+)(e^{(Q)}_k-e^-)}\approx\mx m_1\Biggl [\,1+\frac
{2N_c-N_F}{\nt-N_c}\frac{\langle\Lambda^{SU(\nd-\no)}_{{\cal N}=2\,\,SYM}\rangle^2}{m^2_1}\,\Biggr ]\,,\label{(B.14)}
\eq
this agrees with the {\it independent} calculation in \eqref{(A.24)}.\\

{\bf c)} From \eqref{(B.11)},\eqref{(B.12)} for the dyon condensates in br2 vacua of section 2.1
\bbq
\langle{\ov D}_j D_j\rangle^{SU(N_c)}=\langle{\ov D}_j\rangle^{SU(N_c)}\langle D_j\rangle^{SU(N_c)}= - \mx\sqrt{(e^{(D)}_j-e^+)(e^{(D)}_j-e^-)}\approx
\eeq
\bq
\approx\mx\Biggl [-\omega^{j-1}\lm+\frac{\nt-\no}{2N_c-N_F}\,m_1\Biggl (1+\frac{\bb}{\nt-N_c}\frac{\langle\Lambda^{SU(\nd-\no)}_{{\cal N}=2\,\,SYM}\rangle^2}{m^2_1}\Biggl )\,\Biggr ]\,,\quad j=1...2 N_c-N_F\,,\label{(B.15)}
\eq
this agrees with the {\it independent} calculation in \eqref{(B.8)}.\\

\end{document}